\documentclass[traditabstract]{aa} 
\usepackage{graphicx,longtable,lscape,natbib,amssymb,amsmath,txfonts}
\newcommand{\ltsima} {$\; \buildrel < \over \sim \;$}  
\newcommand{\gtsima} {$\; \buildrel > \over \sim \;$}  
\newcommand{\lta} {\lower.5ex\hbox{\ltsima}}  
\newcommand{\gta} {\lower.5ex\hbox{\gtsima}}  
\newcommand{\Ha} {H$\alpha$}  
\newcommand{\Hb} {H$\beta$}  

\newcommand{\ergs}{\>{\rm erg}\,{\rm s}^{-1}}
\newcommand{\ergsHz}{\>{\rm erg}\,{\rm s}^{-1}\,{\rm Hz}^{-1}}

\begin{document}

\title{Origin of X-shaped radio-sources:\\ further insights from the properties of their host galaxies}
\subtitle{} \titlerunning{Properties of X-shaped radio sources}
\authorrunning{Gillone, Capetti \& Rossi}

\author{M.~Gillone\inst{1,2}
\and A.~Capetti\inst{3}
\and P.~Rossi\inst{3}}

\institute {Dipartimento di Fisica, Universit\`a degli Studi di Torino, via
    Pietro Giuria 1, 10125 Torino, Italy \and School of Physics and Astronomy, University of Birmingham,
  Edgbaston, Birmingham B15 2TT, UK \and INAF - Osservatorio Astrofisico di
  Torino, Via Osservatorio 20, I-10025 Pino Torinese, Italy}

\offprints{capetti@oato.inaf.it} 

\abstract{We analyze the properties of a sample of X-shaped radio-sources
 (XRSs). These objects show, in addition to the main lobes, a pair of wings
  that produce their peculiar radio morphology. We obtain our sample by selecting
  from the initial list of Cheung (2007, AJ, 133, 2097) the 53 galaxies with
  the better defined wings and with available SDSS images. We identify the
  host galaxies and measure their optical position angle, obtaining a
  positive result in 22 cases. The orientation of the secondary radio
  structures shows a strong connection with the optical axis, with all (but
  one) wing forming an angle larger than $40^\circ$ with the host major
  axis. The probability that this is compatible with a uniform distribution is
  $P = 0.9 \times 10^{-4}$.

For all but three sources of the sample, spectroscopic or photometric
redshifts are avaliable. The radio luminosity distribution of XRSs has a high
power cut-off at $L \sim10^{34}$ $\ergsHz$ at 1.4 GHz.  Spectra are available
from the SDSS for 28 XRSs. We modeled them to extract information on their
emission lines and stellar population properties. The sample is formed by
approximately the same number of high and low excitation galaxies (HEGs and
LEGs); this classification is essential for a proper comparison with
non-winged radio-galaxies. XRSs follow the same relations between radio and
line luminosity defined by radio-galaxies in the 3C sample. While in HEGs a
young stellar population is often present, this is not detected in the 13
LEGs, which is, again, in agreement with the properties of non-XRSs. The lack
of young stars in LEGs supports the idea that they have not experienced a
recent gas-rich merger.

  The connection between the optical axis and the wing orientation, as well
  as the stellar population and emission-line properties, provide further
  support for a hydro-dynamic origin of the radio-wings (for example,
  associated with the expansion of the radio cocoon in an asymmetric external
  medium) rather than with a change of orientation of the jet axis. In this
  framework, the high luminosity limit of XRSs can be interpreted as being due to
   high power jets being less affected by the properties of the
  surrounding medium.}

\keywords{Galaxies: active, Galaxies: jets, Galaxies: ISM}
\maketitle

\section{Introduction}
\label{intro}

Historically, extended radio sources have been  classified on the basis of their
radio morphology, the main division being based on their edge-darkened or edge-brightened structure that leads to the identification of the Fanaroff-Riley
classes I and II \citep{fanaroff74}.  The characteristic structure of
FR~II sources is dominated by two hot spots, each located at the edges of the radio
lobes that, in most cases, show bridges of emission, which link the core to the
hot spots. The presence of significant distortions in the bridges has been
recognized since early interferometric imaging of 3C sources (see
e.g., \citealt{leahy84}). Distortion in FR~II can be
classified in two general classes: mirror symmetric (or C-shaped), when the
bridges bend away from the galaxy in the same direction, or centro-symmetric,
when they bend in opposite directions and form an X-shaped or Z-shaped radio
source, depending on the location of the point of insertion of the wings. In
many XRSs the radio emission along the secondary axis, although
more diffuse, is still quite well collimated and can be even more extended
than the main double-lobed structure.

C-shaped morphologies are observed also in FR~I radio-galaxies,
although in these sources the distortions affect their jets rather
than their lobes and they give rise to the typical shape of Narrow
Angle Tails, where the opposite jets bend dramatically and become
almost parallel to one another. Another common morphology for FR~I
is that of centro-symmetric S-shaped sources. Conversely, 
X-shapes among FR~I are extremely rare \citep{saripalli09}.

There is now general agreement that the C-shaped radio sources form when they
are in motion with respect to the external medium: jets or bridges are bent by
the ram pressure of the surrounding gas. Models successfully reproduced the
morphology of FR~I narrow angle tails (see e.g. \citealt{odea86}) and the
extension of this scenario to FR~II bridges appears quite natural.

With regard to X- or Z-shaped sources, several mechanism have been proposed for
their origin (see \citealt{gopalkrishna12} for a review). \citet{ekers78}
suggested that the tails of radio emission in one of these sources, NGC 326,
are the result of the trail caused by a secular jet precession (see also
\citealt{rees78}). A similar model accounts for the morphology of 4C 32.25
\citep{klein95}. In a similar line, \citet{wirth82} noted that a change in the
jet direction can be caused by gravitational interaction with a companion
galaxy. \citet{dennettthorpe02}, from the analysis of spectral variations
along the lobes, proposed that the jet reorientation occurs over short time
scales, a few Myr, and are possibly associated with instabilities in the
accretion disk that cause a rapid change in the jet axis. More recently,
another process of jet reorientation has been suggested, which relates to the sudden
spin change that results from the coalescence of two black holes (see, e.g.,
\citealt{merritt02}).  In all these models, the secondary axis of radio
emission represents a relic of the past activity of the radio source.
An alternative interpretation was suggested by \citet{leahy84} and
\citet{worrall95}. They emphasize the role of the external medium in shaping
radio sources, suggesting that buoyancy forces can bend the back-flowing
material away from the jet axis into the direction of decreasing external gas
pressure. 

In \citet{butterflies}, we re-examine the origin of these extensions that link
the radio morphology to the properties of their host galaxies. We discovered
that the orientation of the wings shows a striking connection with the
structure of the host galaxy since they are preferentially aligned with its
minor axis. Furthermore, wings are only observed in galaxies of high projected
ellipticity (a result confirmed by \citealt{saripalli09}). We concluded that
XRSs naturally form in this geometrical situation: as a jet propagates in a
non-spherical gas distribution, the cocoon surrounding the radio-jets expands
laterally at a high rate, which produces wings of radio emission, in a way
that is reminiscent of the twin-exhaust model for radio-sources.

The importance of the external medium in shaping the morphology of extended
and giant radio-sources has been later strengthened by \citet{saripalli09};
furthermore, \citet{hodgeskluck10} find that the distribution of the hot,
X-ray emitting, gas follows that of the stellar light distribution and
confirm the strong tendency for XRSs to have the wings directed along the
minor axis of the hot gas distribution.

The \citeauthor{butterflies} study is limited by the small size of the XRSs
sample they considered, being formed of only nine objects. Although the connection
between the wings and host orientation is already of high significance, the
study of a larger sample can clearly be used to test this result with a
strongly improved statistical basis. Indeed, by using the list of 100 XRSs
selected by \citet{cheung07}, we were able to explore the radio wings/host
connection in 22 galaxies (none in common with the previous group) and to
fully confirm the results obtained by our initial analysis.

In addition, we can also now include in our study  the analysis the SDSS
optical spectroscopic data. It will then be possible, following the similar
analysis by \citet{landt10} and \citet{mezcua11,mezcua12}, to constraint the
origin of XRSs from the comparison of properties of their stellar population
and emission lines against the general population of radio-galaxies.

In Sect. \ref{sample} we present the selection of the sample of XRSs for which
we measure the optical PA, see Sect. \ref{optical}. The
relationship between radio and optical axis is presented in Sect. \ref{axis}.
In the following two Sections (\ref{redshift} and \ref{spectral}), we study the
radio power distribution and the spectroscopic properties of XRSs. Our summary
and conclusions are given in Sect. \ref{summary}.

\section{Sample selection}
\label{sample}

\begin{figure*}
\centerline{
\includegraphics[width=19mm]{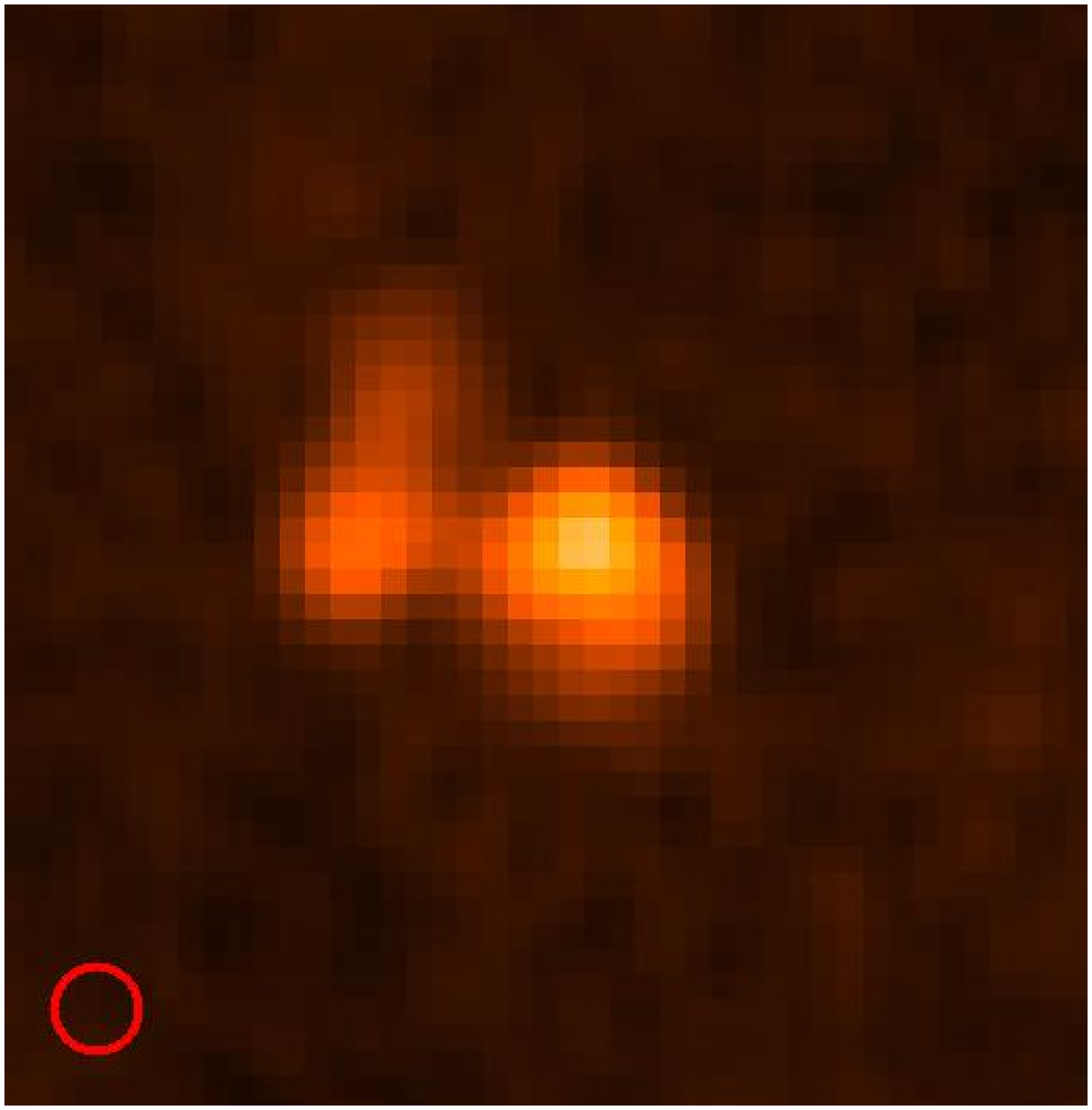}
\includegraphics[width=19mm]{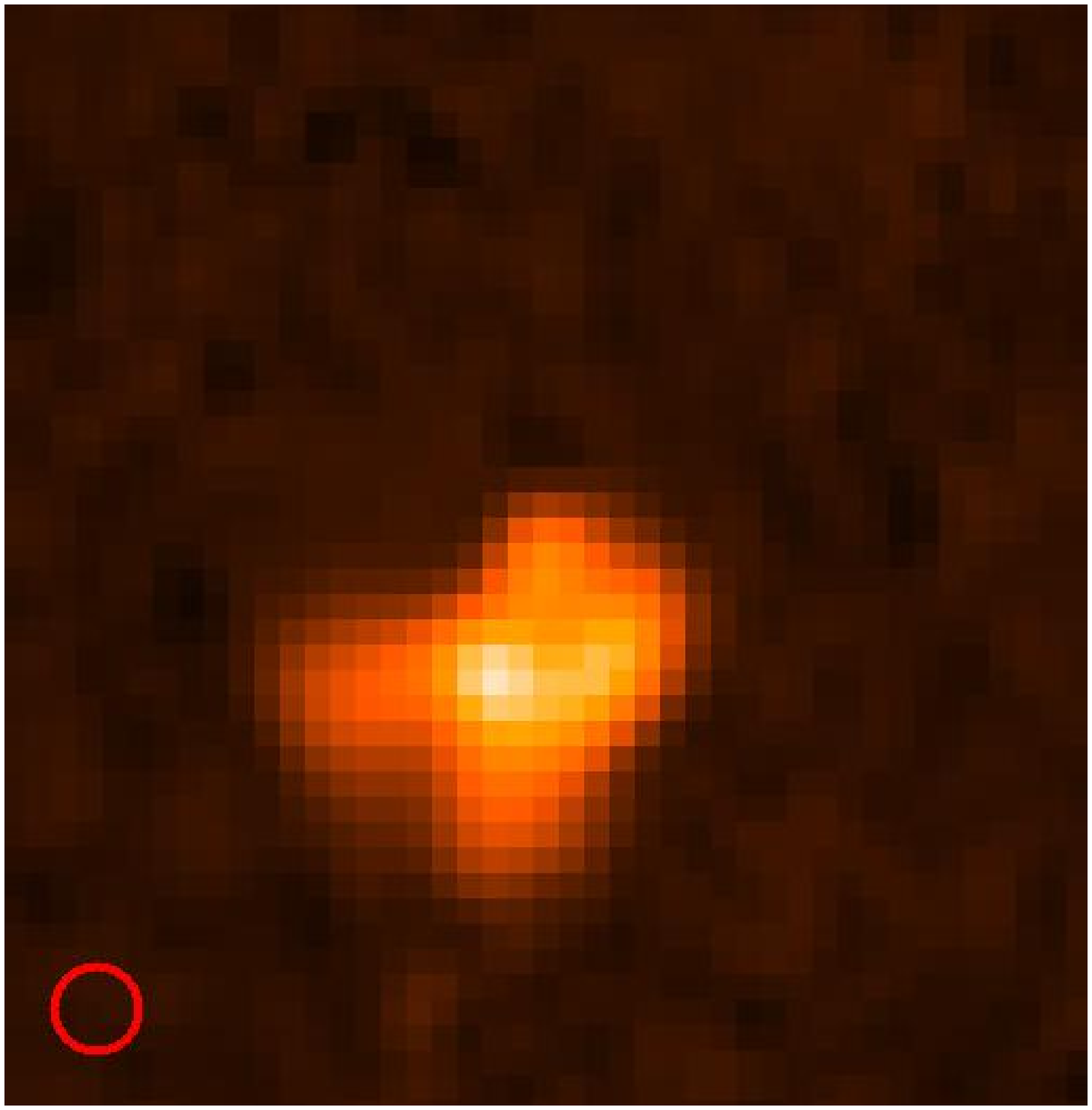}
\includegraphics[width=19mm]{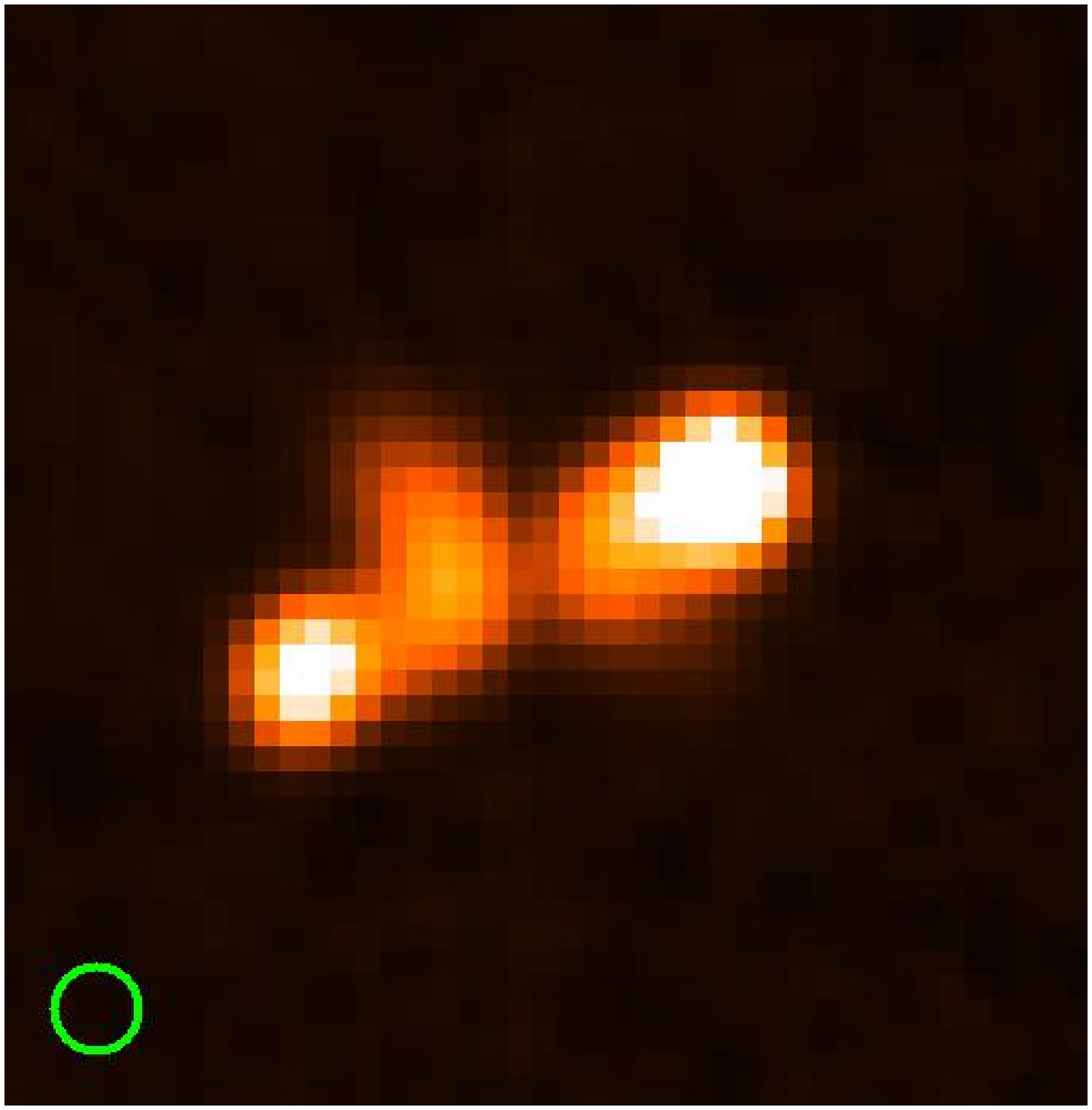}
\includegraphics[width=19mm]{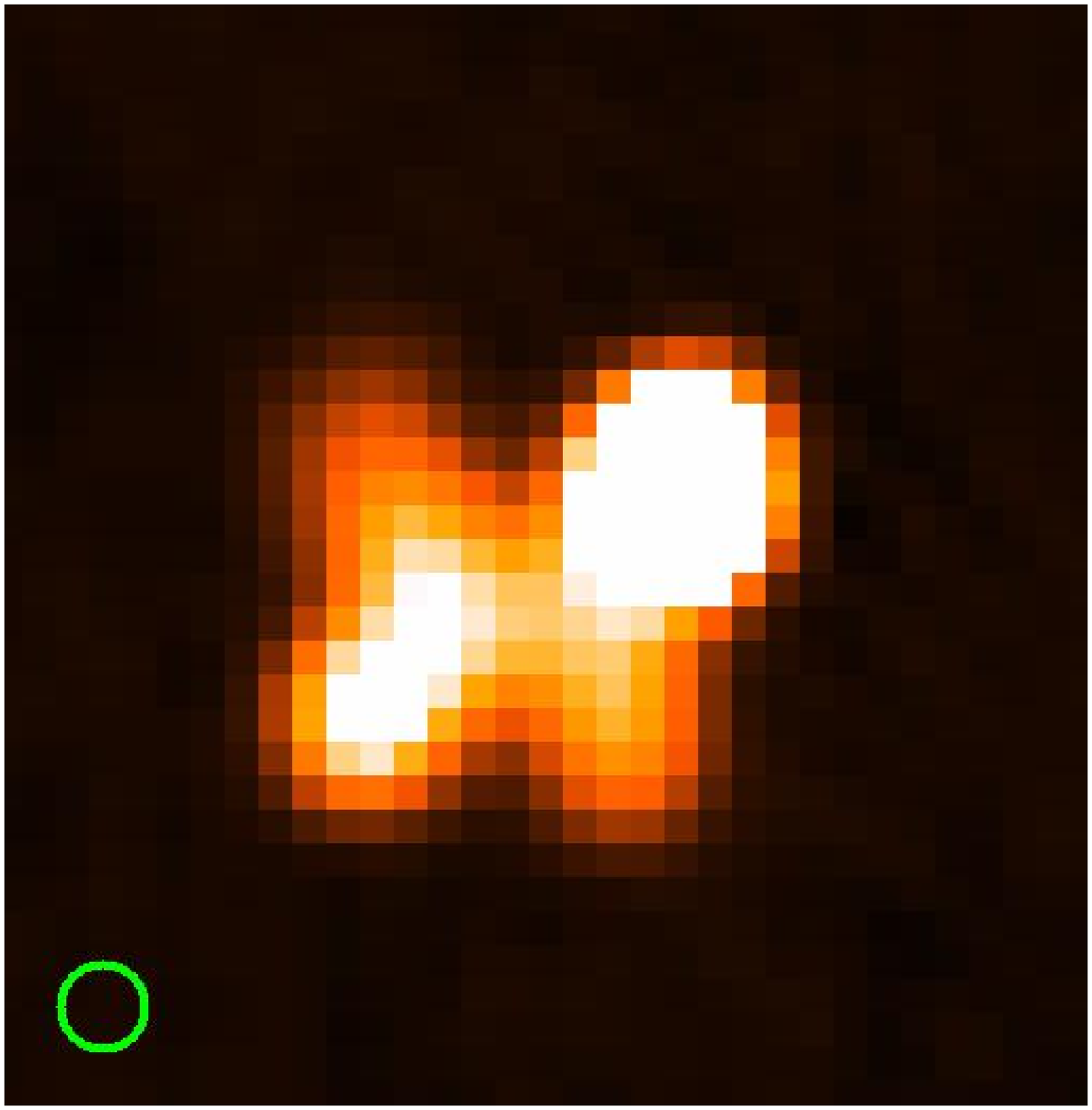}
\includegraphics[width=19mm]{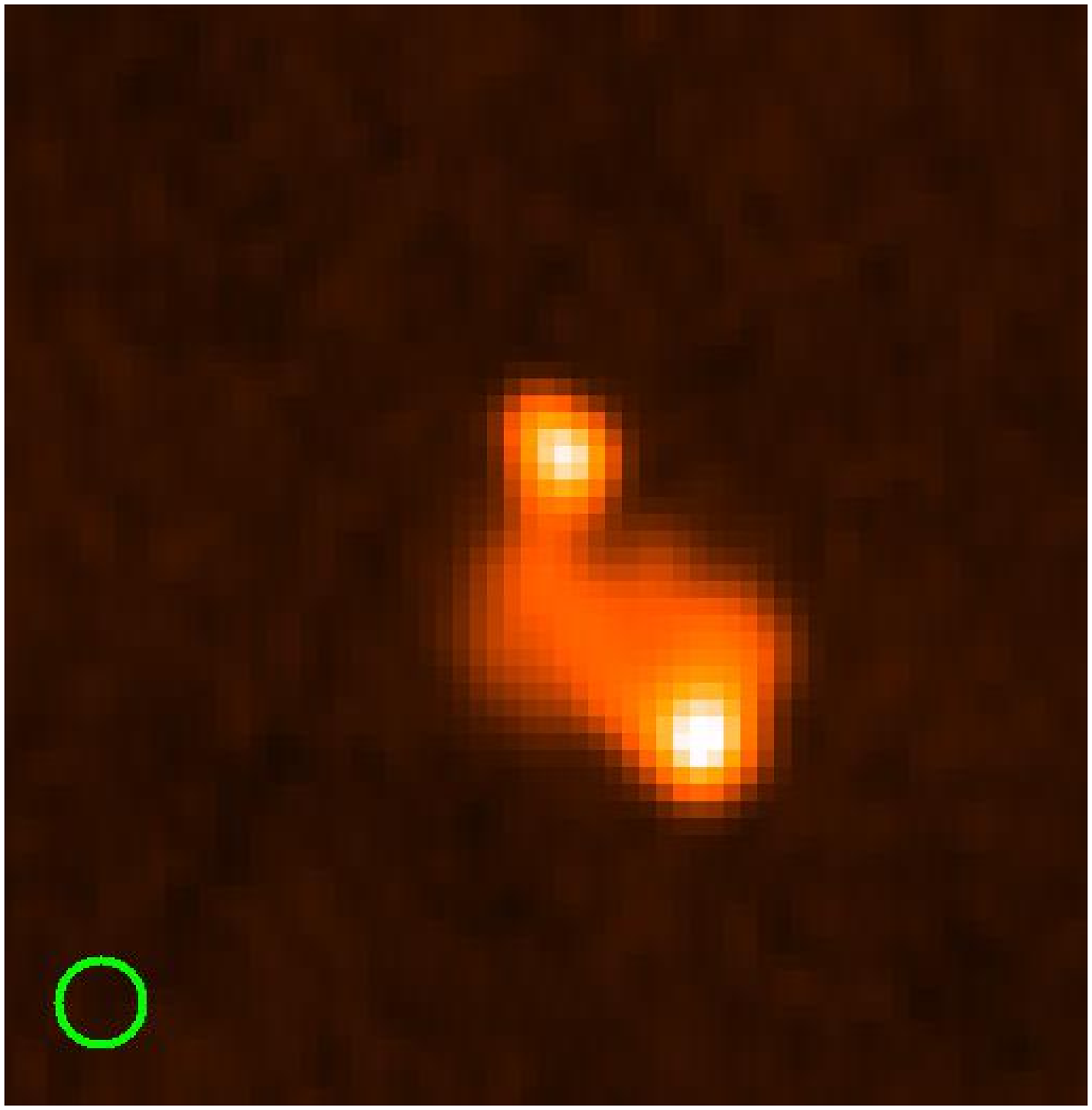}
\includegraphics[width=19mm]{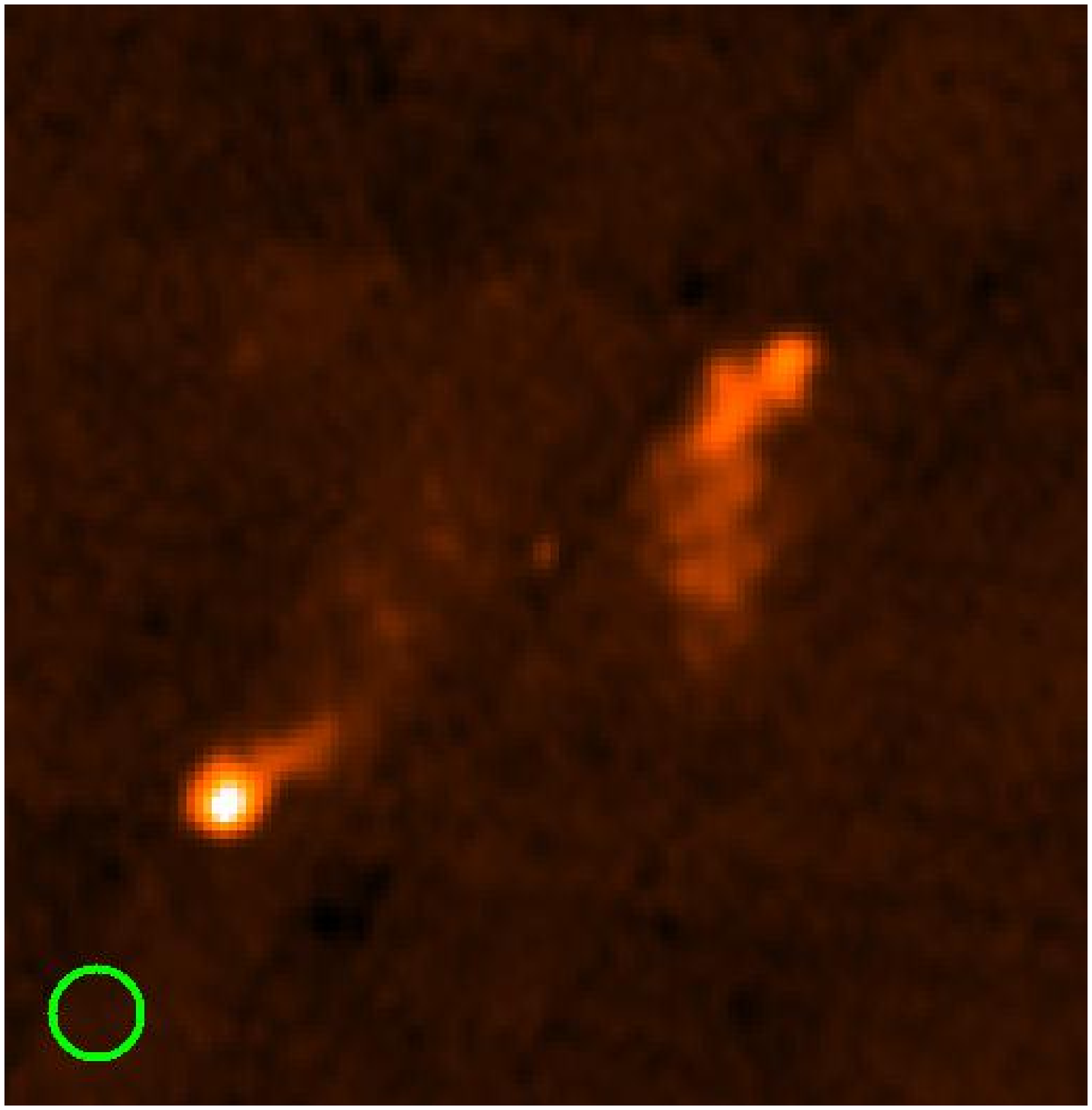}
\includegraphics[width=19mm]{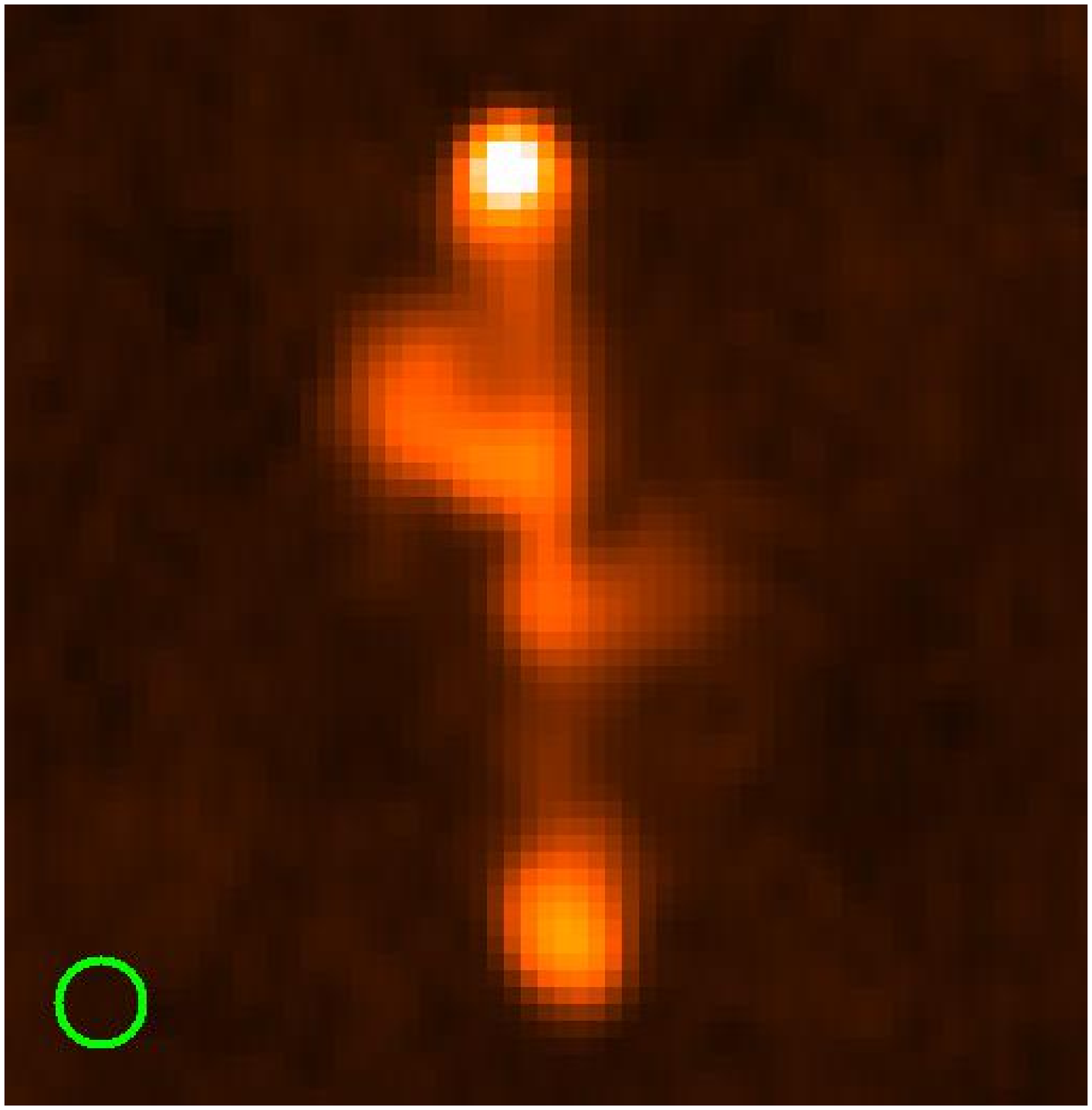}
\includegraphics[width=19mm]{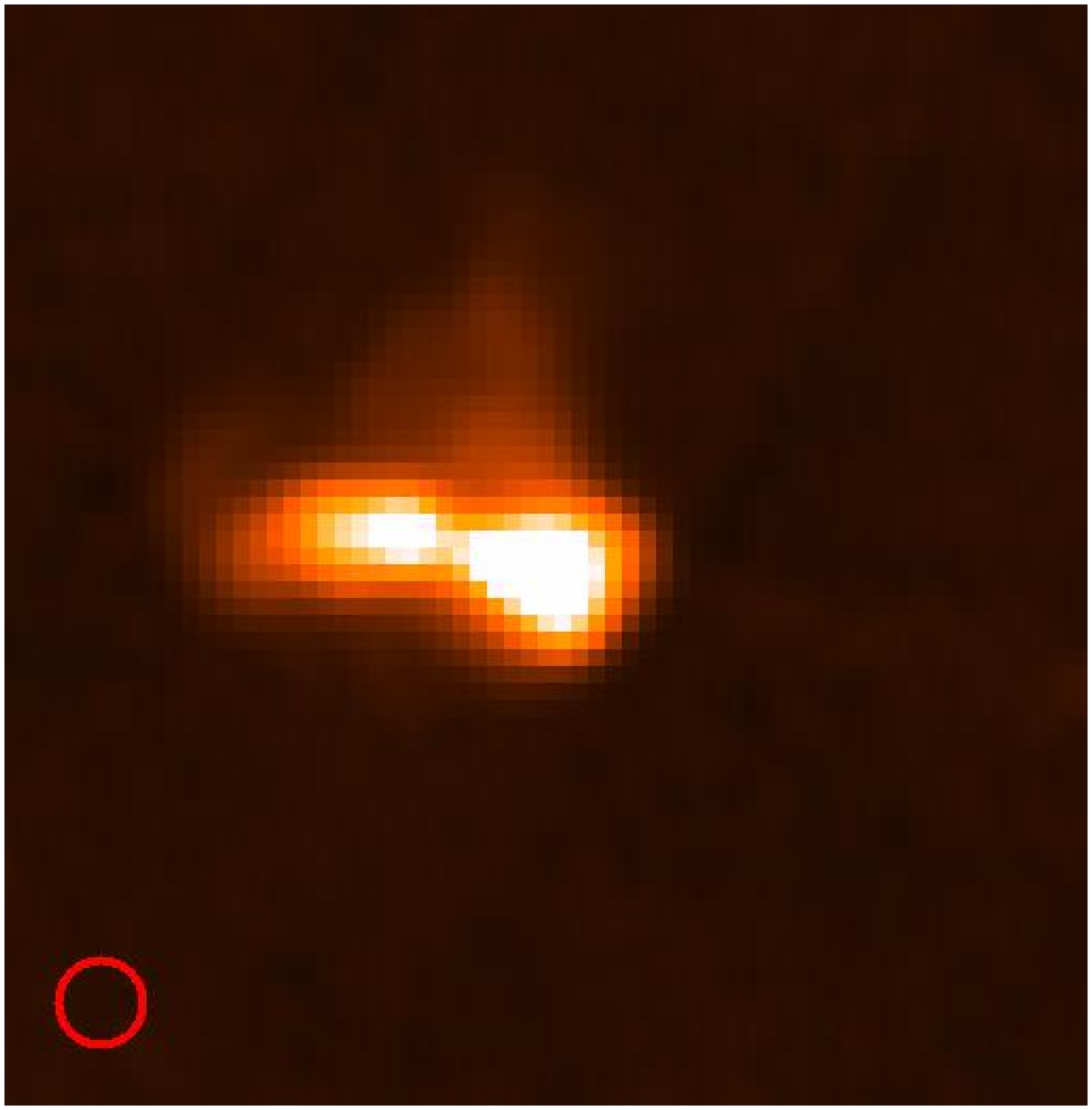}
\includegraphics[width=19mm]{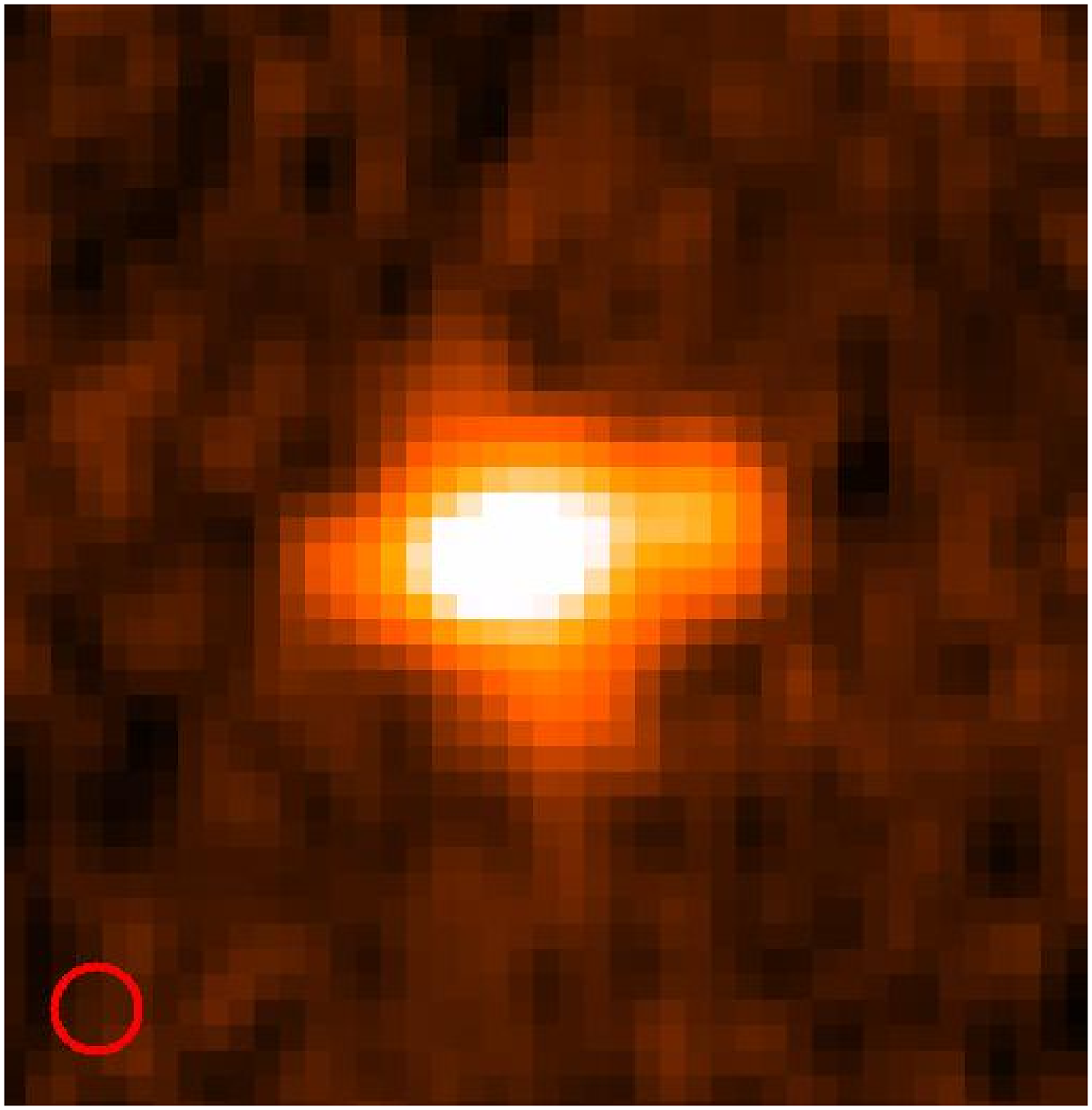}
\includegraphics[width=19mm]{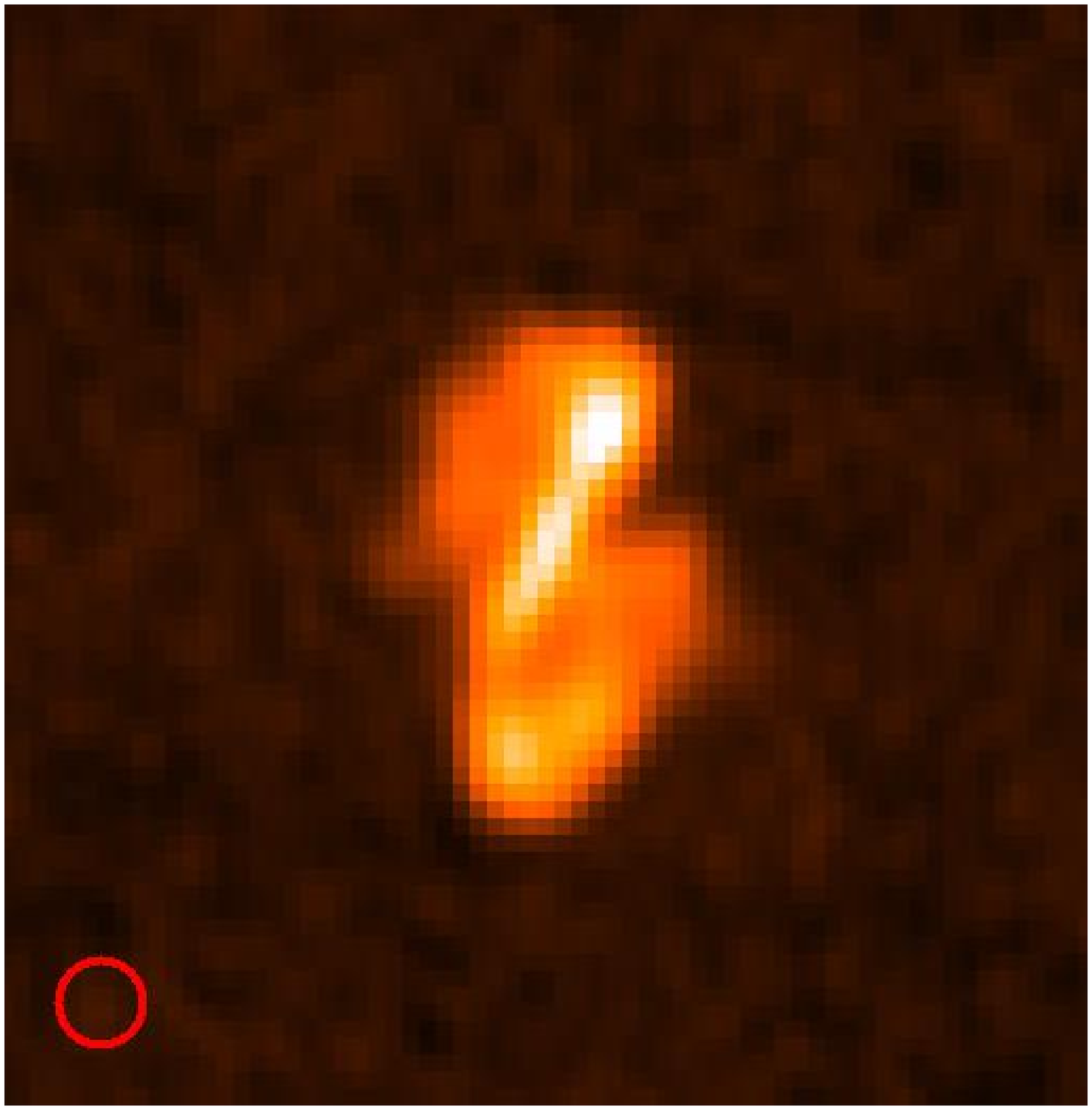}
}
\centerline{
\includegraphics[width=19mm]{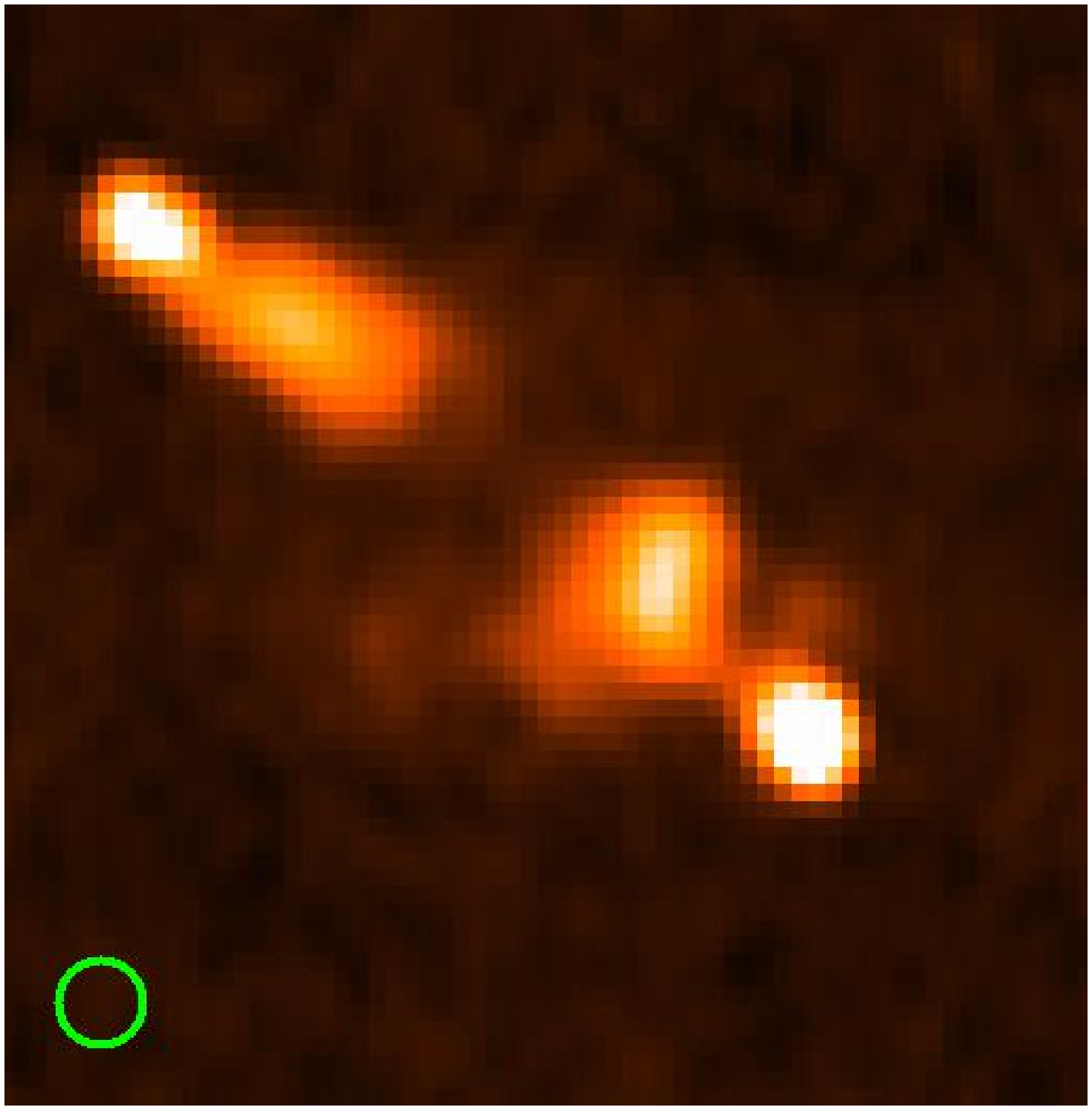}
\includegraphics[width=19mm]{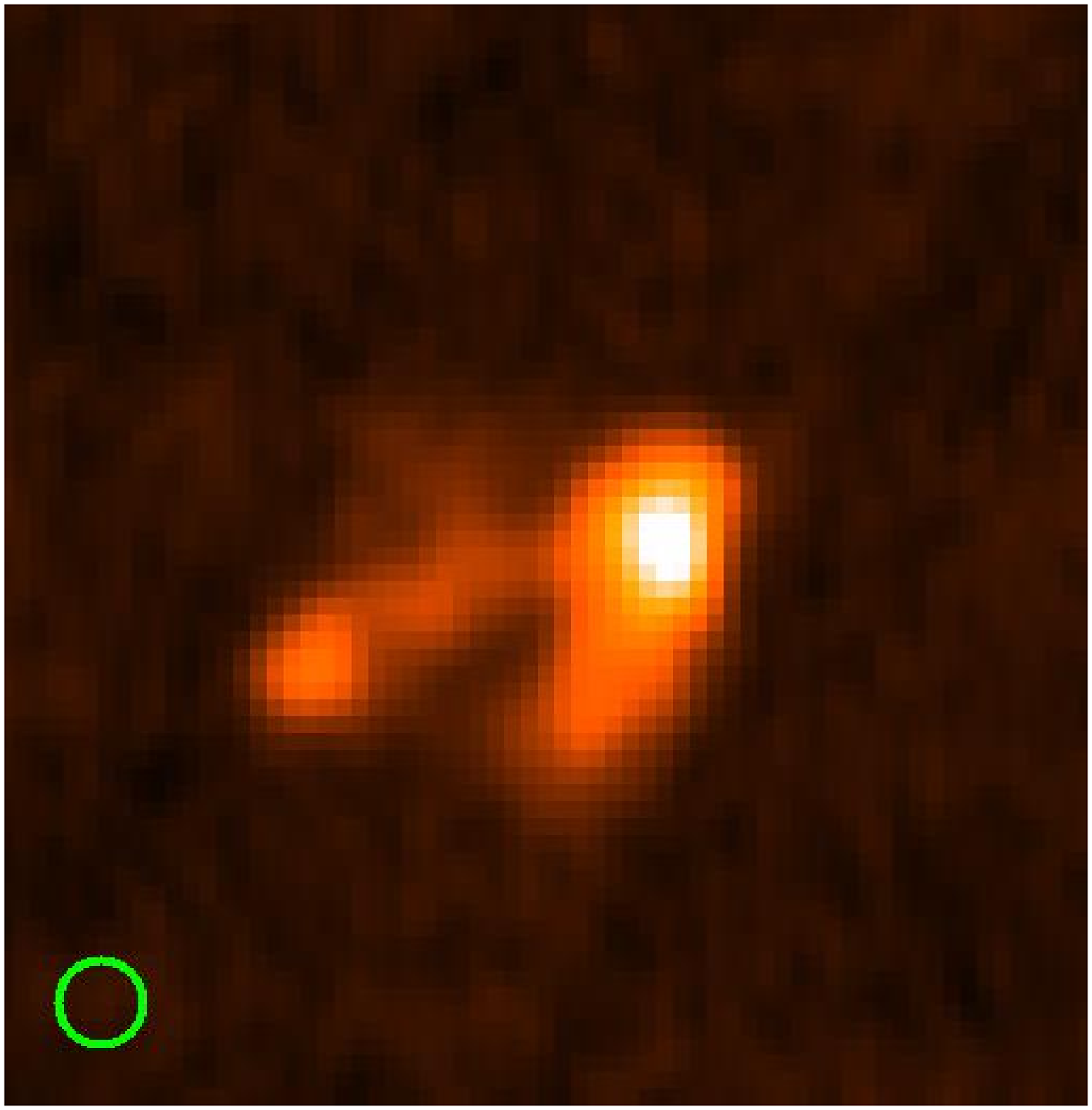}
\includegraphics[width=19mm]{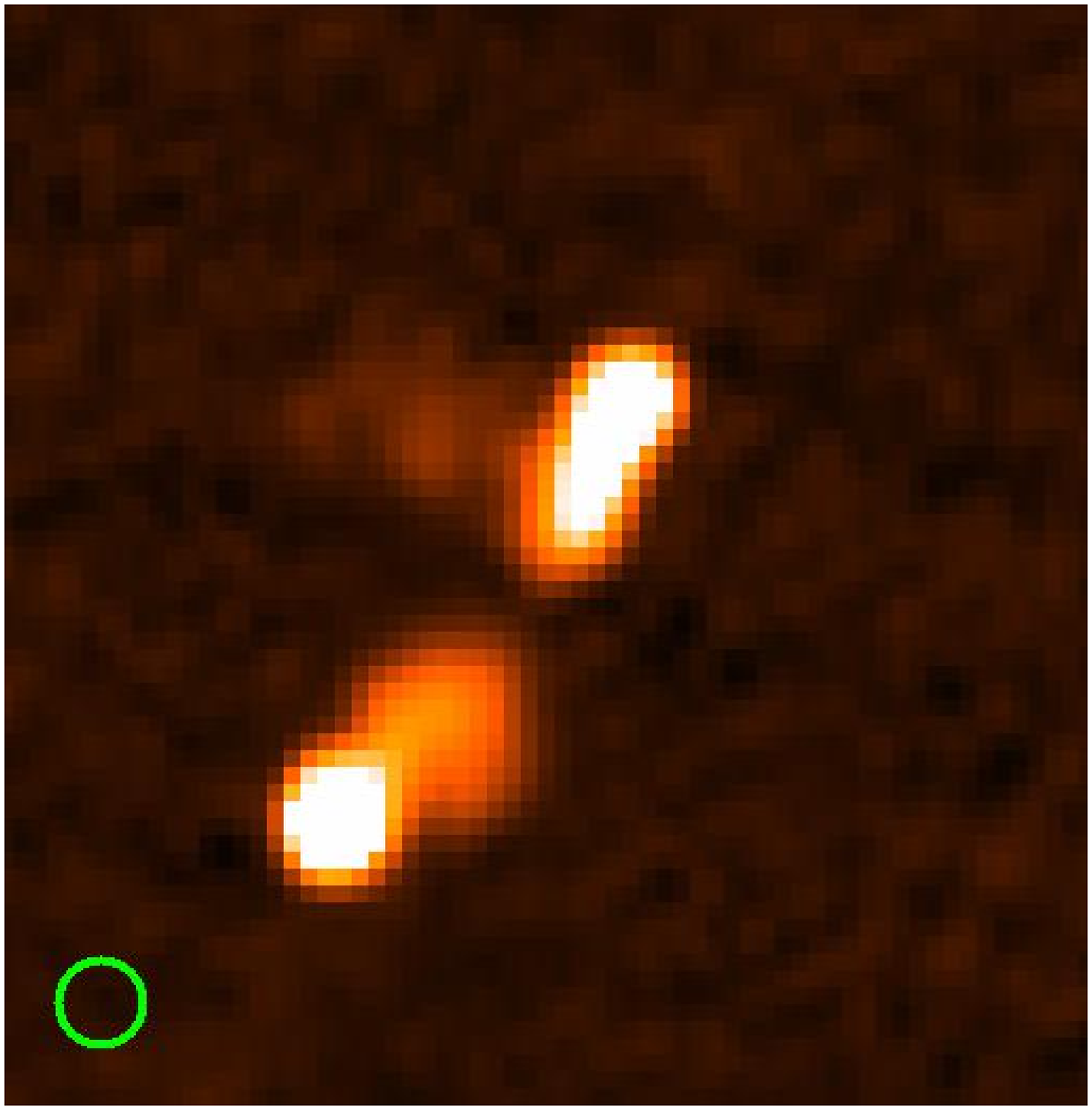}
\includegraphics[width=19mm]{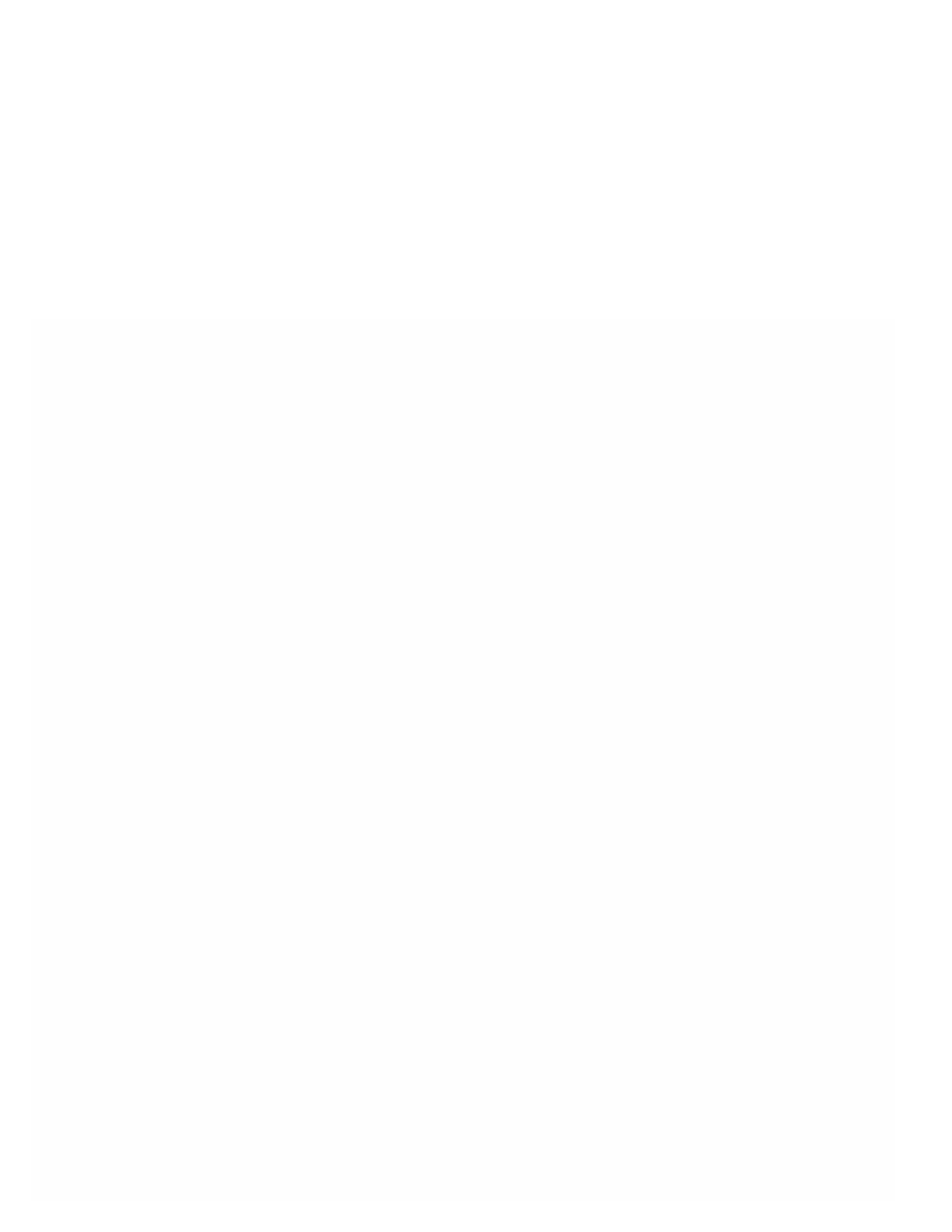}
\includegraphics[width=19mm]{radiowh.ps}
\includegraphics[width=19mm]{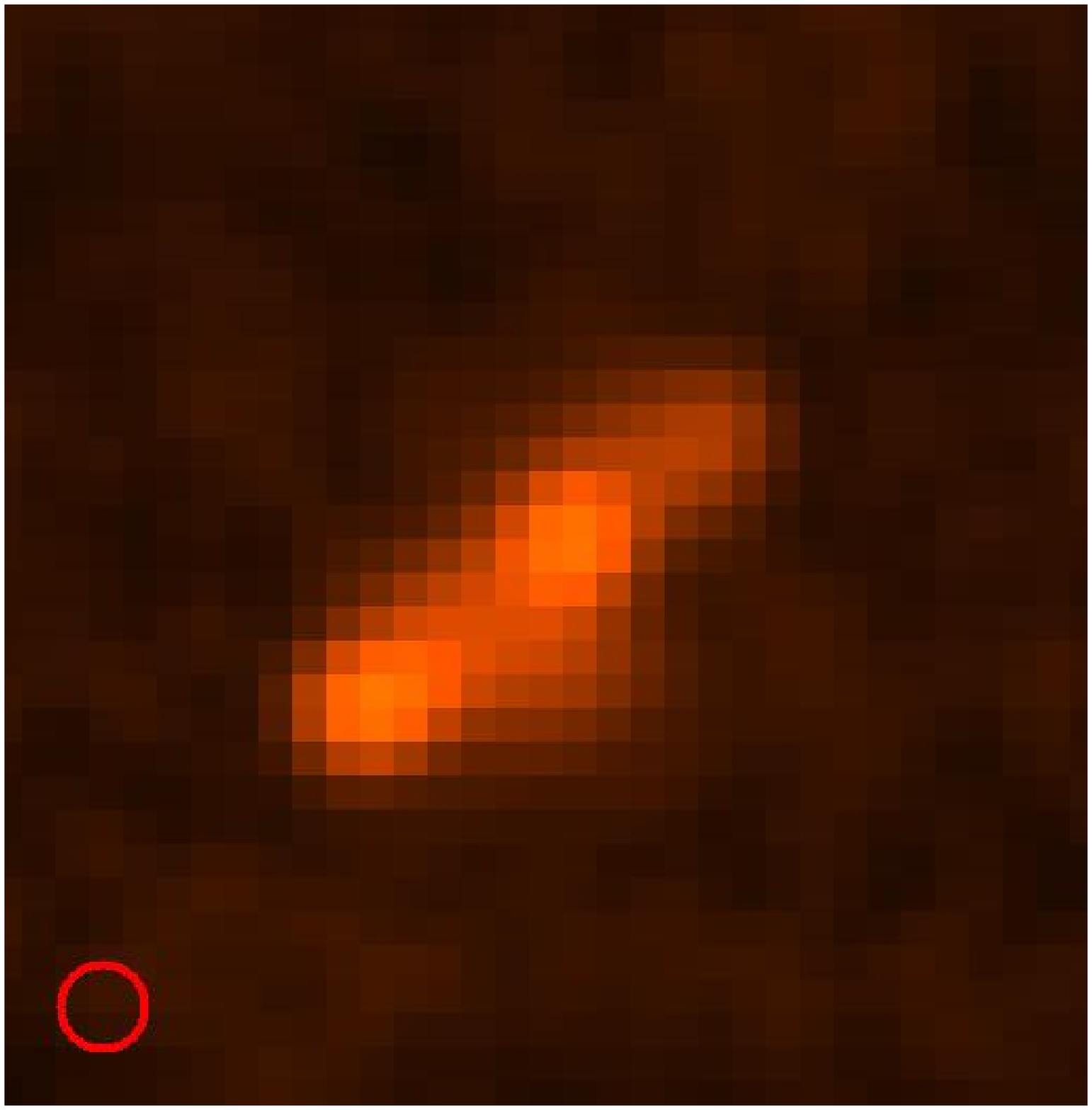}
\includegraphics[width=19mm]{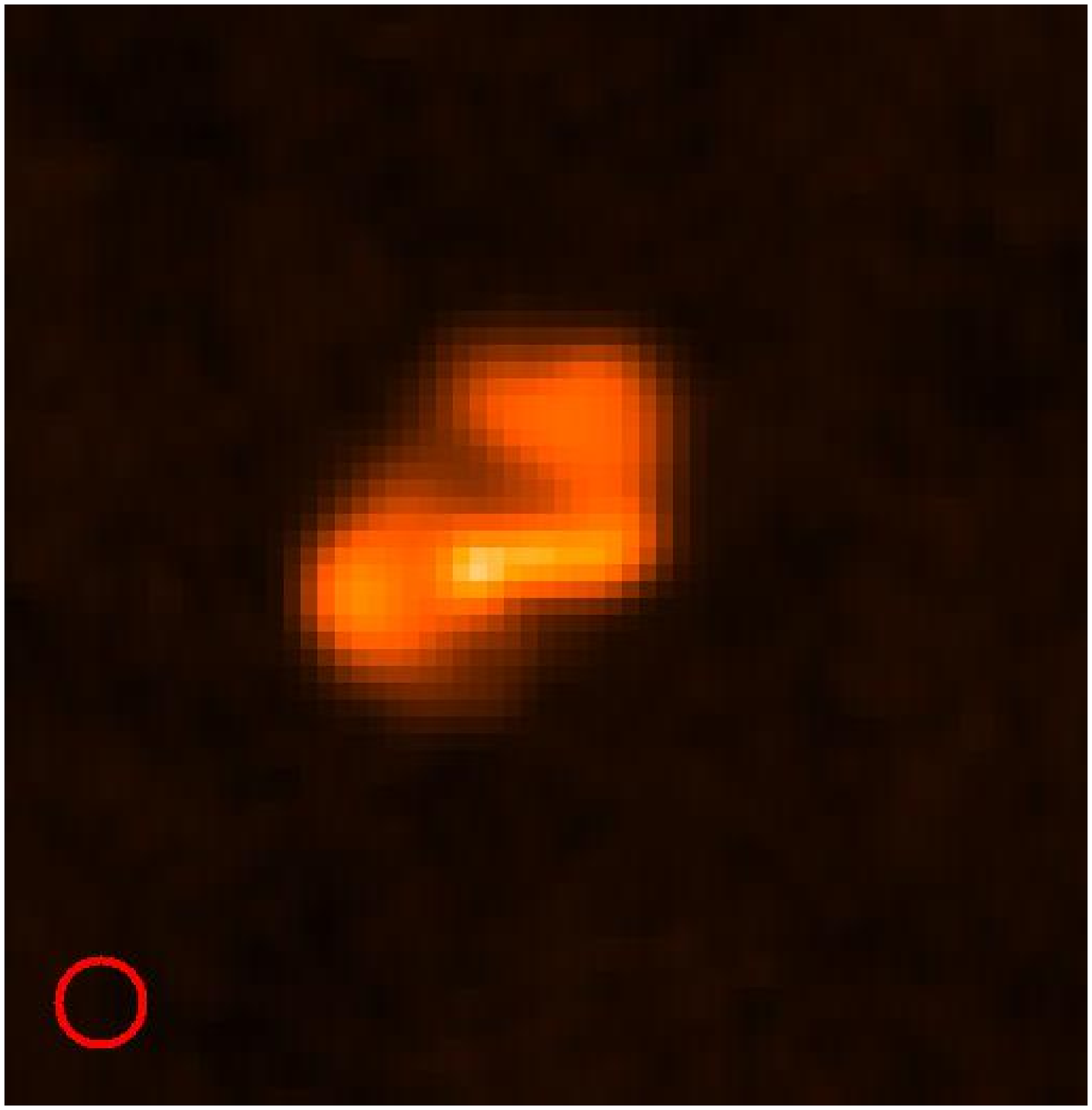}
\includegraphics[width=19mm]{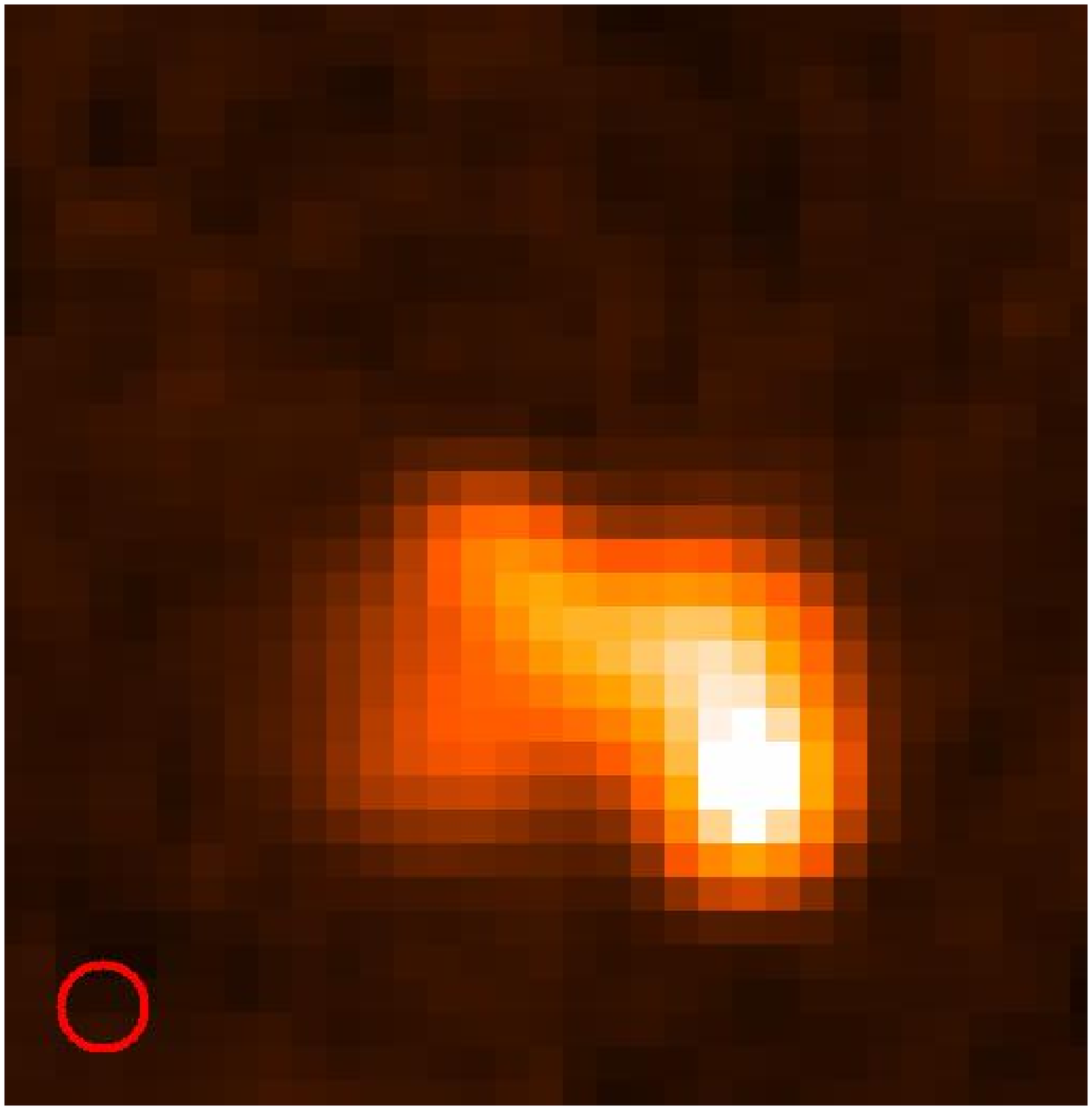}
\includegraphics[width=19mm]{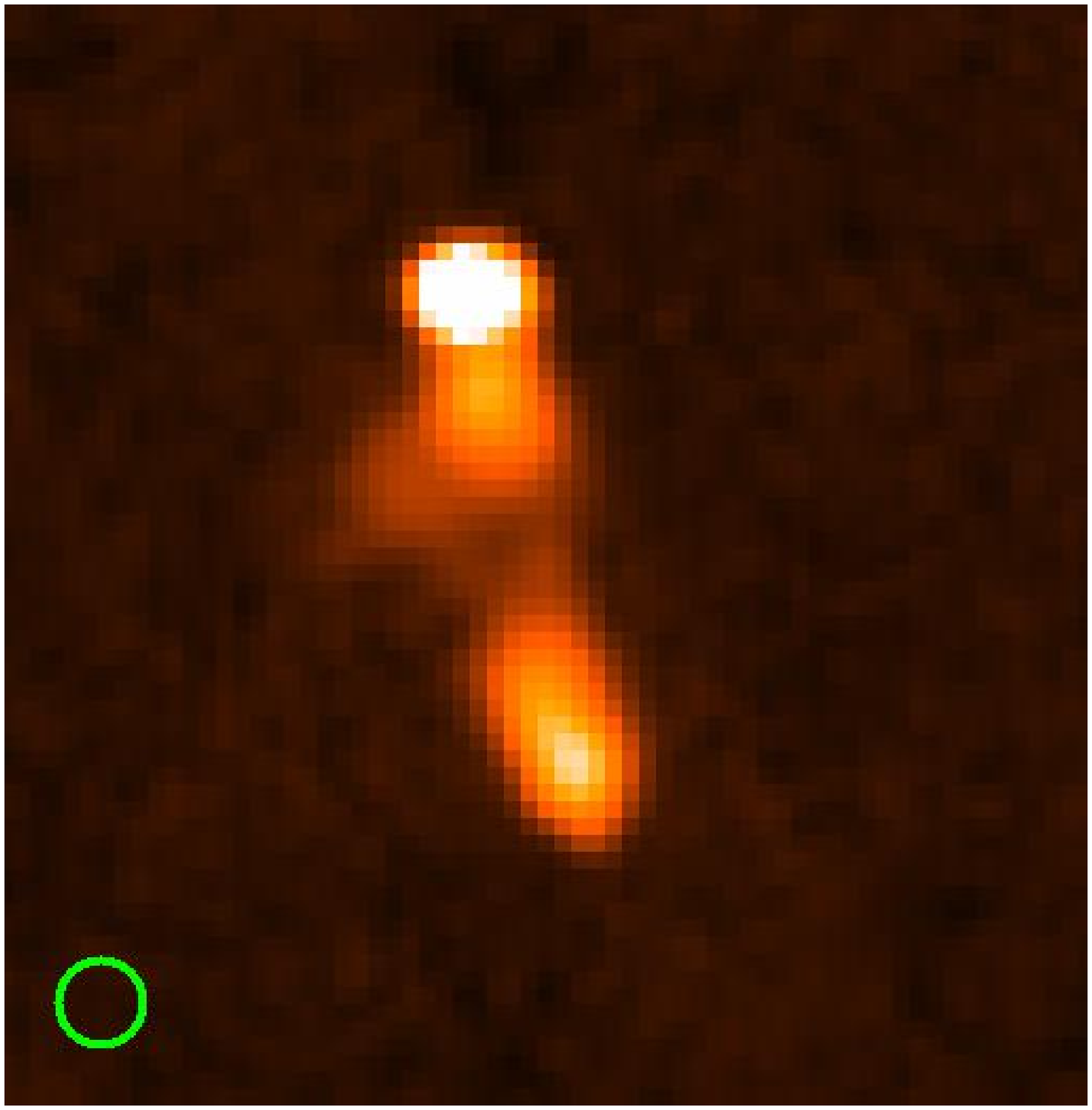}
\includegraphics[width=19mm]{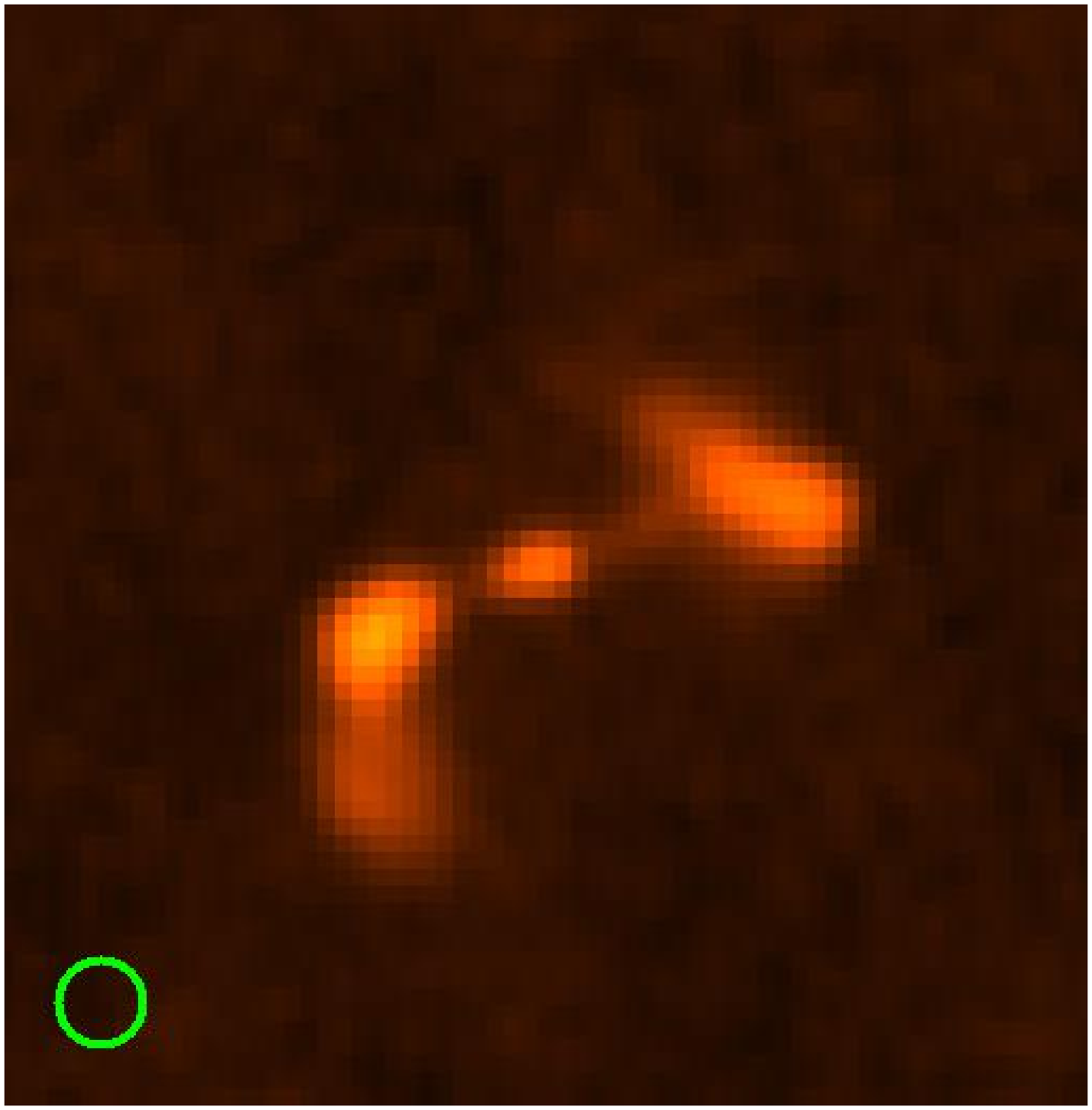}
}
\centerline{
\includegraphics[width=19mm]{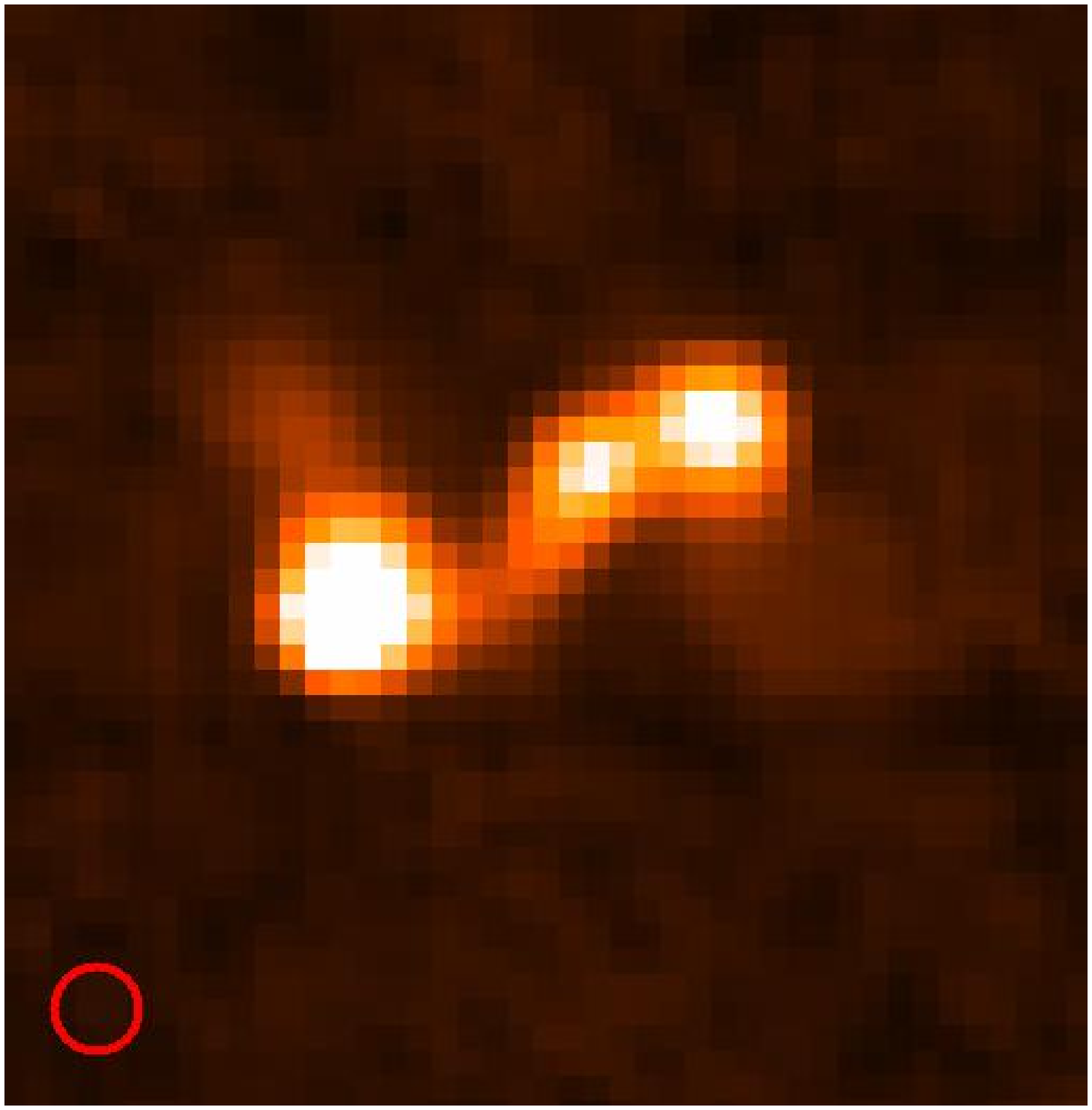}
\includegraphics[width=19mm]{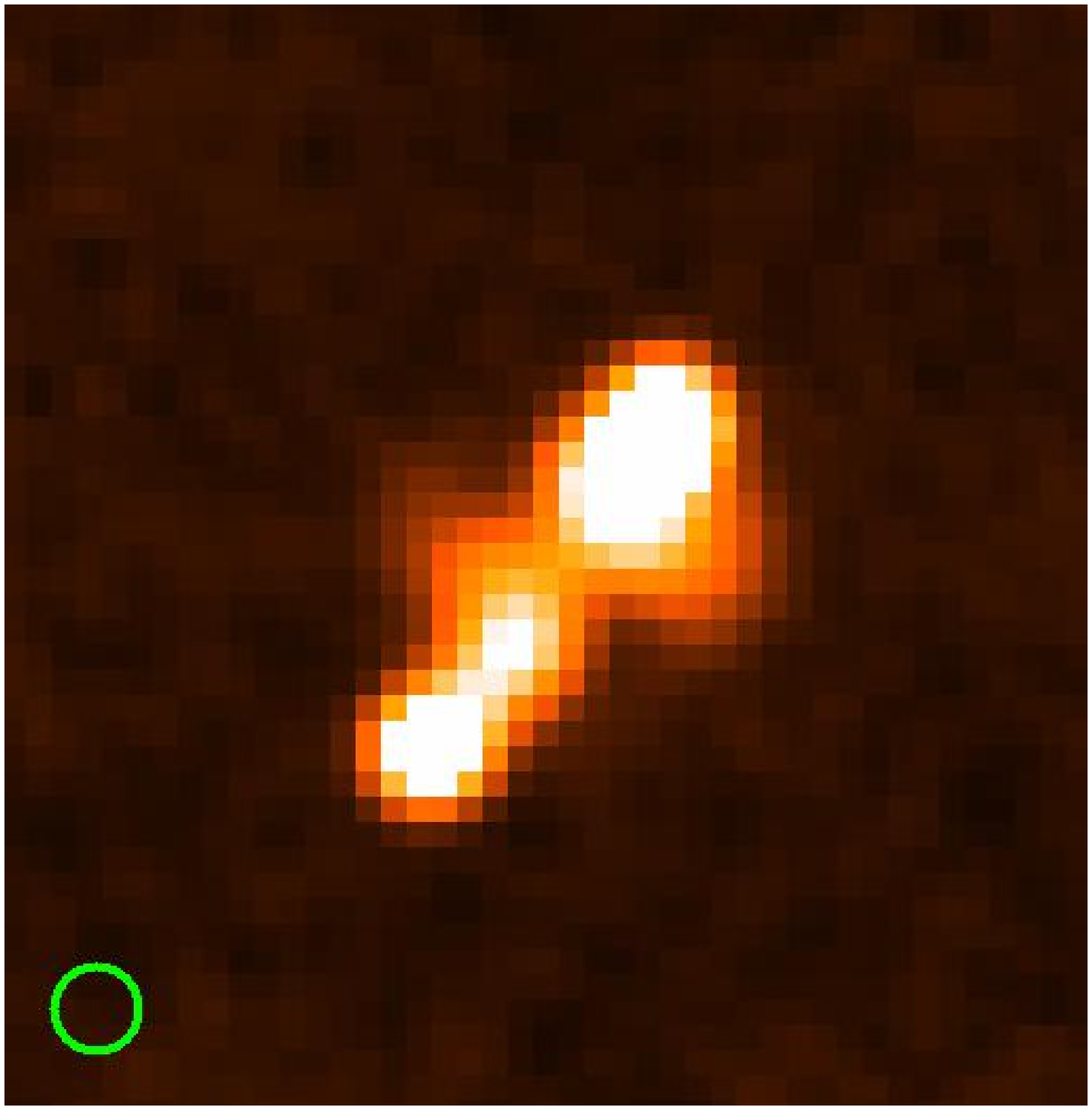}
\includegraphics[width=19mm]{radiowh.ps}
\includegraphics[width=19mm]{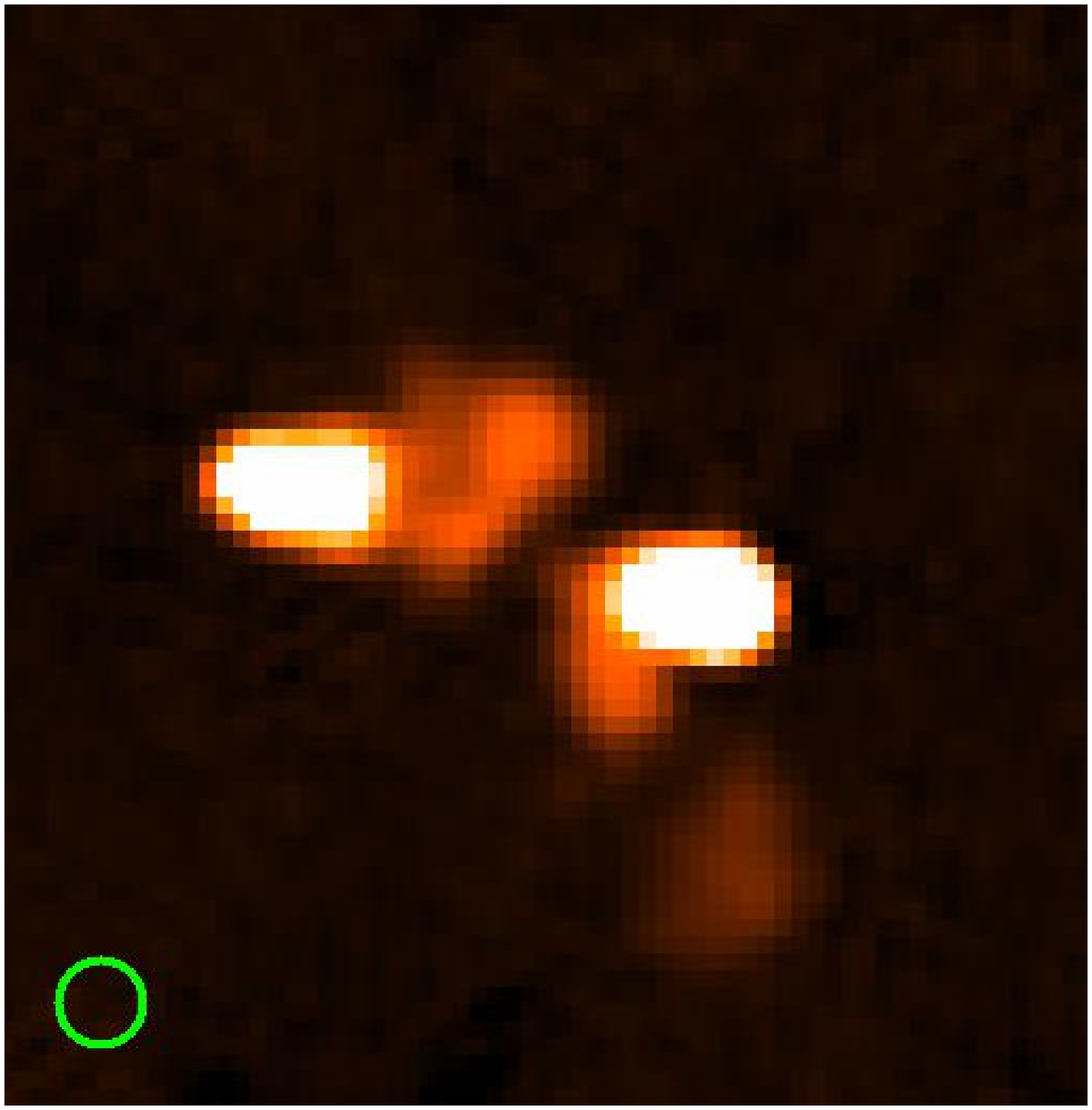}
\includegraphics[width=19mm]{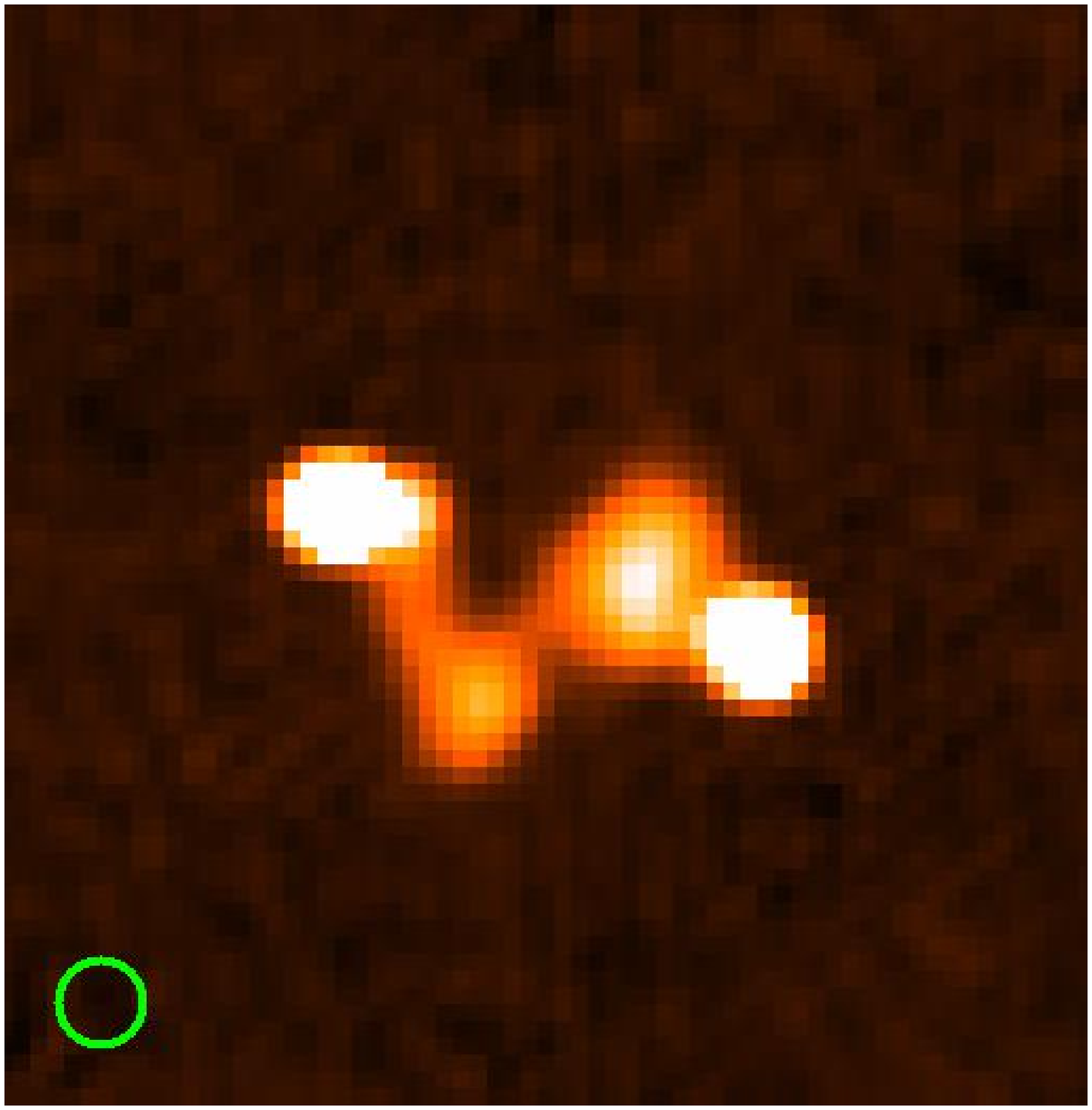}
\includegraphics[width=19mm]{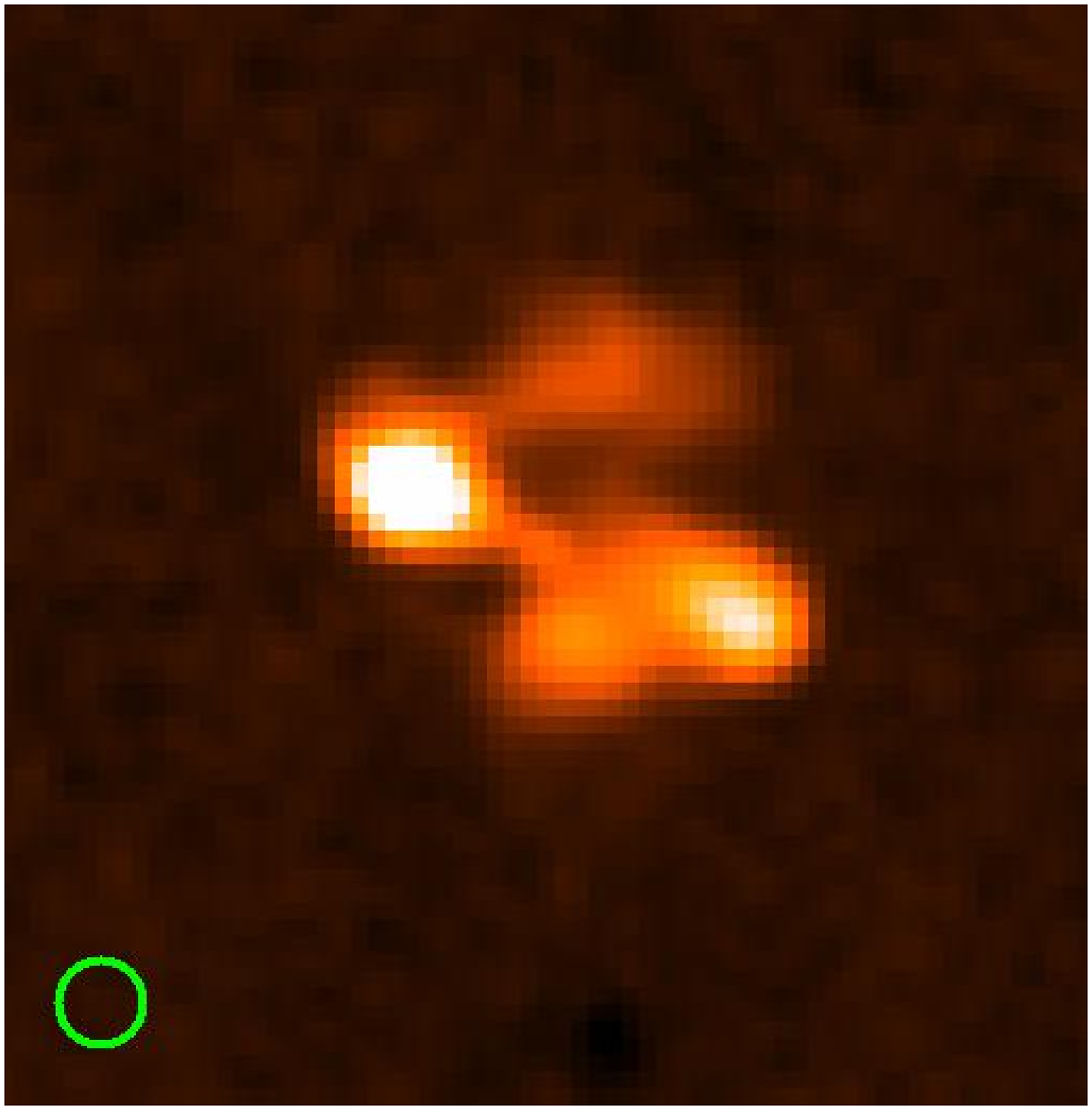}
\includegraphics[width=19mm]{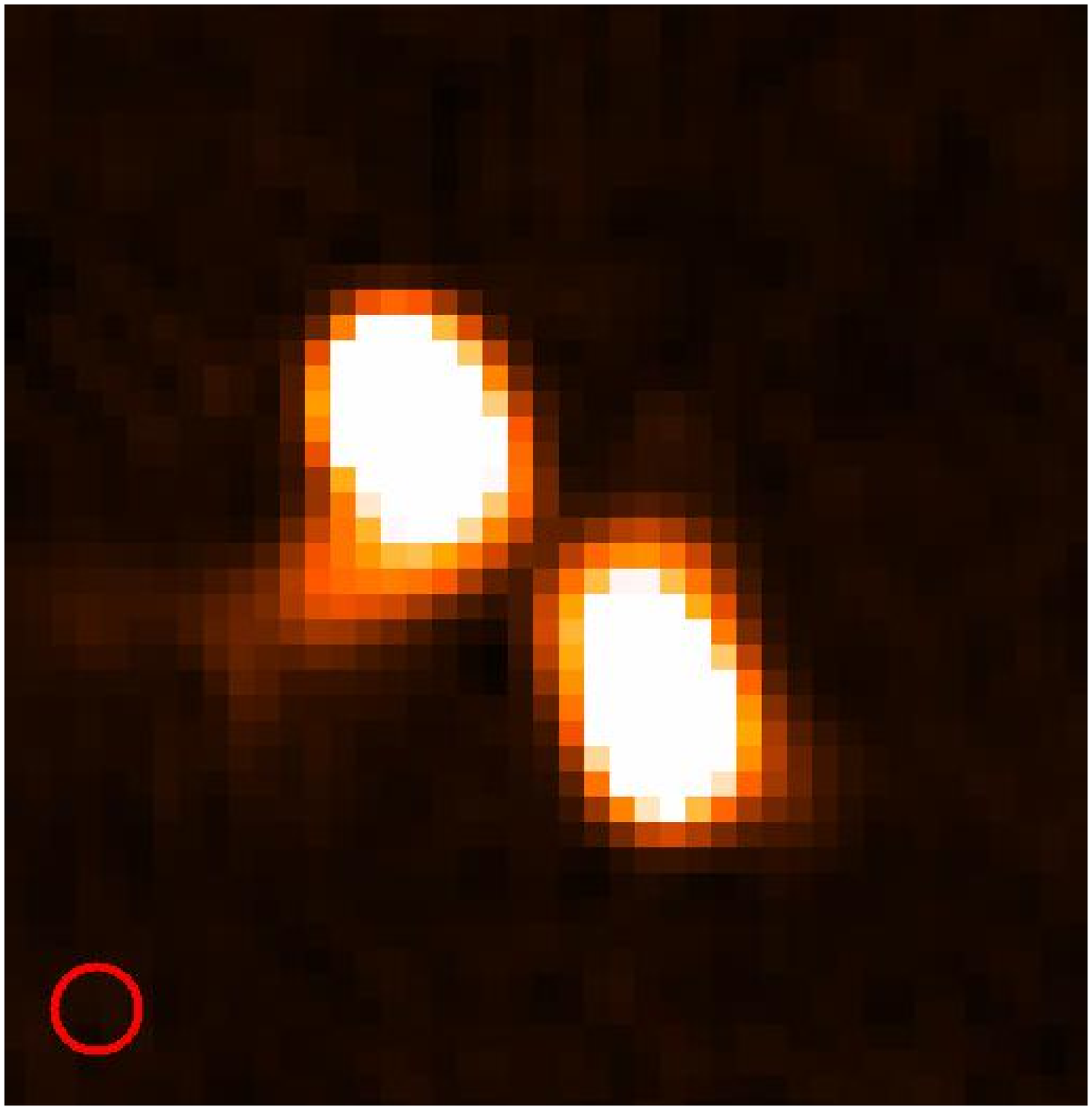}
\includegraphics[width=19mm]{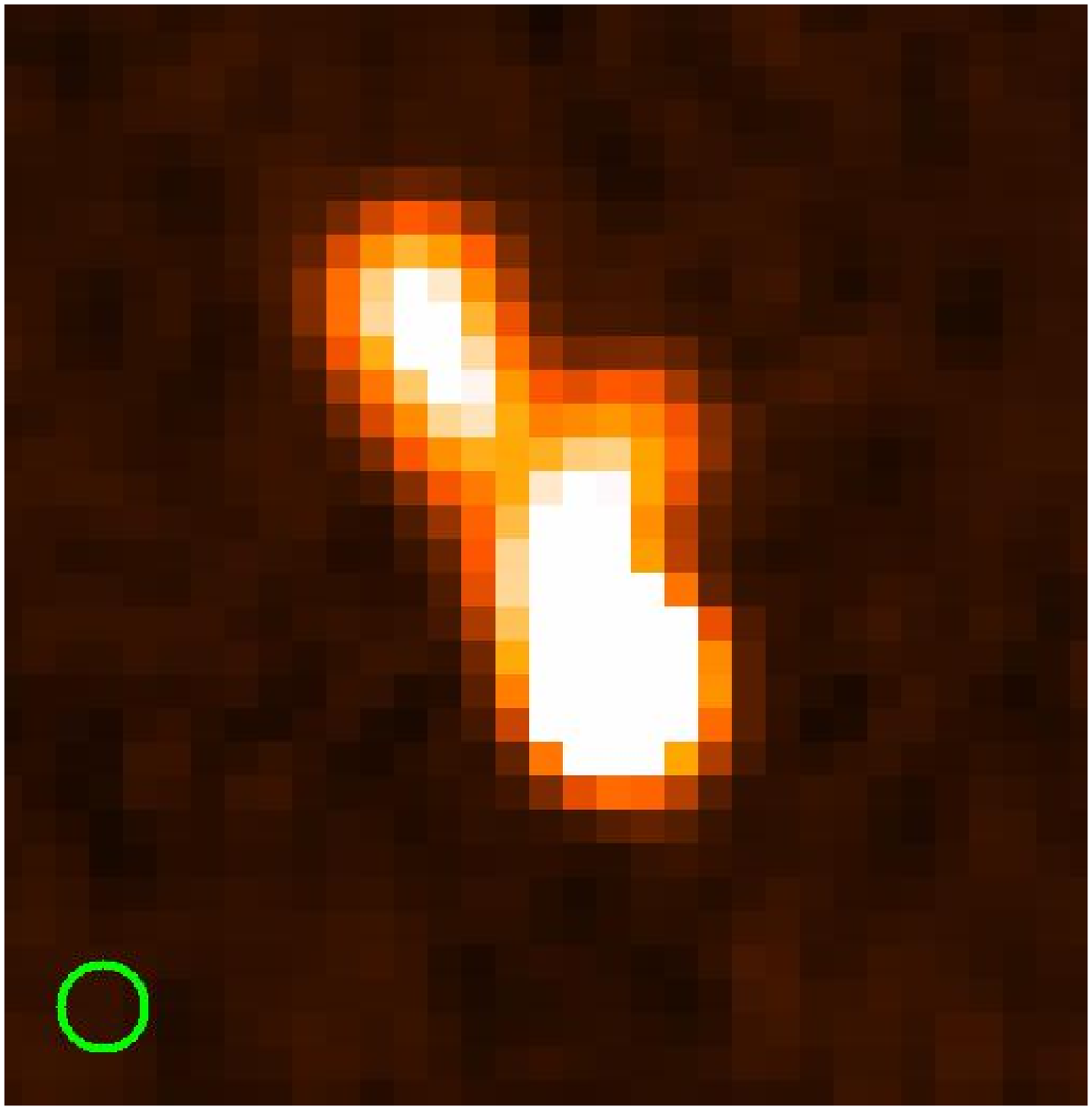}
\includegraphics[width=19mm]{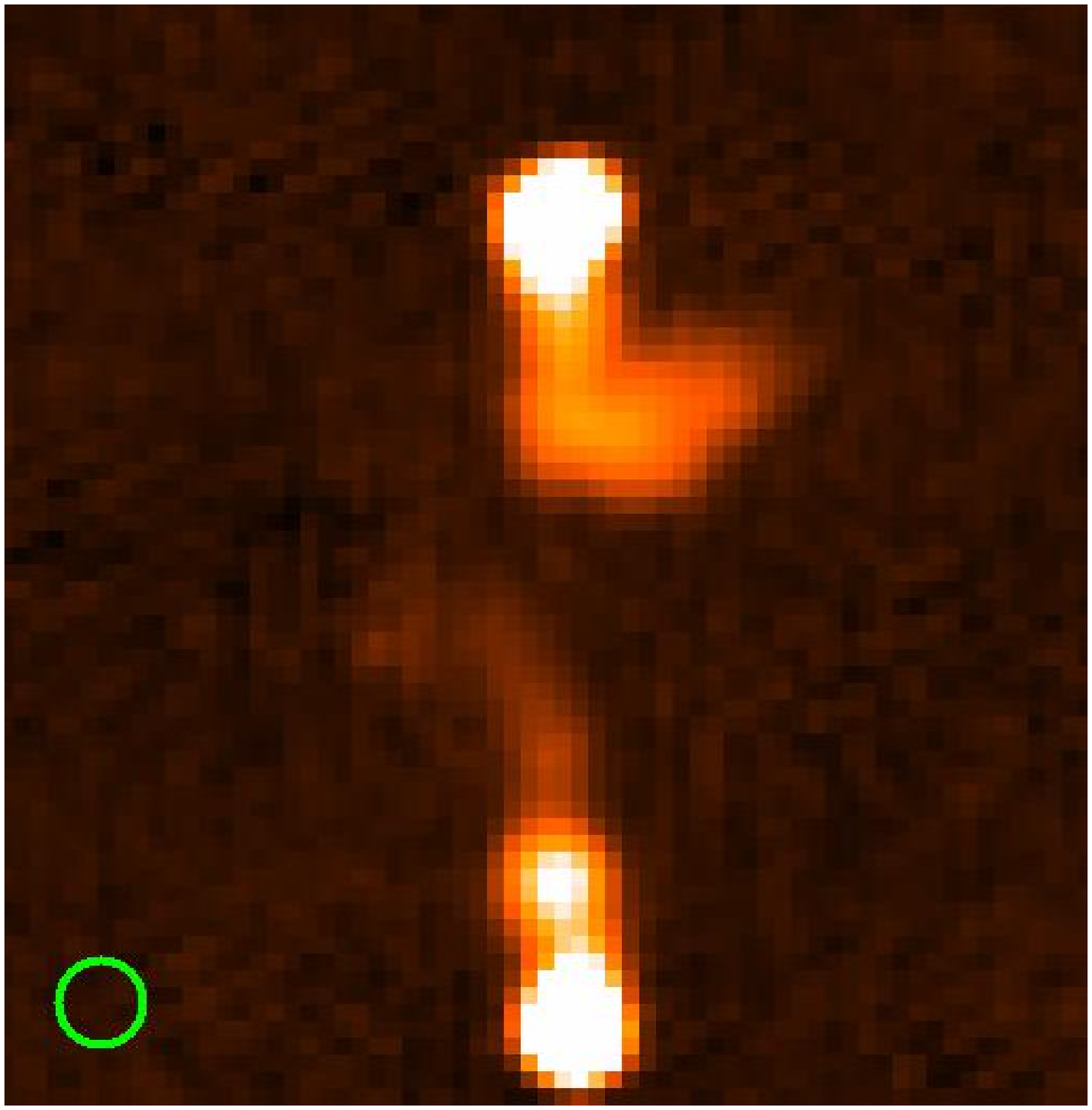}
\includegraphics[width=19mm]{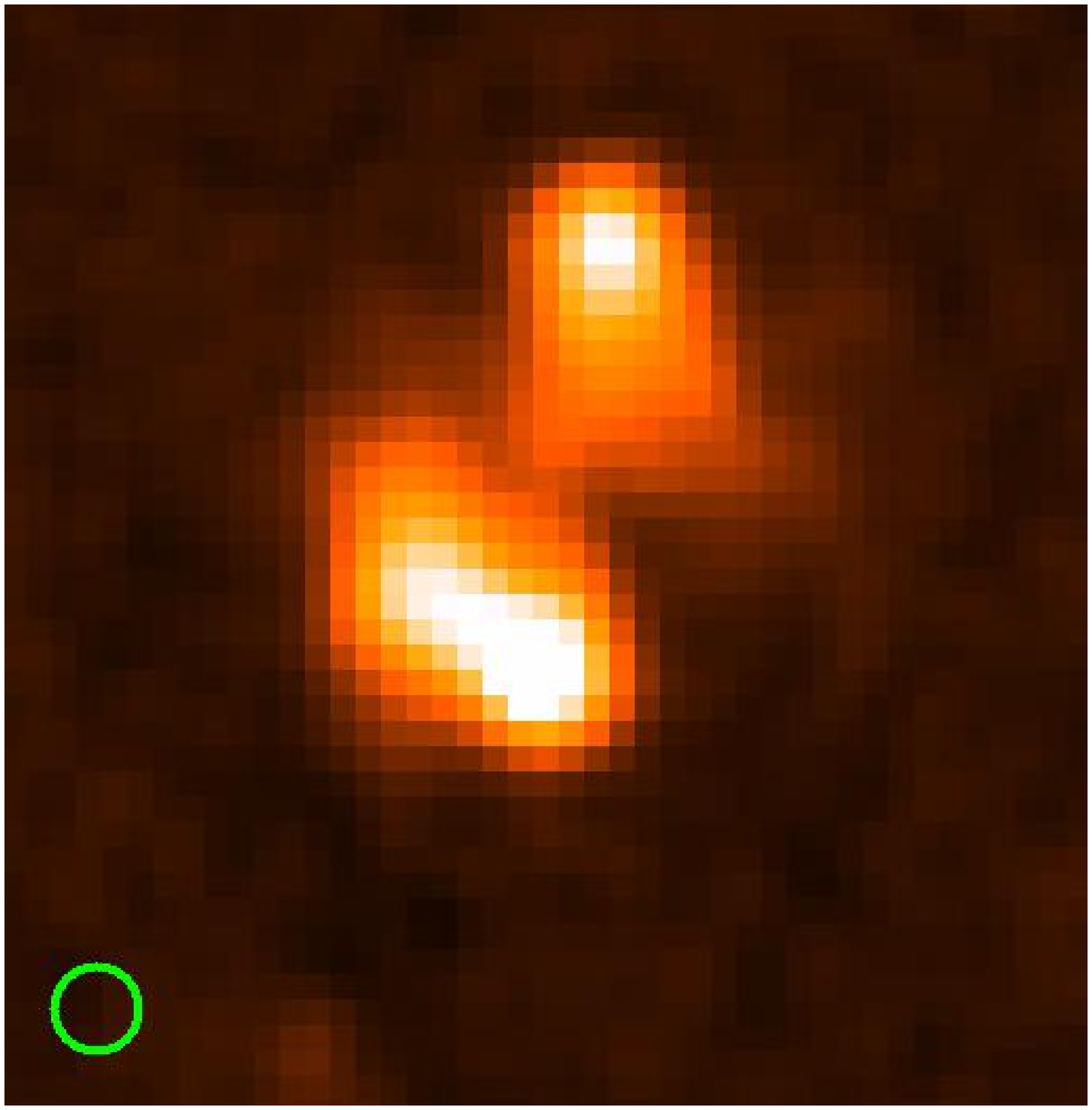}
}
\centerline{
\includegraphics[width=19mm]{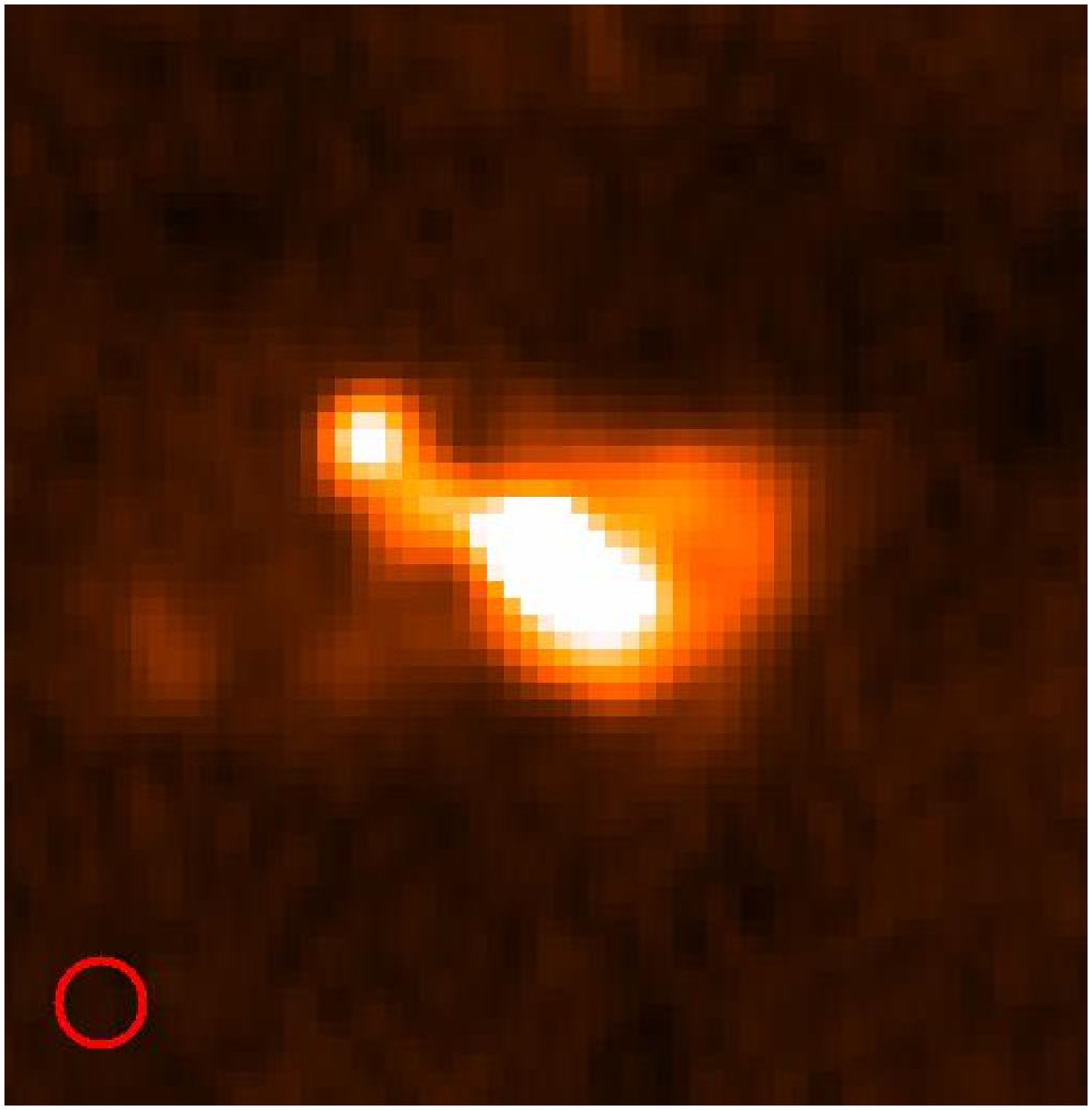}
\includegraphics[width=19mm]{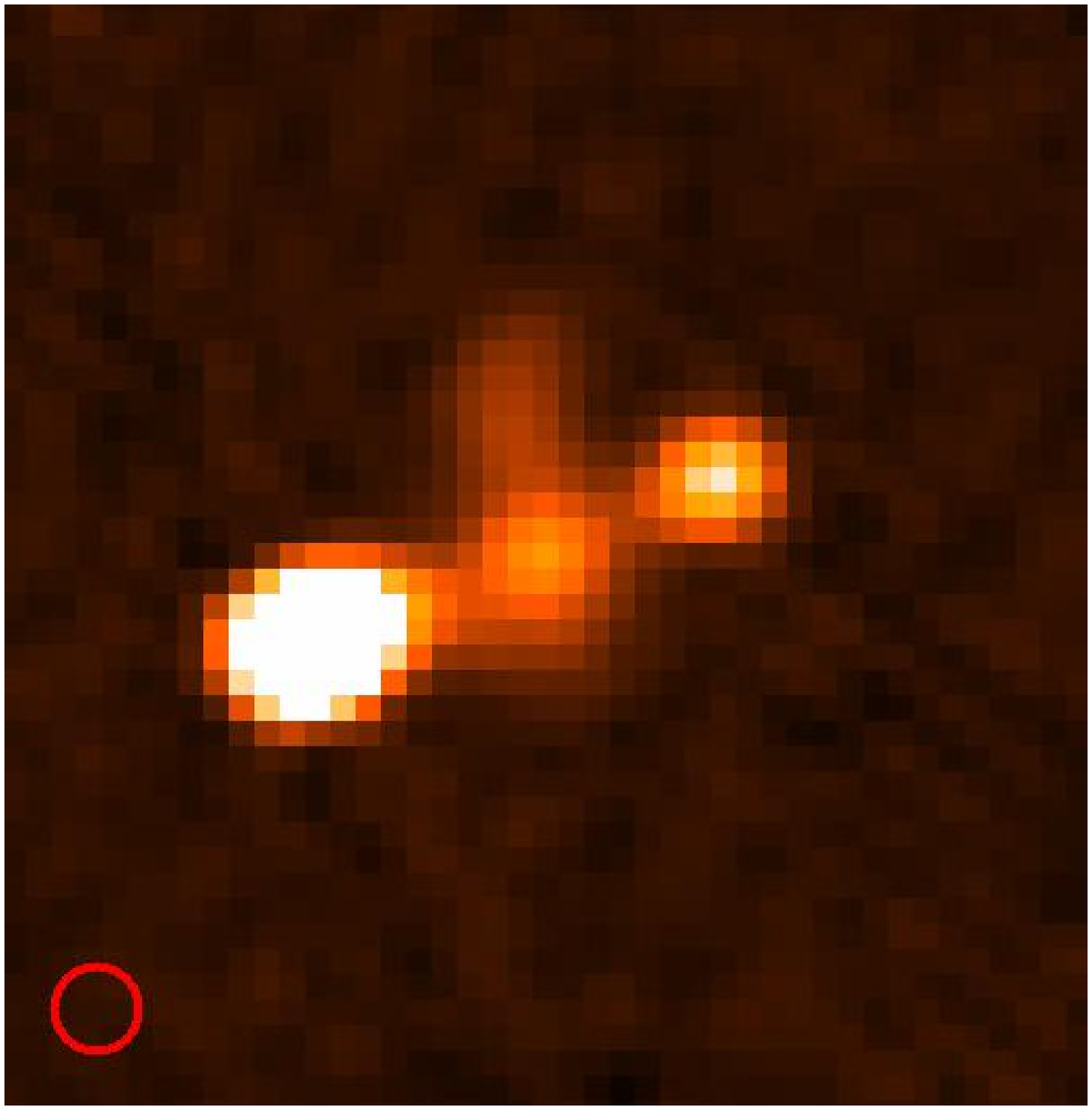}
\includegraphics[width=19mm]{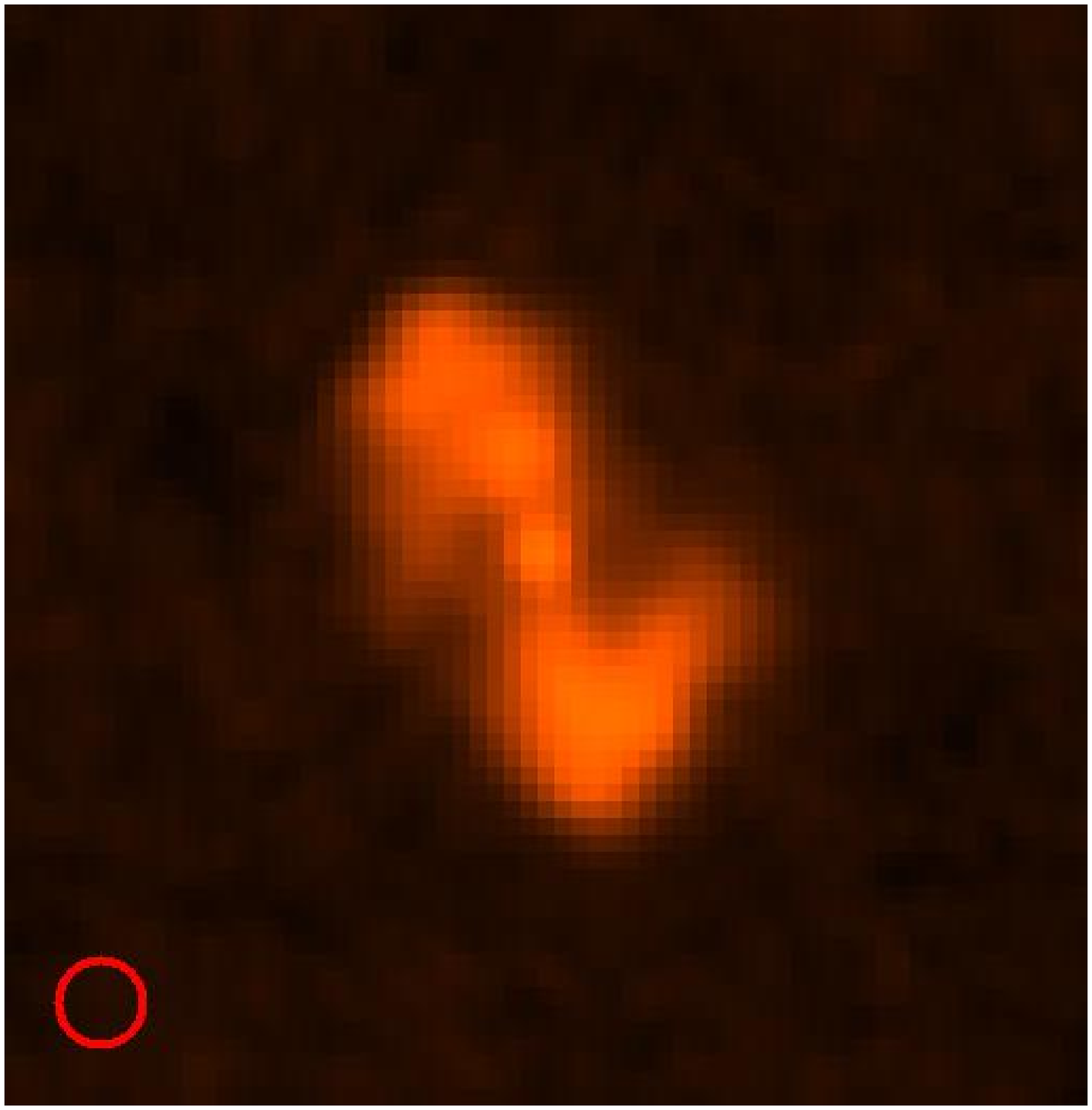}
\includegraphics[width=19mm]{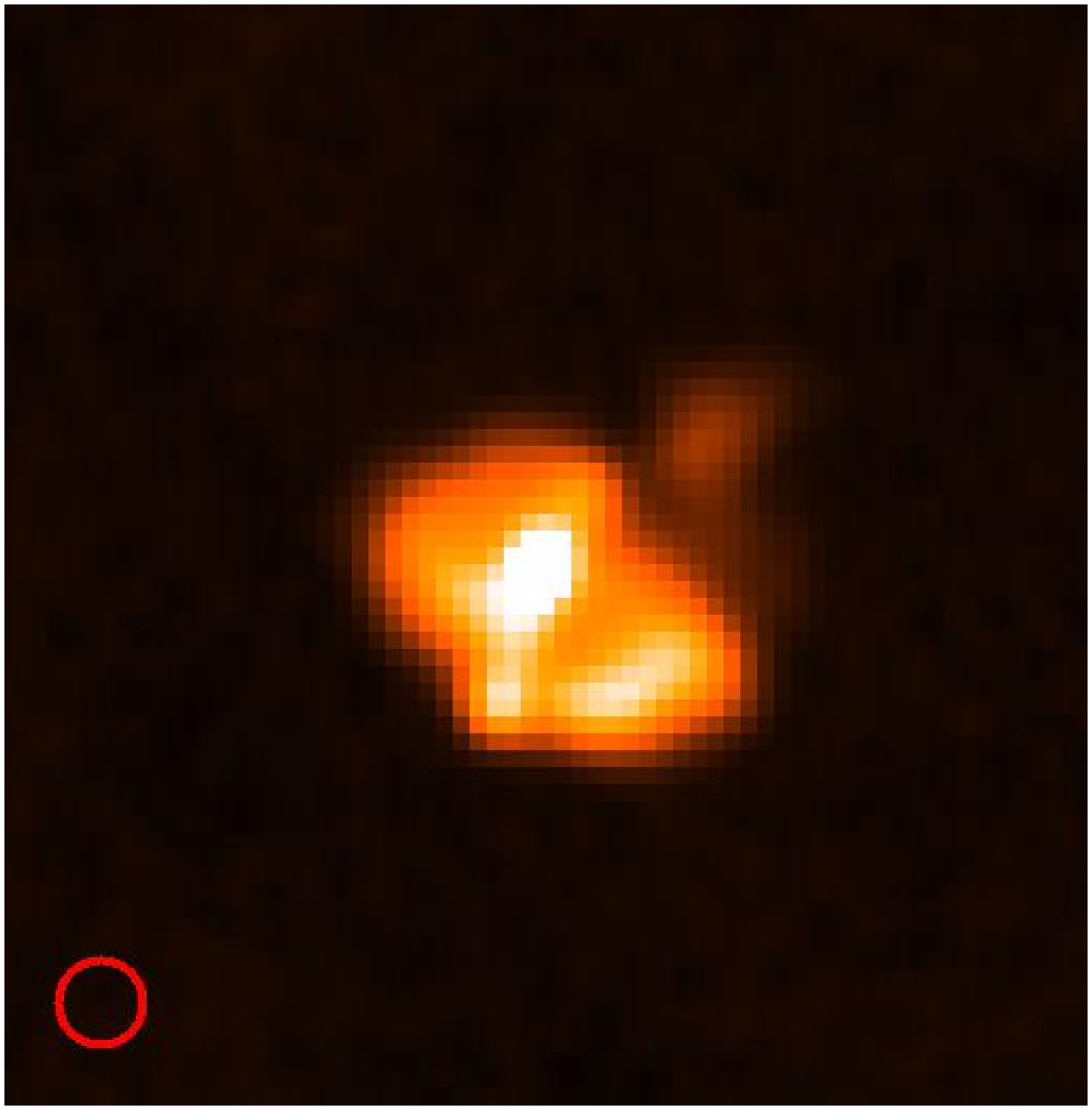}
\includegraphics[width=19mm]{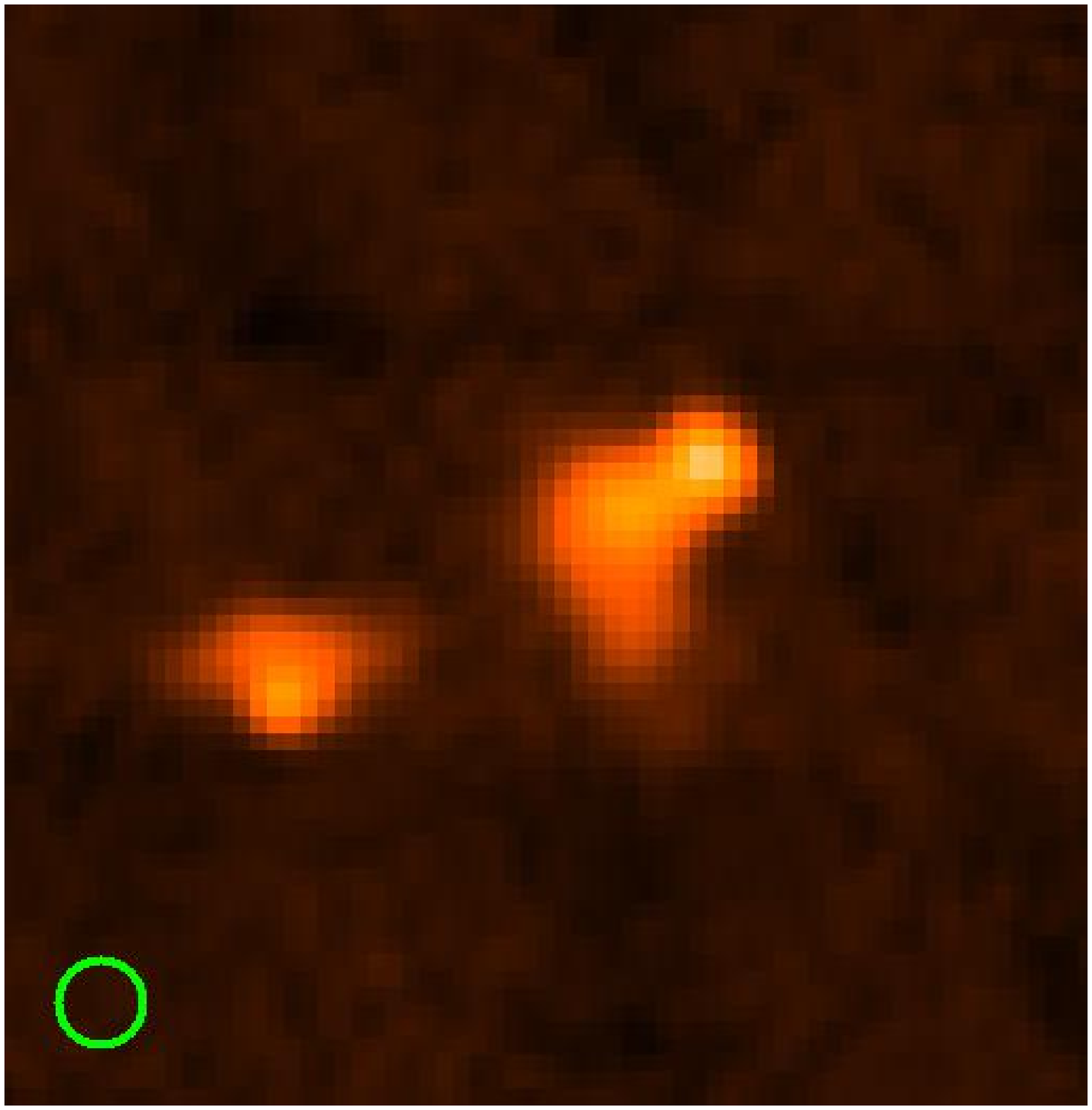}
\includegraphics[width=19mm]{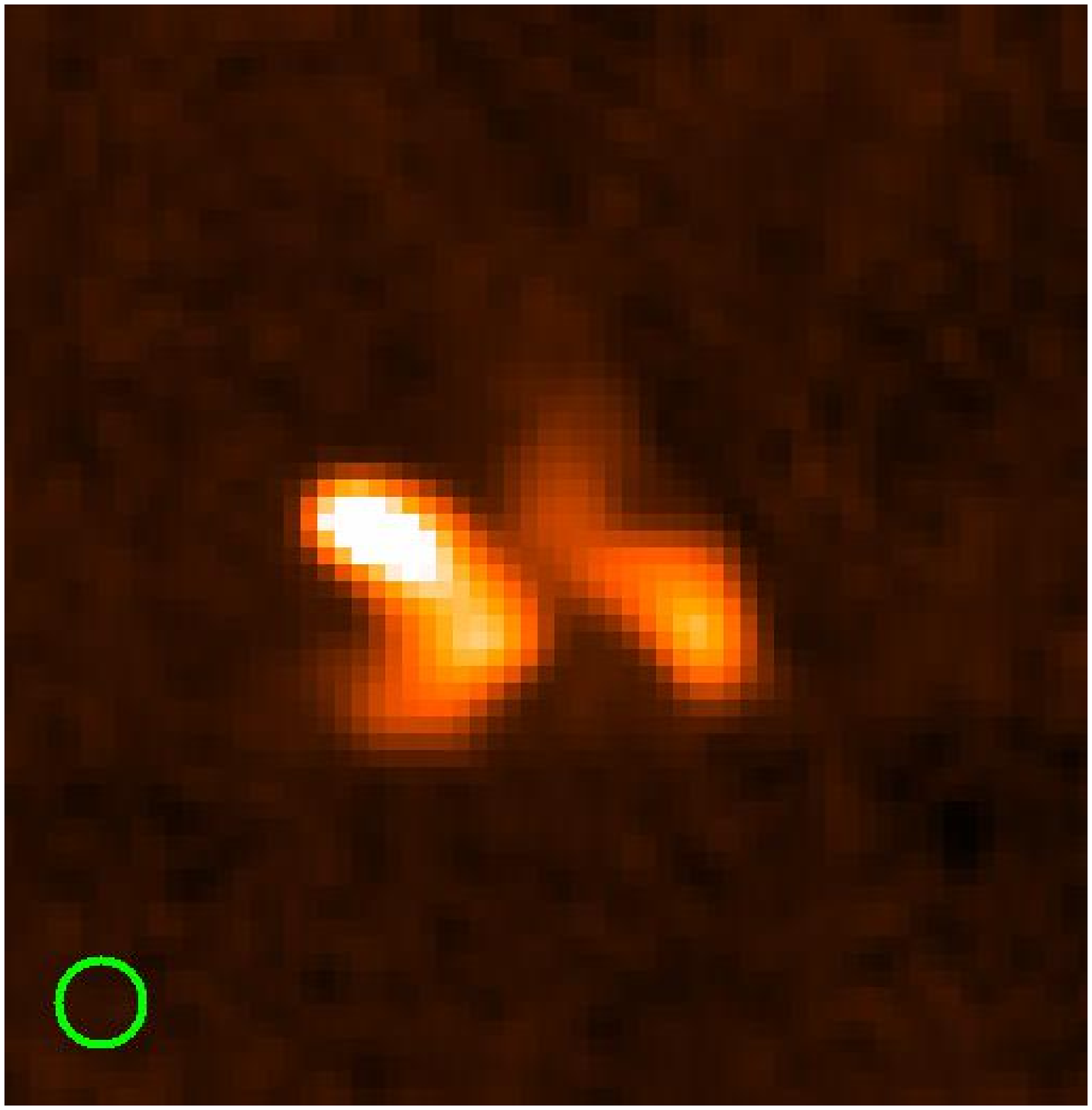}
\includegraphics[width=19mm]{radiowh.ps}
\includegraphics[width=19mm]{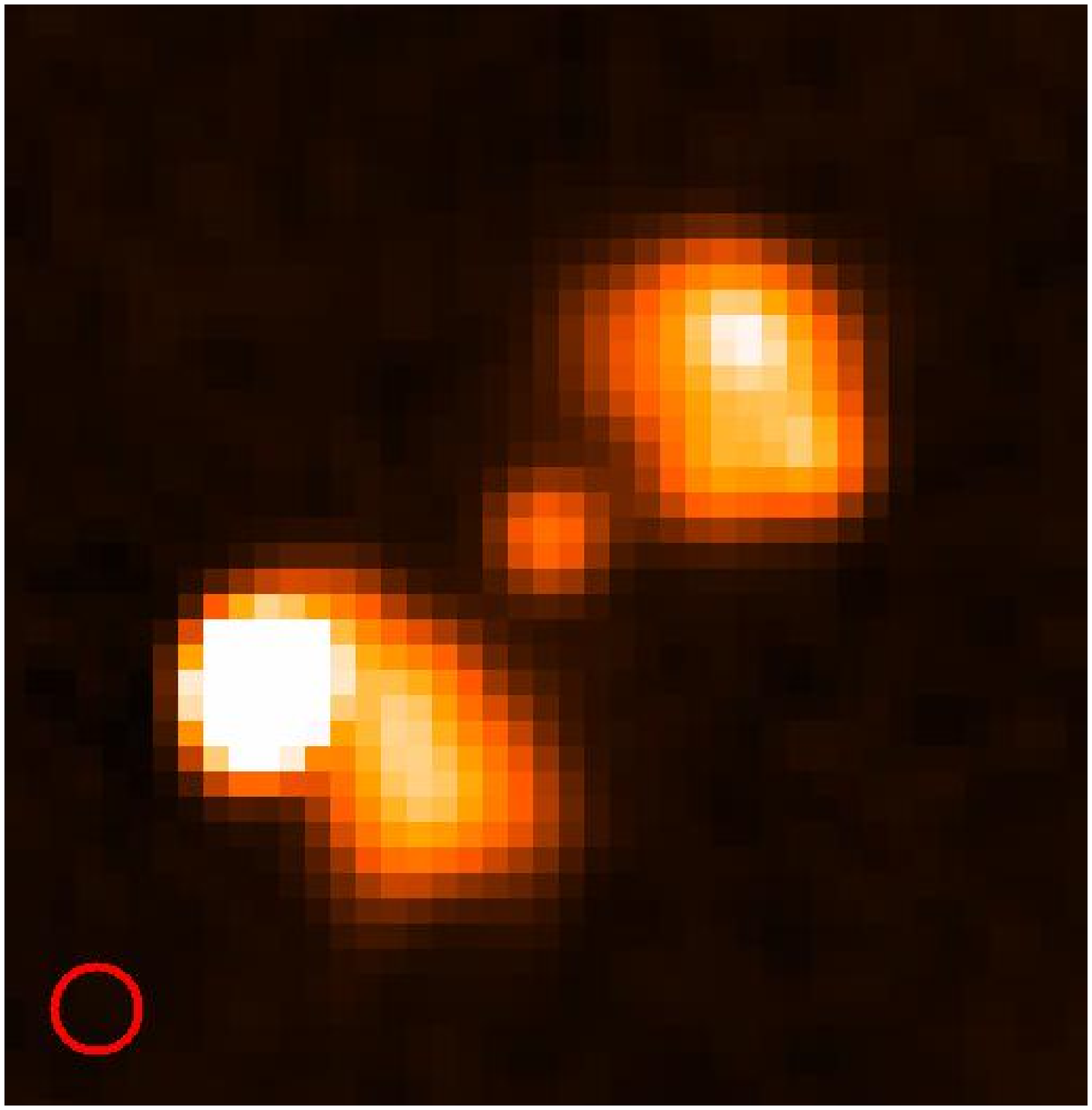}
\includegraphics[width=19mm]{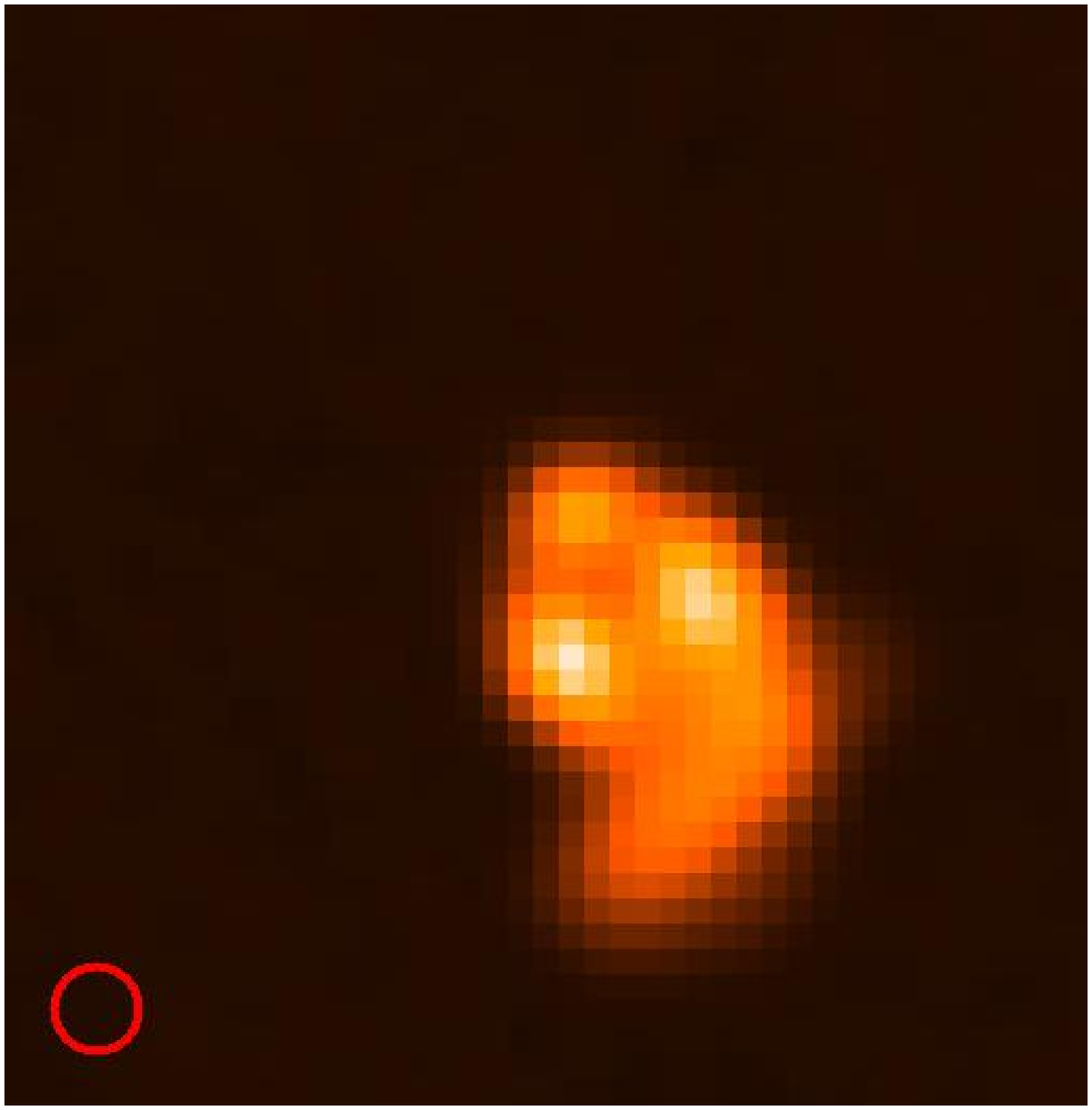}
\includegraphics[width=19mm]{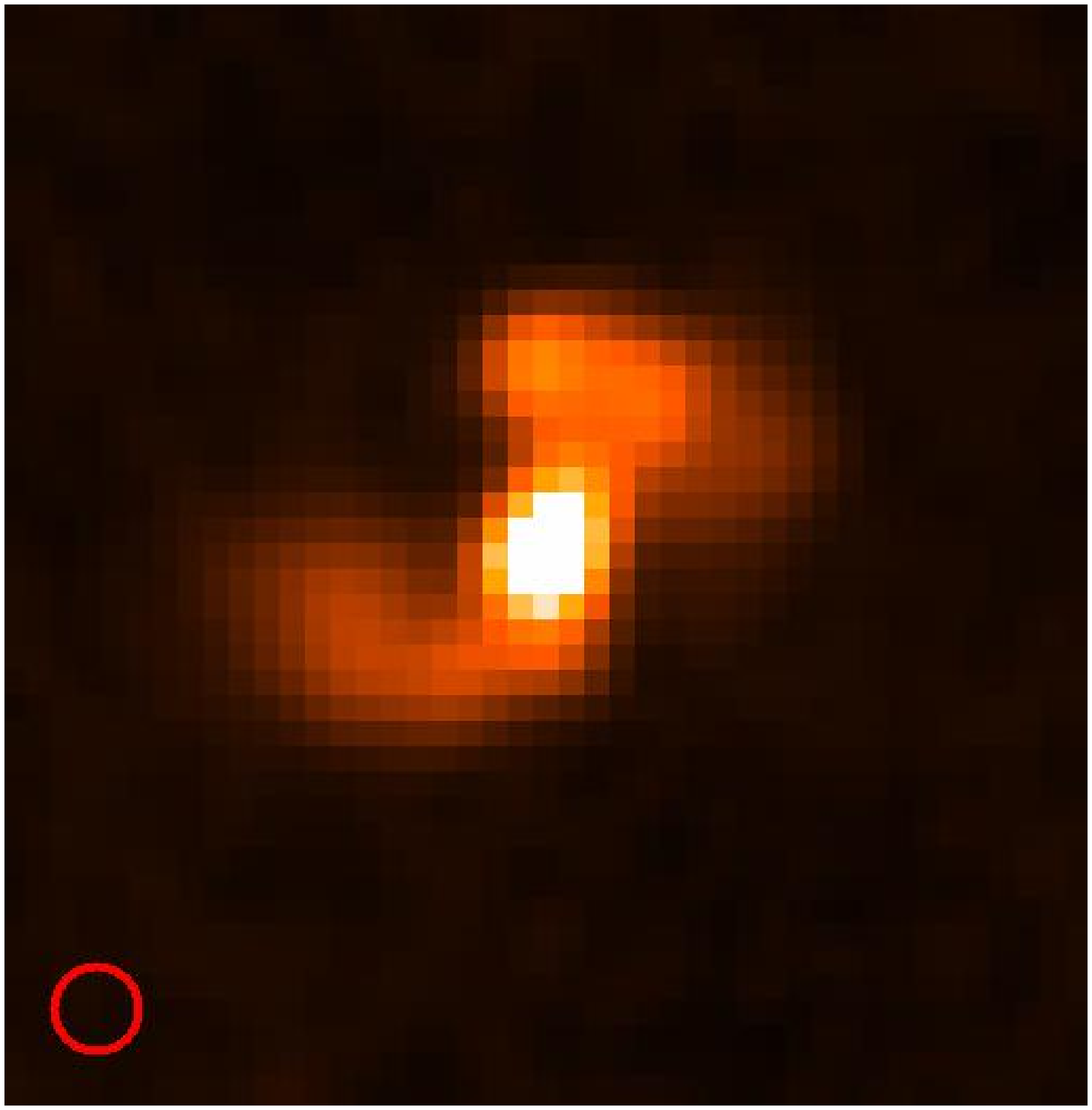}
}
\centerline{
\includegraphics[width=19mm]{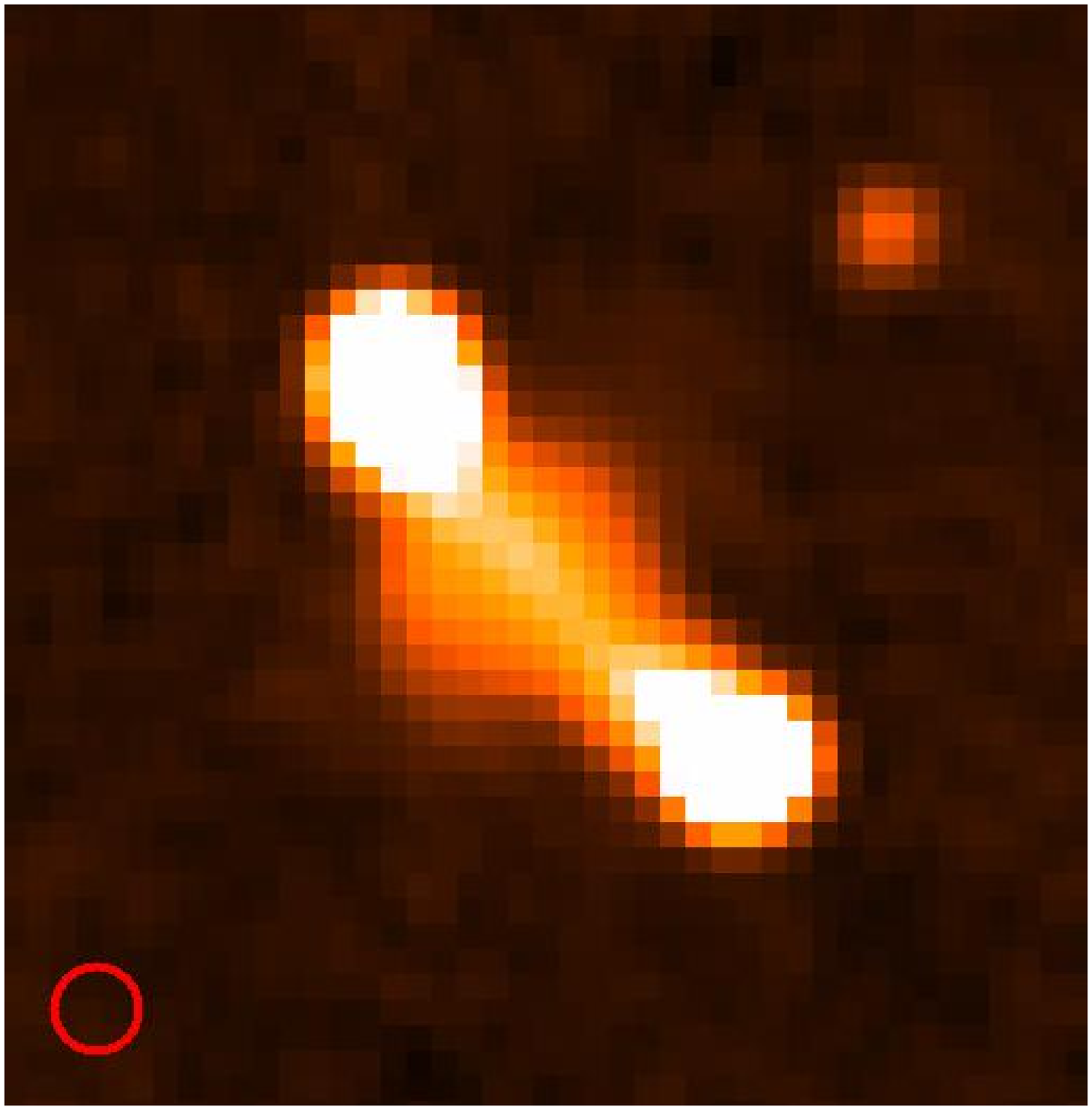}
\includegraphics[width=19mm]{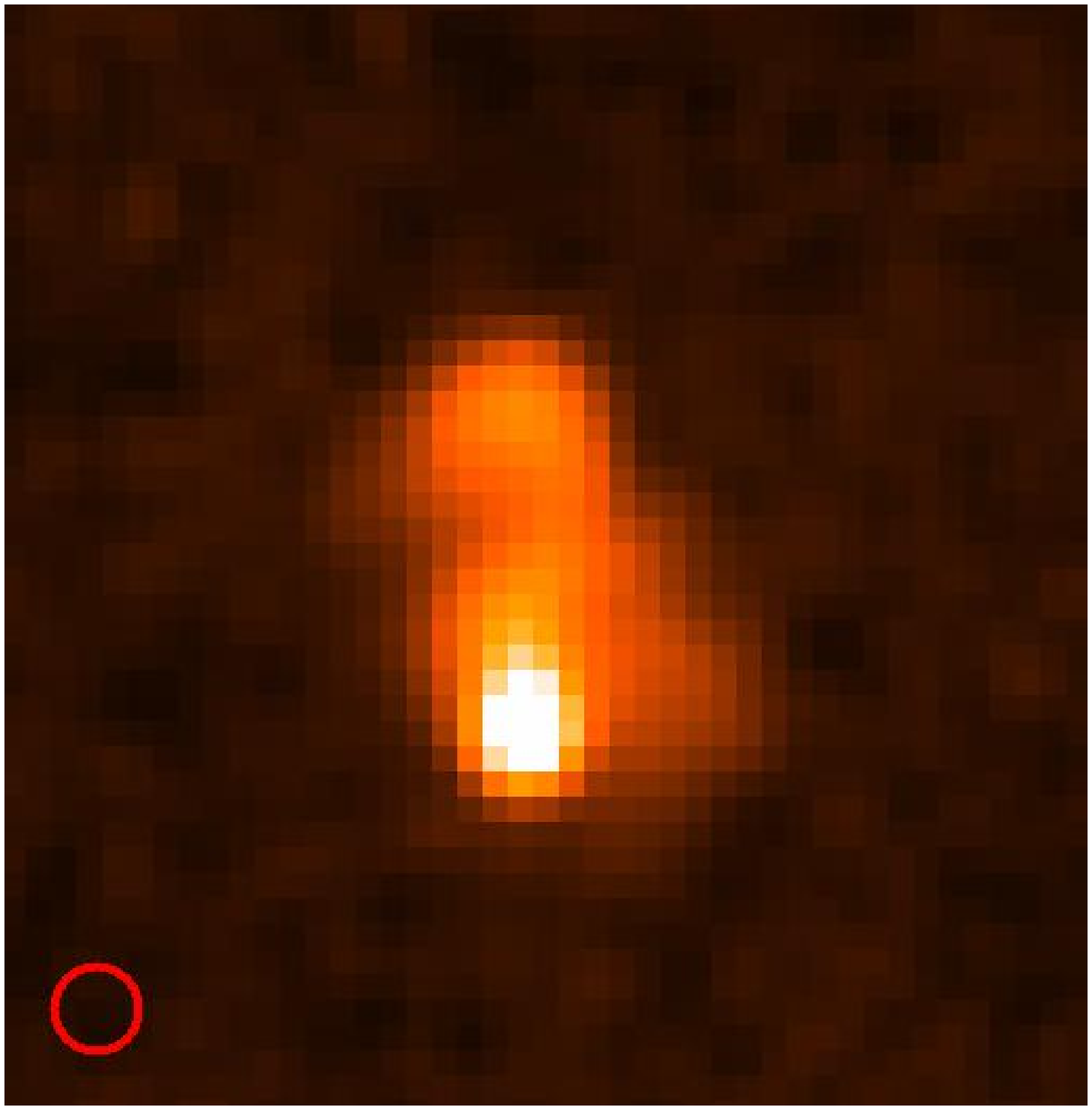}
\includegraphics[width=19mm]{radiowh.ps}
\includegraphics[width=19mm]{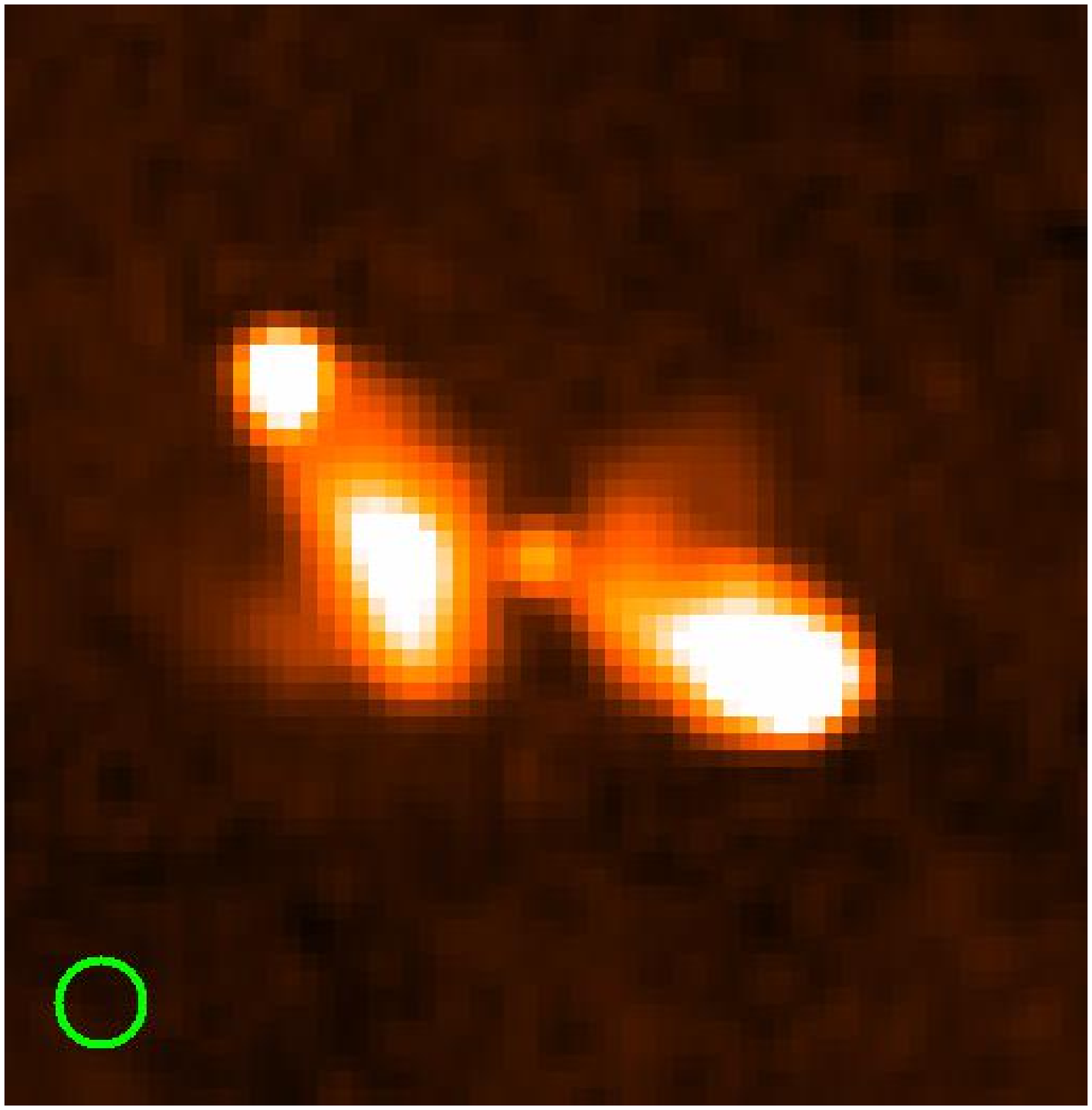}
\includegraphics[width=19mm]{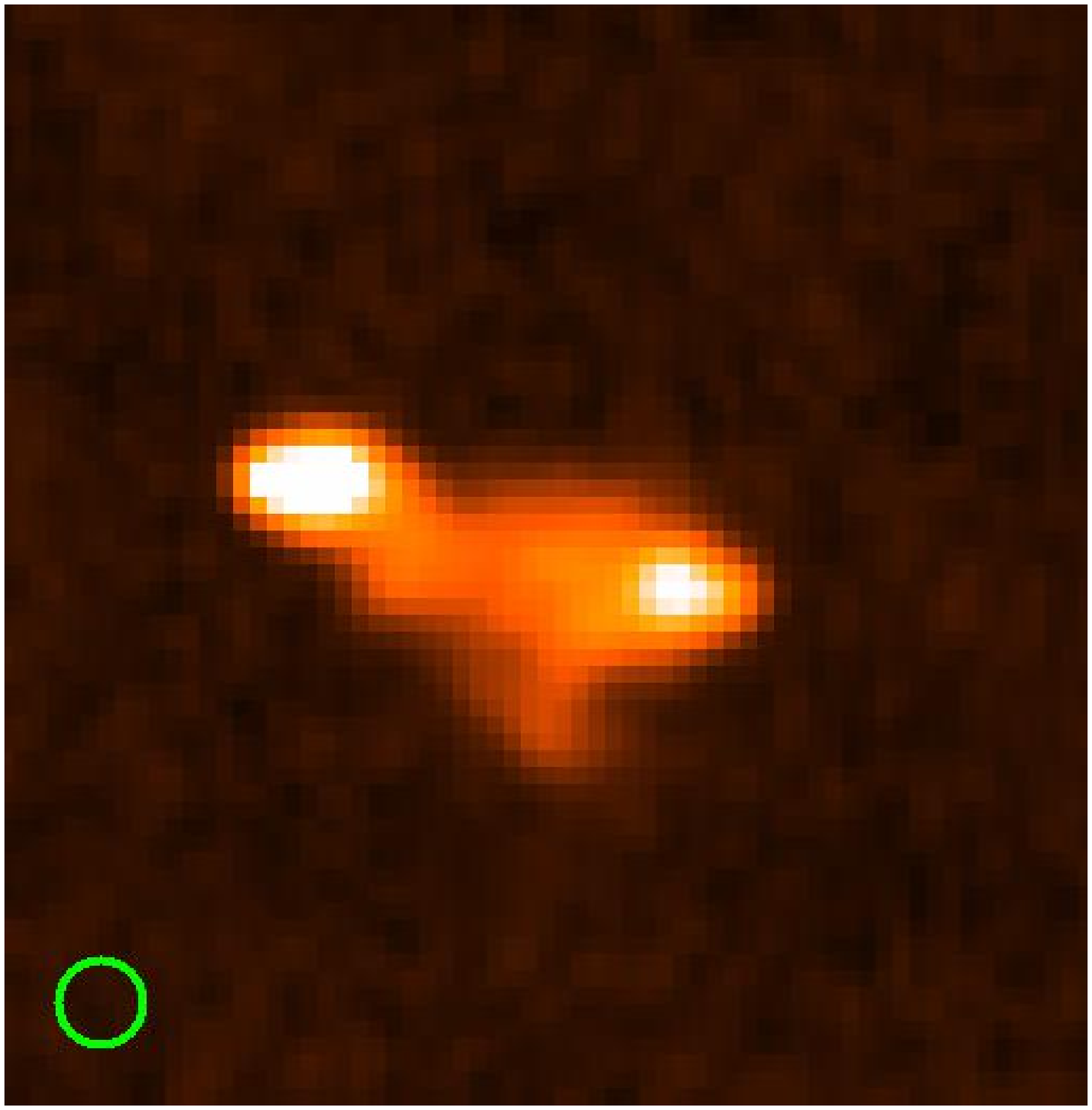}
\includegraphics[width=19mm]{radiowh.ps}
\includegraphics[width=19mm]{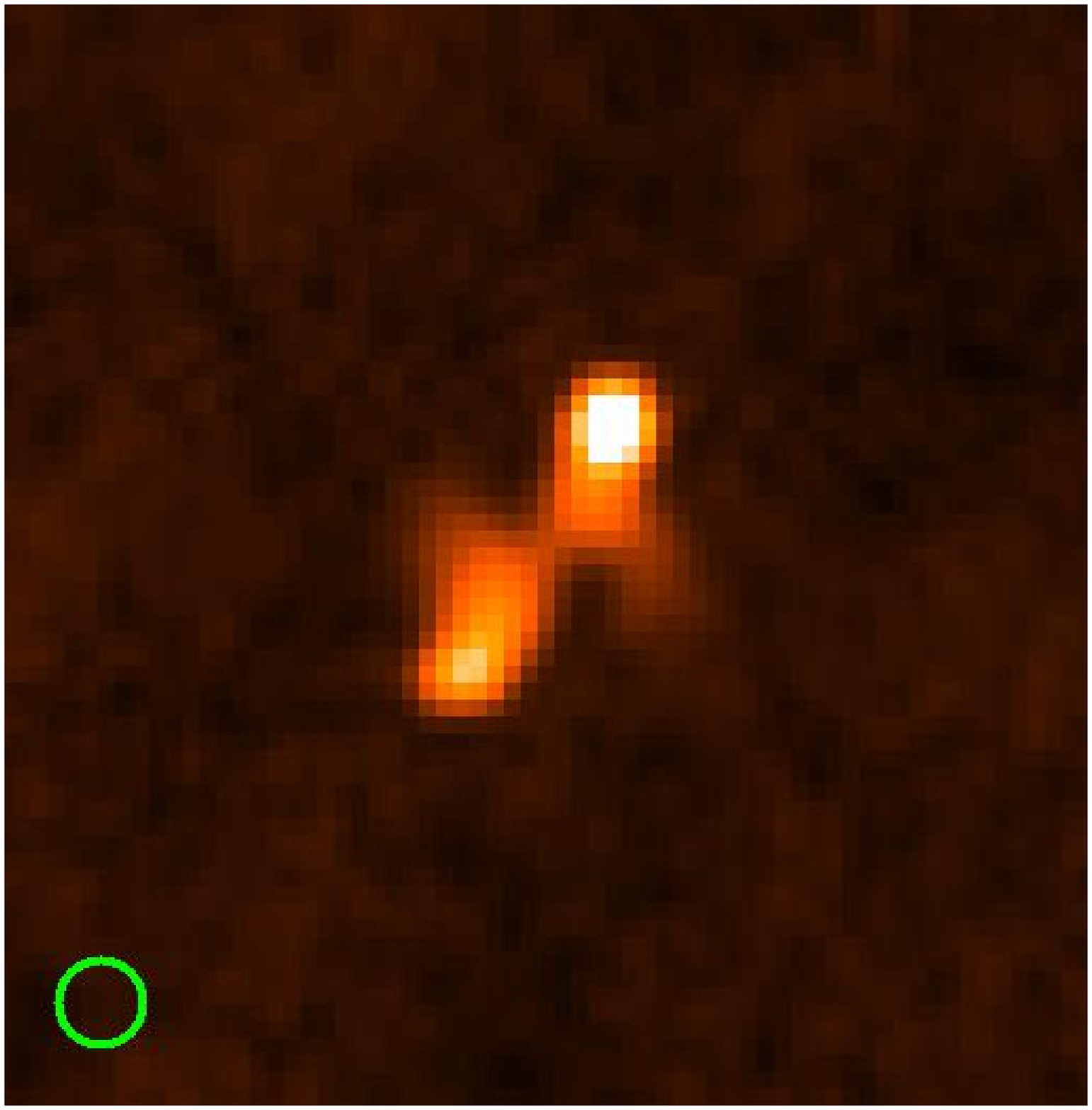}
\includegraphics[width=19mm]{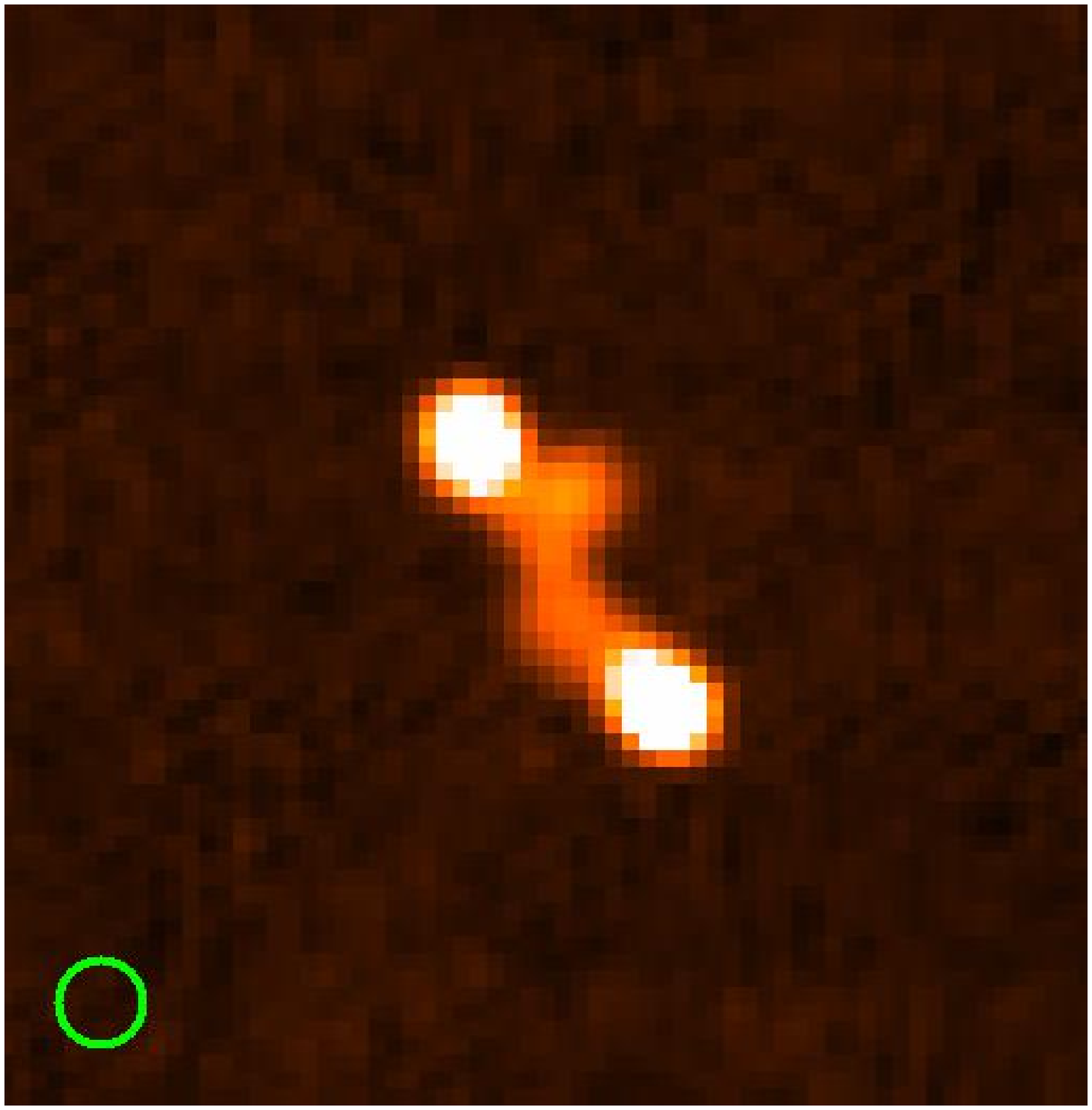}
\includegraphics[width=19mm]{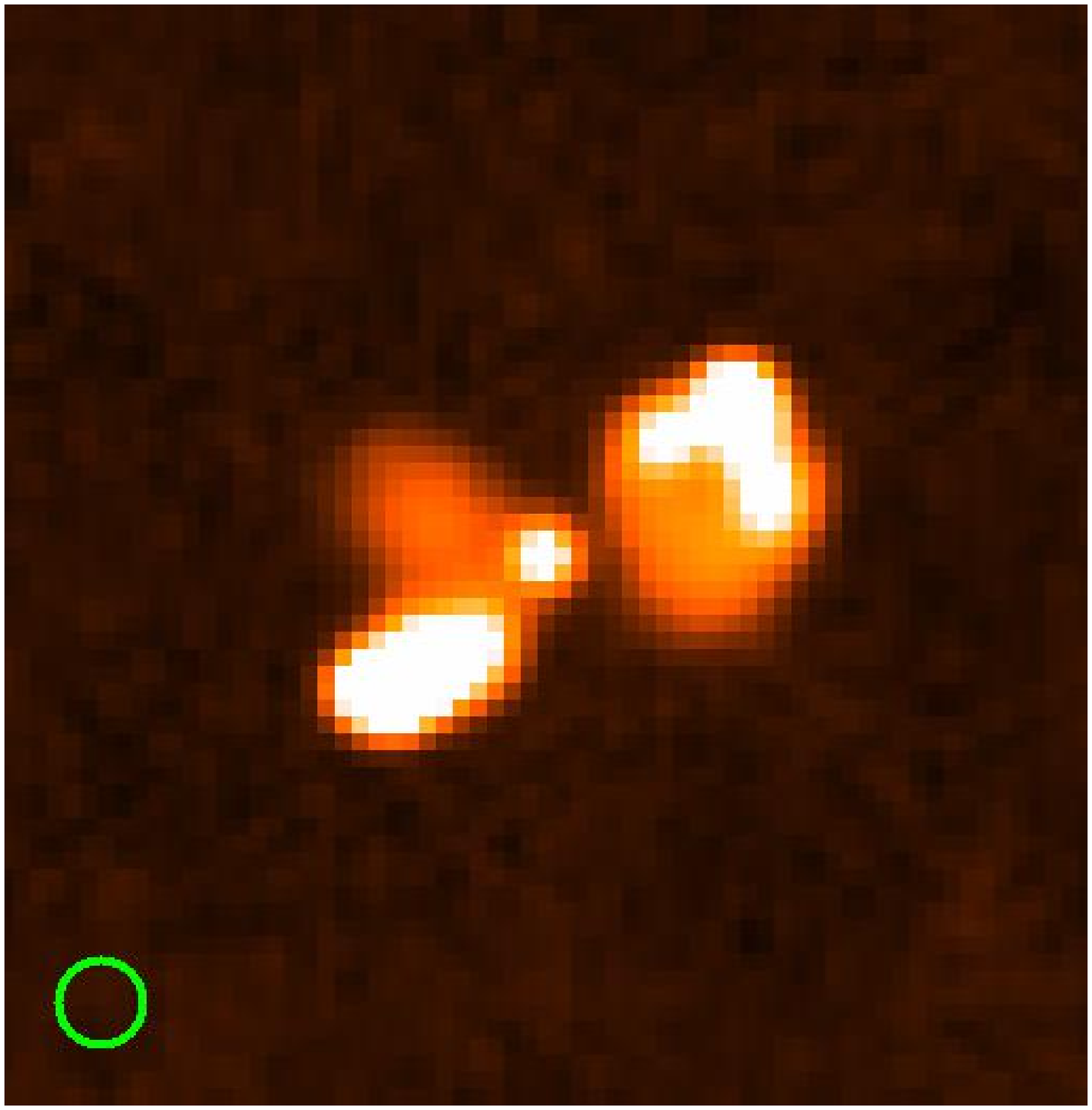}
\includegraphics[width=19mm]{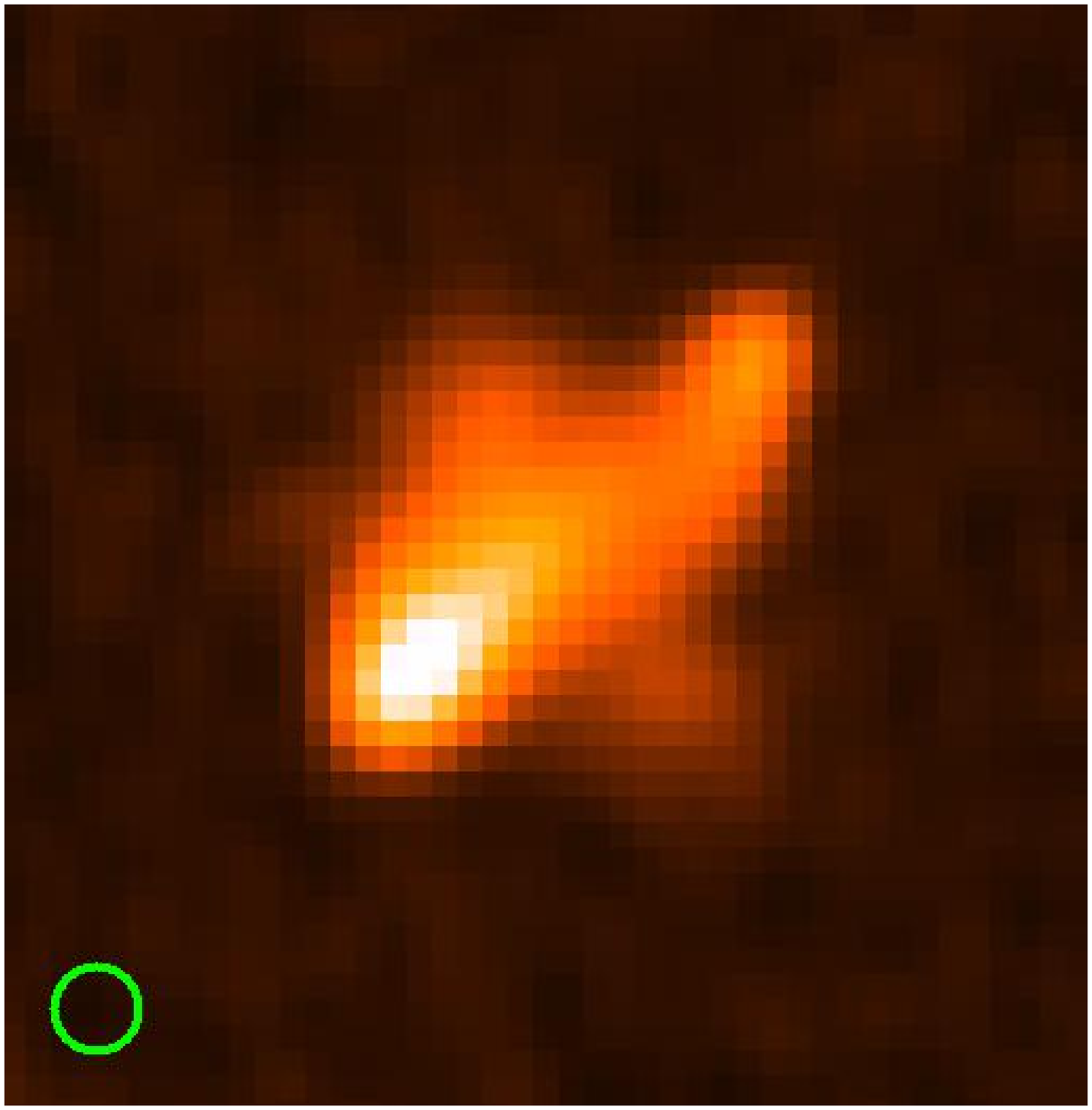}
}
\centerline{
\includegraphics[width=19mm]{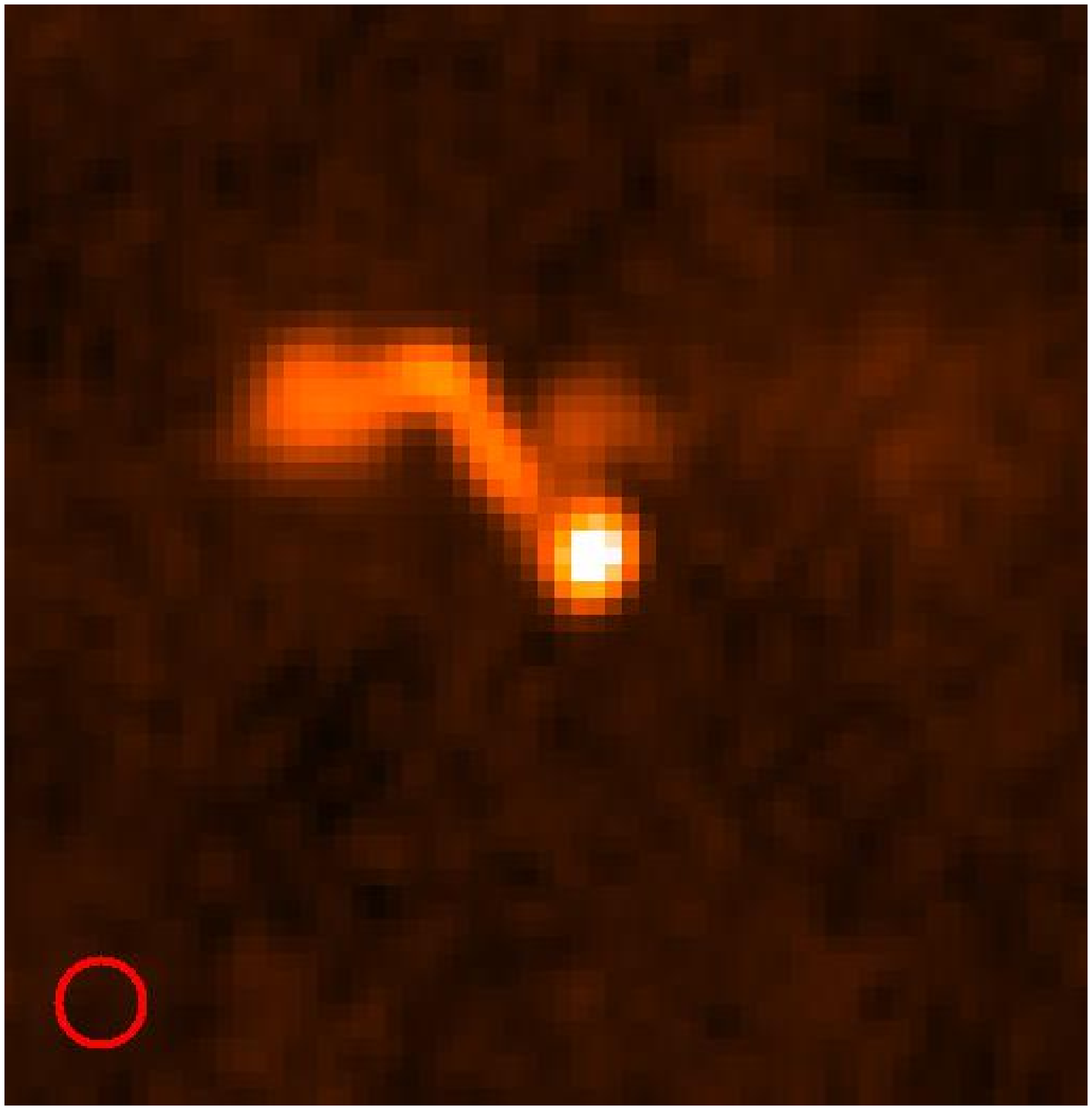}
\includegraphics[width=19mm]{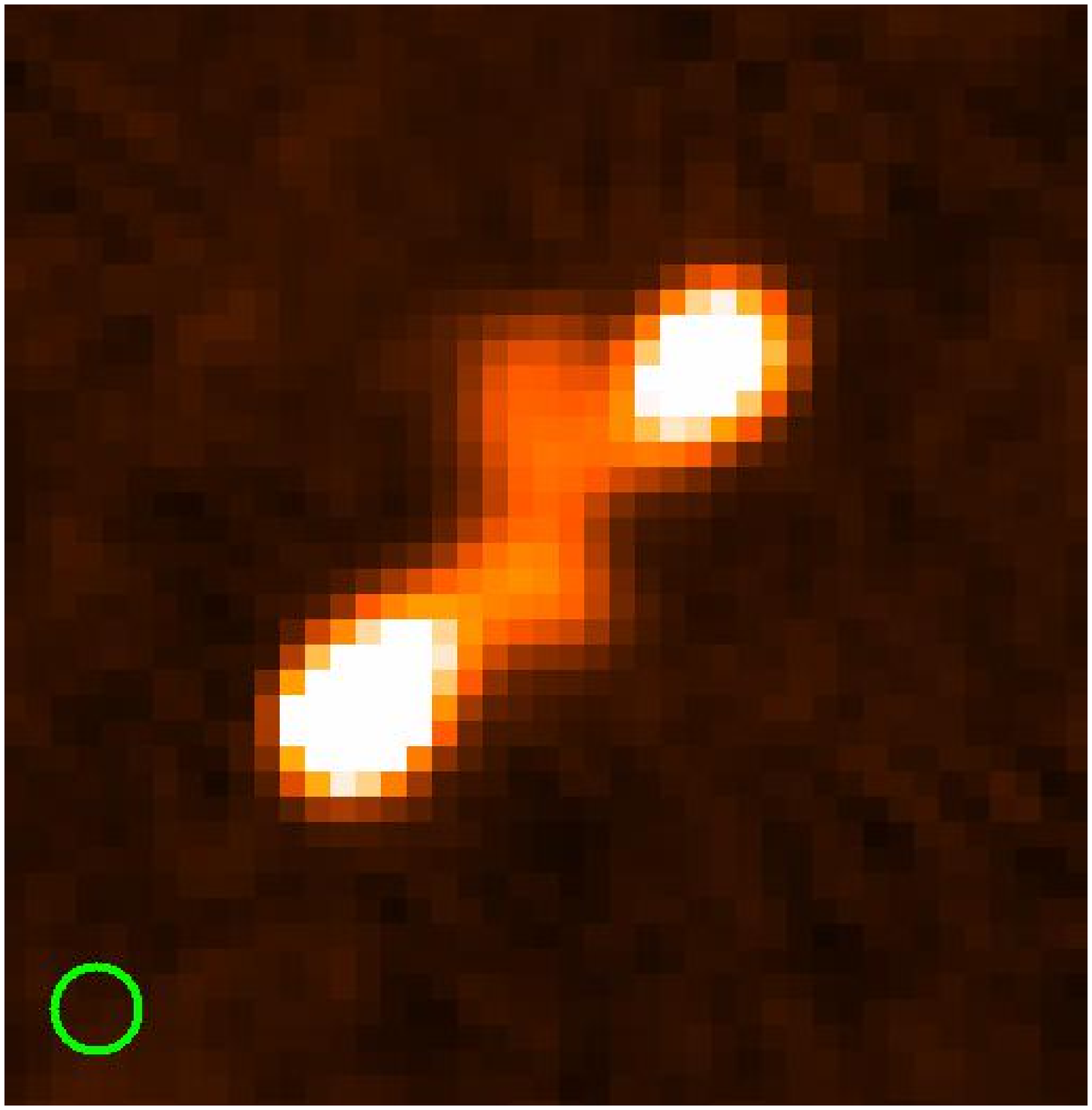}
\includegraphics[width=19mm]{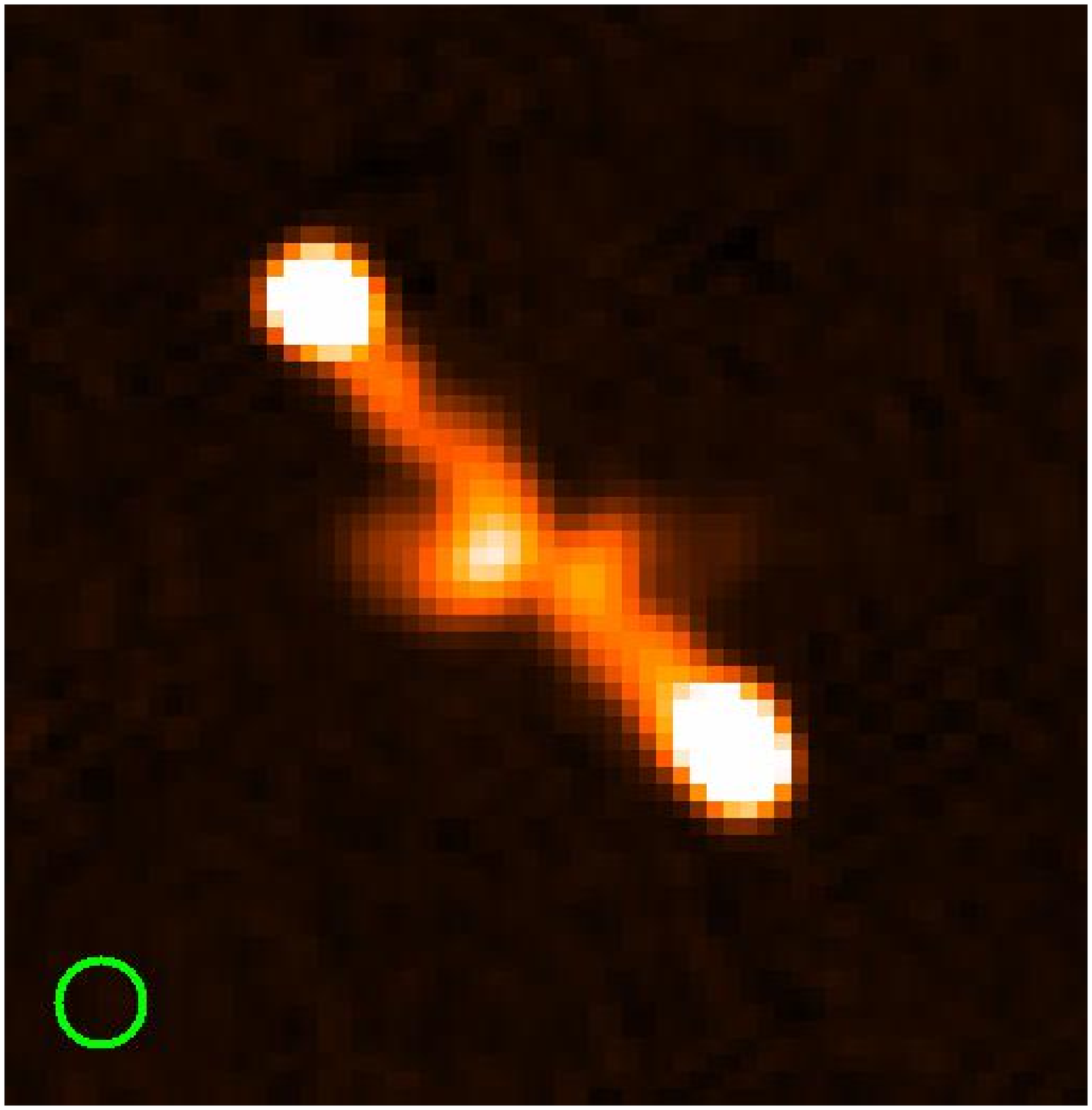}
\includegraphics[width=19mm]{radiowh.ps}
\includegraphics[width=19mm]{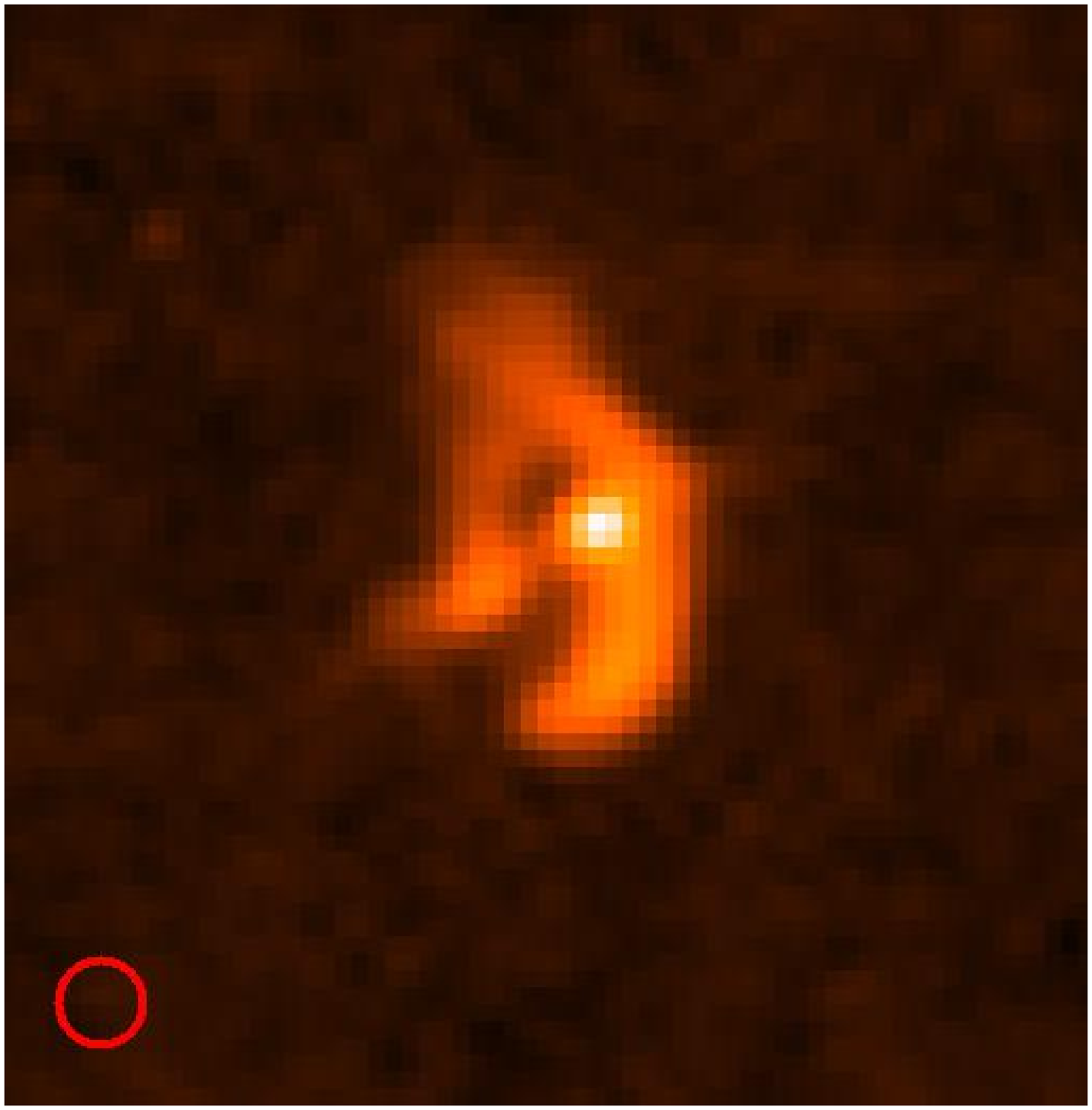}
\includegraphics[width=19mm]{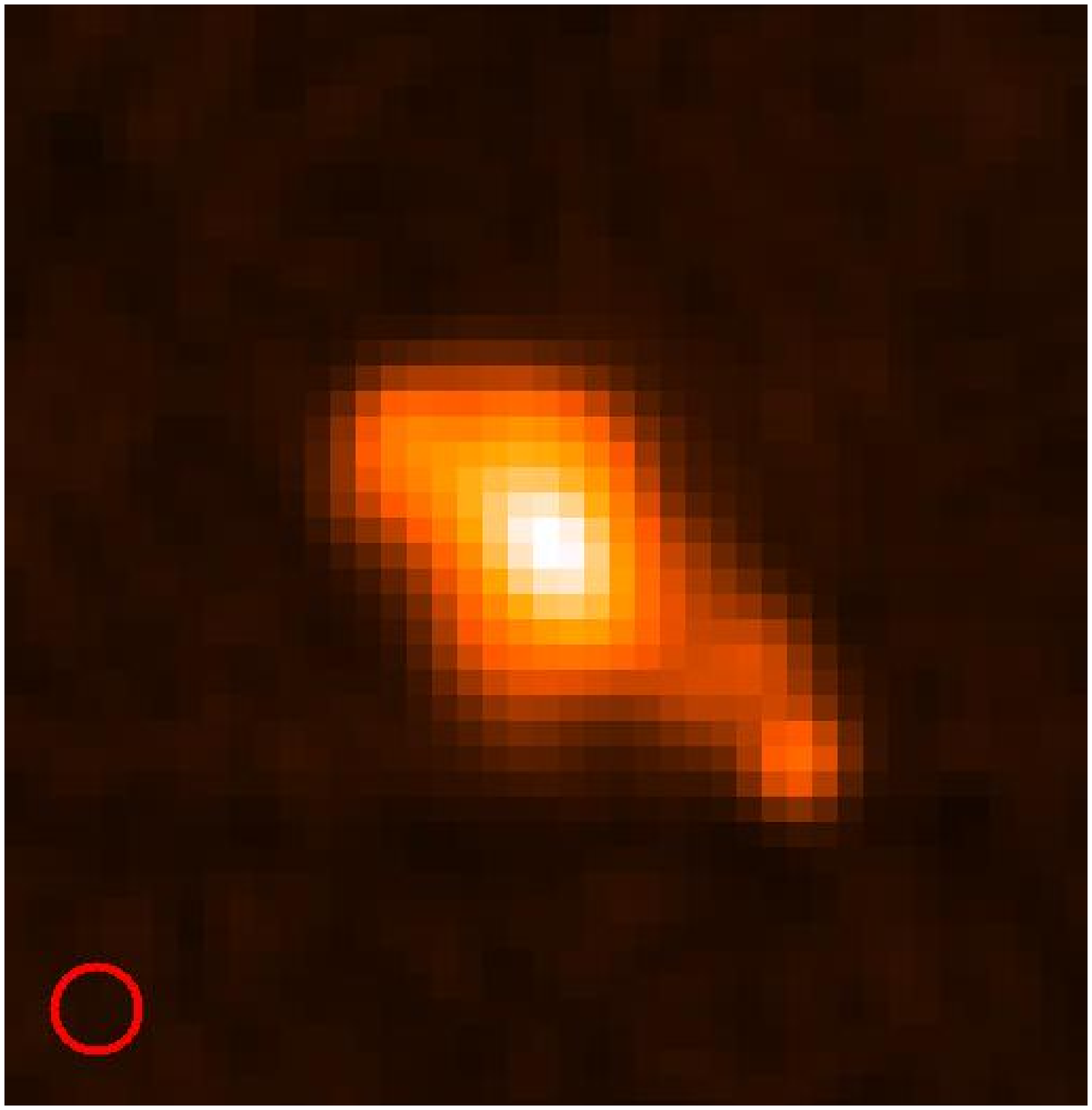}
\includegraphics[width=19mm]{radiowh.ps}
\includegraphics[width=19mm]{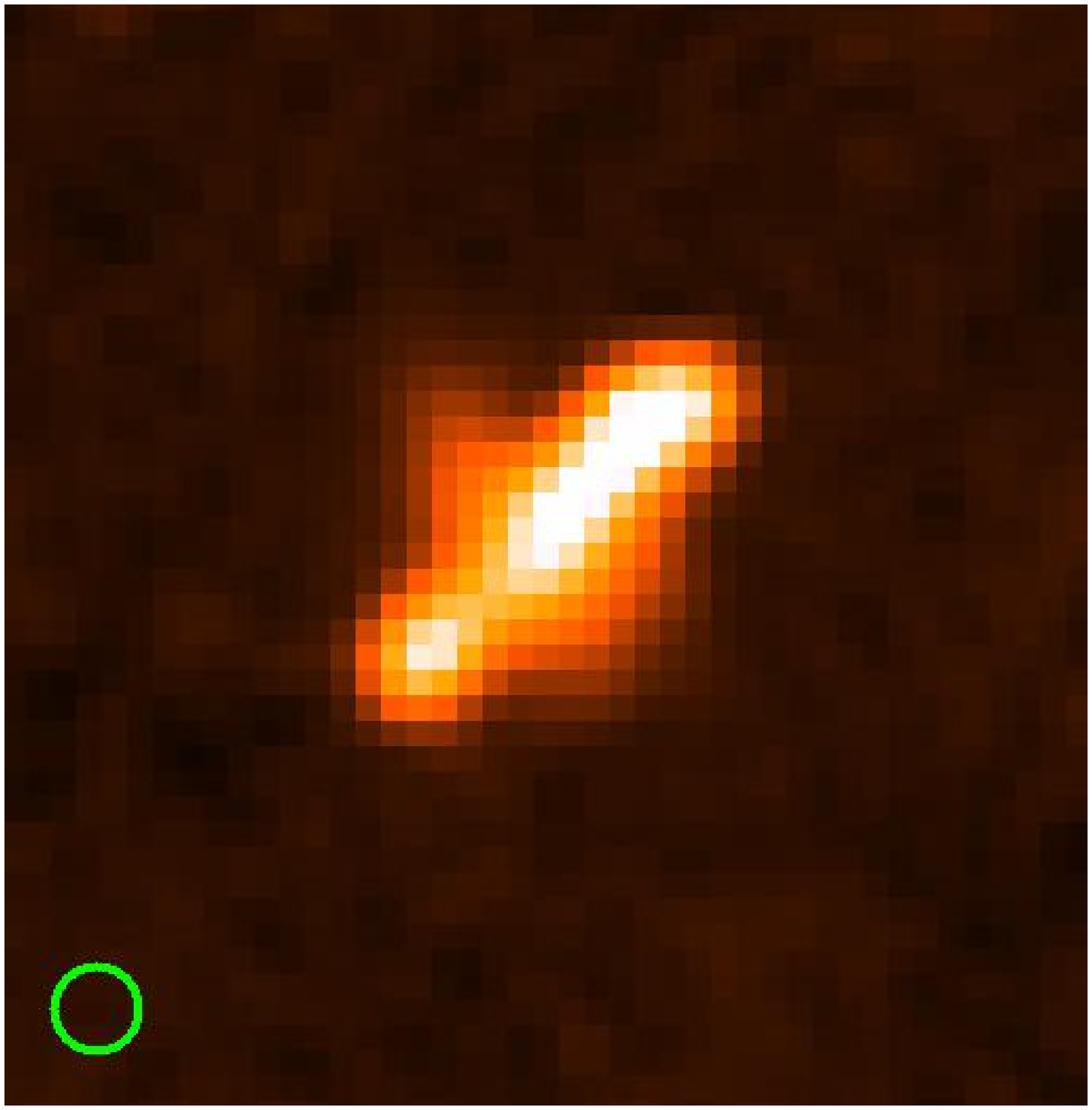}
\includegraphics[width=19mm]{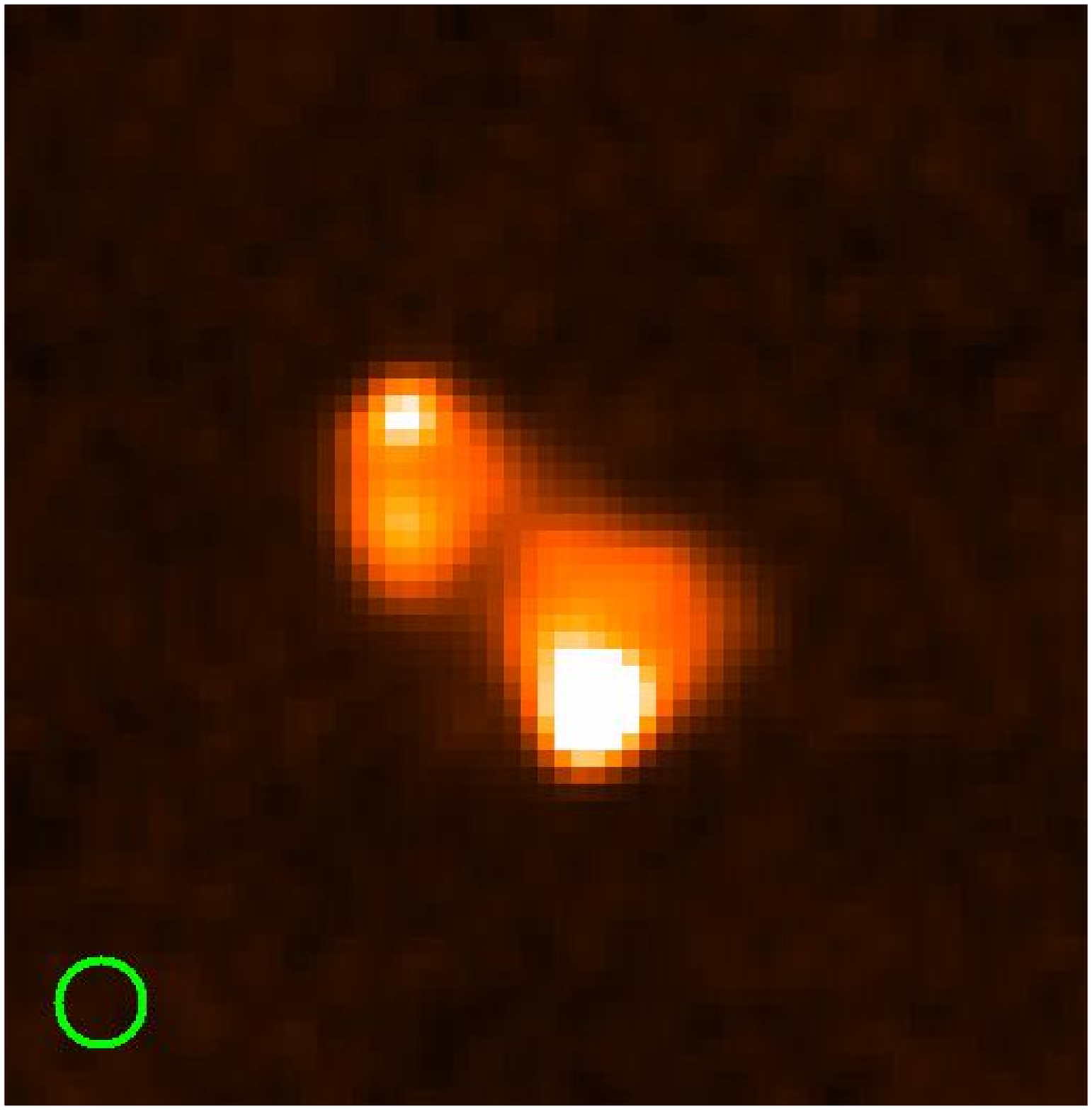}
\includegraphics[width=19mm]{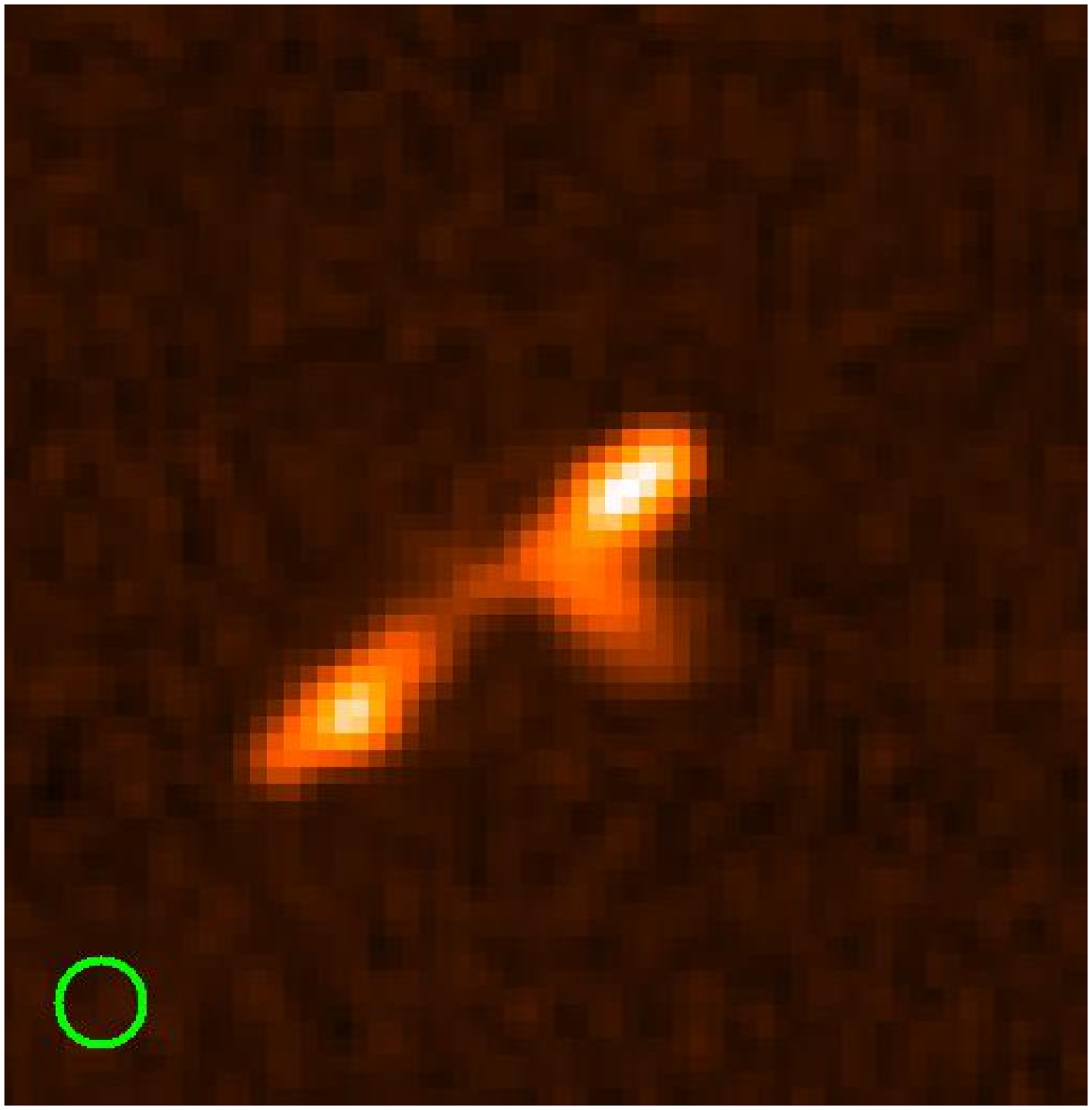}
}
\centerline{
\includegraphics[width=19mm]{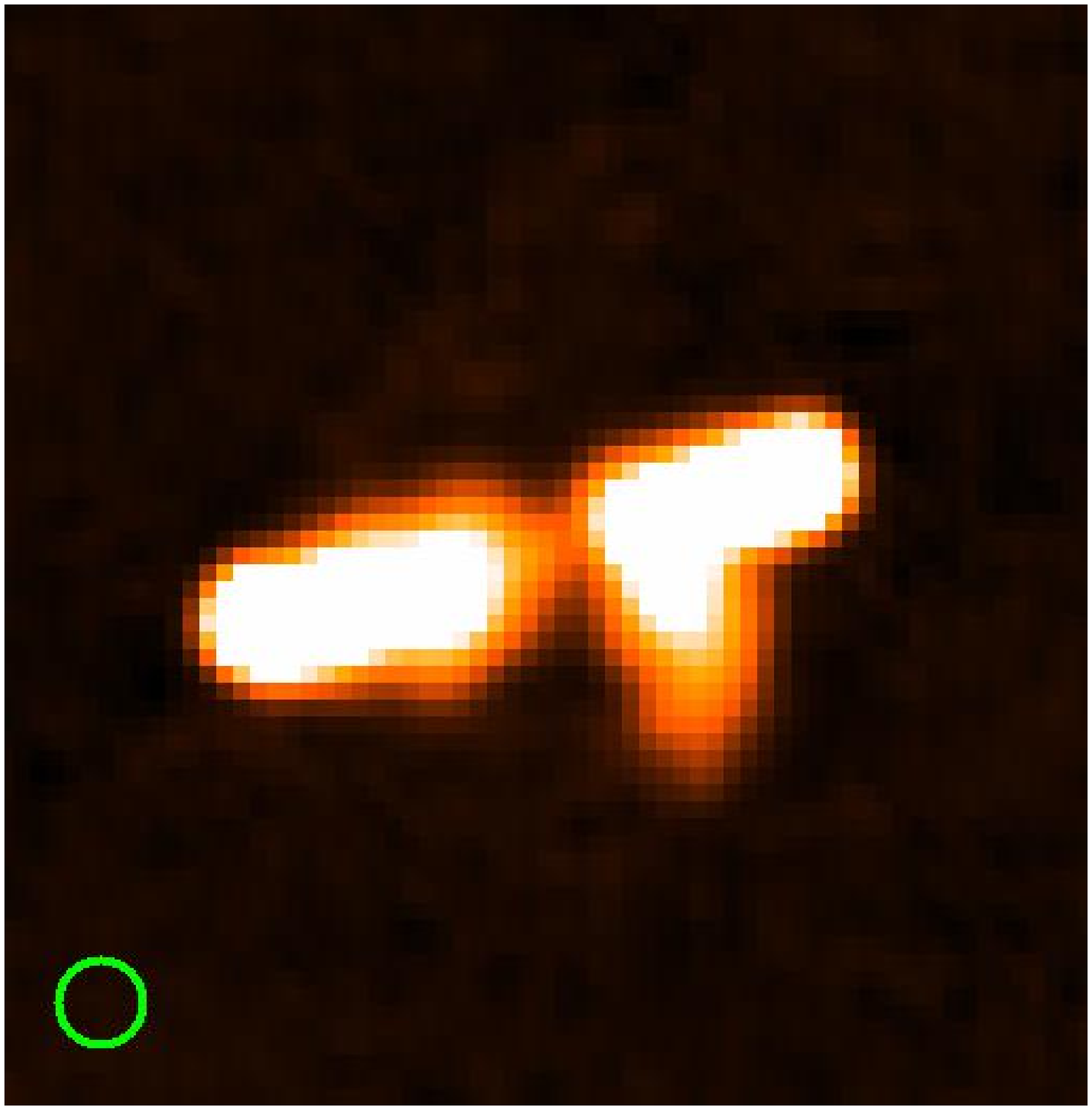}
\includegraphics[width=19mm]{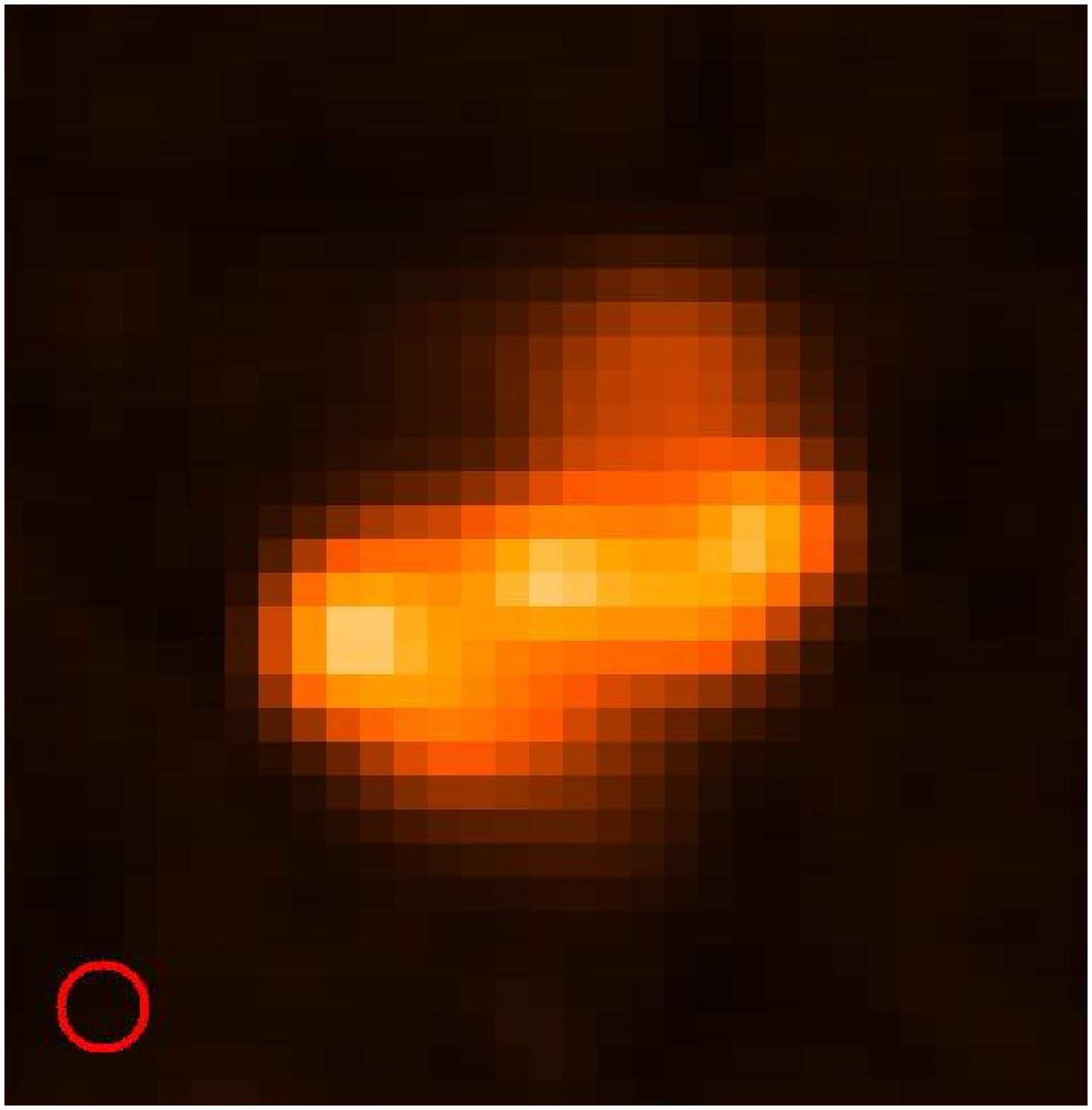}
\includegraphics[width=19mm]{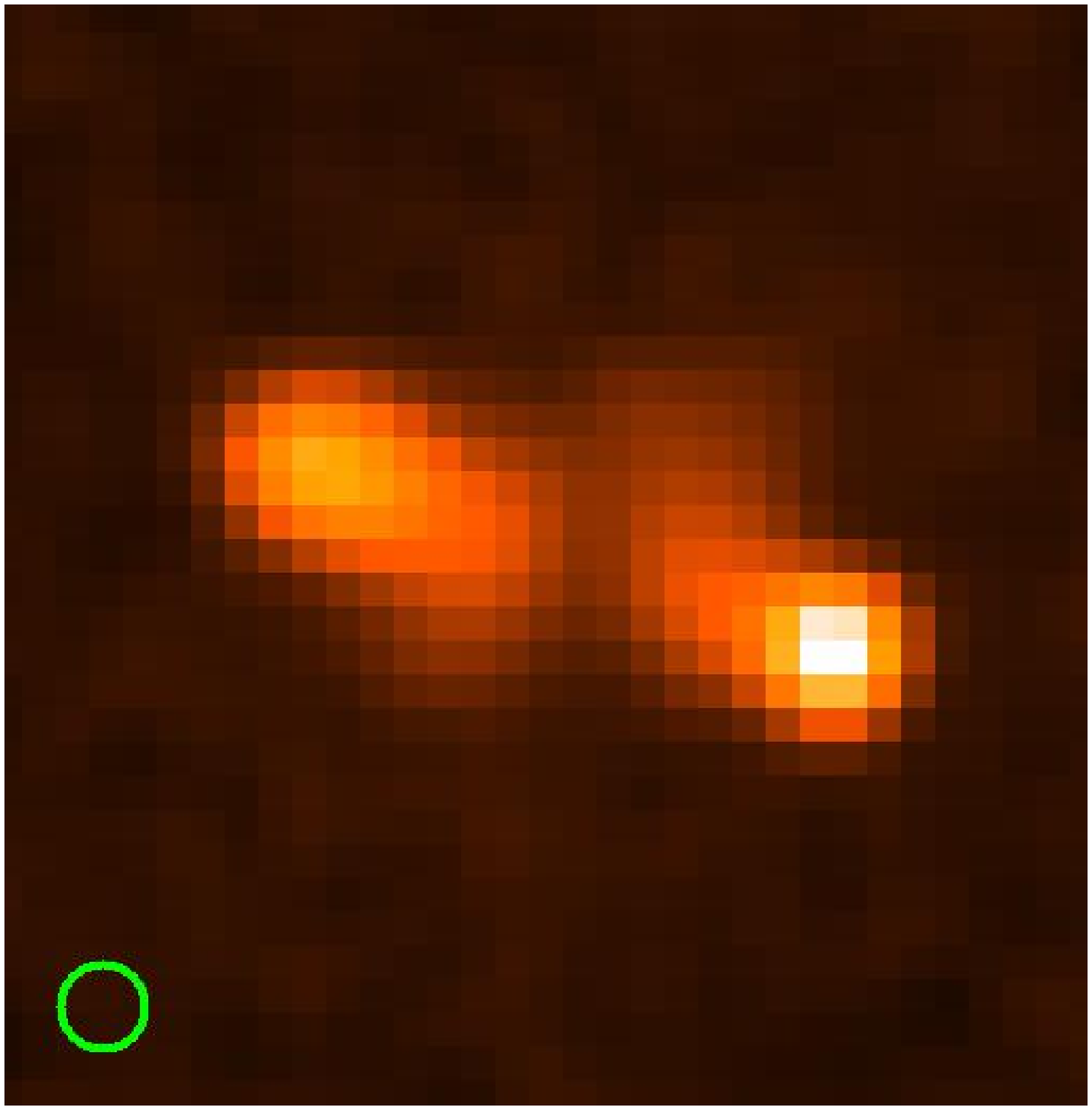}
\includegraphics[width=19mm]{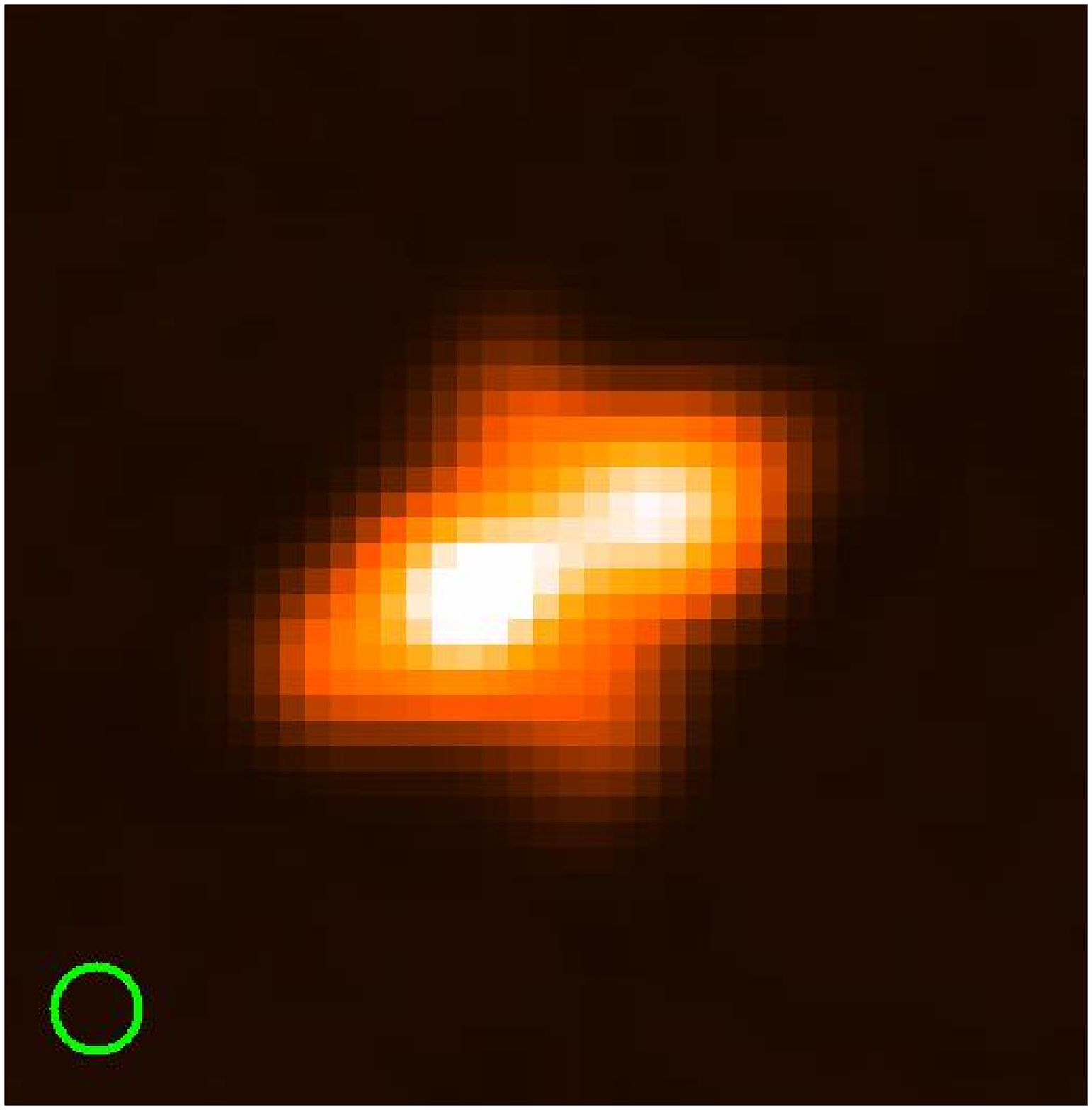}
\includegraphics[width=19mm]{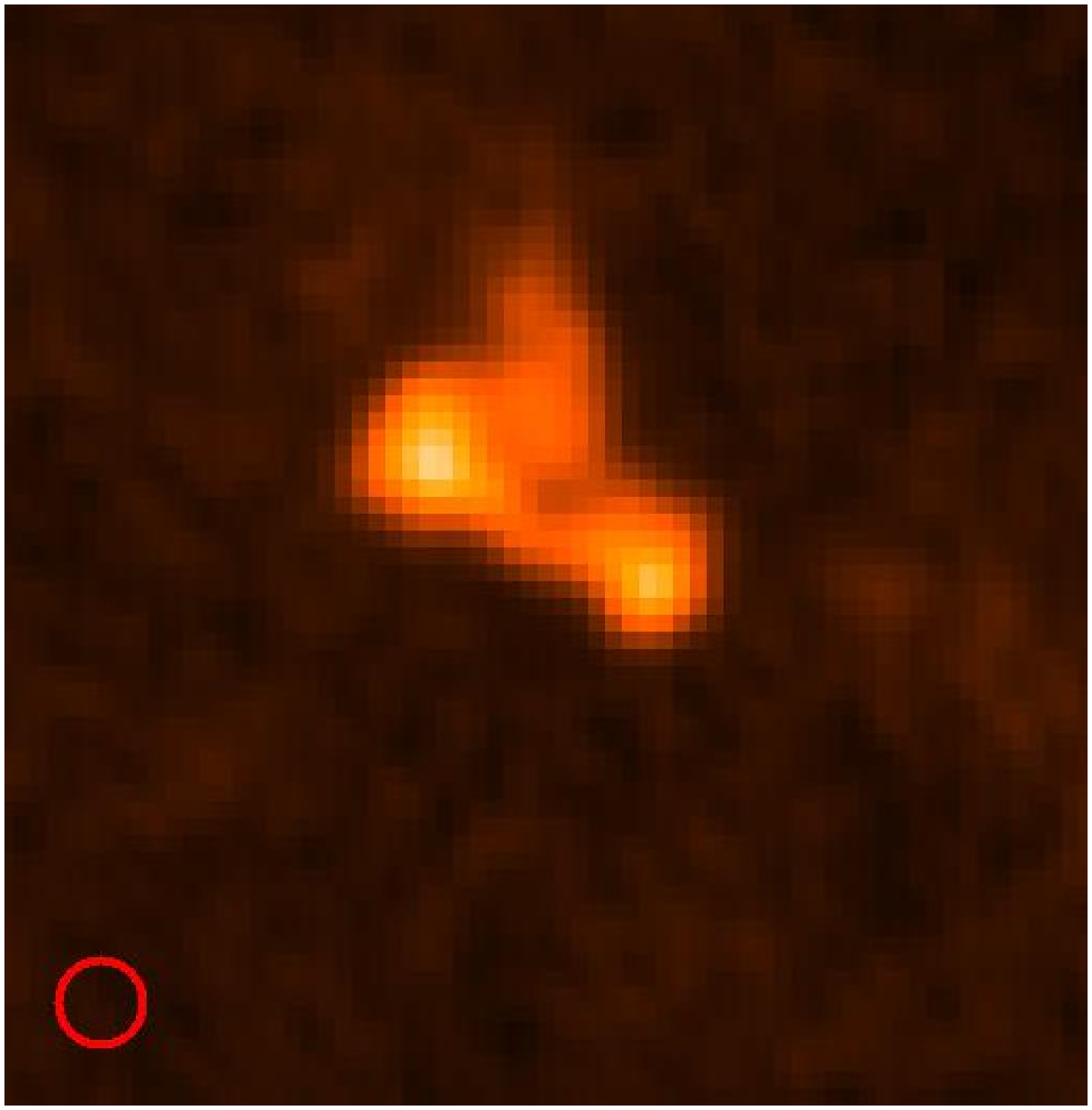}
\includegraphics[width=19mm]{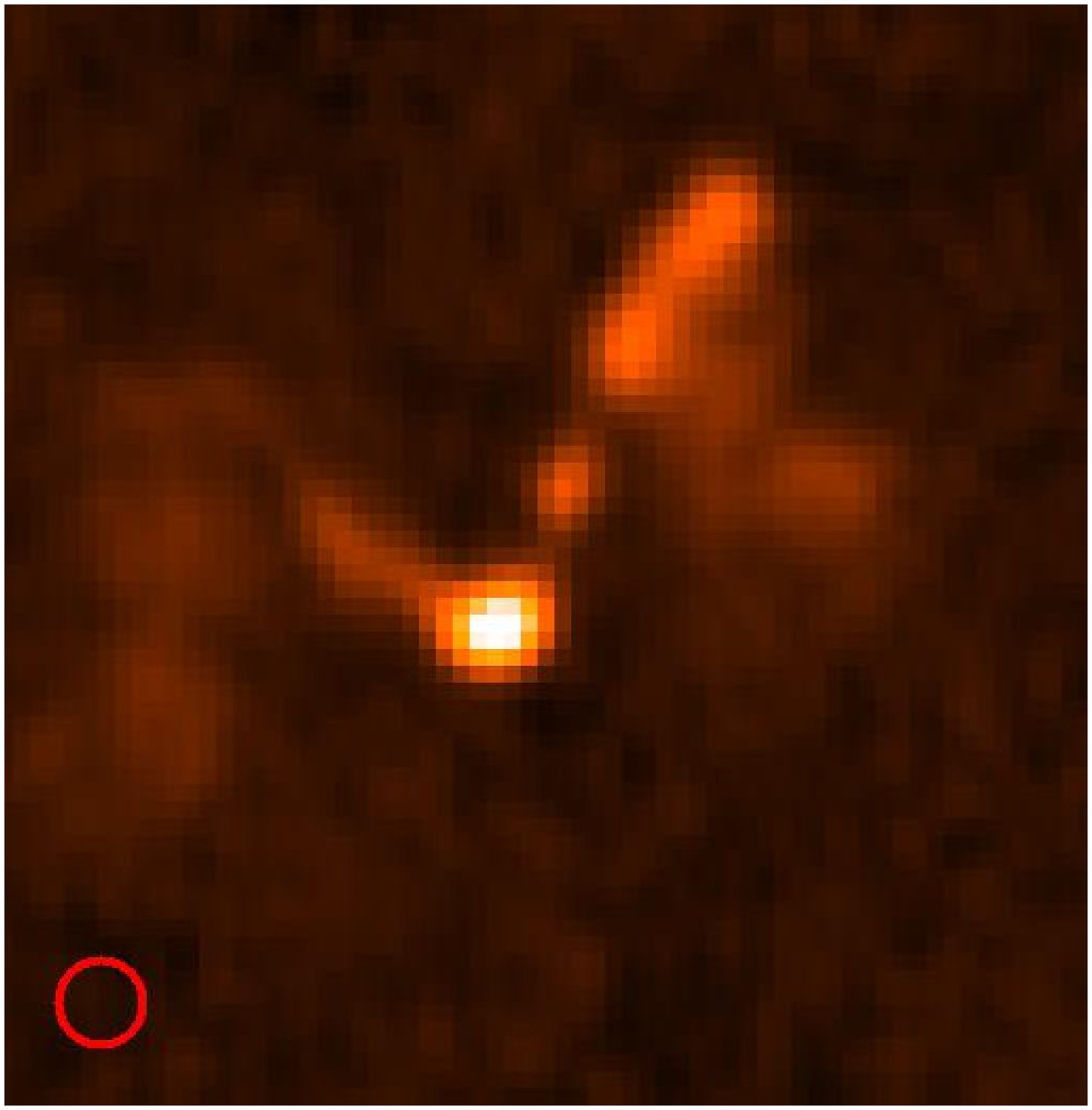}
\includegraphics[width=19mm]{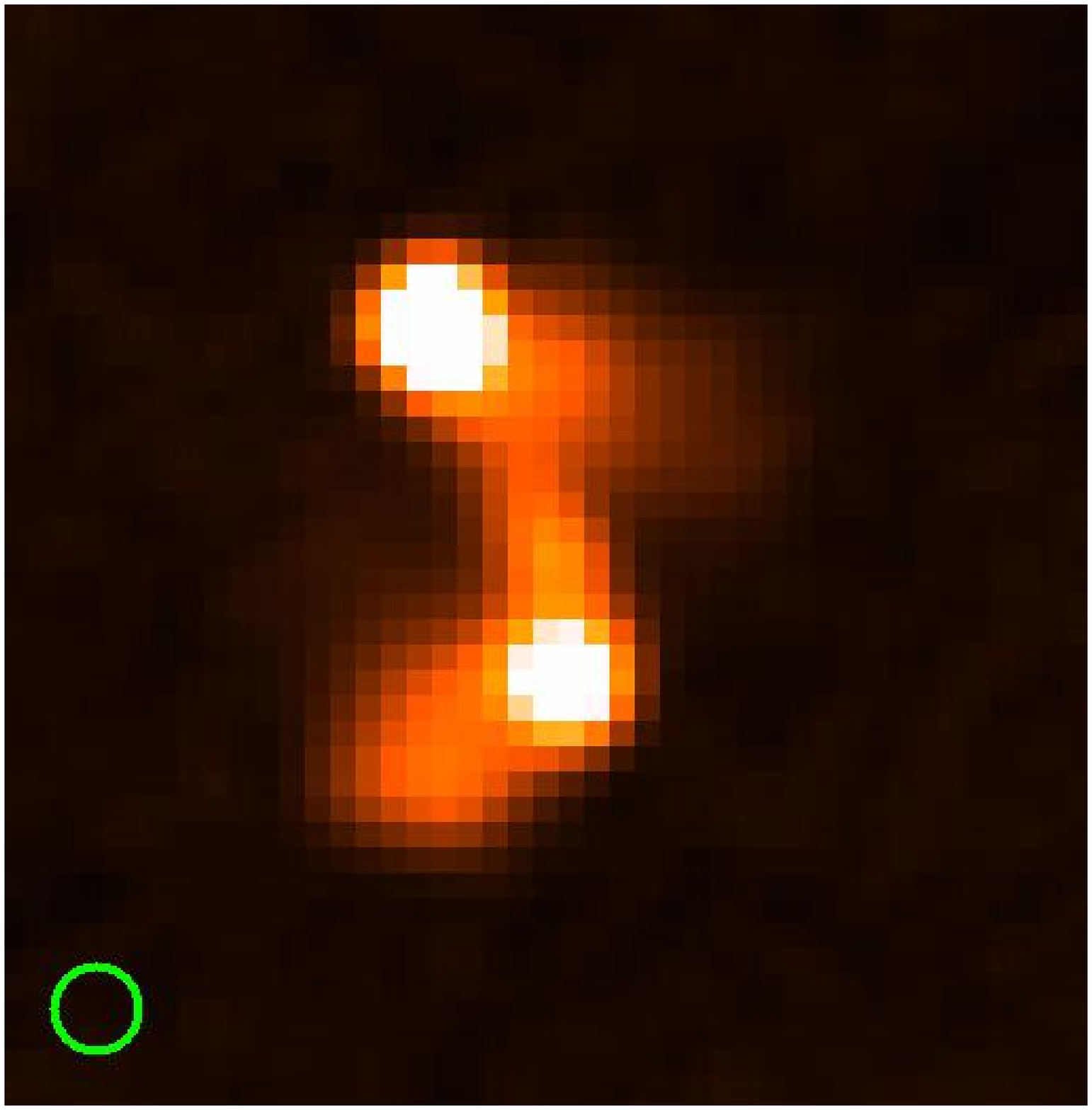}
\includegraphics[width=19mm]{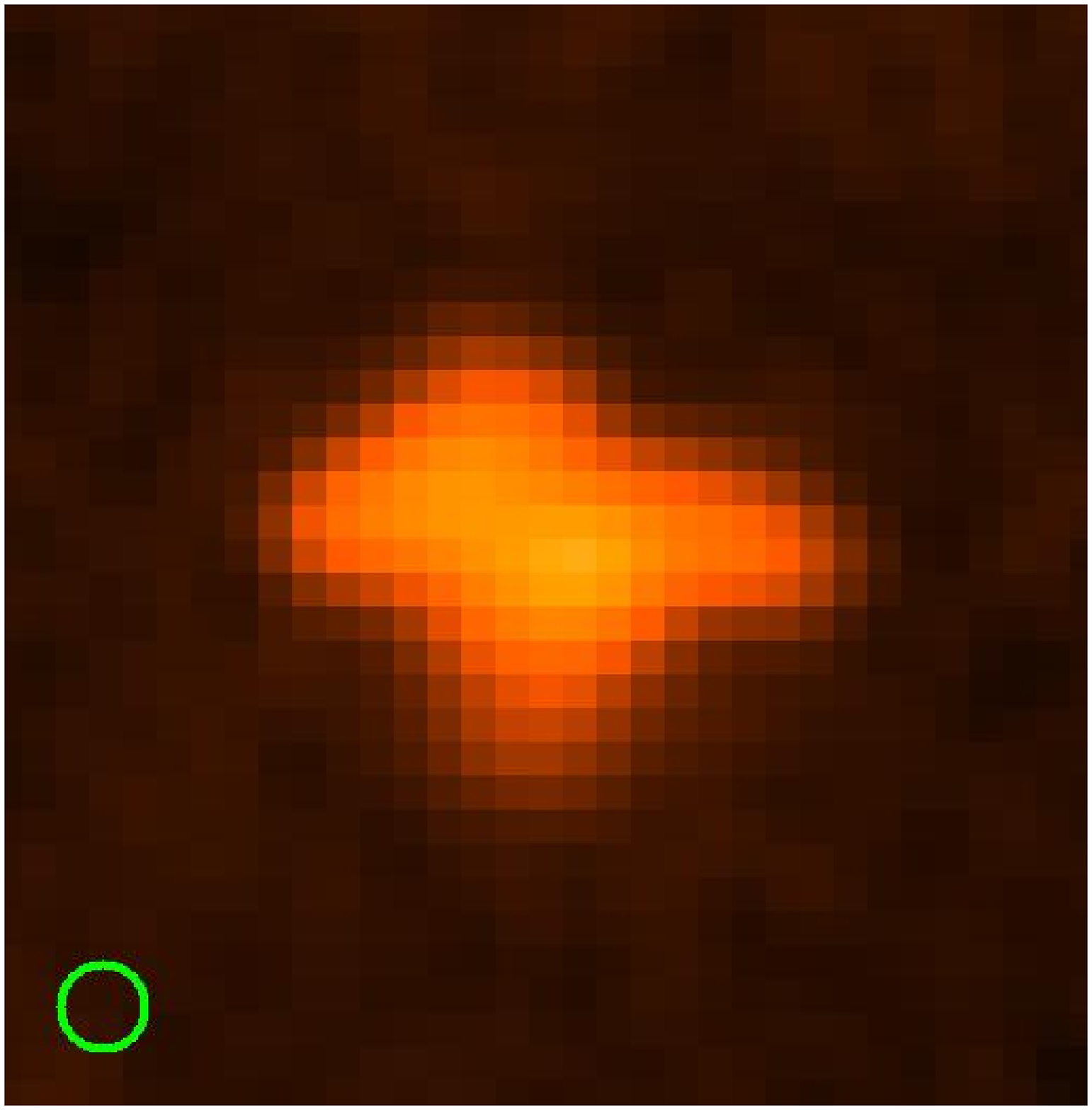}
\includegraphics[width=19mm]{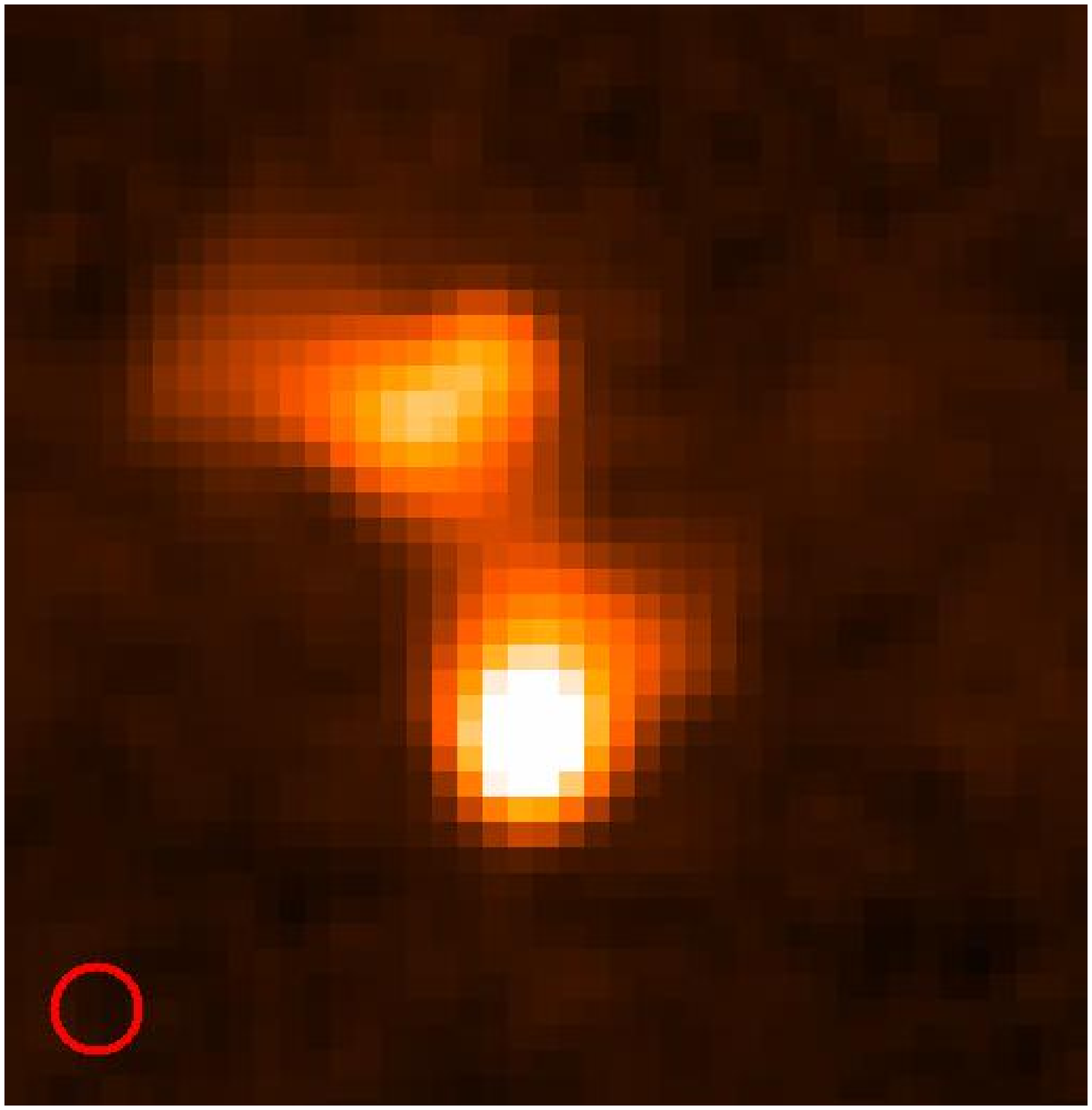}
\includegraphics[width=19mm]{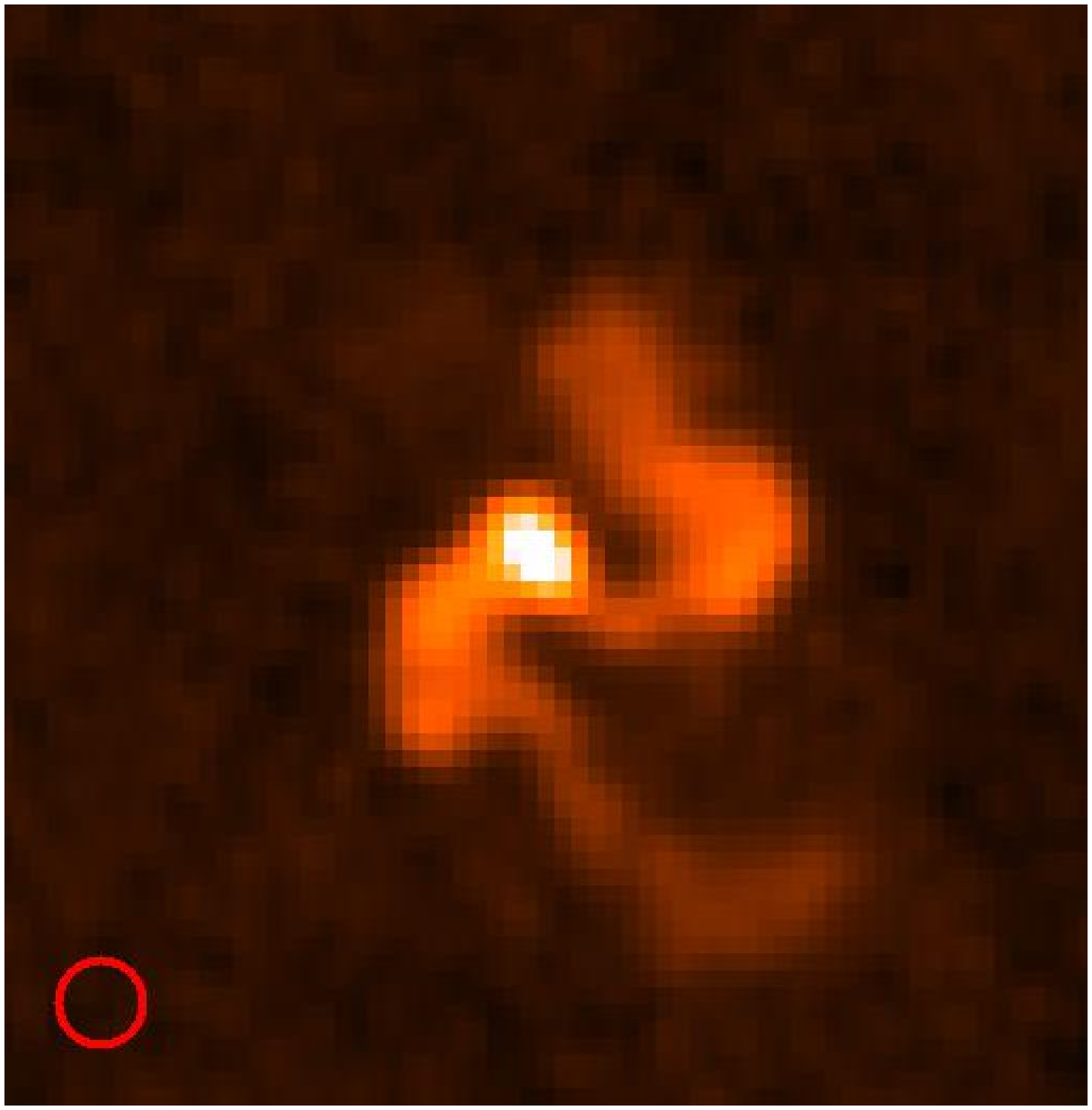}
}
\centerline{
\includegraphics[width=19mm]{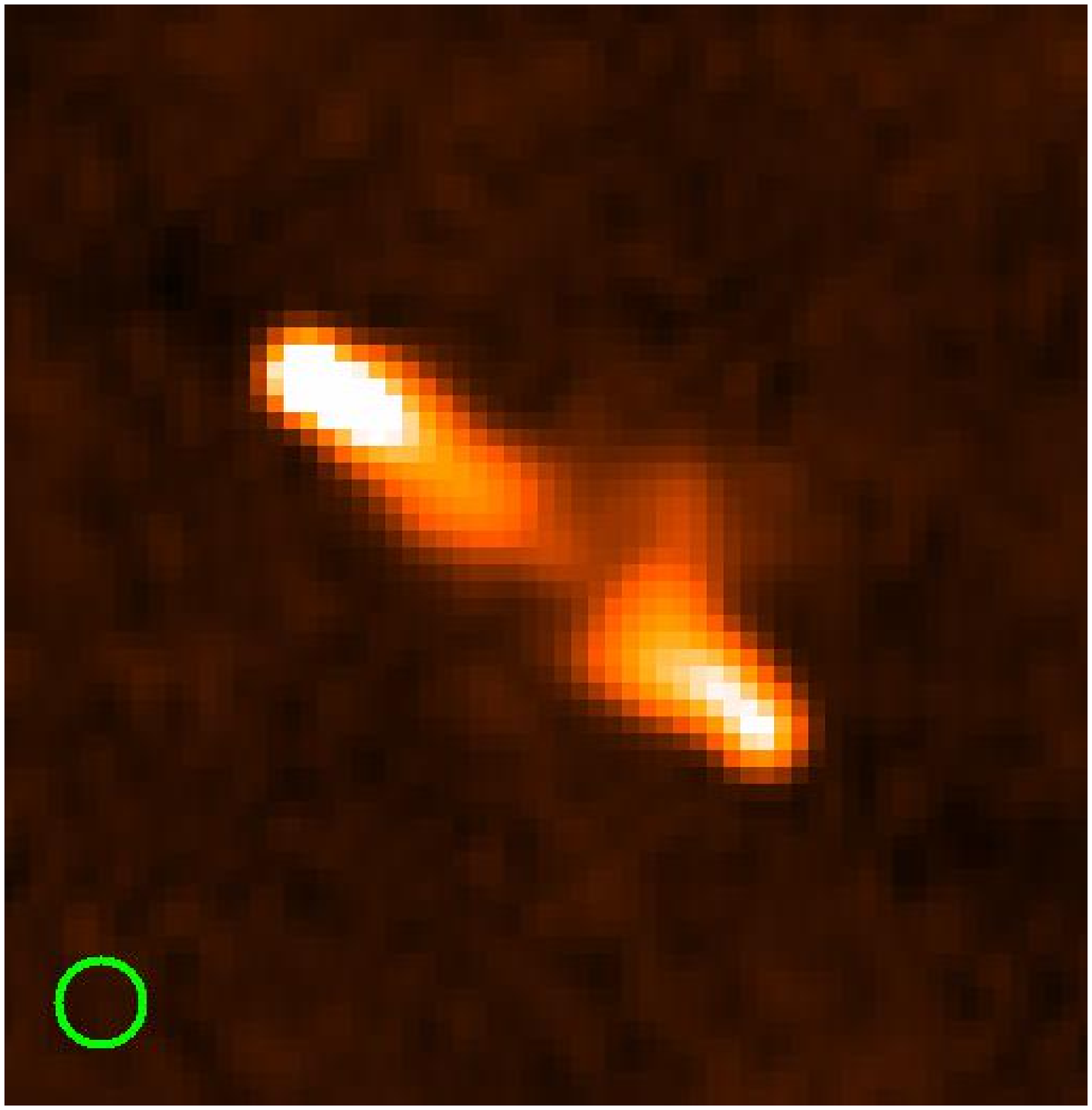}
\includegraphics[width=19mm]{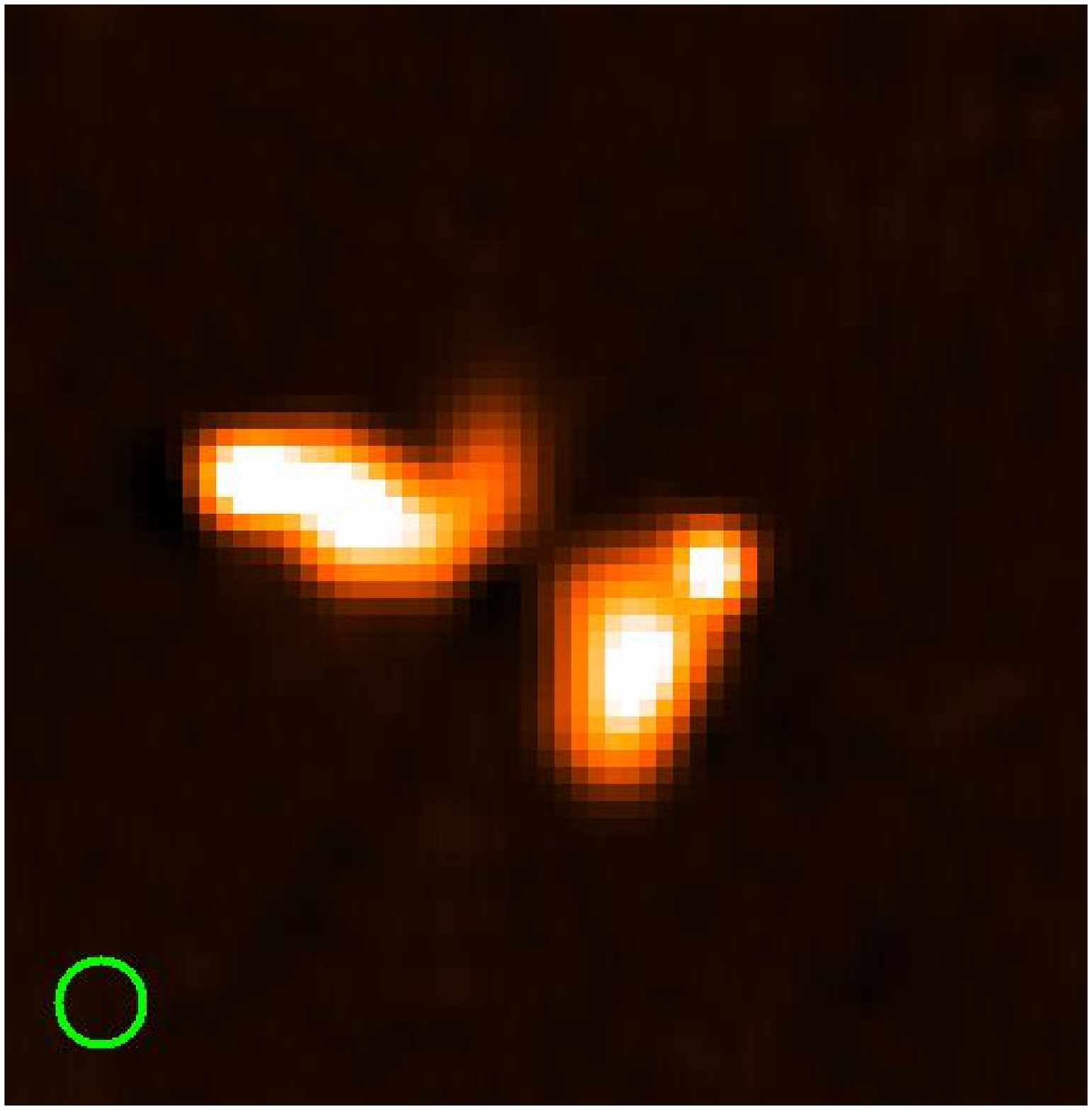}
\includegraphics[width=19mm]{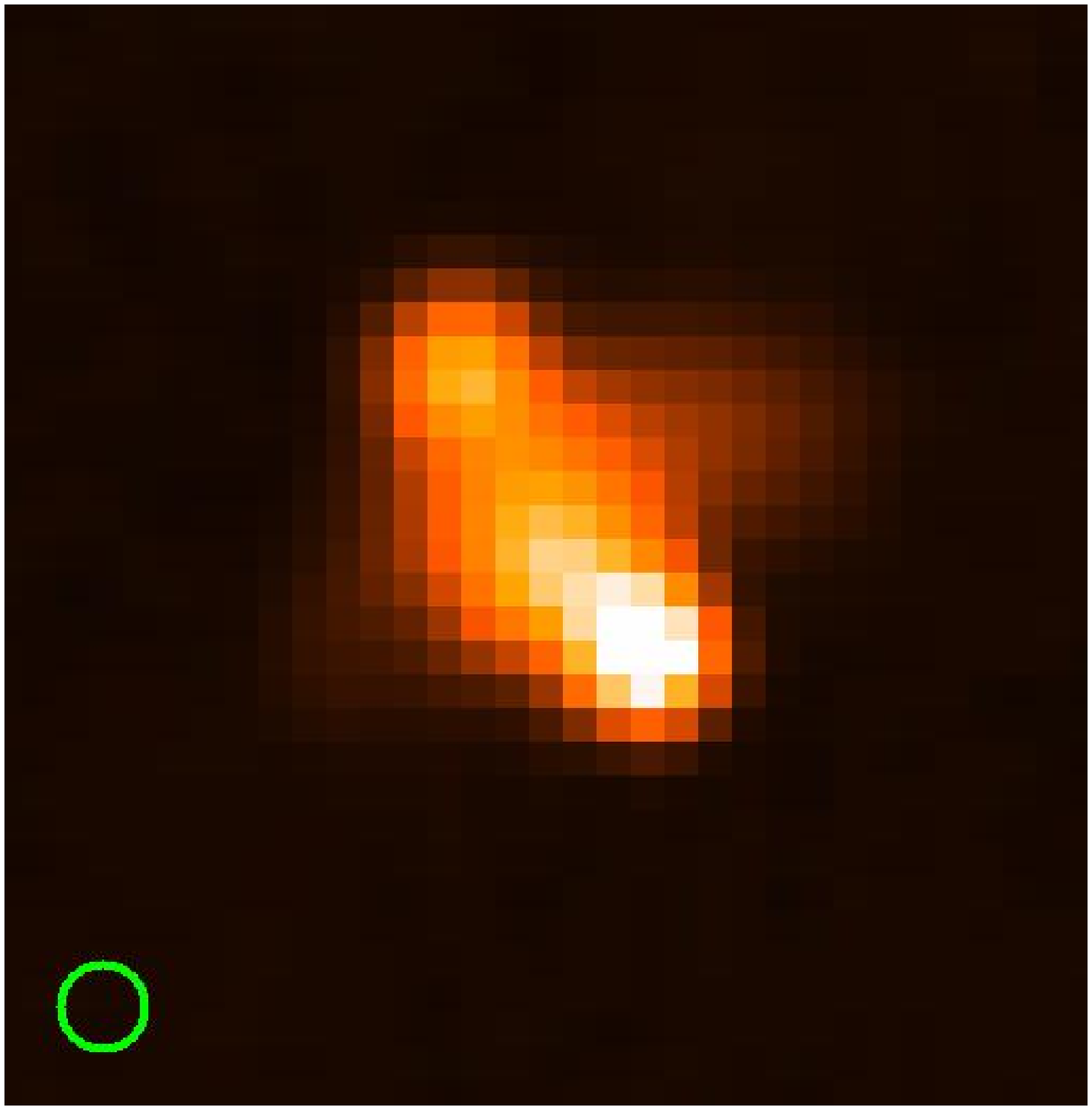}
\includegraphics[width=19mm]{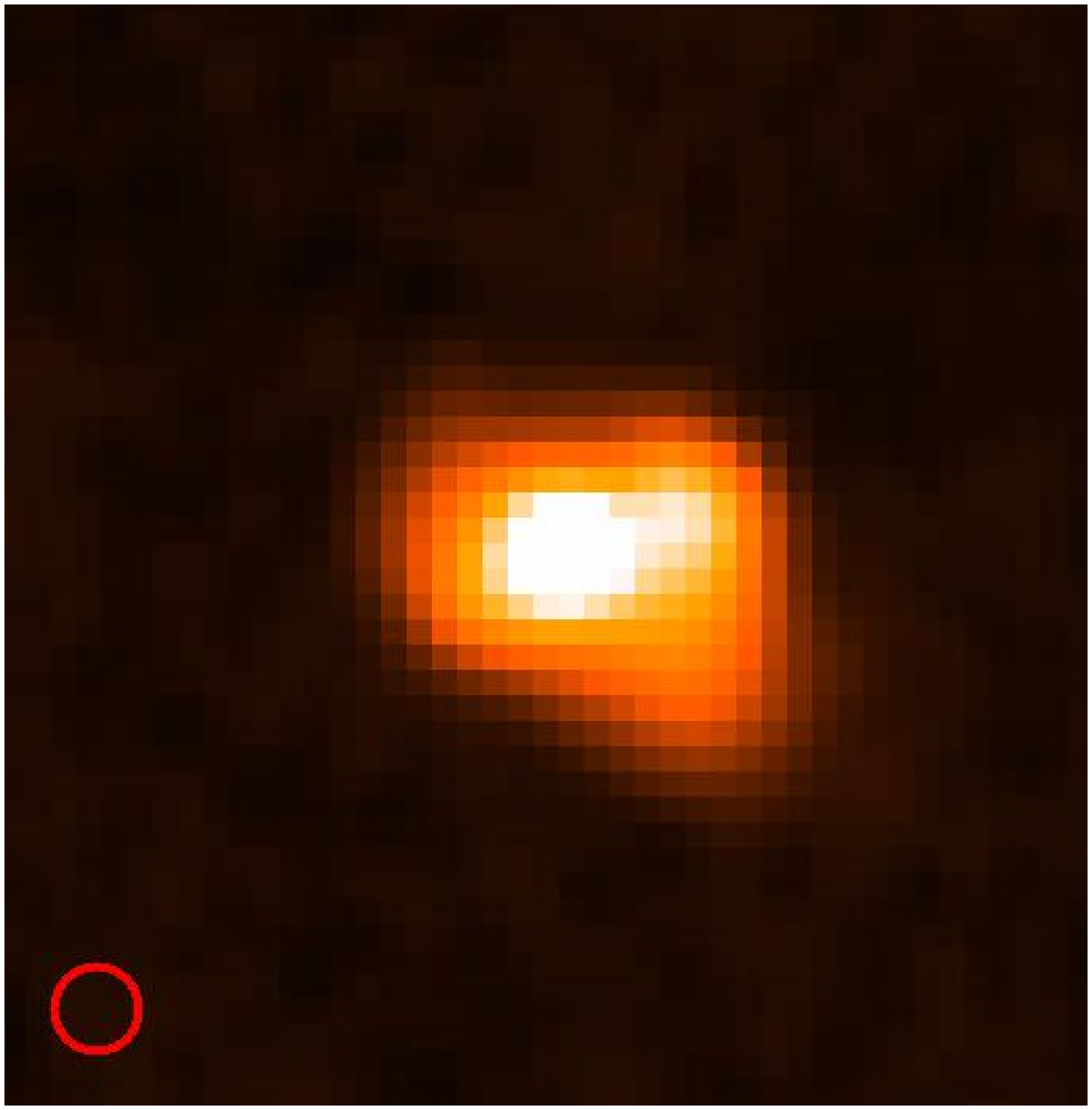}
\includegraphics[width=19mm]{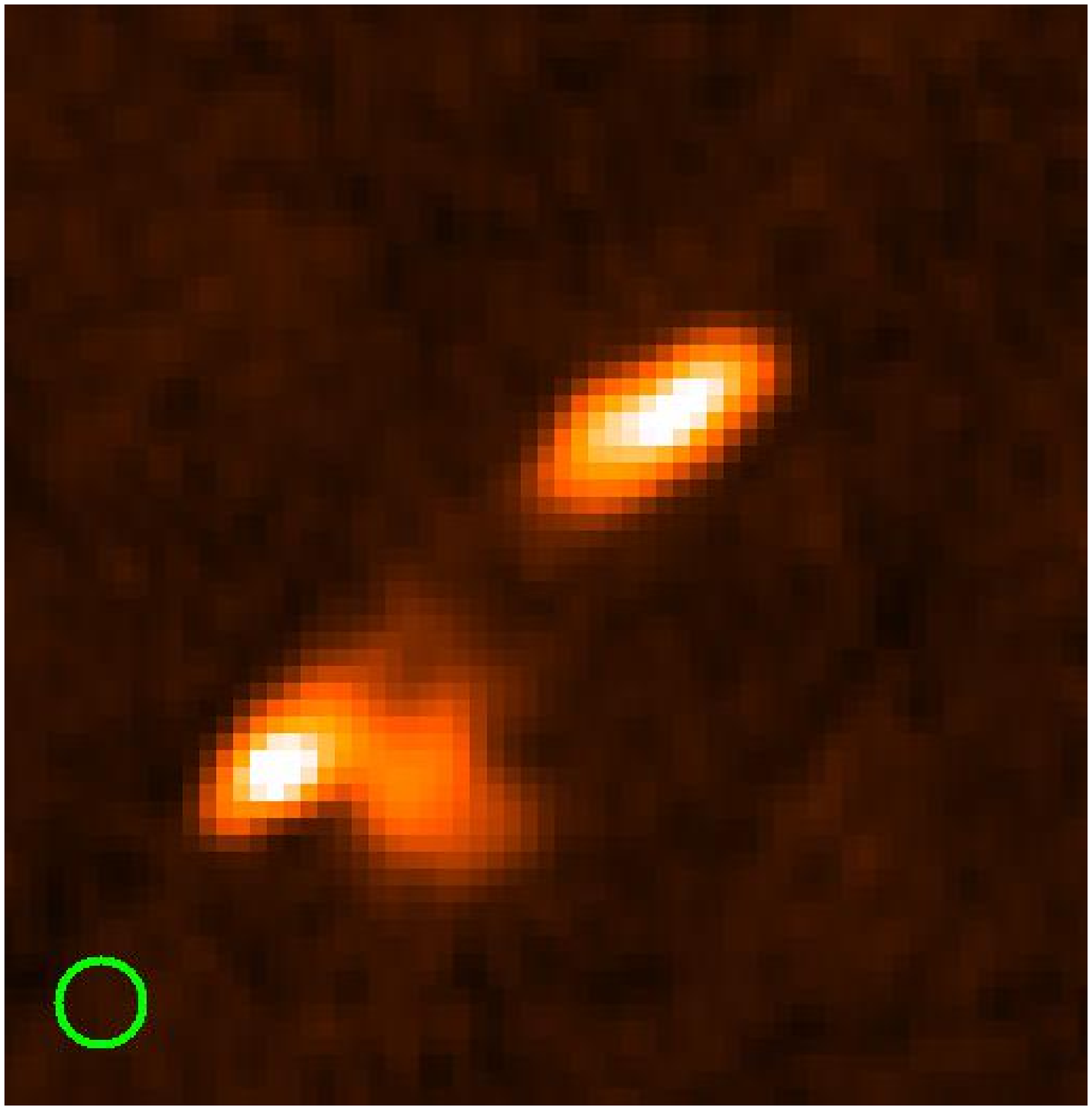}
\includegraphics[width=19mm]{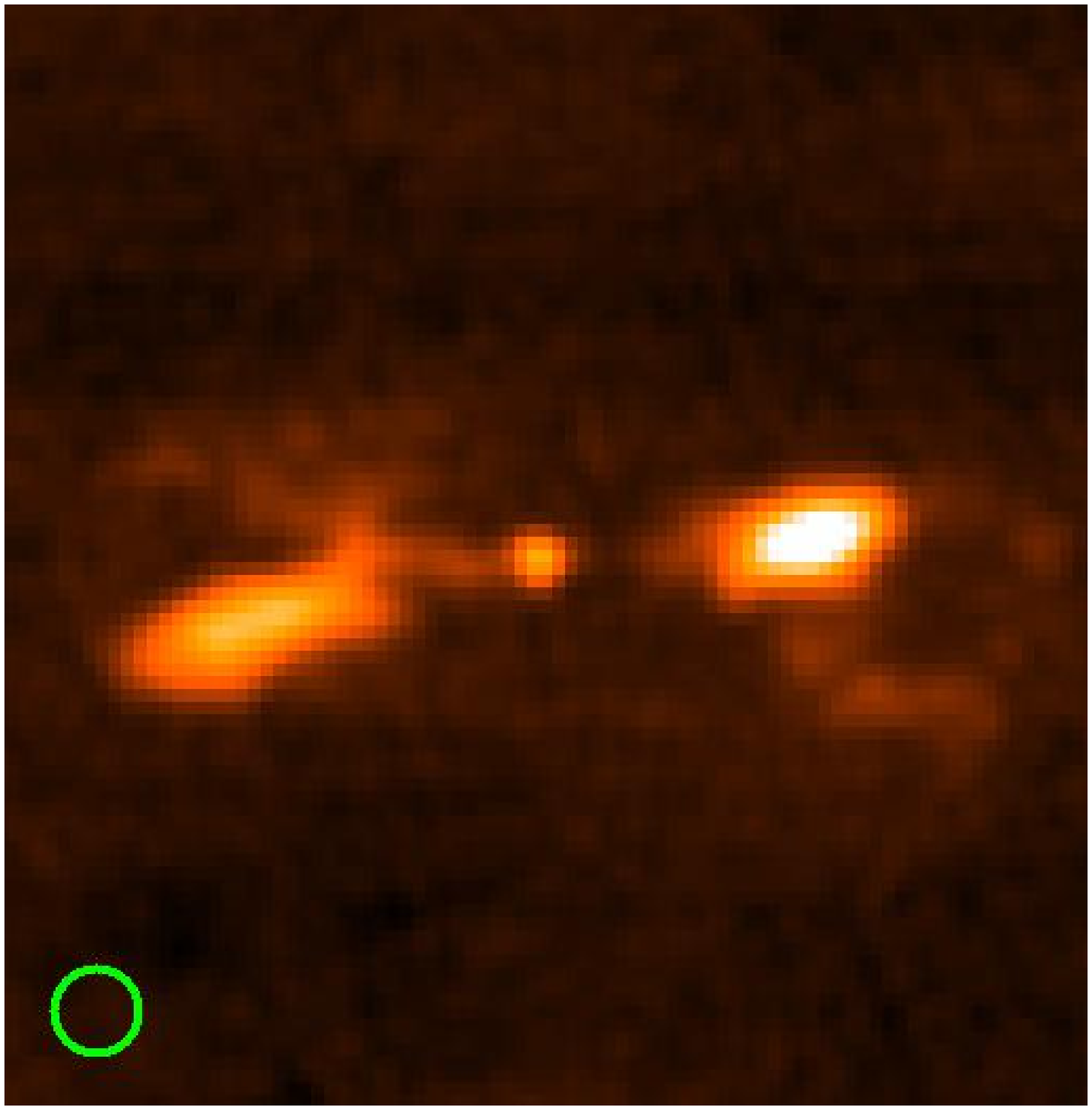}
\includegraphics[width=19mm]{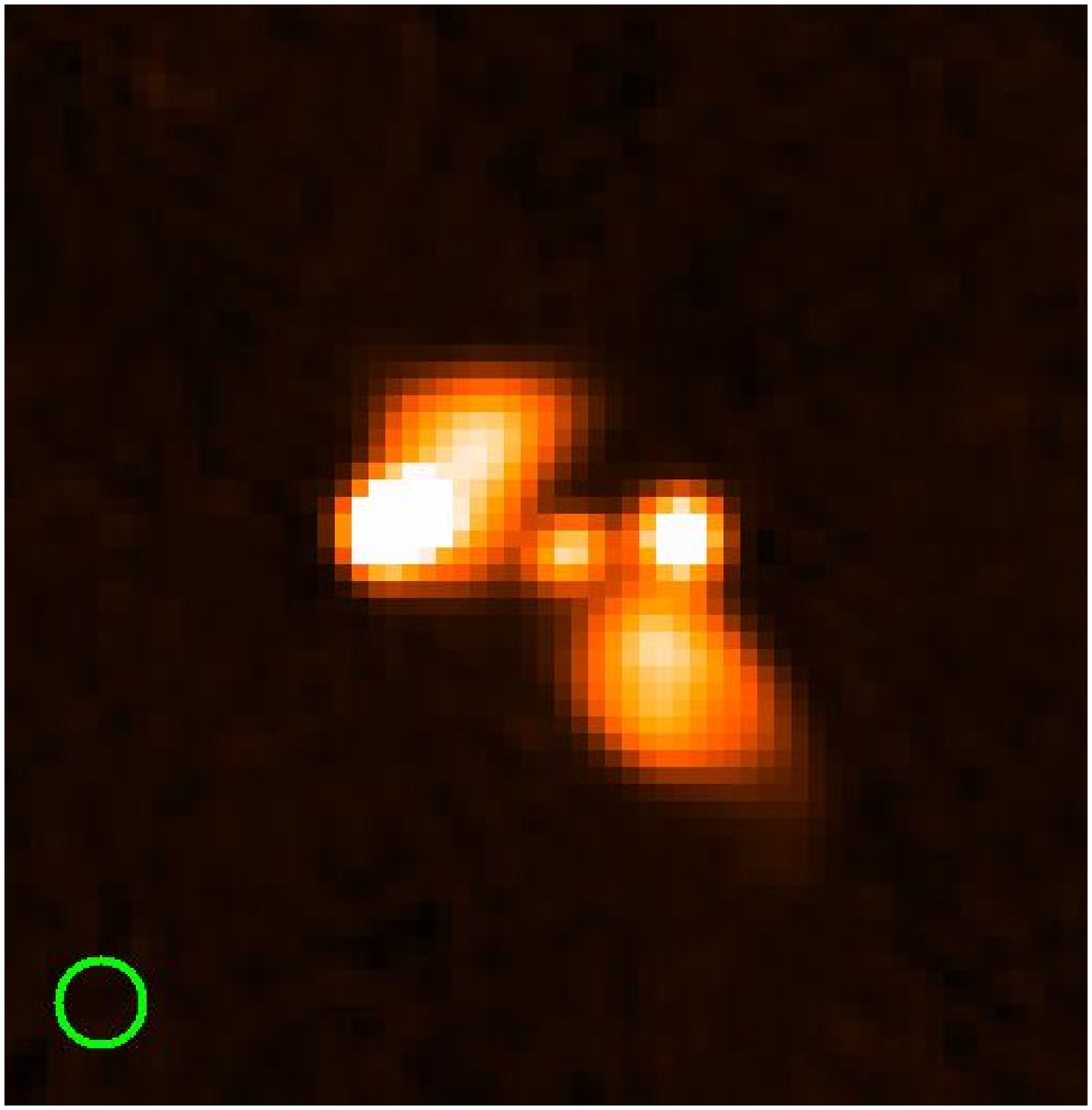}
\includegraphics[width=19mm]{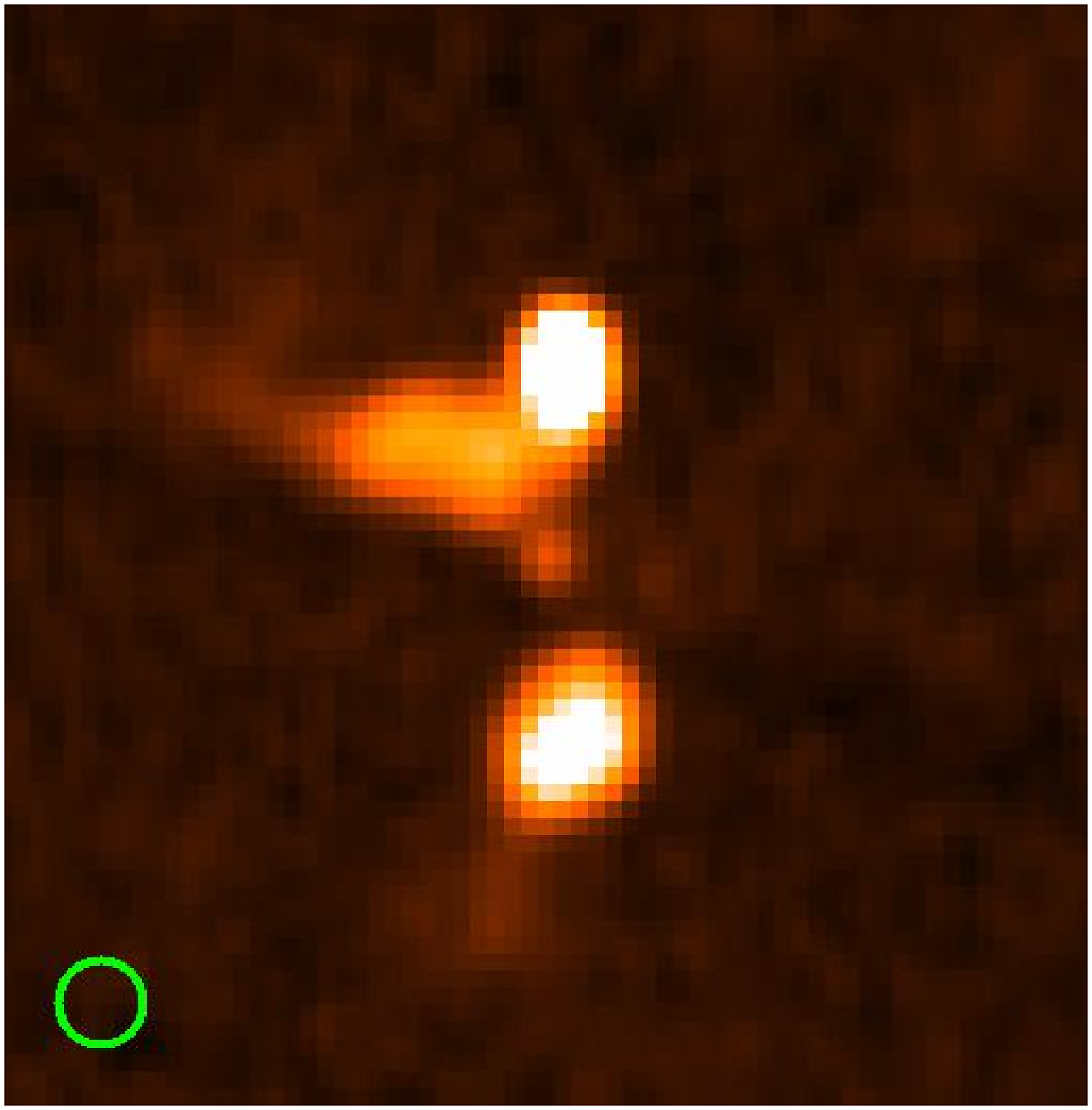}
\includegraphics[width=19mm]{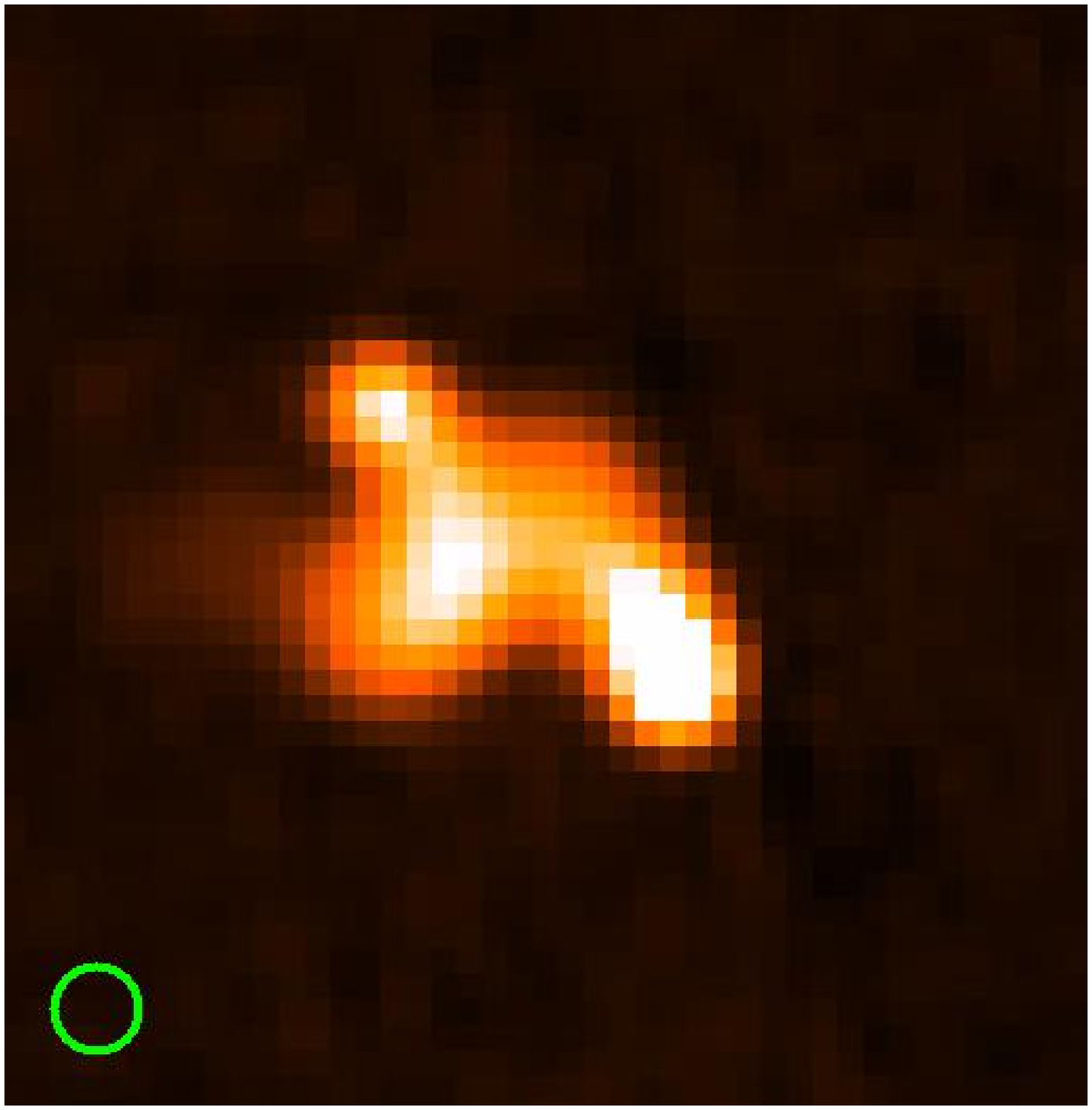}
\includegraphics[width=19mm]{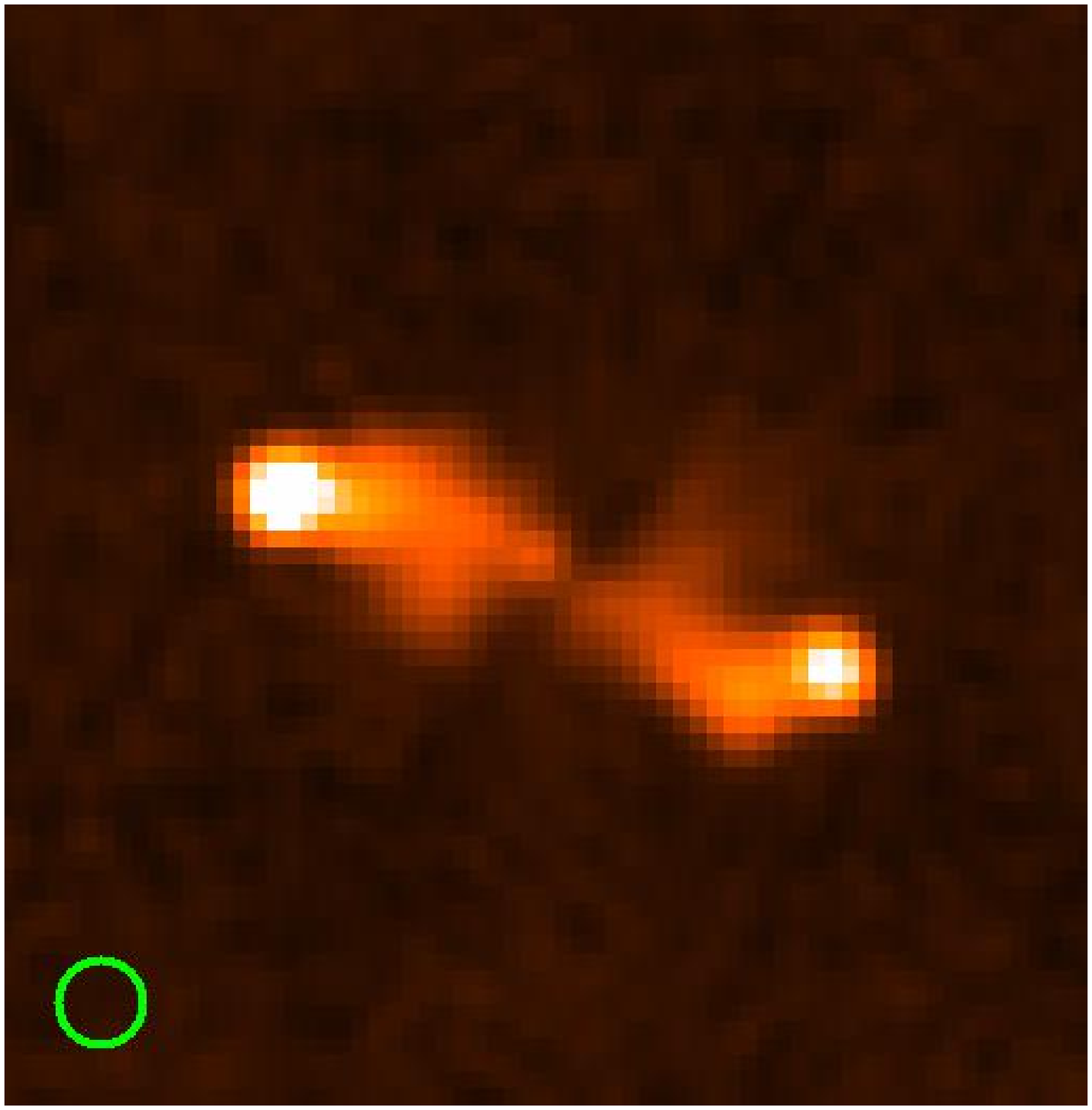}
}
\centerline{
\includegraphics[width=19mm]{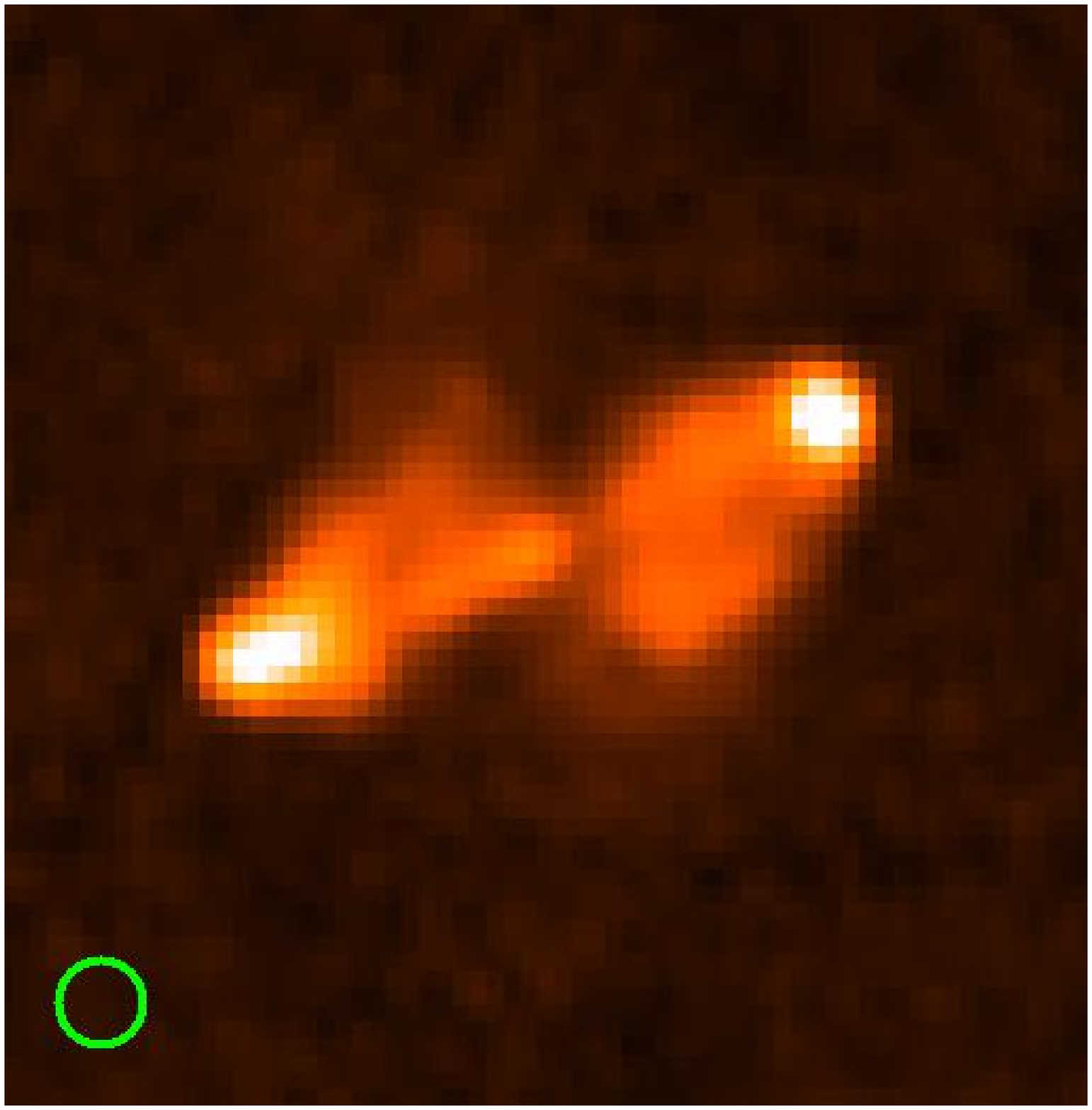}
\includegraphics[width=19mm]{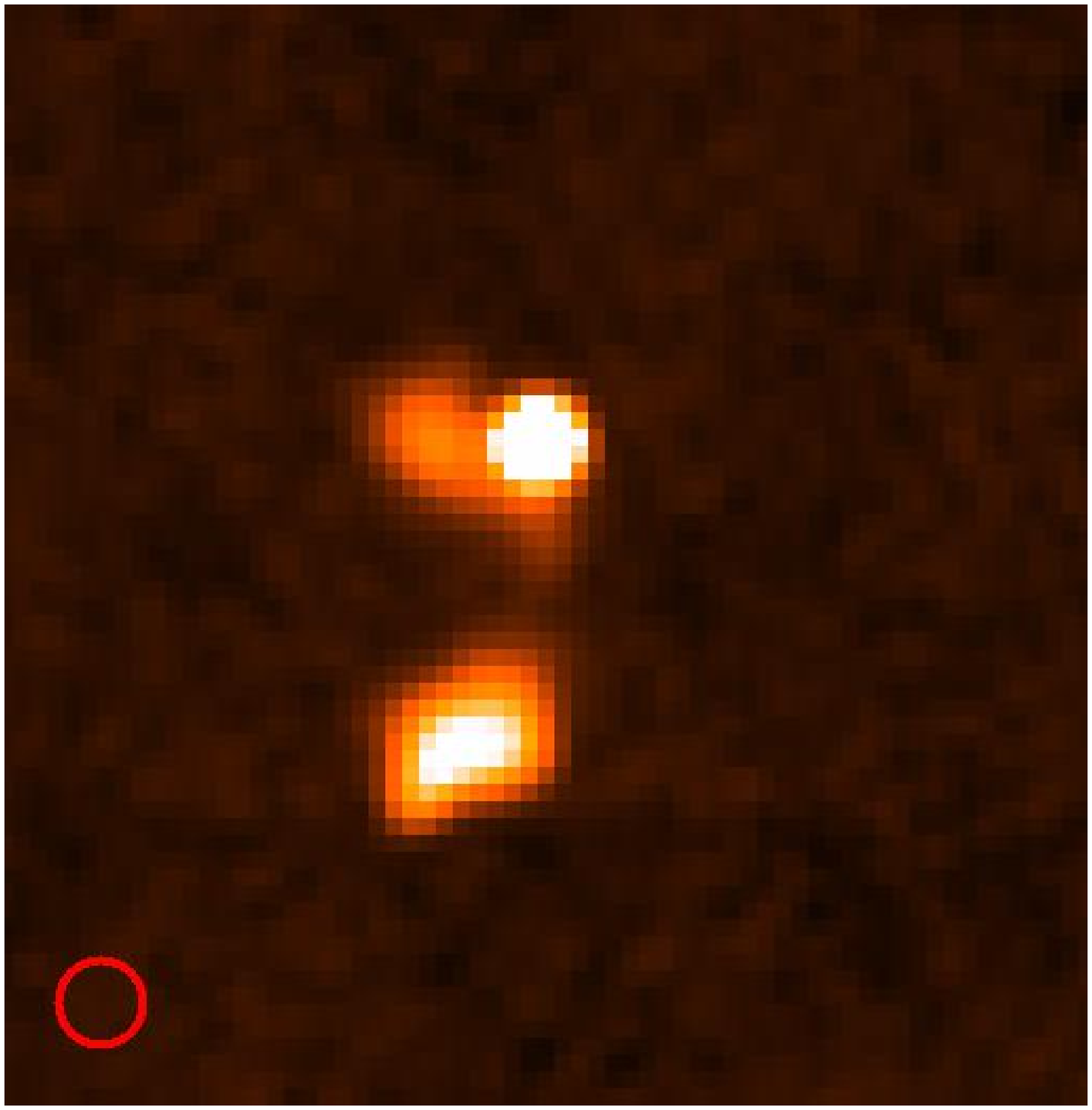}
\includegraphics[width=19mm]{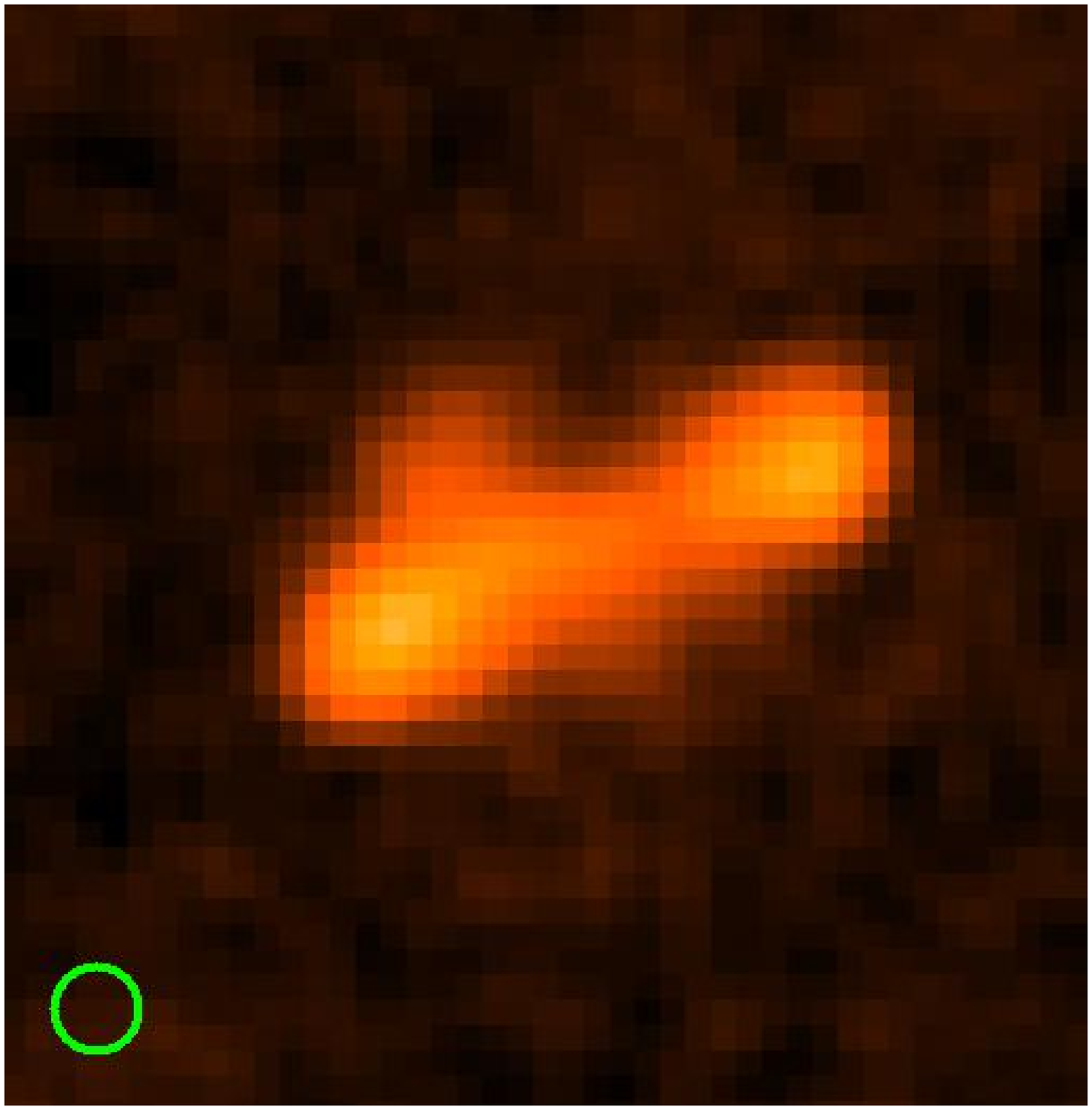}
\includegraphics[width=19mm]{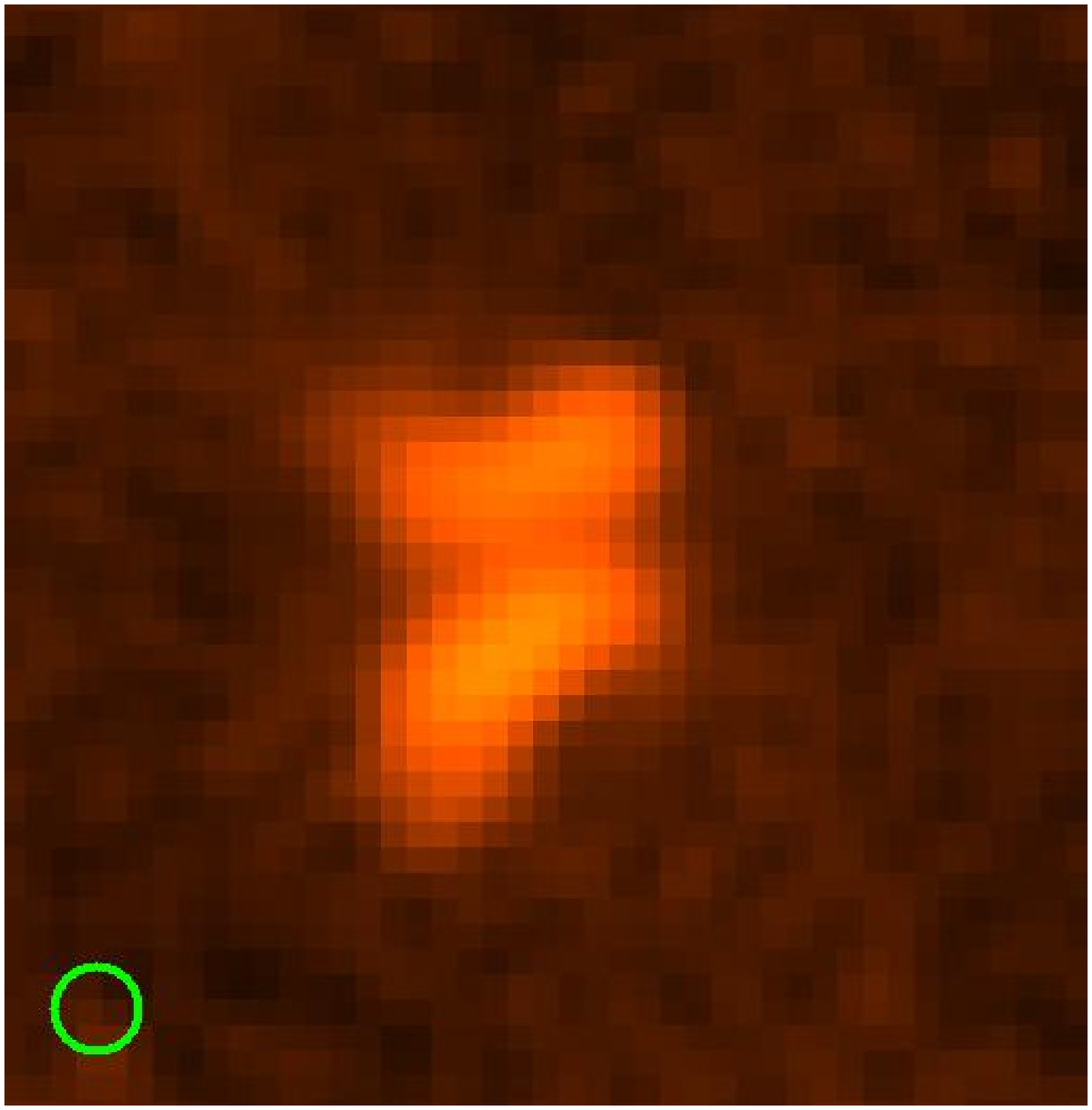}
\includegraphics[width=19mm]{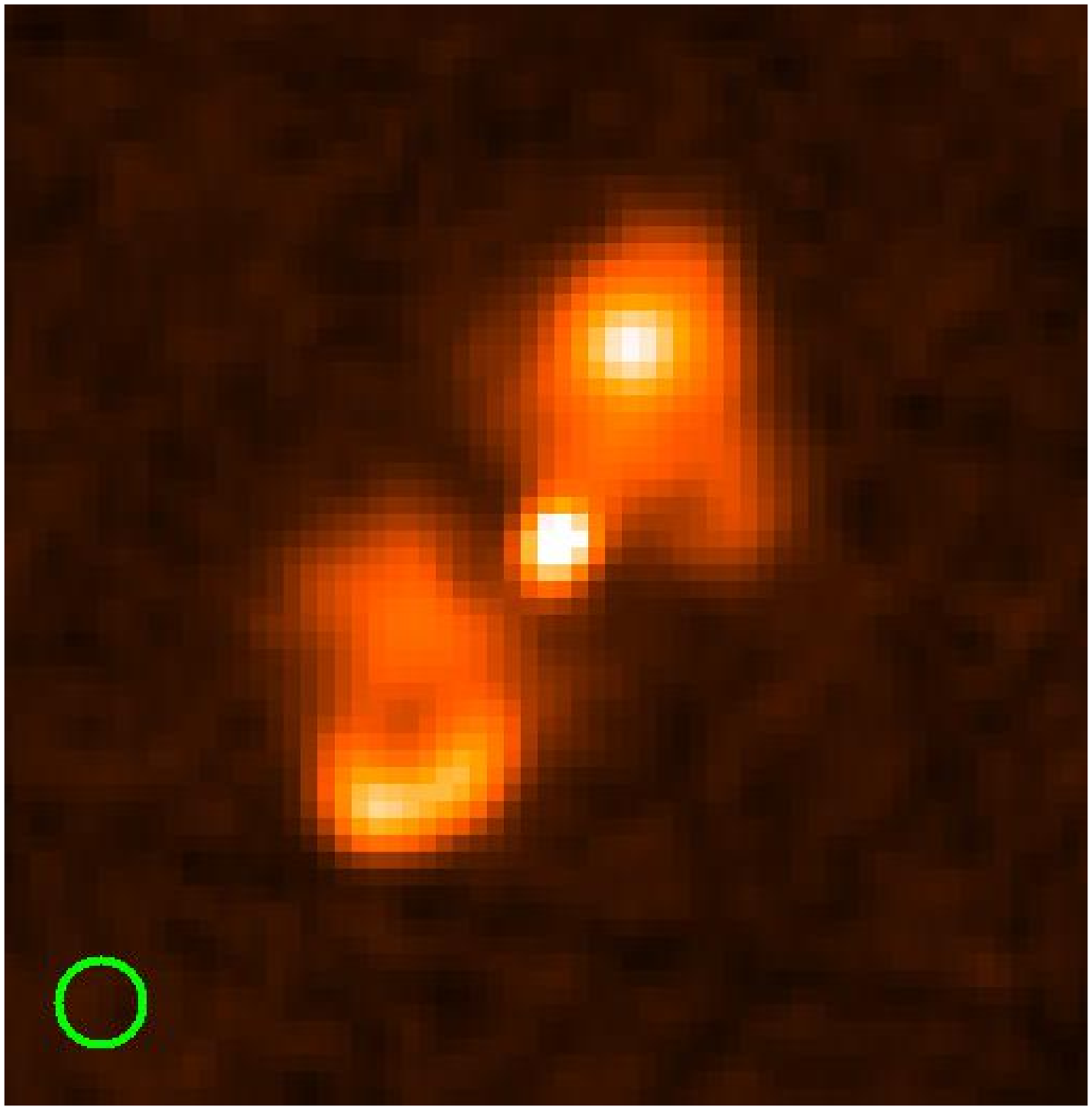}
\includegraphics[width=19mm]{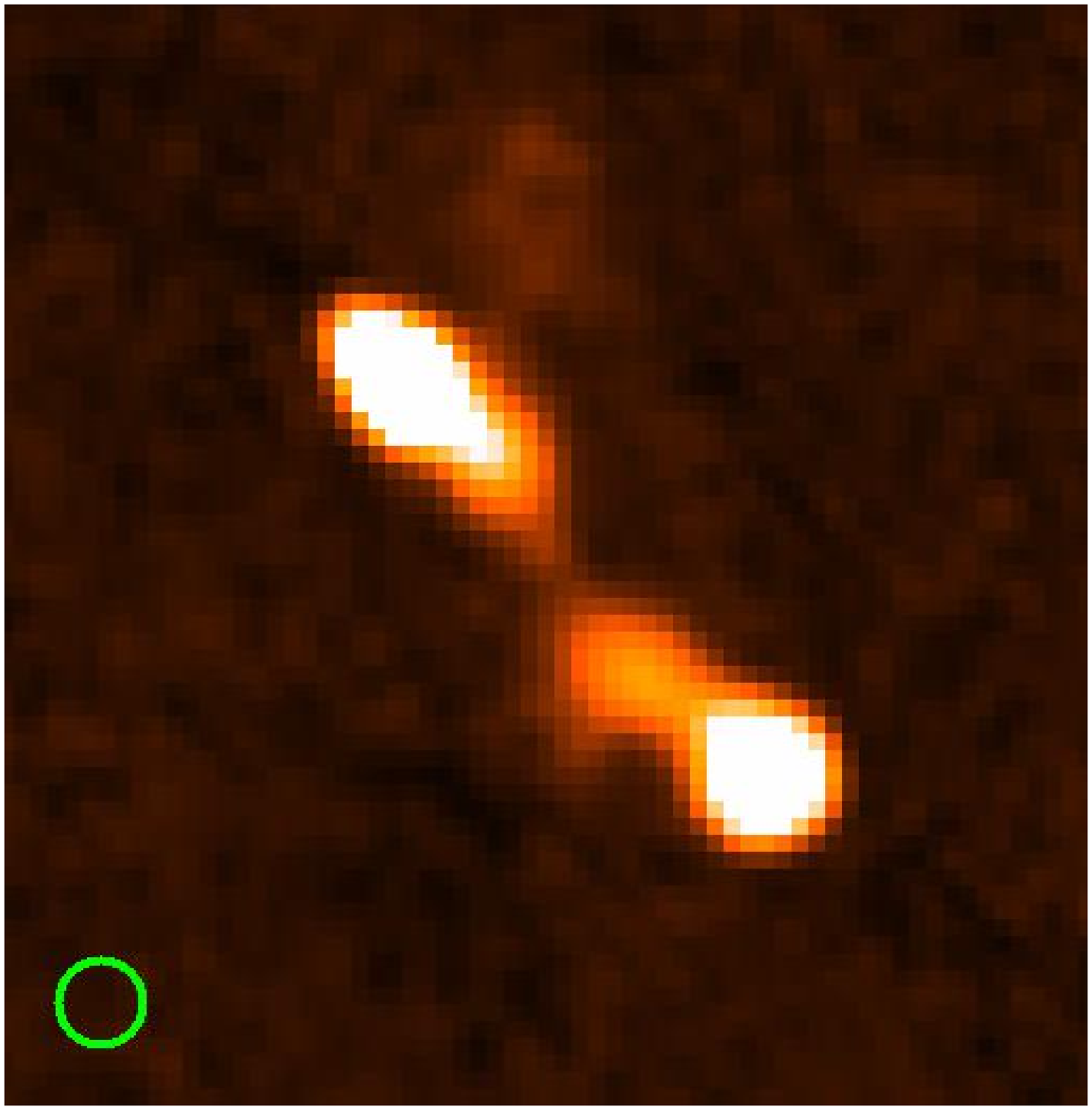}
\includegraphics[width=19mm]{radiowh.ps}
\includegraphics[width=19mm]{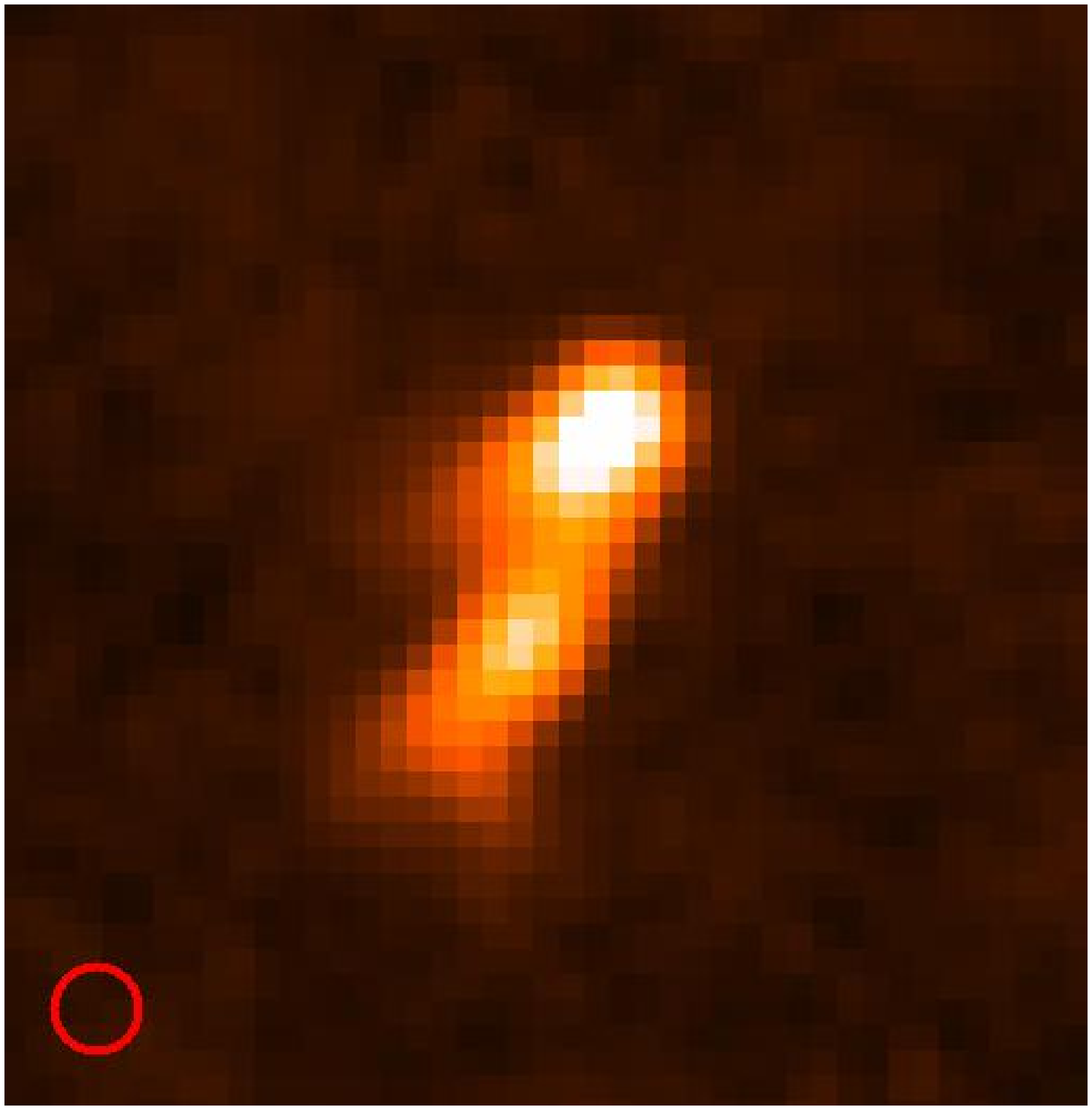}
\includegraphics[width=19mm]{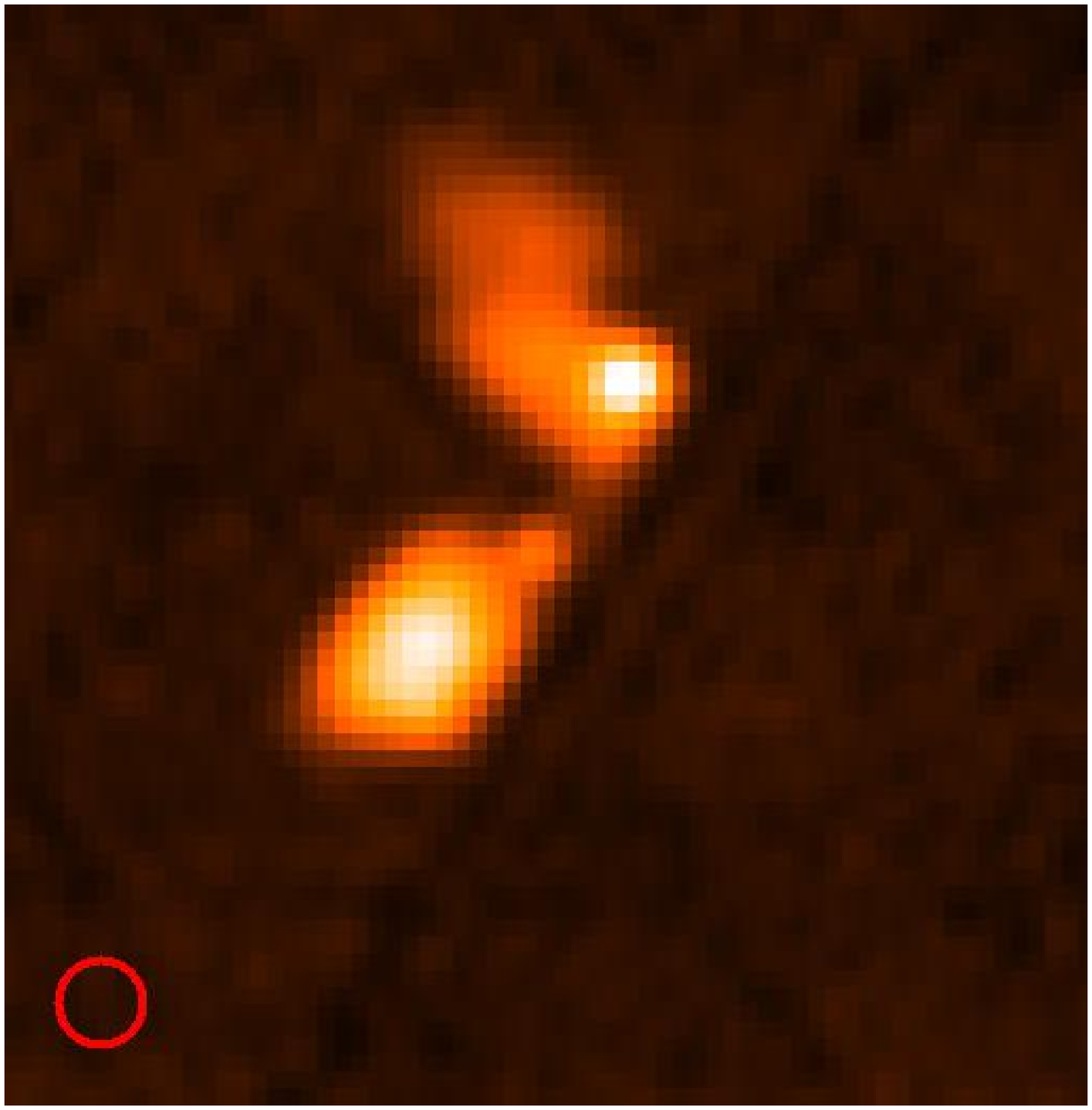}
\includegraphics[width=19mm]{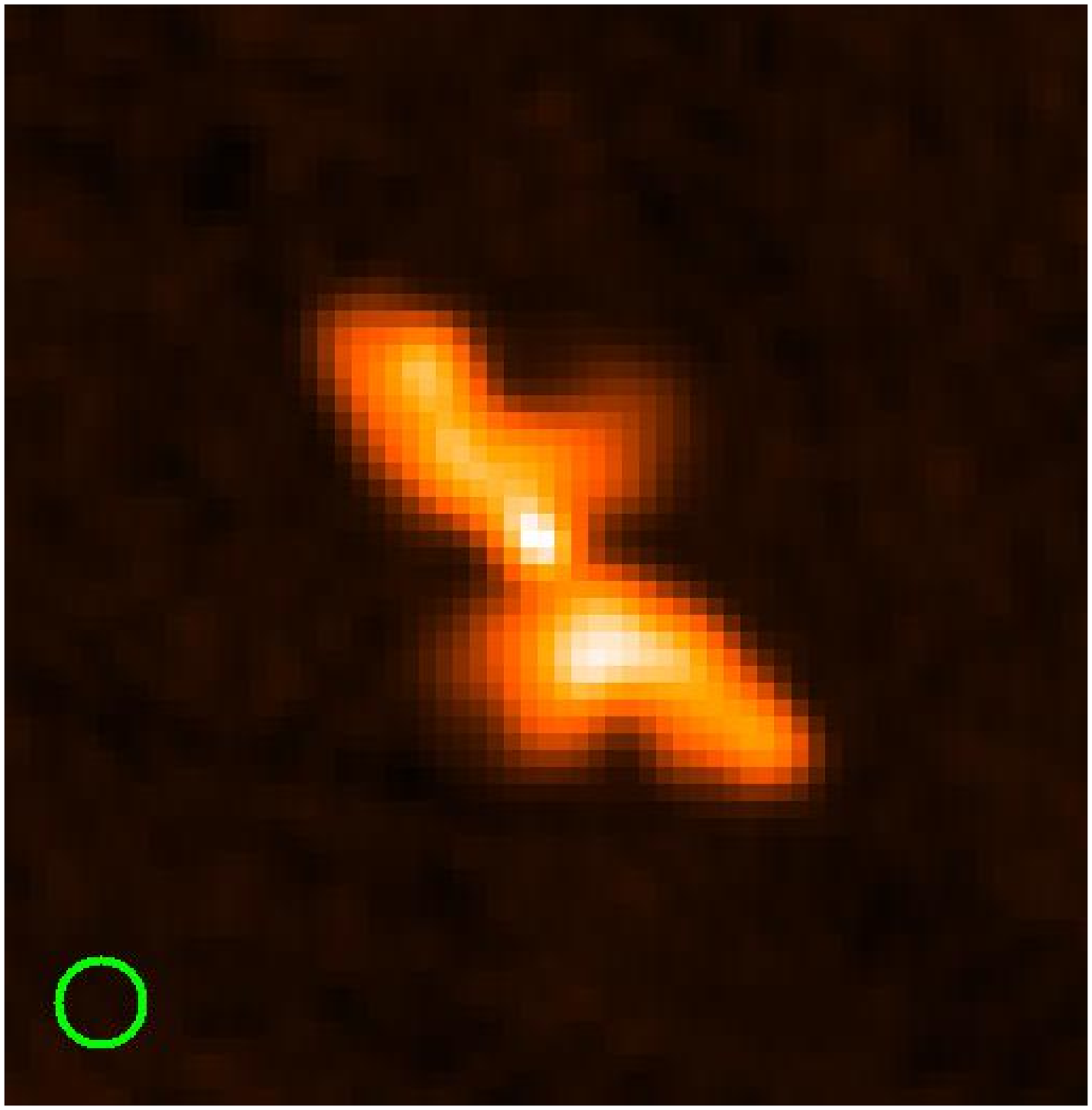}
}
\centerline{
\includegraphics[width=19mm]{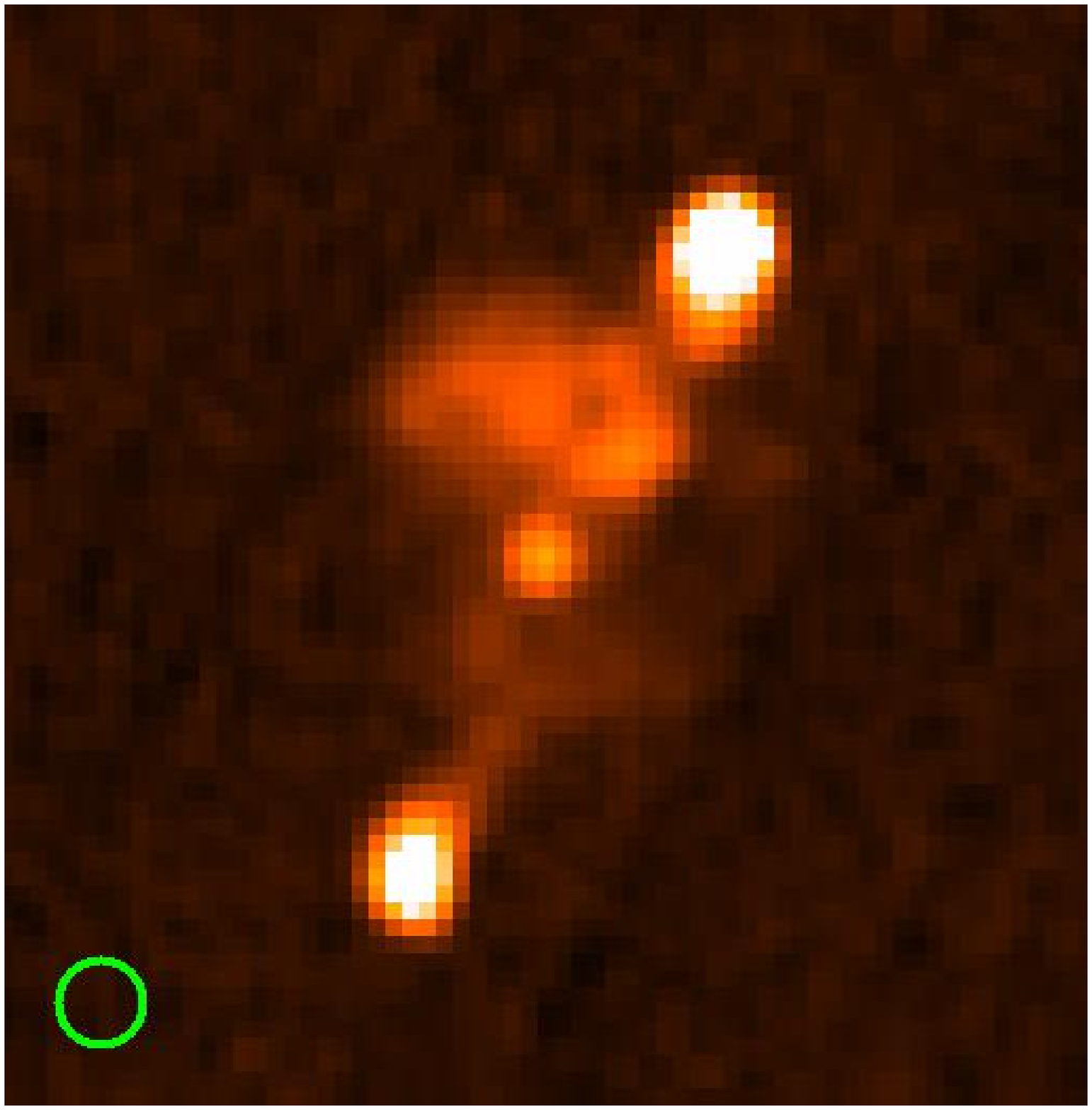}
\includegraphics[width=19mm]{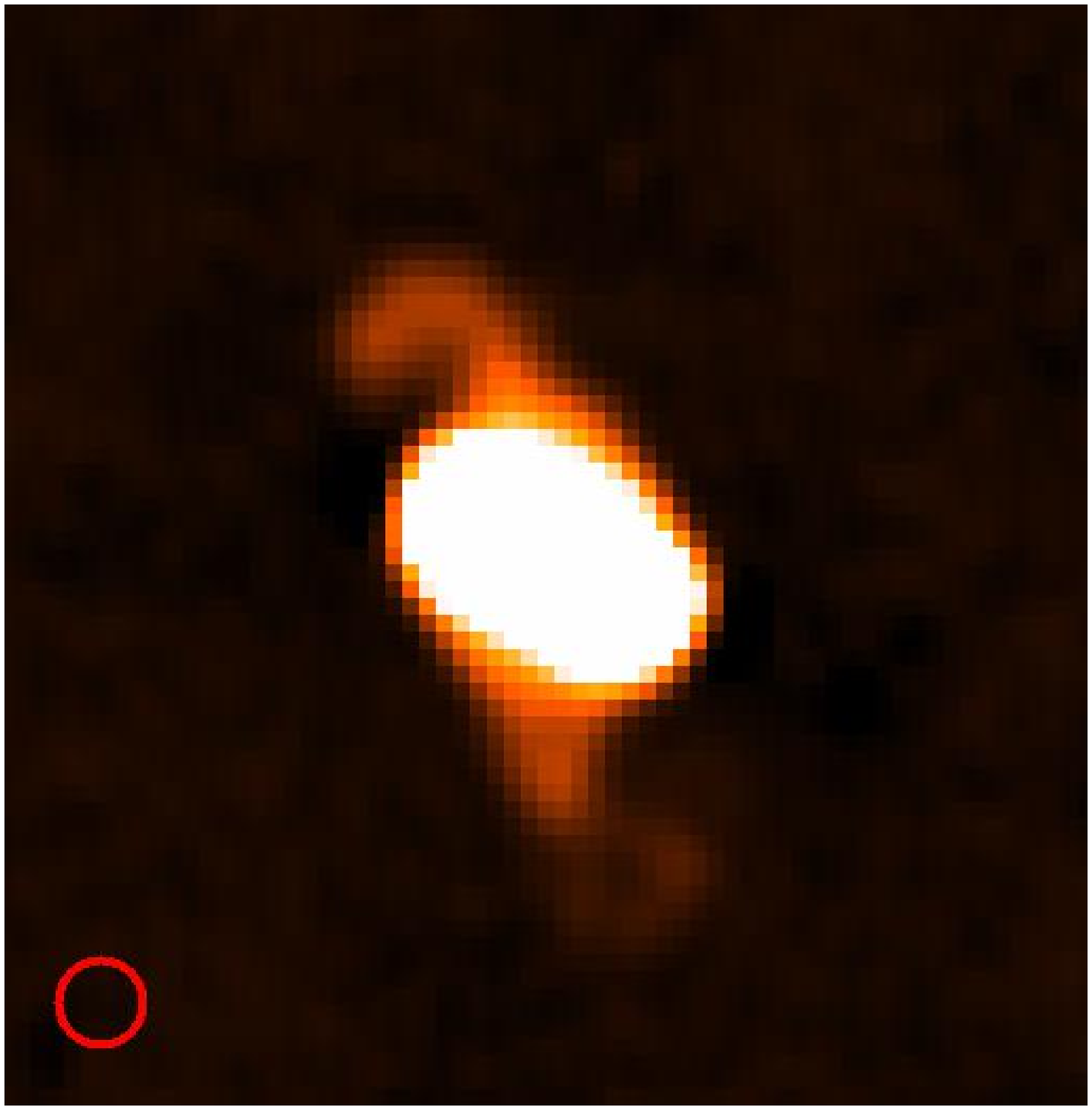}
\includegraphics[width=19mm]{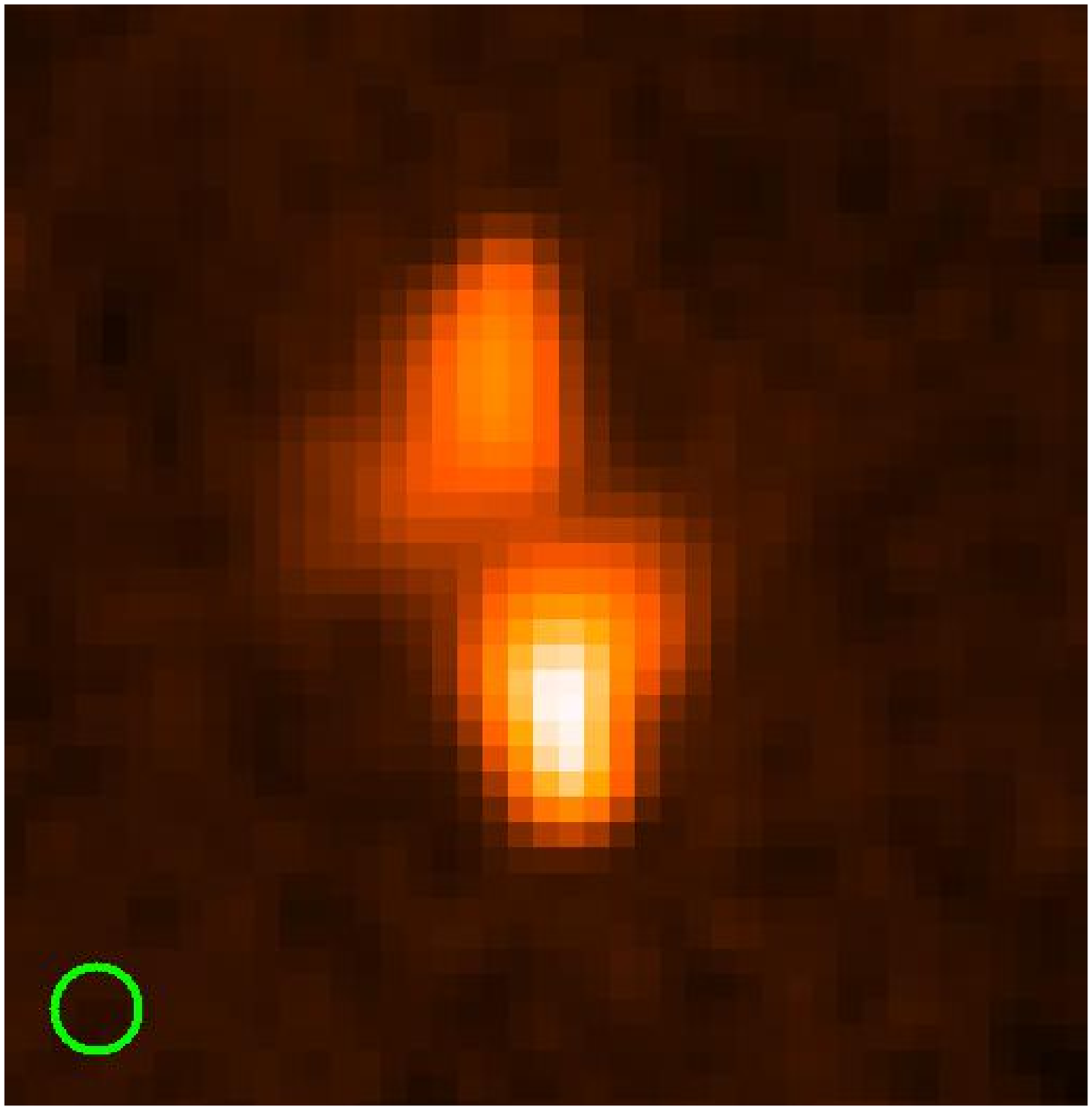}
\includegraphics[width=19mm]{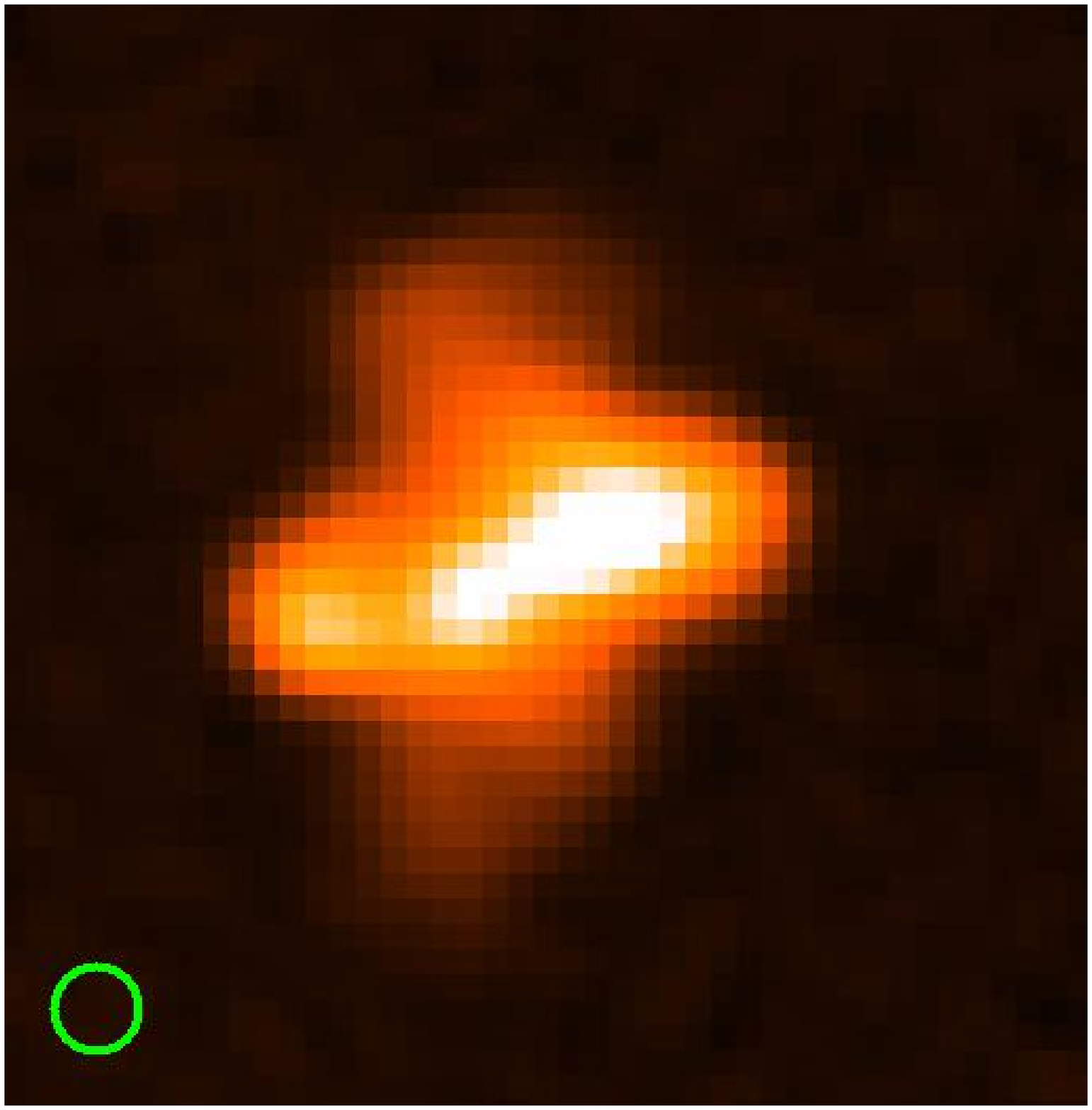}
\includegraphics[width=19mm]{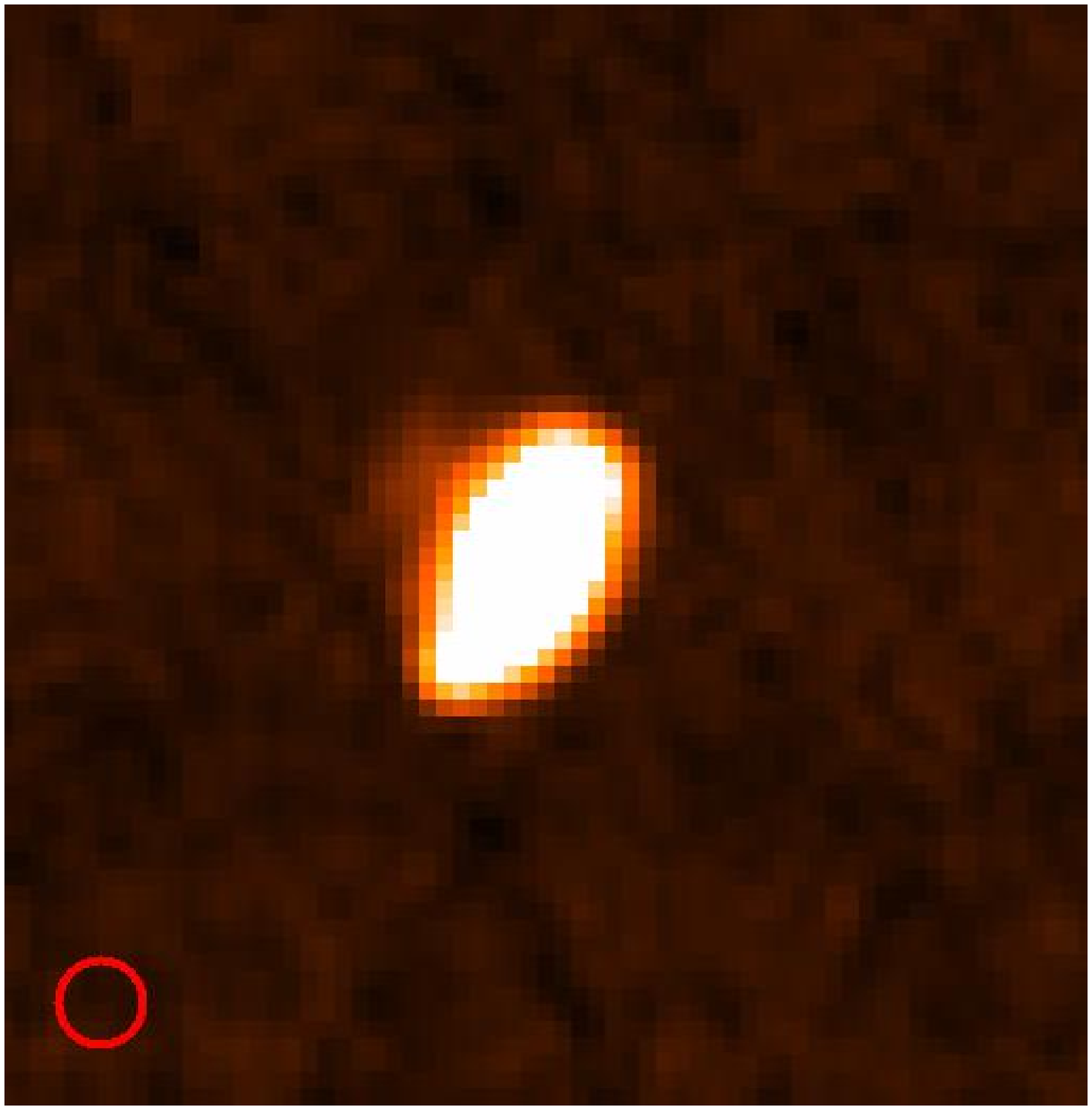}
\includegraphics[width=19mm]{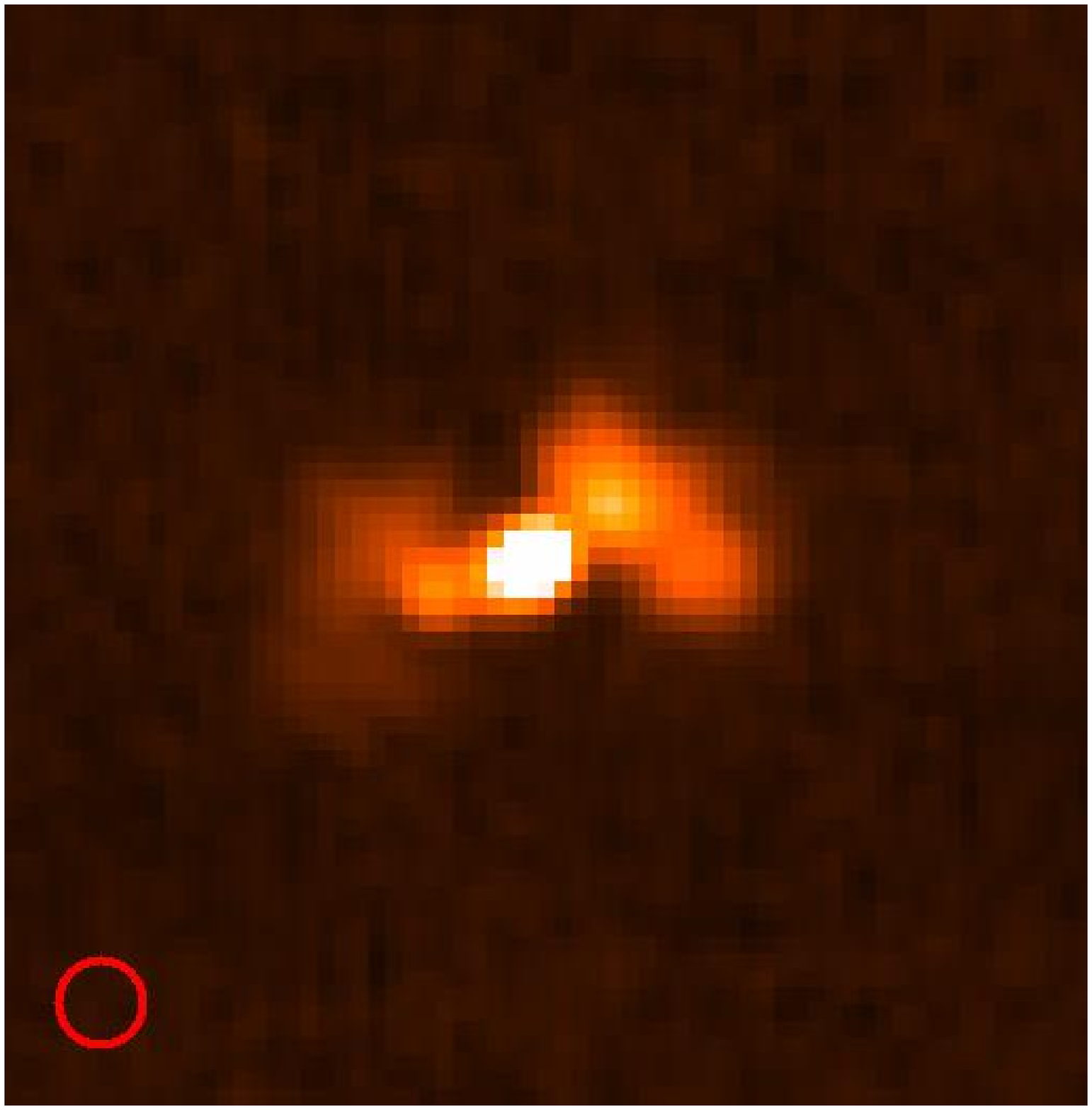}
\includegraphics[width=19mm]{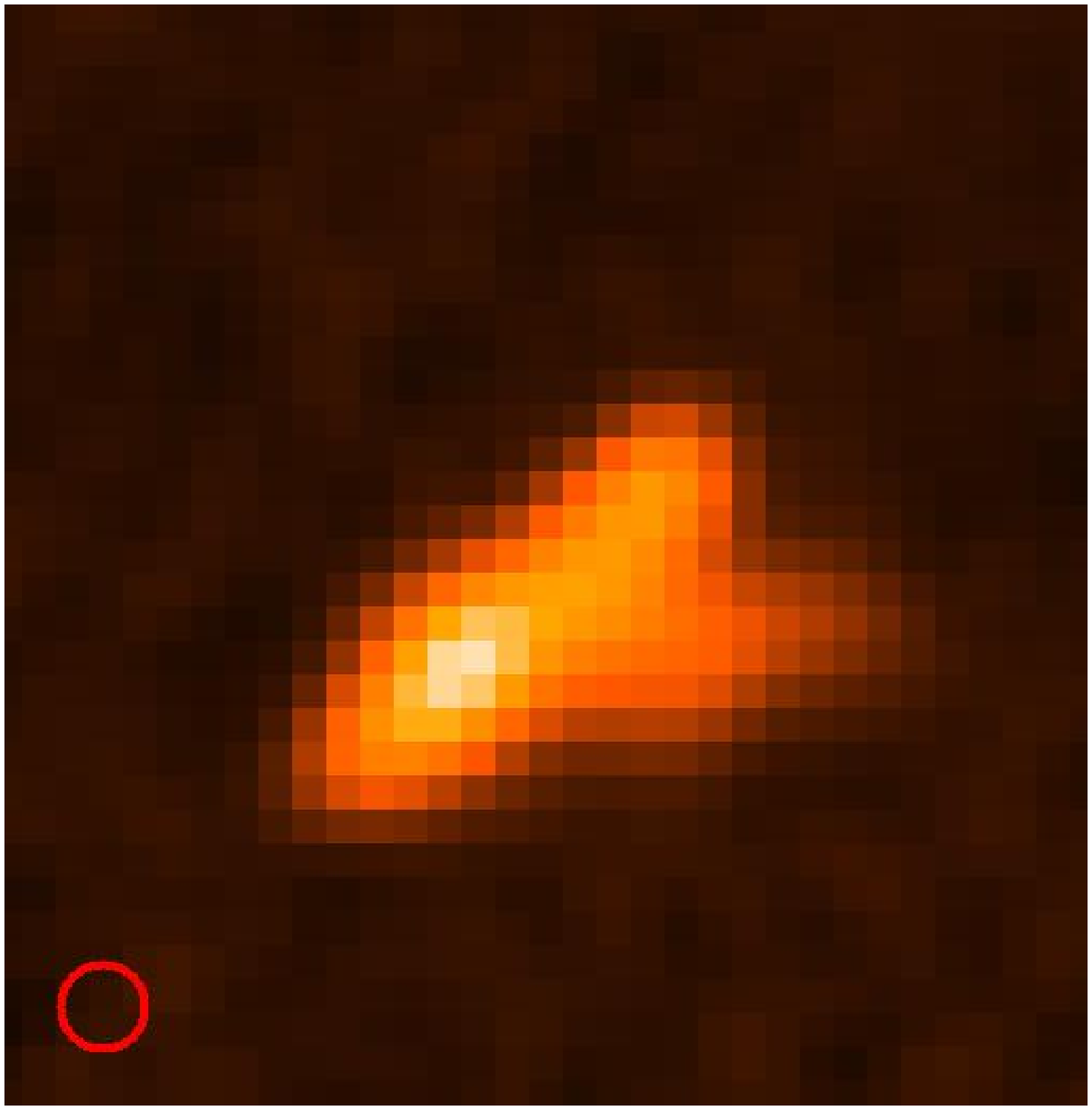}
\includegraphics[width=19mm]{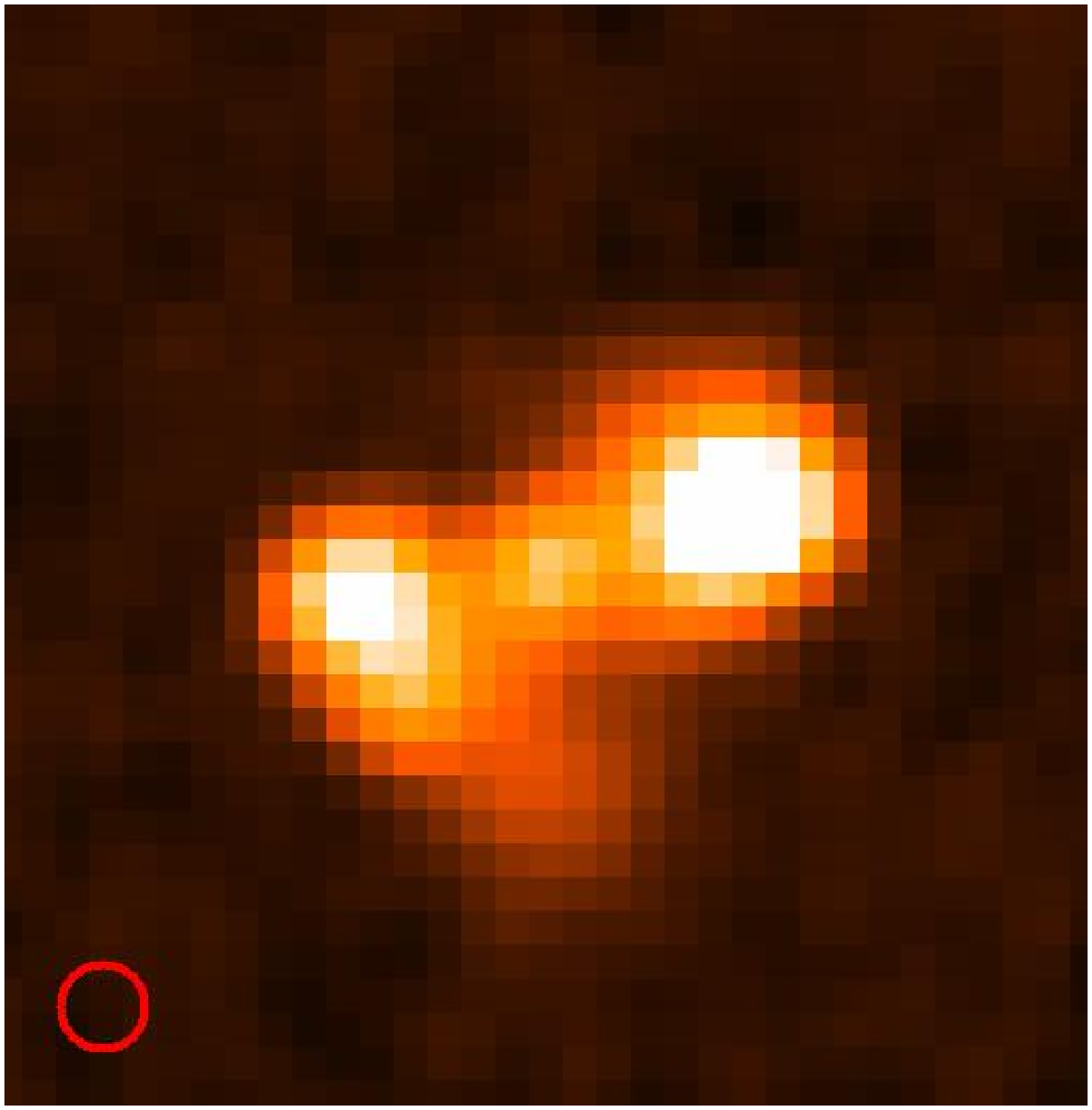}
\includegraphics[width=19mm]{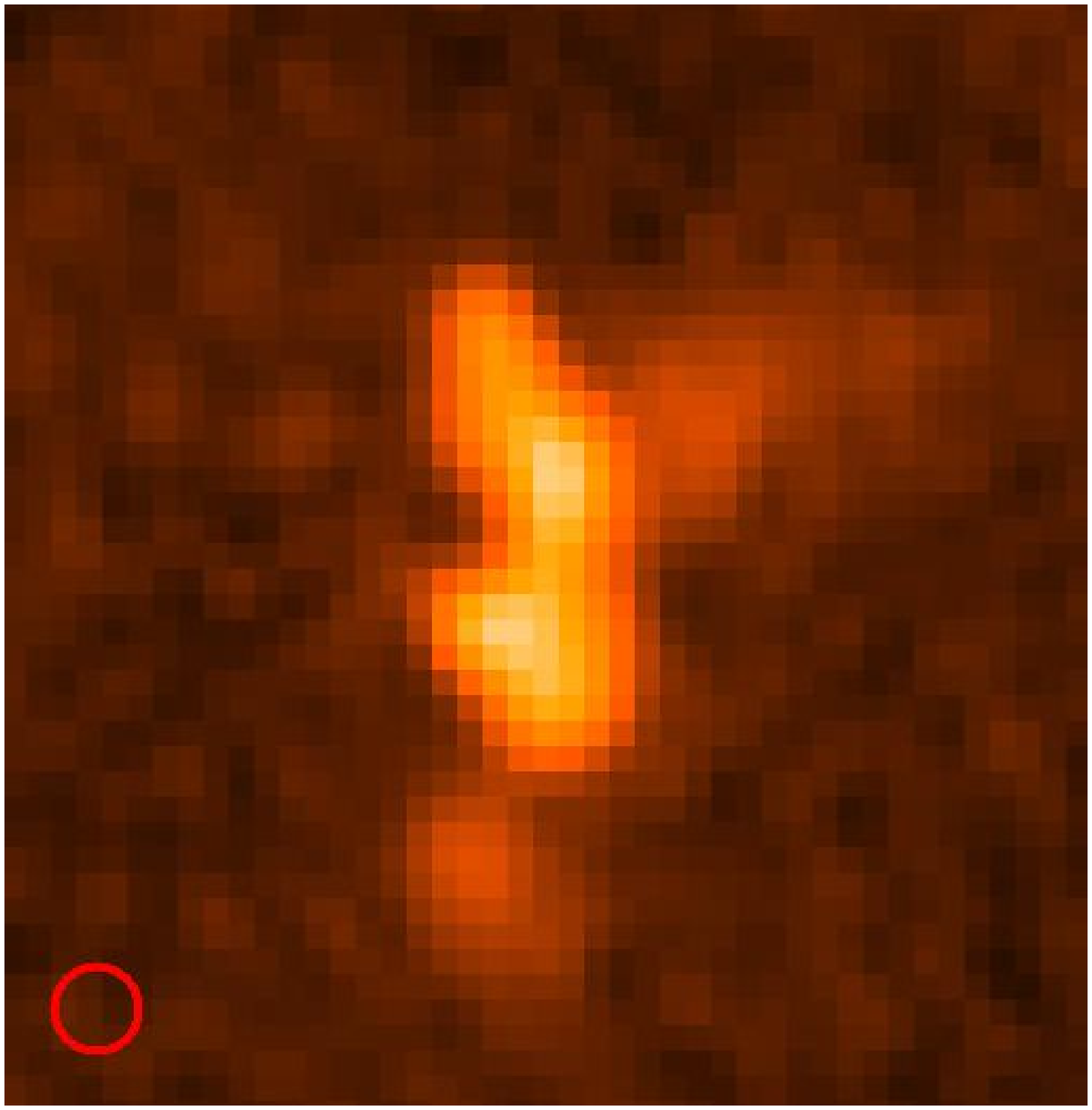}
\includegraphics[width=19mm]{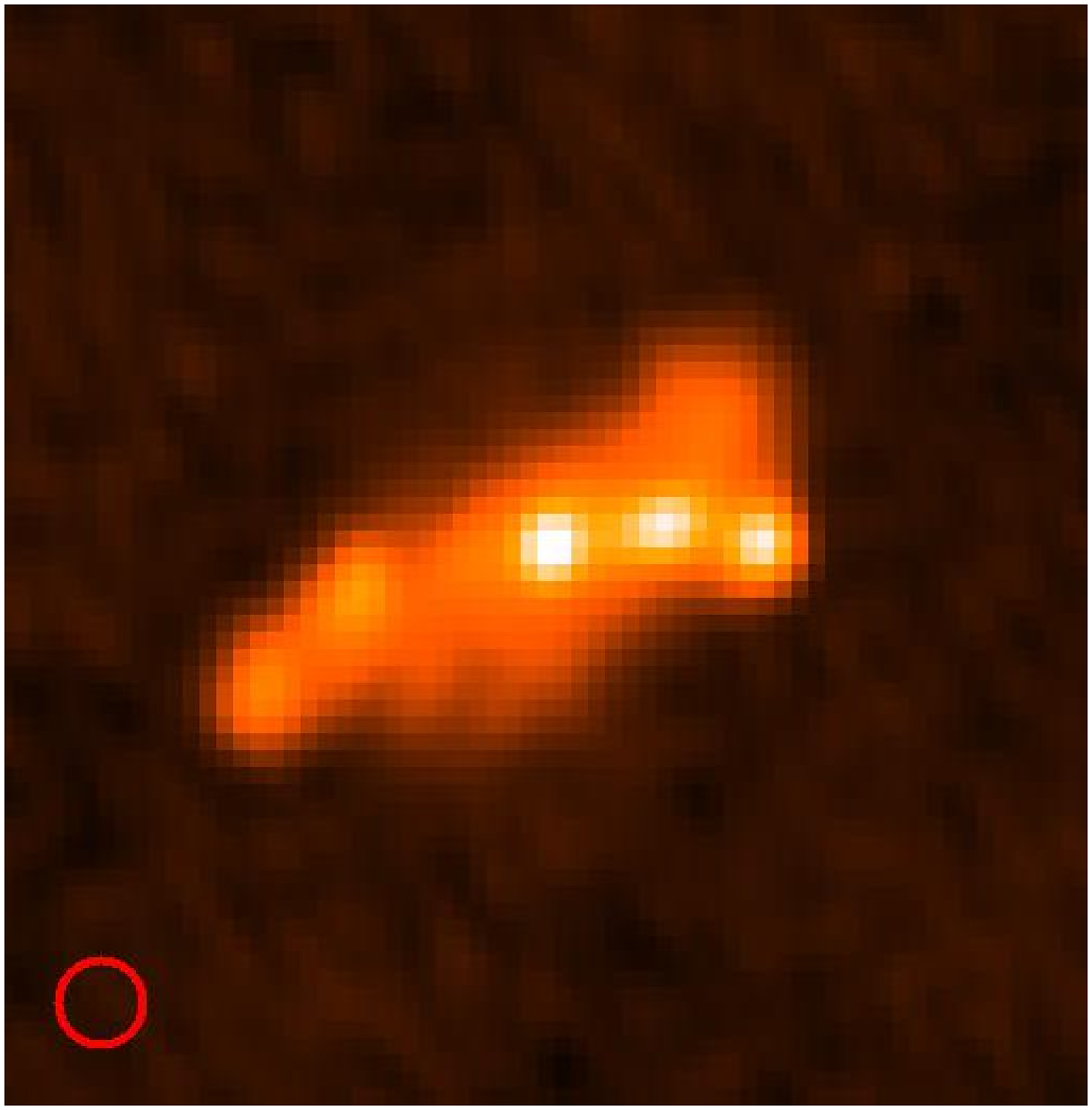}
}
\caption{FIRST images of the 91 X-shaped galaxies selected by \citet{cheung07}
  with available SDSS images. The color scale and the image size (ranging from
  50 to 200 $\arcsec$) have been selected to emphasize the wing
  structure. The 53 galaxies with well-defined wings selected for 
  analysis are shown with a green circle (those rejected with a red
  circle). }
\label{all}
\end{figure*}

\begin{table}
  \begin{center}
 \caption{Average morphological score of the 91 radio-sources.}
\begin{tabular}{cll|cll}
\hline
ID & Name       &       Marks           &       ID      &Name           &Marks            \\
\hline
  1 &J0001-0033 &2.67  $\bullet$        &       53      &J1218+1955     &1.33                   \\
  2 &J0033-0149 &2.33                   &       55      &J1227+2155     &4.00 $\bullet$   \\
  3 &J0036+0048 &2.00                   &       56      &J1228+2642     &3.33 $\bullet$   \\
  4 &J0045+0021 &1.00  $\bullet$        &       58      &J1247+4646     &1.67                   \\
  5 &J0049+0059 &2.00 \text{\sffamily X}&       59      &J1253+3435     &1.33 $\bullet$   \\
  6 &J0113+0106 &1.00  $\bullet$        &       60      &J1258+3227     &1.33                   \\
  7 &J0115-0000 &1.33                   &       61      &J1309-0012     &1.33 $\bullet$   \\
  8 &J0143-0119 &2.33  $\bullet$        &       62      &J1310+5458     &2.33 $\checkmark$\\
  9 &J0144-0830 &2.33  $\bullet$        &       63      &J1316+2427     &1.00             \\
 10 &J0145-0159 &2.66  $\bullet$        &       64      &J1327-0203     &1.00 $\bullet$   \\
 11 &J0147-0851 &2.00                   &       65      &J1330-0206     &2.33           \\
 12 &J0211-0920 &2.00 \text{\sffamily X}&       66      &J1339-0016     &2.33           \\
 13 &J0225-0738 &1.33                   &       67      &J1342+2547     &1.67 $\bullet$   \\
 16 &J0805+4854 &3.33                   &       68      &J1345+5233     &1.00 $\bullet$   \\
 17 &J0813+4347 &3.67  $\bullet$        &       69      &J1348+4411     &2.33 $\checkmark$\\
 18 &J0821+2922 &3.33  $\bullet$        &       70      &J1351+5559     &3.67           \\
 19 &J0836+3125 &2.00                   &       71      &J1353+0724     &1.67           \\
 20 &J0838+3253 &2.00                   &       72      &J1406-0154     &1.00 $\bullet$   \\
 21 &J0845+4031 &2.33  $\bullet$        &       73      &J1406+0657     &1.67           \\
 22 &J0846+3956 &1.67  $\bullet$        &       74      &J1408+0225     &3.33 $\bullet$   \\
 24 &J0914+1715 &1.67                   &       75      &J1411+0907     &2.00           \\
 25 &J0917+0523 &1.67  $\bullet$        &       76      &J1424+2637     &1.33           \\
 26 &J0924+4233 &1.33  $\bullet$        &       77      &J1430+5217     &1.00 $\bullet$   \\
 27 &J0941-0143 &2.33  $\checkmark$     &       78      &J1433+0037     &1.67           \\
 28 &J0941+2147 &2.00                   &       79      &J1434+5906     &1.33 $\bullet$   \\
 29 &J0943+2834 &1.33                   &       80      &J1437+0834     &1.33             \\
 30 &J1005+1154 &2.00  $\bullet$        &       81      &J1444+4147     &1.33           \\
 31 &J1008+0030 &3.00  $\bullet$        &       82      &J1454+2732     &2.33           \\
 32 &J1015+5944 &2.33  $\checkmark$     &       83      &J1455+3237     &2.00           \\
 33 &J1040+5056 &3.00                   &       84      &J1456+2542     &1.33 $\bullet$   \\
 34 &J1043+3131 &2.67  $\bullet$        &       85      &J1459+2903     &1.67 $\bullet$   \\
 35 &J1049+4422 &2.00                   &       86      &J1501+0752     &2.00           \\
 36 &J1054+5521 &1.00  $\bullet$        &       88      &J1522+4527     &4.00           \\
 38 &J1102+0250 &2.33                   &       89      &J1537+2648     &2.67           \\
 39 &J1111+4050 &4.00  $\bullet$        &       90      &J1600+2058     &1.00 $\bullet$   \\
 40 &J1114+2632 &3.00                   &       91      &J1603+5242     &1.33           \\
 41 &J1120+4354 &2.33                   &       92      &J1606+0000     &3.33 $\checkmark$\\
 42 &J1128+1919 &2.33                   &       93      &J1606+4517     &1.00 $\bullet$   \\
 44 &J1140+1057 &1.00                   &       94      &J1614+2817     &1.00 $\bullet$   \\
 45 &J1200+6105 &2.00                   &       95      &J1625+2705     &4.00 $\bullet$   \\
 47 &J1202+4915 &1.00  $\bullet$        &       96      &J1653+3115     &3.00           \\
 48 &J1206+3812 &1.67  $\bullet$        &       97      &J1655+4551     &3.33           \\
 49 &J1207+3352 &2.00  $\bullet$        &       98      &J1656+3952     &2.33 $\bullet$   \\
 50 &J1210-0341 &1.00  $\bullet$        &       99      &J2226+0125     &3.00             \\
 51 &J1210+1121 &3.33                   &      100      &J2359-1041     &2.33             \\
 52 &J1211+4539 &1.33  $\bullet$        &               &               &                 \\
 \hline
\end{tabular}
  \end{center}

\medskip
The sources marked with a $\bullet$ symbol following the score have radio
images from \citet{roberts15}; a $\checkmark$ indicates the objects for which
these images suggest an XRS classification that contrasts with the score
assigned by us based on the FIRST images, while an \text{\sffamily X}
indicates the opposite outcome.
\label{tab1}
\end{table}

Among the 100 XRSs selected by \citet{cheung07}, we only considered  the 91
sources covered by the SDSS, thus excluding nine objects, namely XRS~14, 15, 23,
37, 43, 46, 54, 57, and 87. The FIRST images of the 91 objects we considered are
shown in Fig. \ref{all}.

The presence of distortions and/or radio structures, in addition to the
classical main double-lobed morphology, are often present in extended radio
sources, including the X-shaped morphology, but they also show S-shaped
structures, a ``bottle-neck'', and overall bendings and asymmetries of the
radio lobes. The initial \citeauthor{cheung07} sample is based on the presence
of an X-shaped radio morphology. However, classifying the morphology of
radio-galaxies into the various subclasses relies on subjective choices,
further complicated by projection effects. In addition, the depth and
resolution of the FIRST images does not always allow us to properly inspect
the structure of a given source. Indeed an object-by-object analysis shows
that, in some cases, such a structure is not sufficiently well defined. We
therefore prefer to filter the objects, preserving only those with the
clearest X-shaped morphology. This is also necessary to restrict the analysis
to the objects in which the wings are sufficiently well developed and defined
to allow us to derive their geometrical parameters (one of the main points of
our study) with a good degree of accuracy. Operatively we graded all sources
by using a score ranging from one (the best XRSs examples) to four (the less
convincing ones).  The grades are based on various morphological aspects, the
most important of which is that the lateral extension in the radio images
should not originate from the lobe end; in this case, rather common in the
sample (including, for example, XRS 01, XRS 10, XRS 21, and XRS 40), we
consider such objects as Z- or C-shaped sources. Furthermore, the lateral
extensions of bona fide XRSs must be located on the opposite sides of the main
radio axis. In several sources, however, the morphology is very complex and it
is difficult to recognize even the main lobes (such, as, for example, XRS 02,
XRS 39, and XRS 55). In a many objects, the radio structure suggests an
X-shaped morphology, but the FIRST resolution is insufficient to explore in
detail their structures (e.g., XRS 74 and XRS 95); these objects are not
included in the clean XRSs sample. The grades were associated with all 91
sources independently by the three authors, and we only kept the 53 galaxies
in the sample where the average grade was $\leq$ 2 (see Table \ref{tab1}).
While we were completing our analysis, \citet{roberts15} published an atlas of
radio maps of 52 sources from the \citeauthor{cheung07} sample (44 in common
with our subsample) mostly obtained at 1.4 GHz with a resolution of $\sim
1\arcsec$. These images allow us to test our initial classification based on
the lower resolution FIRST images, in particular for the objects with the
least extended radio emission. Although in several of the
\citeauthor{roberts15} images the low brightness extensions are resolved out,
they often improve our ability to recognize XRSs. The result of this
comparison is reported in Table 1. In two sources the XRS classification based
on the high-resolution images (and on the same criteria listed above) is less
secure; the opposite outcome occurs for five objects. We note that, with only
one exception, these are all sources with scores of 2.00 or 2.33, i.e.,
borderline objects. Since high resolution images are available for less than
half the sample, we prefer to maintain the classification based on the FIRST
images for the following analysis; nonetheless we will discuss the effects of
including/excluding the objects of uncertain nature.

\begin{figure*}
\includegraphics[width=90mm,angle=0]{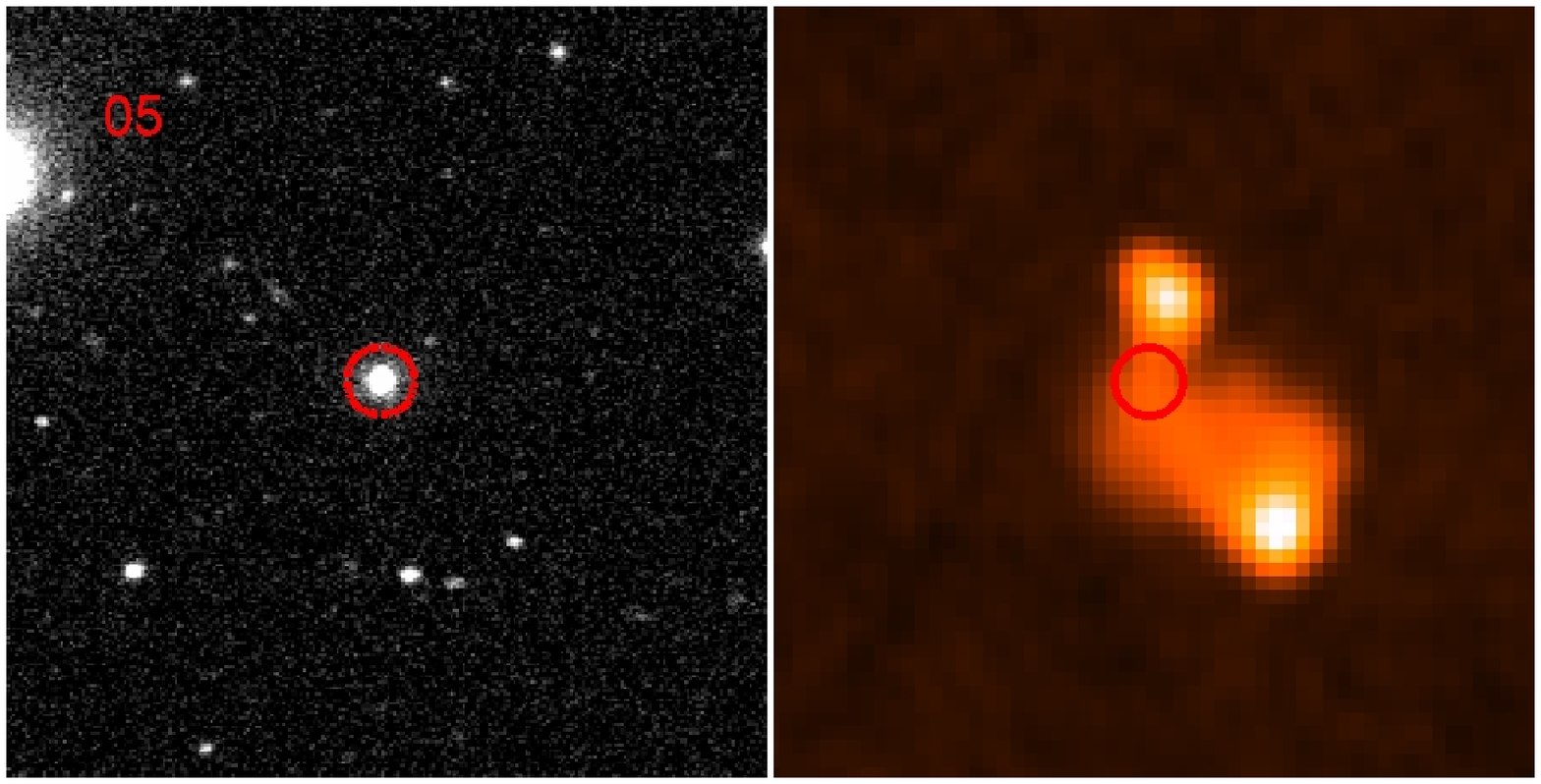}
\includegraphics[width=90mm,angle=0]{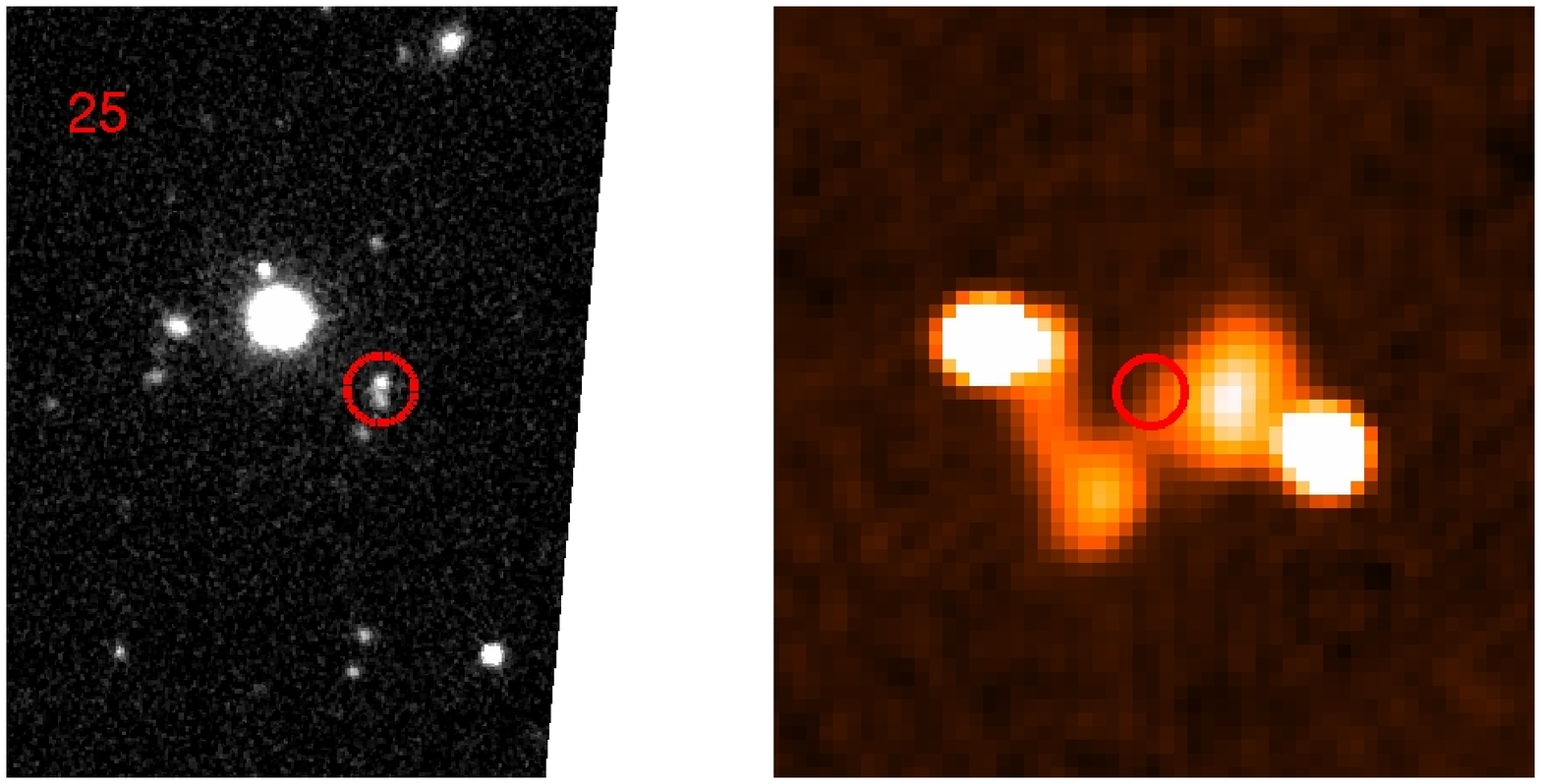}

\includegraphics[width=90mm,angle=0]{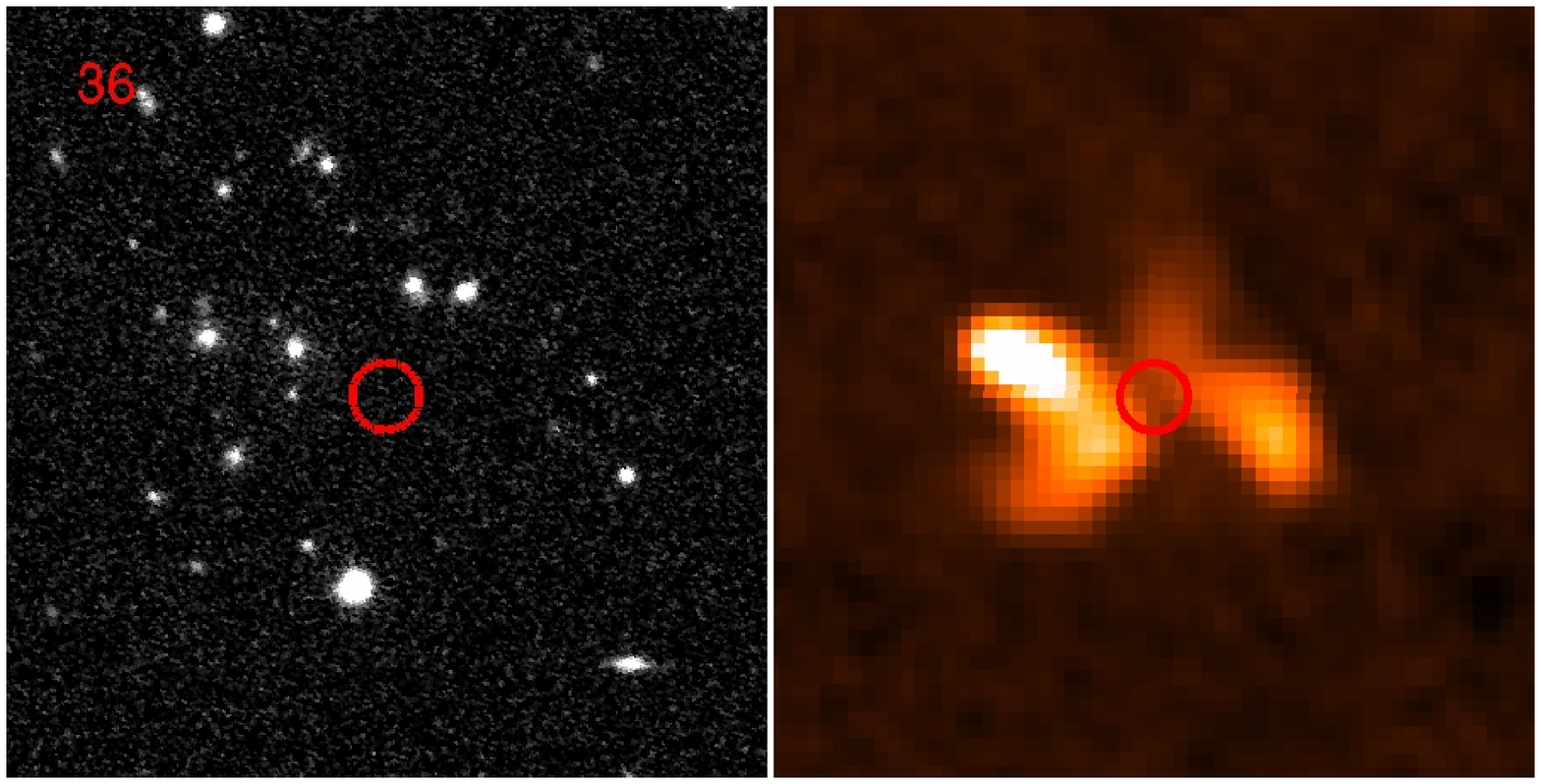}
\includegraphics[width=90mm,angle=0]{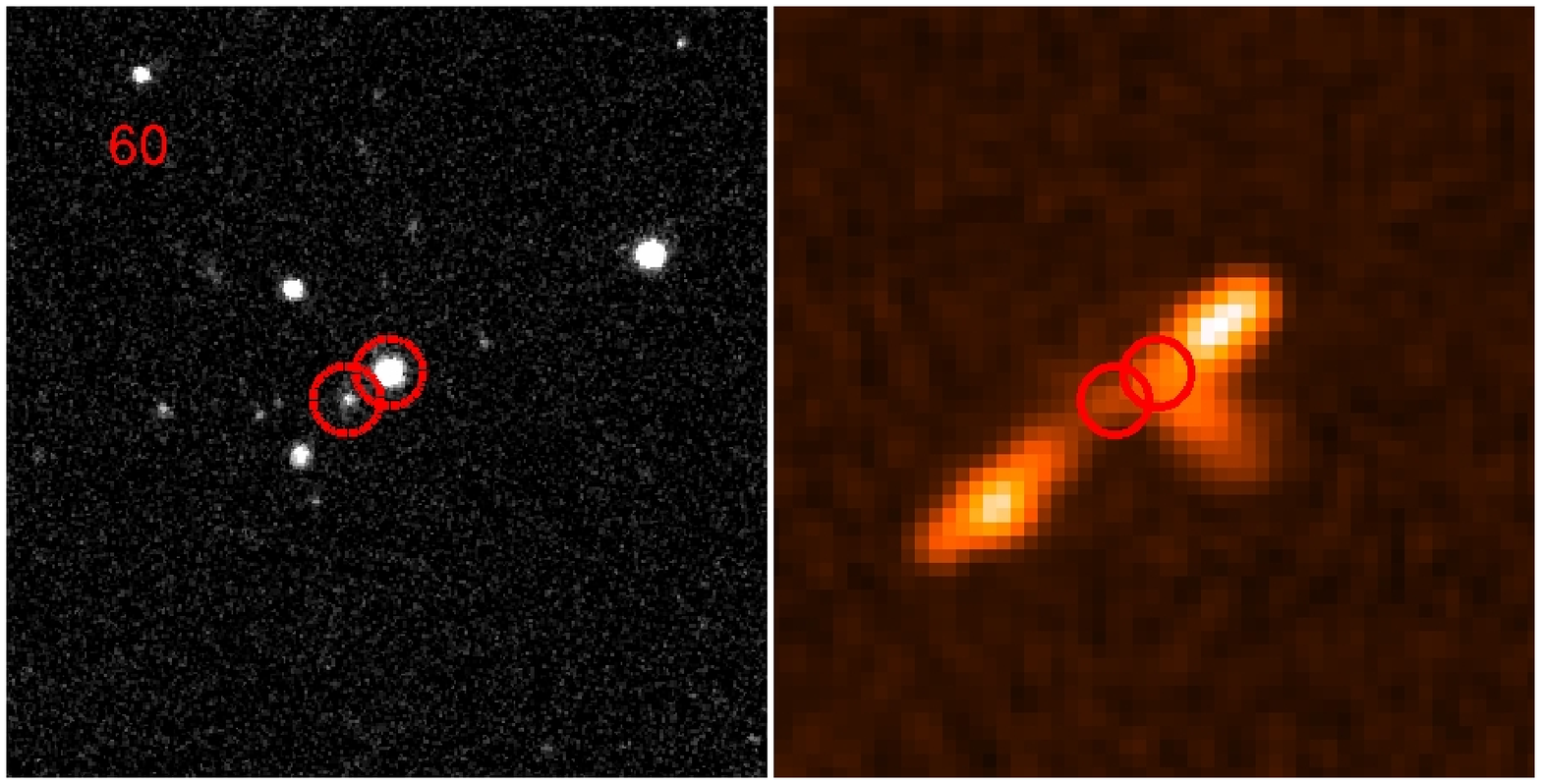}

\includegraphics[width=90mm,angle=0]{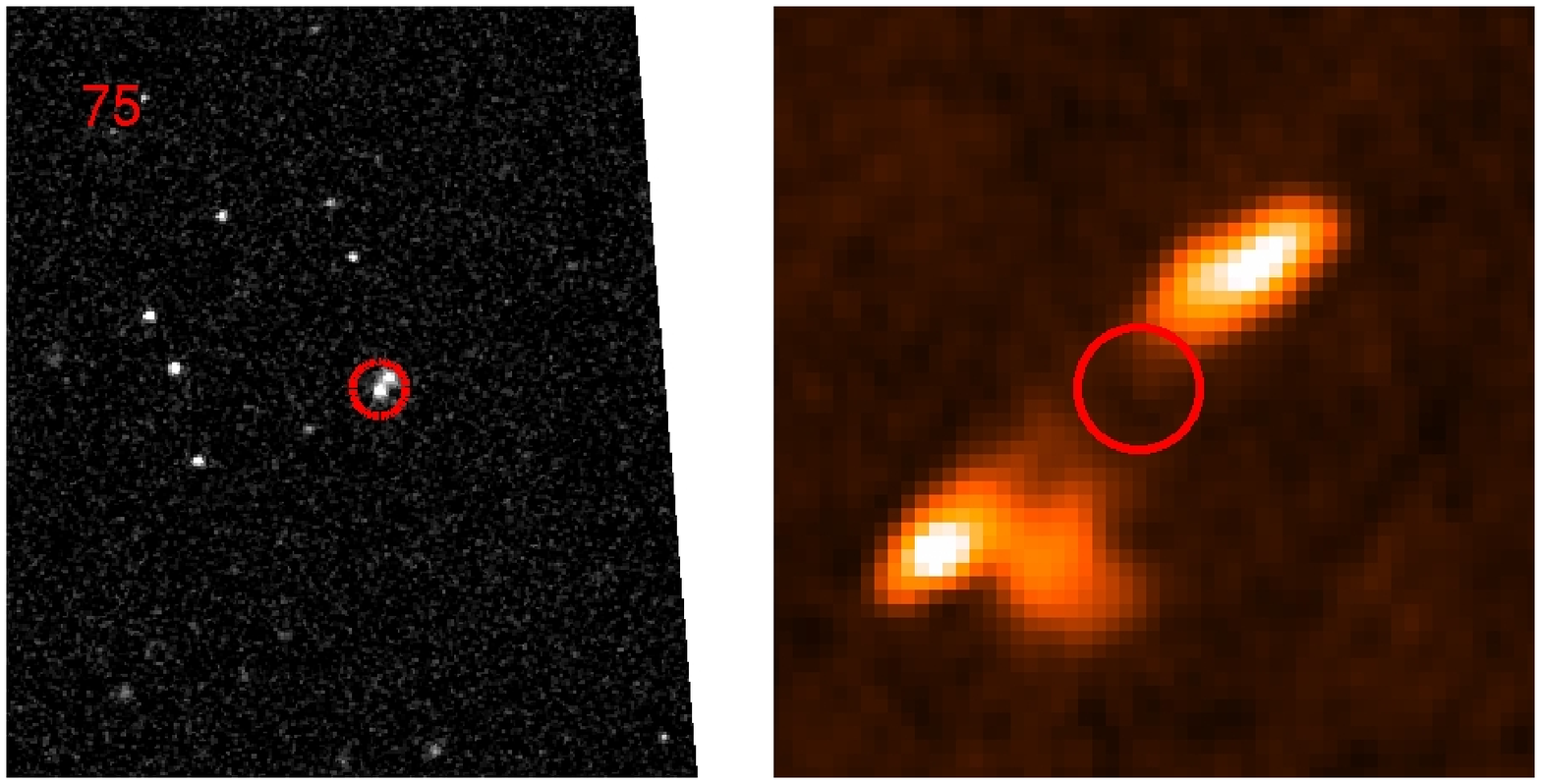}
\caption{The five cases of uncertain or unsuccessful optical counterpart
  identifications. The circle(s) indicate the location of the possible host(s) in
  both the optical SDSS (left) and radio FIRST (right) images.}
\label{identifications}
\end{figure*}

We then proceeded at the identification of the host of the XRSs in the SDSS
images. The host was sought close to the midpoint of the line joining the
peaks of the two main radio lobes. In most cases the association is
straightforward. Nonetheless, in XRS~05, the apparent host is significantly
offset from the radio axis, in three cases (XRS~25, XRS~60, and XRS~75) there is
more than one plausible optical identification (we then tentatively adopt the
closest to the center of the radio sources), while for XRS~36, no optical
emission is seen close to the center of the radio source (see
Fig. \ref{identifications}).

\section{Measurement of the optical position angle}
\label{optical}
We estimated the optical position angle, $\theta_{\rm {opt}}$, of the XRS's
host by using the SDSS images. In most cases, the $i$ band image was used
because of its higher signal-to-noise ratio (S/N). Only for XRS~06 did we instead use
the $r$-band image because its $i$-band image is strongly contaminated by an
emission line feature. 

\begin{figure*}
\centerline{
\includegraphics[width=65mm,angle=0]{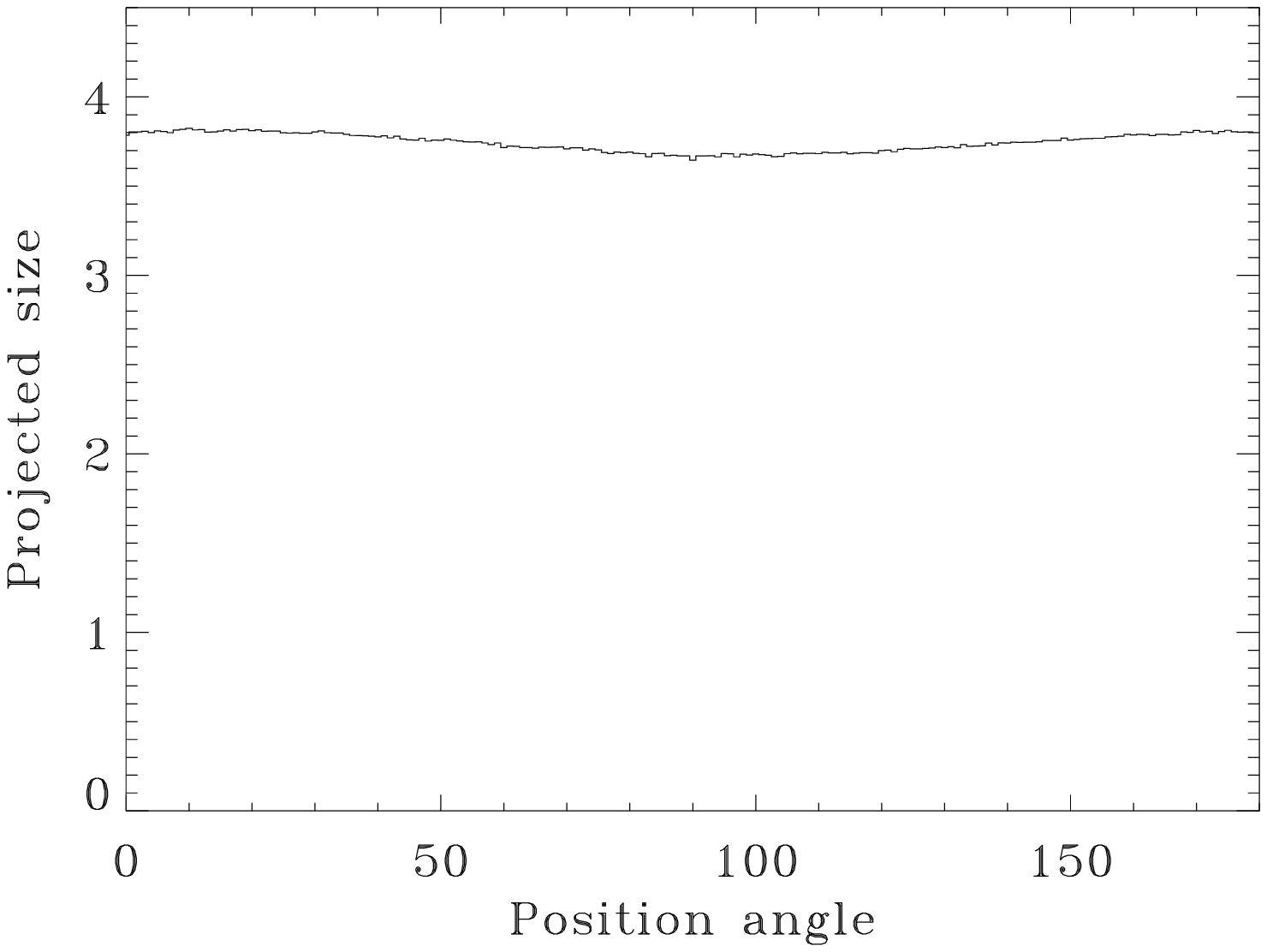}
\includegraphics[width=65mm,angle=0]{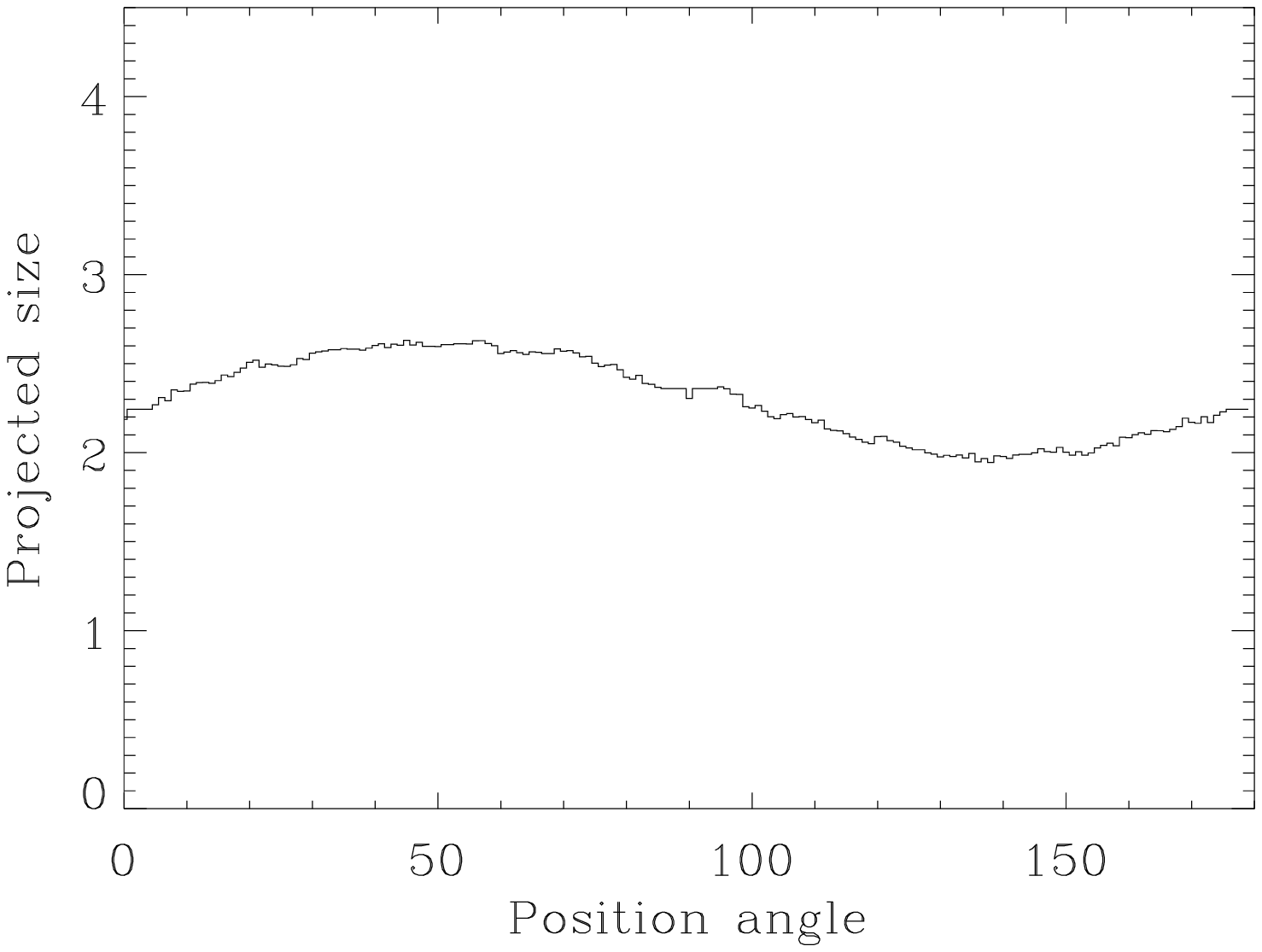}
\includegraphics[width=65mm,angle=0]{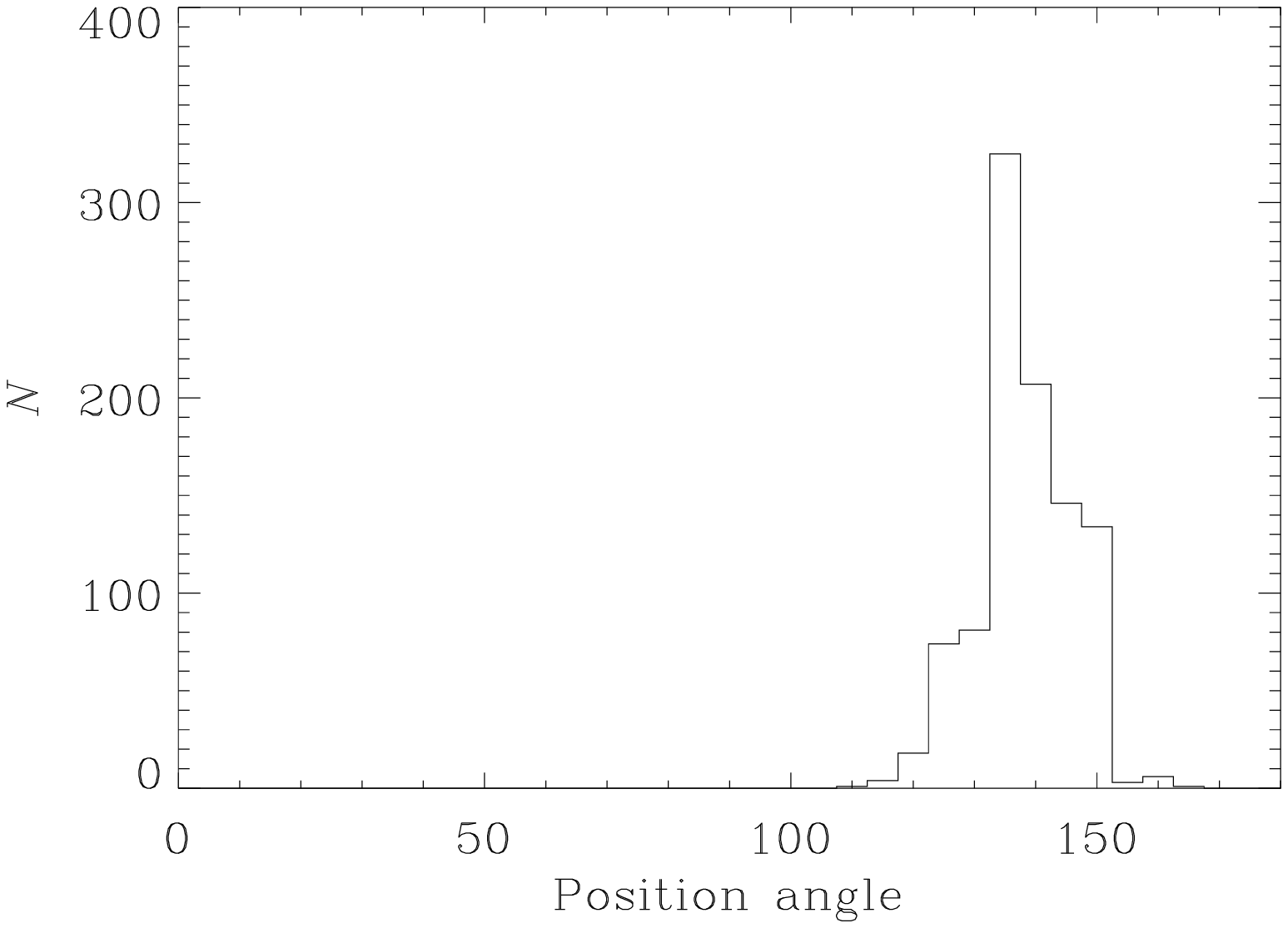}}
\caption{Left and central panels: galaxy's size (defined as 2.35 times the
  standard deviation in arcseconds of the projected profile) at varying angle
  of the projection axis for objects XRS~05 and XRS~07. Right panel:
  distribution of optical position angle, derived for XRS~07 from the
  bootstrap technique (see text for details).}
\label{varianza}
\end{figure*}

The classical approach to estimate $\theta_{\rm {opt}}$ is based on the
fitting of elliptical isophotes to the optical images. We followed this method
by using the IRAF task {\sl ellipse} \citep{jedrzejewski87}. However, in the
case of the XRS's hosts, this is successful for only a minority of objects,
mostly because of their rather small angular size. As a consequence, we prefer
to adopt a different strategy, free from any prior assumption on the galaxy's
shape, which is more robust and effective for small objects. For each galaxy,
we projected its optical image onto an axis whose orientation varies from
0$^{\circ}$ to 180$^{\circ}$. The orientation of the galaxy's major axis is
defined as the angle at which the width of the projected profile (measured as
the standard deviation of the one-dimensional image) reaches its maximum. In
Fig. \ref{varianza} we show two examples.

To estimate the error on the optical PA we used the bootstrap
technique. To each original image we added a random noise image (whose
amplitude is defined from a blank area close to the region of interest) and
re-estimated the PA. This procedure was repeated 1000 times. The PA error
is defined as the range  that includes 95 \% of the resulting values,
corresponding approximatively to a 2$\sigma$ level. The distribution of
PA derived for one galaxy is shown in Fig. \ref{varianza}, right panel. By
using the same method, we estimated the central value (and the error) of the
host's size (defined as 2.35 times the standard deviation of the projected
profile), its axial ratio, and the difference between the major and minor
axis.

In nine galaxies the optical source has an insufficient S/N to return
a useful measurement of its geometrical parameters and, as a result, the error
on their optical PA exceeds 45$^\circ$. These sources are marked as ``faint''
(or ``undetected'') in Table \ref{tab2}. 

For a successful measurement of the optical position angle we adopted the
following criteria: 1) the axial ratio of the source must be higher than 1.05
and 2) the difference between the width along the minor and major axis must be
greater than 0\farcs15. In both cases, the requirements must be met at least
at the 2 $\sigma$ level. The results are tabulated in Table \ref{tab2}. We
have a positive outcome for 22 galaxies, for which we report the optical PA
and its error in Table \ref{tab3}. We note that i) all galaxies that meet the
first requirement also meet the second, indicating that the check based on the
axial ratio is actually more restrictive and ii) by adopting lower thresholds
(1.03 for the axial ratio and 0\farcs1 for the axis difference) no additional
source would be found. We now consider the XRSs with uncertain optical
counterparts, as discussed in Sect. \ref{sample}, namely XRS~25, XRS~60, and
XRS~75. Because it is impossible to measure $\theta_{\rm {opt}}$ for any of
them, they do not represent an issue for the analysis from this point of view.

The method adopted to measure $\theta_{\rm {opt}}$, although efficient and
robust, has the drawback of providing measurements of the galaxy's axis and
ellipticity that are not straightforward to compare with previous
analysis. Consequently, we present two tests to better understand our results.

We produced images of a gaussian source with a FWHM of 1$\arcsec$ and applied
the same procedure used for the galaxies and obtained a size of $0\farcs76$,
significantly smaller than the input width. This is understandable given that
the one-dimensional projection returns a more concentrated profile with
respect to the radial profile than is normally used to measure the FWHM.

We also test model galaxies at various distances. More specifically, we
produce images of elliptical galaxies following a De Vaucouleur's law, with an
effective radius of 4.6 kpc, the value measured by \citet{donzelli07} for 3C
radio-galaxies. The model is then convolved with a gaussian with FWHM of
$1\farcs5$ to simulate the effects of seeing. For a galaxy at $z=0.4$ (as we
will see, typical of our sample) and with an ellipticity of 0.2, the galaxy's
size is $\sim 2\farcs2$, the difference between the two axis is $0\farcs23$,
and the axial ratio is $1.11$. Such a galaxy, provided that the
signal-to-noise ratio of its image is sufficient, would correspond to a
positive outcome. However, for an ellipticity of 0.1, the axial ratio is only
1.06 and the axis difference is $0\farcs12$; the analysis of such a source
would not return an optical position angle.

\begin{table*}
\begin{center}
\caption{Parameters for the 53 X-shaped RG with SDSS images.}
\begin{tabular}{c|c|c|c|c|c|c}
\hline
\hline
ID  & redshift &  $F_{\rm FIRST}$ (mJy) &  A ($\arcsec$)  & A/a  & A-a ($\arcsec$)      & Outcome  \\
\hline                      
XRS~03 &0.591                           & 280& 2.91 $\pm$0.06  & 1.12 $\pm$ 0.04 &  0.31 $\pm$0.09 & $\checkmark$ \\    
XRS~04 &0.51$\pm$0.17 / 0.59$\pm$0.28   & 509&                 &                 &                 & faint  \\       
XRS~05 &0.304                           & 155& 3.85 $\pm$0.03  & 1.04 $\pm$ 0.01 &  0.15 $\pm$0.05 & \text{\sffamily X}  \\       
XRS~06 &0.281                           & 391& 2.80 $\pm$0.02  & 1.20 $\pm$ 0.01 &  0.47 $\pm$0.03 & $\checkmark$ \\    
XRS~07 &0.381$^a$                       & 222& 2.62 $\pm$0.09  & 1.29 $\pm$ 0.09 &  0.59 $\pm$0.16 & $\checkmark$ \\    
XRS~11 &0.68$\pm$0.10 / 0.61$\pm$0.16   & 306& 2.60 $\pm$0.06  & 1.22 $\pm$ 0.05 &  0.46 $\pm$0.10 & $\checkmark$ \\    
XRS~12 &0.22$\pm$0.02 / 0.21$\pm$0.03   & 180& 2.18 $\pm$0.01  & 1.03 $\pm$ 0.01 &  0.06 $\pm$0.01 & \text{\sffamily X}  \\     
XRS~13 &0.659                           & 318& 2.46 $\pm$0.09  & 1.10 $\pm$ 0.07 &  0.25 $\pm$0.13 & \text{\sffamily X} \\       
XRS~19 &0.376                           & 291& 2.70 $\pm$0.04  & 1.06 $\pm$ 0.03 &  0.14 $\pm$0.06 & \text{\sffamily X} \\       
XRS~20 &0.213                           & 121& 3.37 $\pm$0.01  & 1.02 $\pm$ 0.01 &  0.06 $\pm$0.01 & \text{\sffamily X}  \\       
XRS~22 &0.34$\pm$0.12 / 0.07$\pm$0.03   & 197&                 &                 &                 & faint  \\       
XRS~24 &0.520                           & 527& 3.25 $\pm$0.07  & 1.16 $\pm$ 0.05 &  0.45 $\pm$0.12 & $\checkmark$ \\    
XRS~25 &0.591                           & 612&                 &                 &                 & faint  \\       
XRS~26 &0.227                           & 292& 3.53 $\pm$0.03  & 1.03 $\pm$ 0.01 &  0.11 $\pm$0.05 & \text{\sffamily X}  \\       
XRS~28 &---                             & 206&                 &                 &                 & faint  \\       
XRS~29 &0.73$\pm$0.09 / 0.65$\pm$0.20   & 647& 2.64 $\pm$0.07  & 1.06 $\pm$ 0.04 &  0.16 $\pm$0.10 & \text{\sffamily X} \\       
XRS~30 &0.166                           & 205& 3.34 $\pm$0.01  & 1.04 $\pm$ 0.01 &  0.13 $\pm$0.01 & \text{\sffamily X}  \\       
XRS~35 &0.42$\pm$0.08 / 0.33$\pm$0.16   & 144& 2.48 $\pm$0.07  & 1.01 $\pm$ 0.05 &  0.04 $\pm$0.11 & \text{\sffamily X}  \\       
XRS~36 &---                             & 222&                 &                 &                 & undet.  \\  
XRS~44 &0.081                           & 537& 8.08 $\pm$0.02  & 1.20 $\pm$ 0.01 &  1.35 $\pm$0.04 & $\checkmark$ \\    
XRS~45 &0.30$\pm$0.05 / 0.29$\pm$0.09   & 278& 3.13 $\pm$0.02  & 1.07 $\pm$ 0.04 &  0.21 $\pm$0.03 & \text{\sffamily X} \\       
XRS~47 &0.72$\pm$0.08 / 0.60$\pm$0.20   & 104&                 &                 &                 & faint  \\       
XRS~48 &0.838                           & 247& 1.80 $\pm$0.03  & 1.01 $\pm$ 0.02 &  0.02 $\pm$0.04 & \text{\sffamily X}  \\       
XRS~49 &0.079                           & 490& 6.44 $\pm$0.01  & 1.11 $\pm$ 0.01 &  0.64 $\pm$0.01 & $\checkmark$ \\    
XRS~50 &0.178$^a$                       & 151& 3.89 $\pm$0.03  & 1.09 $\pm$ 0.01 &  0.32 $\pm$0.05 & $\checkmark$ \\    
XRS~52 &0.75$\pm$0.10 / 0.56$\pm$0.25   & 232&                 &                 &                 & faint  \\       
XRS~53 &0.424                           &  47& 2.45 $\pm$0.05  & 1.05 $\pm$ 0.03 &  0.12 $\pm$0.06 & \text{\sffamily X} \\       
XRS~58 &0.61$\pm$0.13 / 0.61$\pm$0.11   & 135&                 &                 &                 & faint  \\       
XRS~59 &0.358$^a$                       & 362& 2.82 $\pm$0.04  & 1.07 $\pm$ 0.03 &  0.17 $\pm$0.07 & \text{\sffamily X} \\       
XRS~60 &---                             & 149& 1.71 $\pm$0.01  & 1.01 $\pm$ 0.01 &  0.02 $\pm$0.01 & \text{\sffamily X} \\       
XRS~61 &0.419$^a$                       & 637& 2.49 $\pm$0.03  & 1.09 $\pm$ 0.02 &  0.20 $\pm$0.04 & $\checkmark$ \\    
XRS~63 &0.447                           &  75& 2.55 $\pm$0.03  & 1.15 $\pm$ 0.03 &  0.34 $\pm$0.05 & $\checkmark$ \\    
XRS~64 &0.183                           & 179& 3.52 $\pm$0.01  & 1.03 $\pm$ 0.01 &  0.09 $\pm$0.01 & \text{\sffamily X}  \\       
XRS~67 &0.585$^a$                       & 365& 1.89 $\pm$0.03  & 1.09 $\pm$ 0.03 &  0.15 $\pm$0.05 & \text{\sffamily X} \\       
XRS~68 &0.38$\pm$0.08 / 0.35$\pm$0.12   &  71& 2.68 $\pm$0.14  & 1.08 $\pm$ 0.08 &  0.20 $\pm$0.20 & \text{\sffamily X} \\       
XRS~71 &0.26$\pm$0.06 / 0.26$\pm$0.08   & 294& 3.18 $\pm$0.02  & 1.49 $\pm$ 0.03 &  1.05 $\pm$0.04 & $\checkmark$ \\    
XRS~72 &0.641                           & 143& 2.63 $\pm$0.07  & 1.08 $\pm$ 0.05 &  0.20 $\pm$0.11 & \text{\sffamily X} \\       
XRS~73 &0.550                           & 339& 1.85 $\pm$0.04  & 1.05 $\pm$ 0.03 &  0.09 $\pm$0.05 & \text{\sffamily X}  \\       
XRS~75 &0.43$\pm$0.15 / 0.49$\pm$0.08   & 260&                 &                 &                 & faint  \\       
XRS~76 &0.037                           & 857& 5.10 $\pm$0.01  & 1.06 $\pm$ 0.01 &  0.27 $\pm$0.01 & $\checkmark$ \\    
XRS~77 &0.367                           & 671& 2.56 $\pm$0.04  & 1.19 $\pm$ 0.04 &  0.41 $\pm$0.07 & $\checkmark$ \\    
XRS~78 &0.504                           & 275& 2.66 $\pm$0.06  & 1.01 $\pm$ 0.03 &  0.02 $\pm$0.09 & \text{\sffamily X}  \\       
XRS~79 &0.538$^a$                       & 315& 2.77 $\pm$0.08  & 1.16 $\pm$ 0.05 &  0.39 $\pm$0.12 & $\checkmark$ \\    
XRS~80 &0.38$\pm$0.07 / 0.26$\pm$0.11   & 208& 2.75 $\pm$0.04  & 1.06 $\pm$ 0.02 &  0.16 $\pm$0.06 & \text{\sffamily X}  \\       
XRS~81 &0.188                           & 372& 2.67 $\pm$0.01  & 1.14 $\pm$ 0.01 &  0.33 $\pm$0.01 & $\checkmark$ \\    
XRS~83 &0.084                           & 107& 4.32 $\pm$0.01  & 1.16 $\pm$ 0.01 &  0.60 $\pm$0.01 & $\checkmark$ \\    
XRS~84 &0.536                           &  36& 2.52 $\pm$0.07  & 1.05 $\pm$ 0.05 &  0.11 $\pm$0.11 & \text{\sffamily X} \\       
XRS~85 &0.146$^a$                       & 367& 4.86 $\pm$0.01  & 1.17 $\pm$ 0.01 &  0.70 $\pm$0.02 & $\checkmark$ \\    
XRS~86 &0.659                           & 509& 2.86 $\pm$0.07  & 1.21 $\pm$ 0.06 &  0.49 $\pm$0.12 & $\checkmark$ \\    
XRS~90 &0.174                           & 523& 2.80 $\pm$0.01  & 1.13 $\pm$ 0.01 &  0.31 $\pm$0.01 & $\checkmark$ \\    
XRS~91 &0.30$\pm$0.13 / 0.37$\pm$0.10   & 348& 2.59 $\pm$0.01  & 1.09 $\pm$ 0.01 &  0.21 $\pm$0.02 & $\checkmark$ \\    
XRS~93 &0.556                           & 116& 2.65 $\pm$0.07  & 1.26 $\pm$ 0.06 &  0.55 $\pm$0.11 & $\checkmark$ \\    
XRS~94 &0.107                           & 424& 3.33 $\pm$0.01  & 1.06 $\pm$ 0.01 &  0.18 $\pm$0.01 & $\checkmark$ \\    
\hline
\end{tabular}
\label{tab2}
\end{center}
\medskip
\noindent
\small{1) Galaxy ID; 2) spectroscopic redshift ($^a$ mark objects with
  redshift from \citealt{landt10}) or, when not available, photometric
  redshifts from the KD and RF methods, respectively; 3) radio flux from the
  FIRST catalog; 4) optical major axis (all errors quoted are at the 2
  $\sigma$ level; 5) axial ratio; 6) axis difference; 7) outcome of the
  measurement of the optical axis.}
\end{table*}

\begin{figure*}
\centerline{
\includegraphics[width=78mm,angle=0]{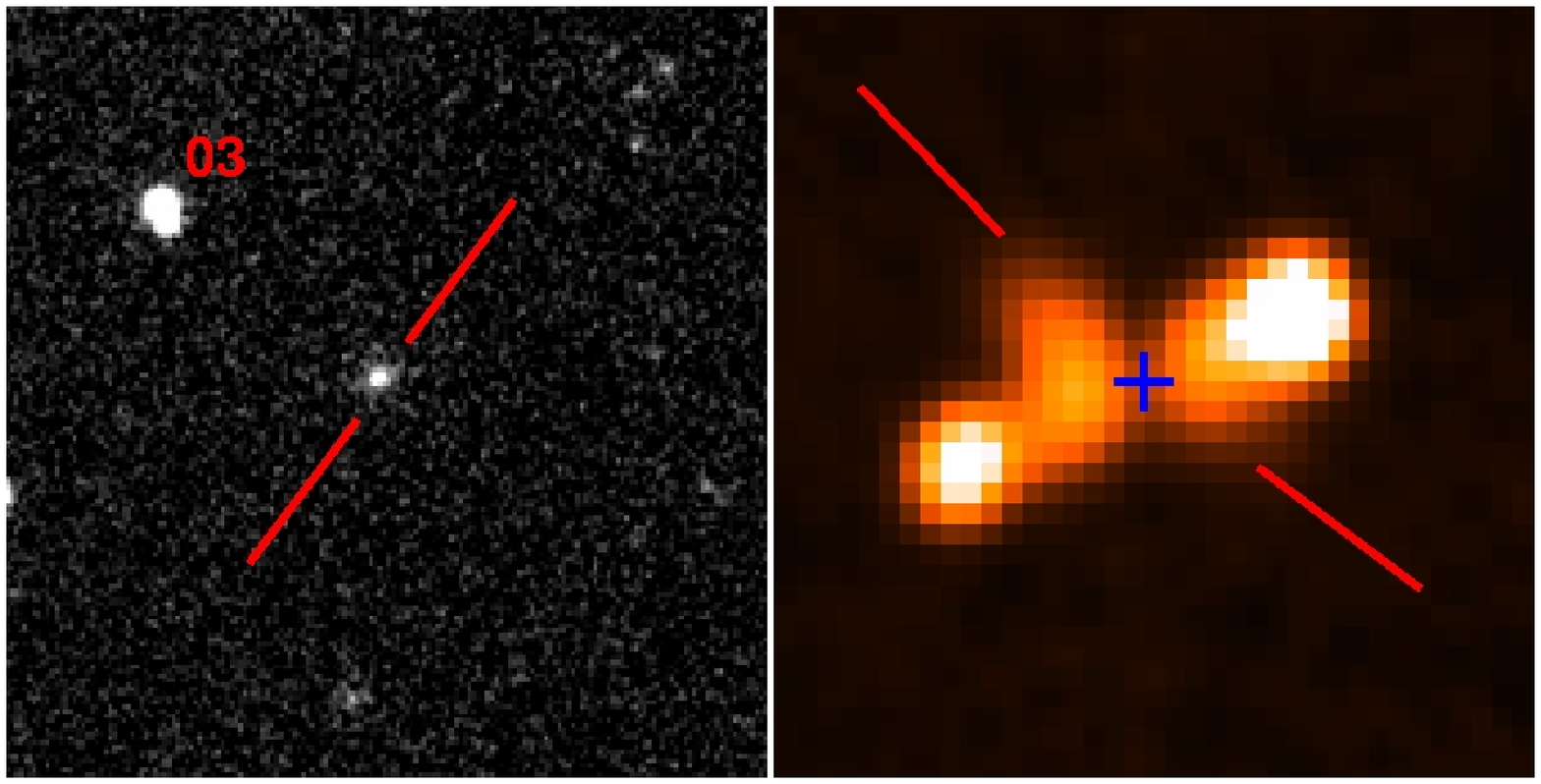}
\includegraphics[width=78mm,angle=0]{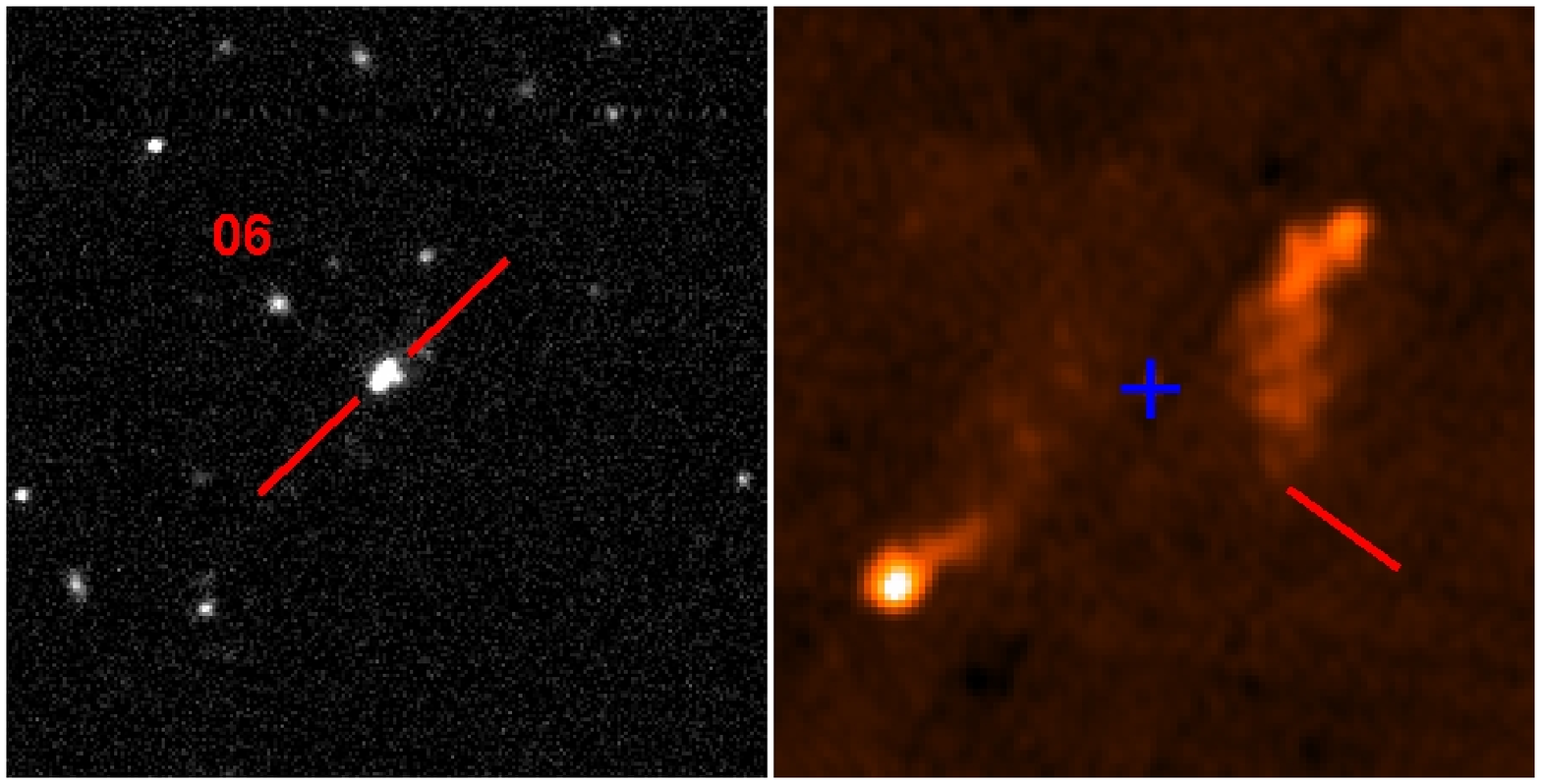}}
\centerline{
\includegraphics[width=78mm,angle=0]{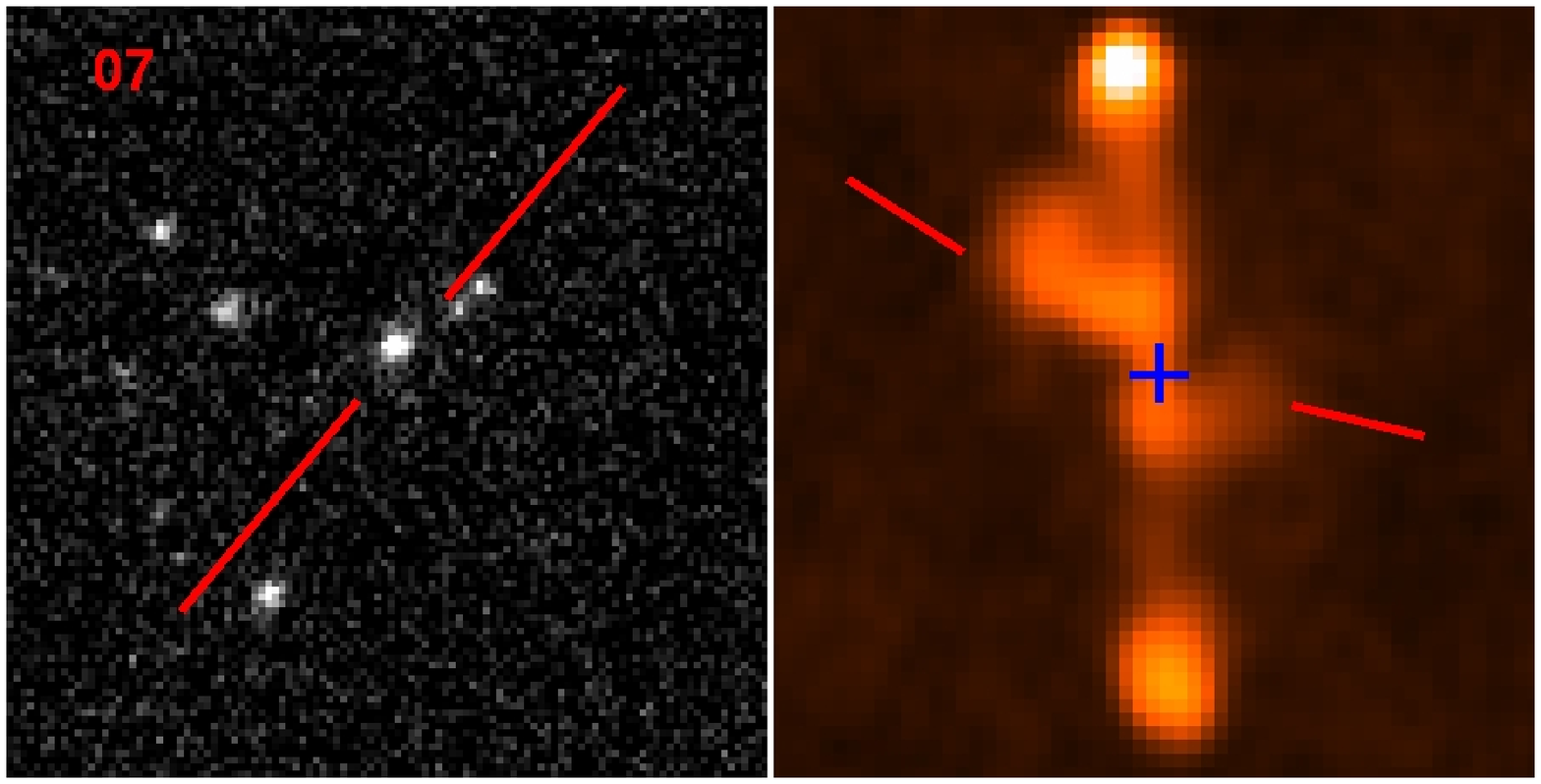}
\includegraphics[width=78mm,angle=0]{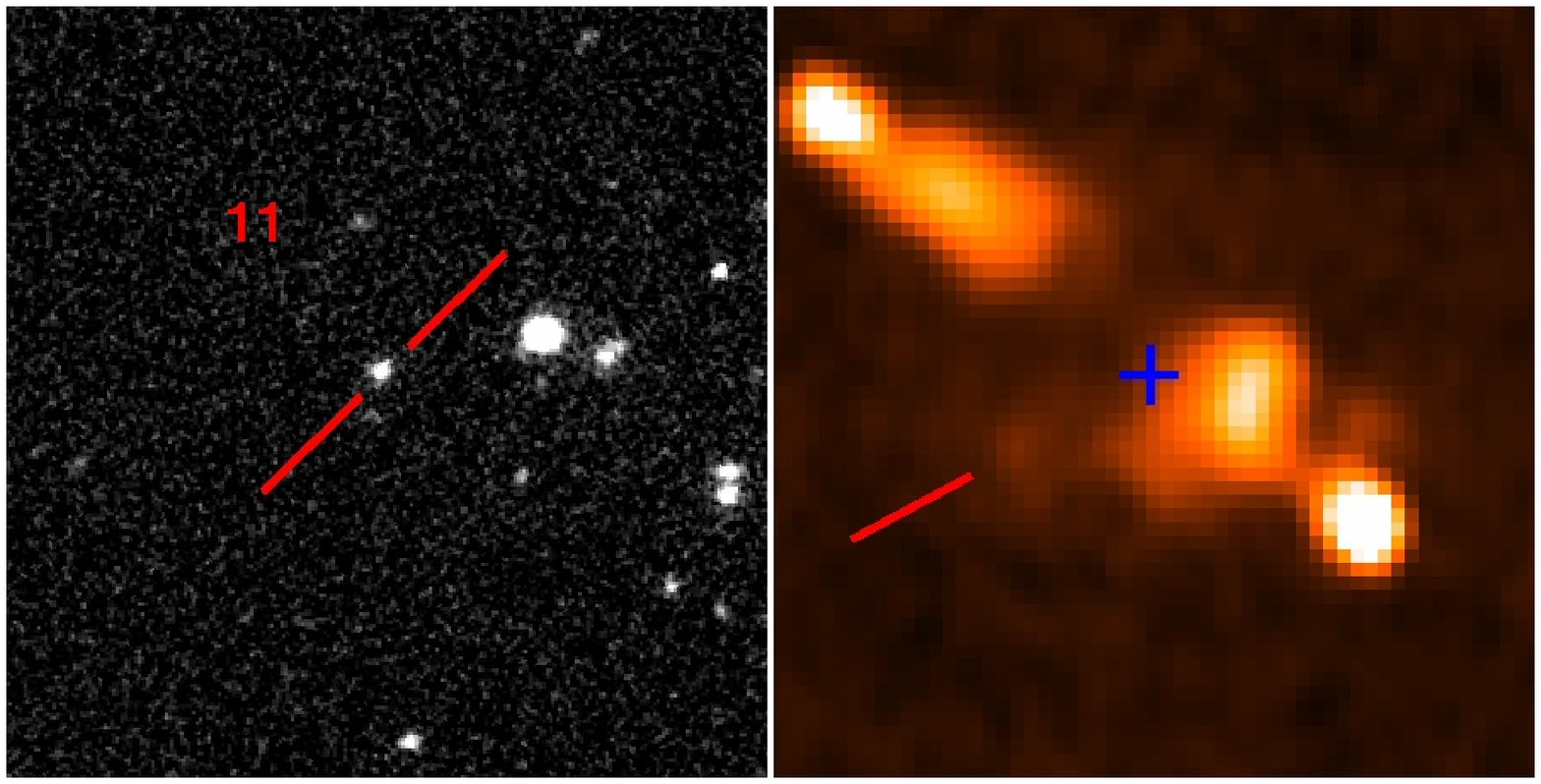}}
\centerline{
\includegraphics[width=78mm,angle=0]{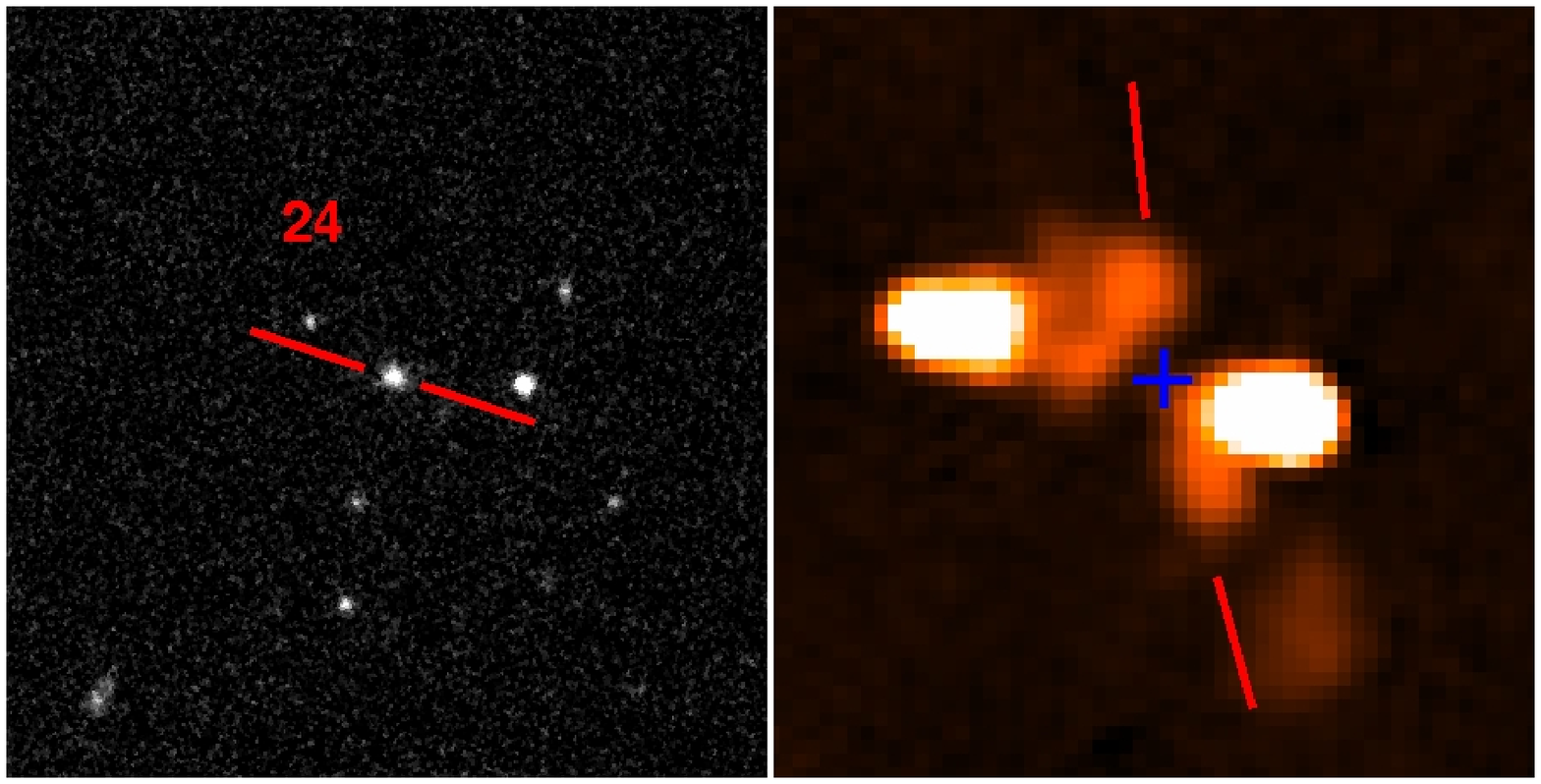}
\includegraphics[width=78mm,angle=0]{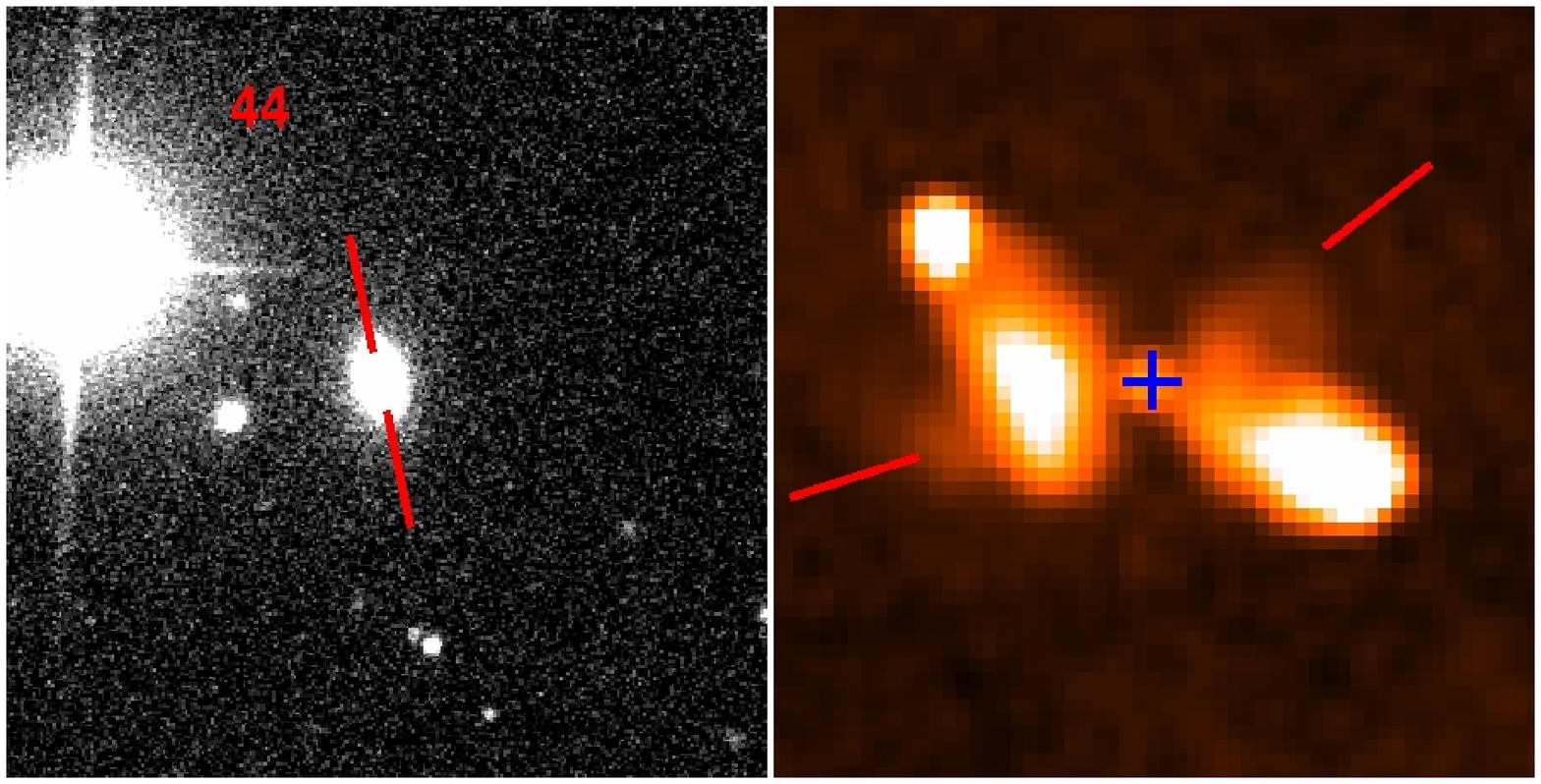}}
\centerline{
\includegraphics[width=78mm,angle=0]{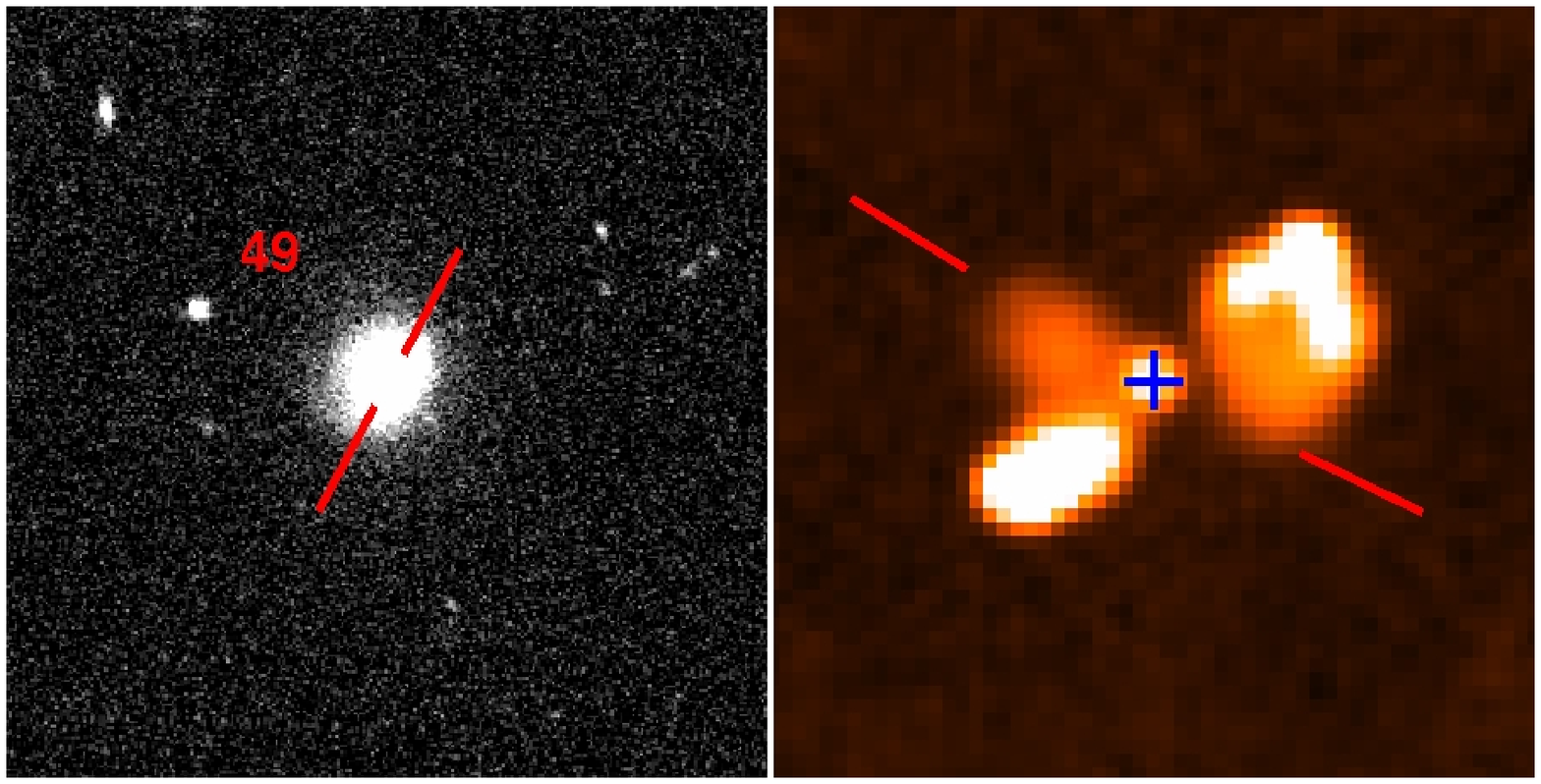}
\includegraphics[width=78mm,angle=0]{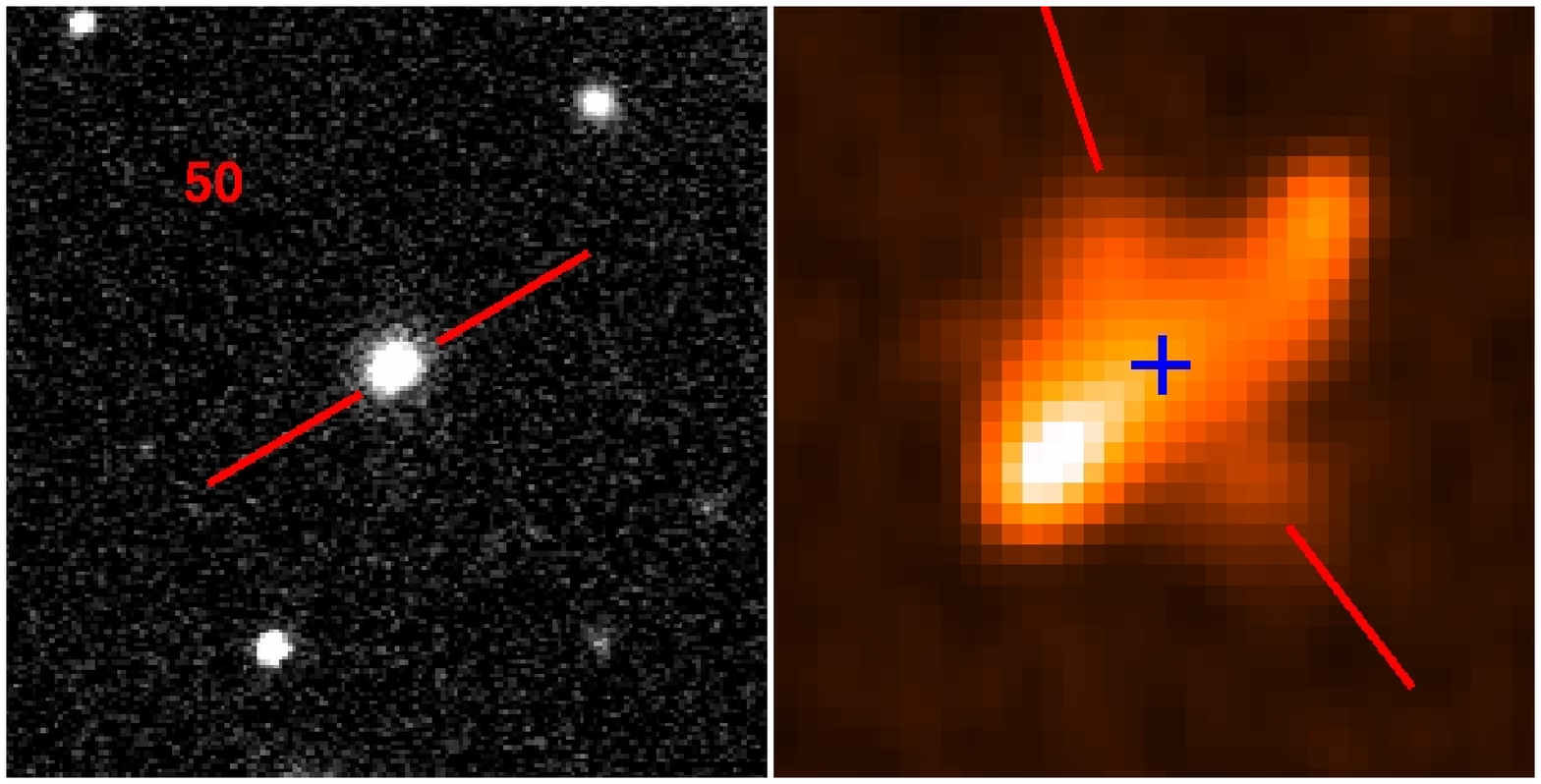}}
\centerline{
\includegraphics[width=78mm,angle=0]{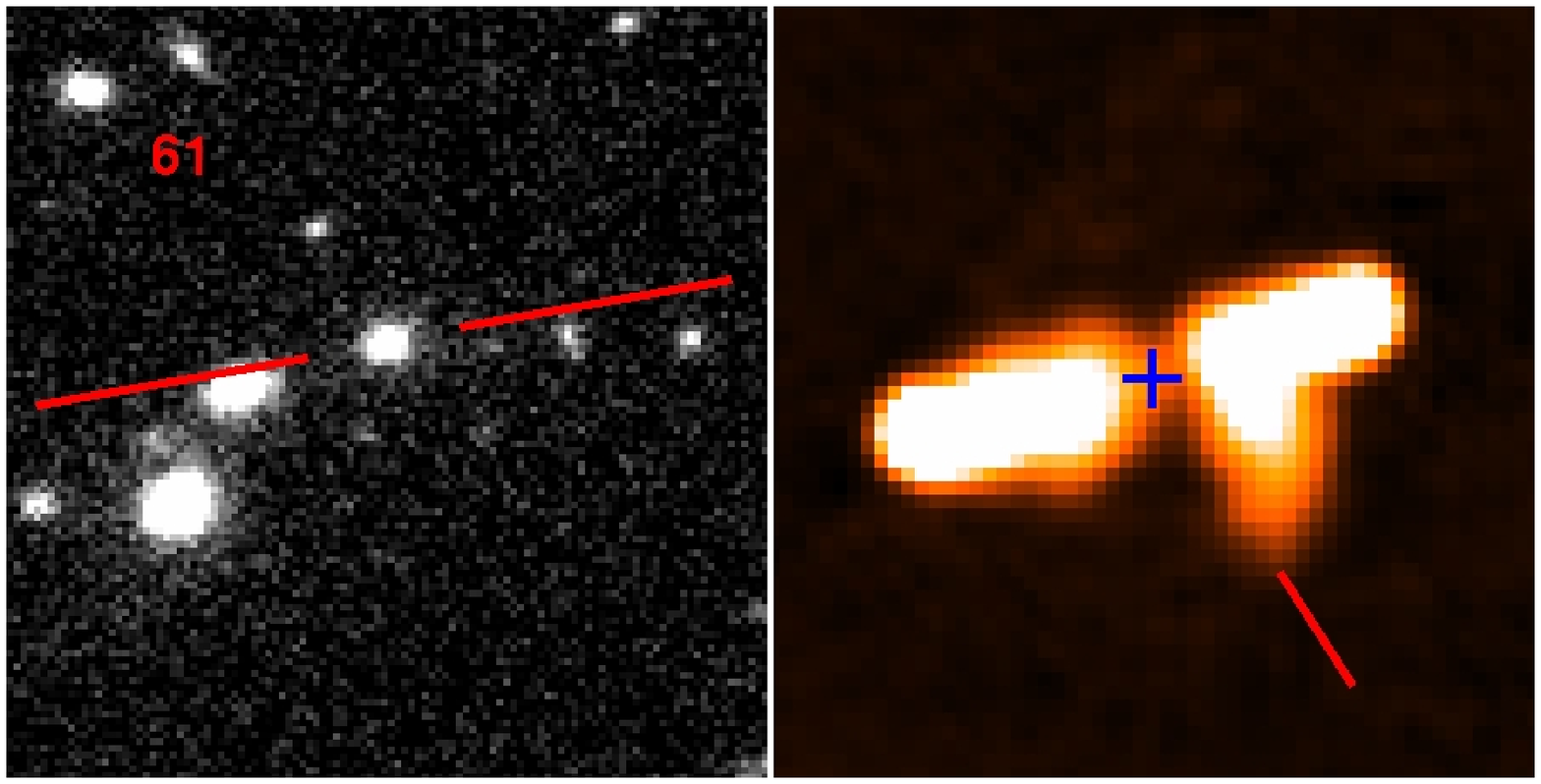}
\includegraphics[width=78mm,angle=0]{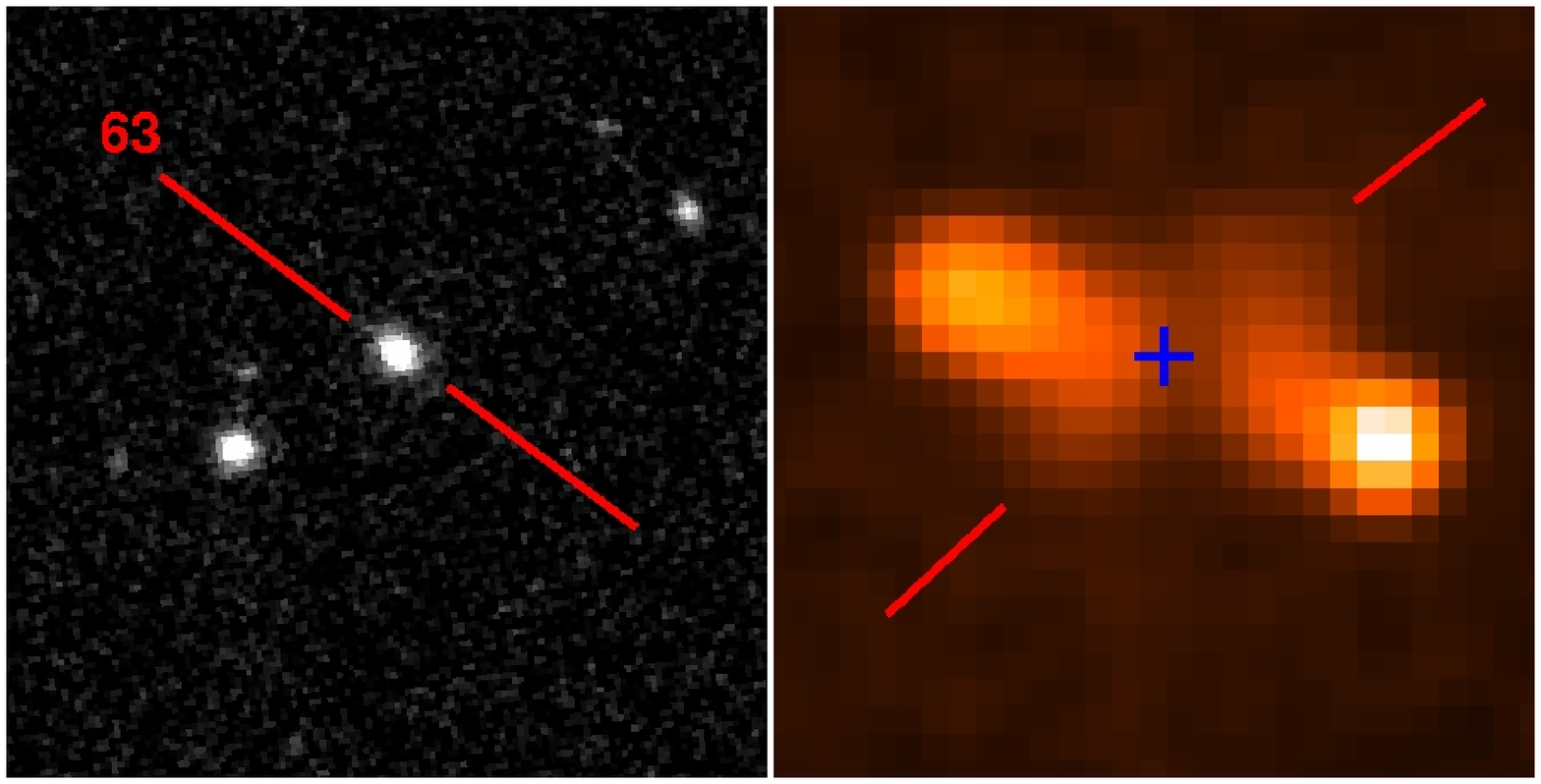}}
\centerline{
\includegraphics[width=78mm,angle=0]{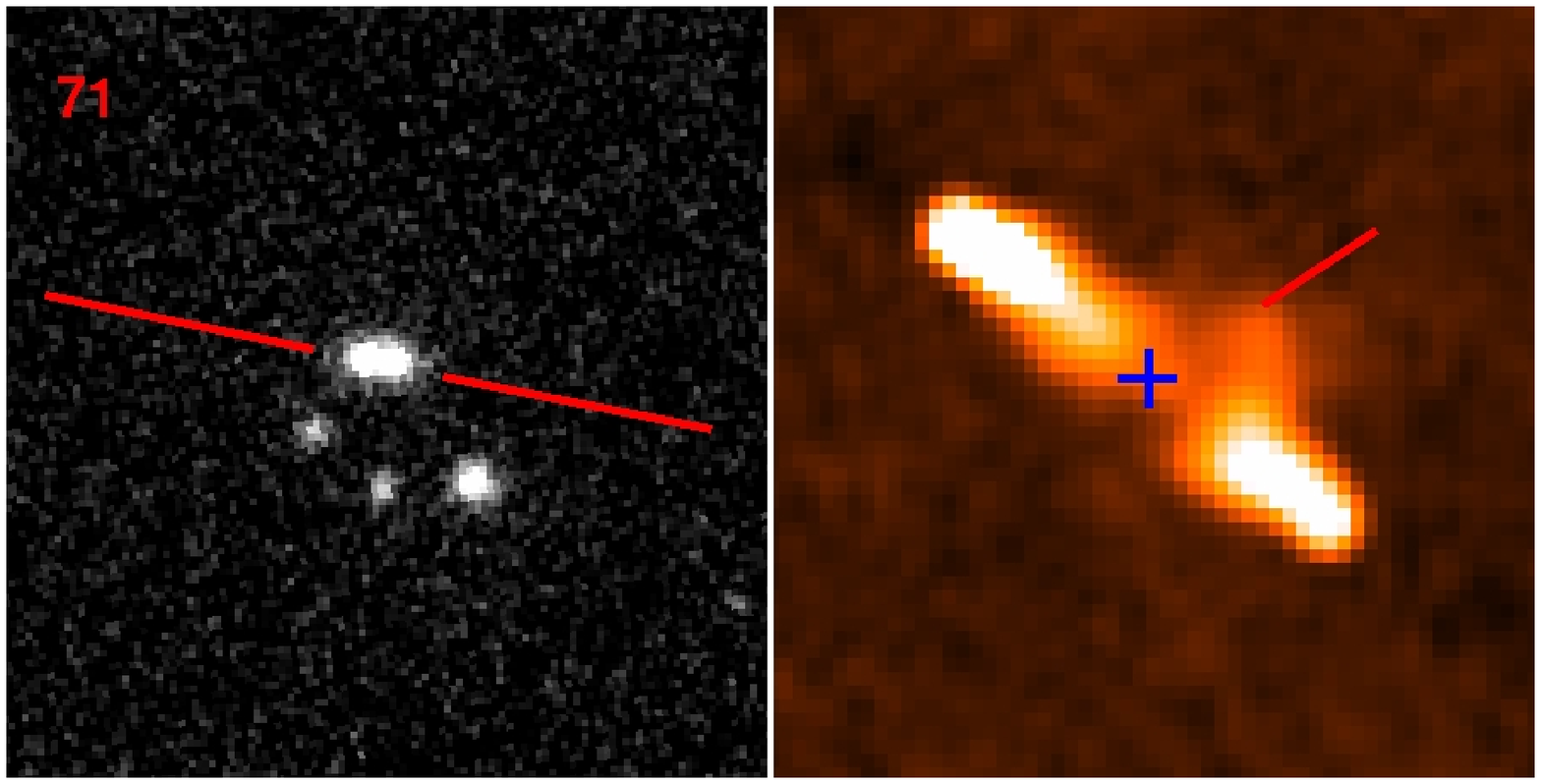}
\includegraphics[width=78mm,angle=0]{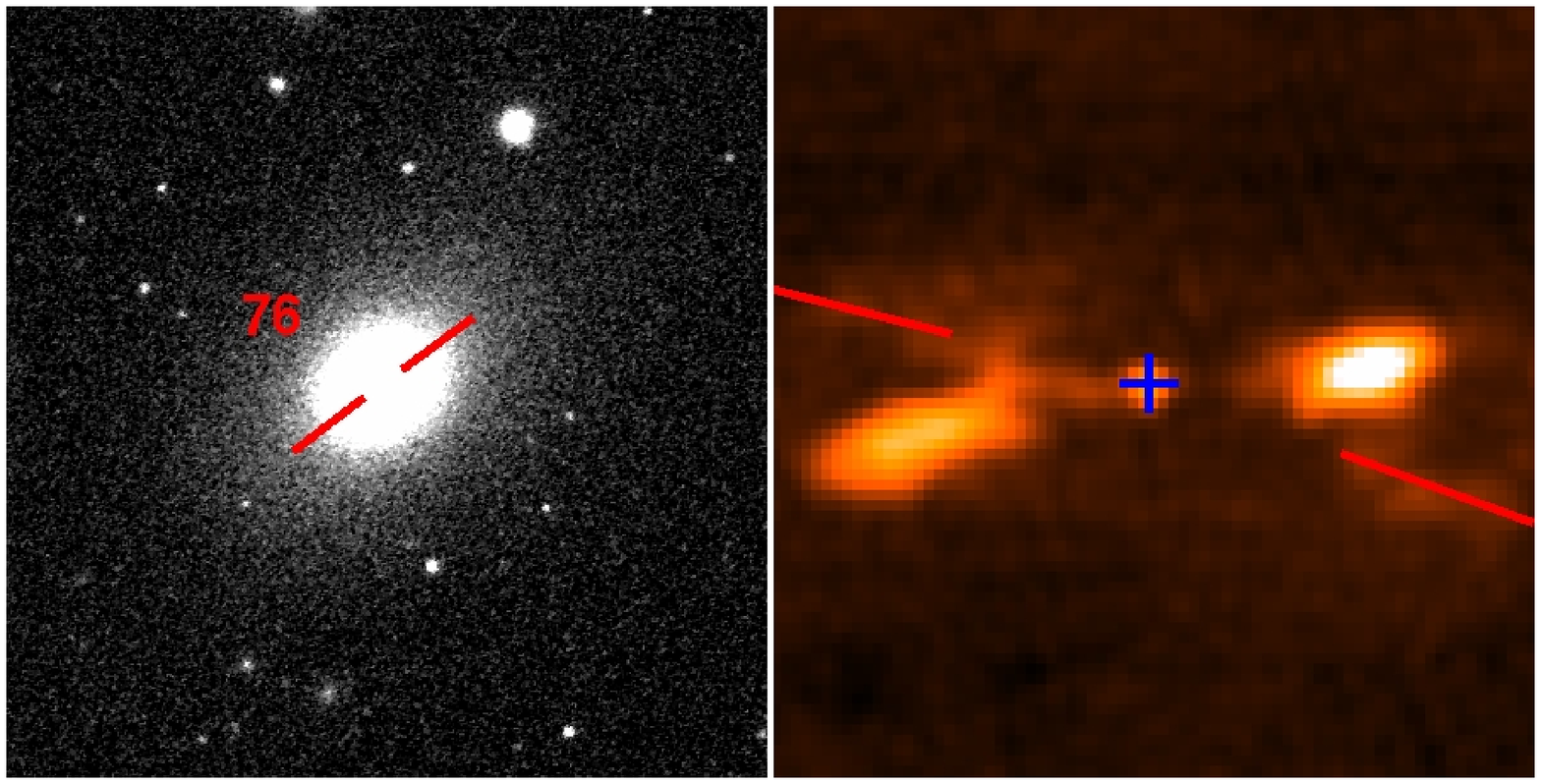}}
\caption{SDSS (left) and FIRST (right) images of the 22 X-shaped galaxies for
  which the optical position angle can be measured. In the SDSS images the
  host is shown by two red ticks, each of them 16$\arcsec$ long and separated
  by 8$\arcsec$. The tick(s) overplotted onto the FIRST image indicated the PA of the
  radio wing(s), while the blue cross is at the location of the optical
  counterpart.}
\label{cfr}
\end{figure*}

\addtocounter{figure}{-1}
\begin{figure*}
\centerline{
\includegraphics[width=78mm,angle=0]{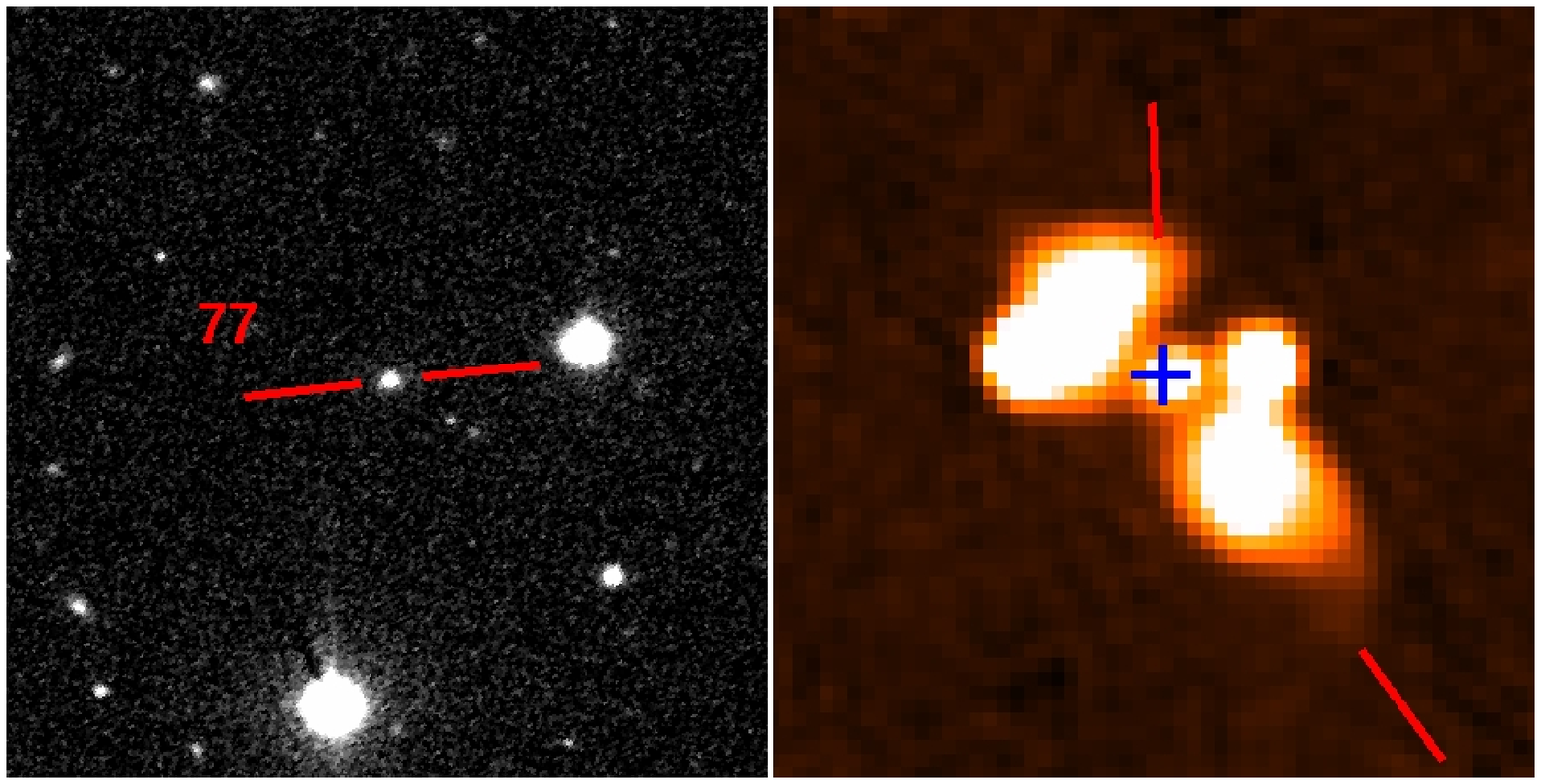}
\includegraphics[width=78mm,angle=0]{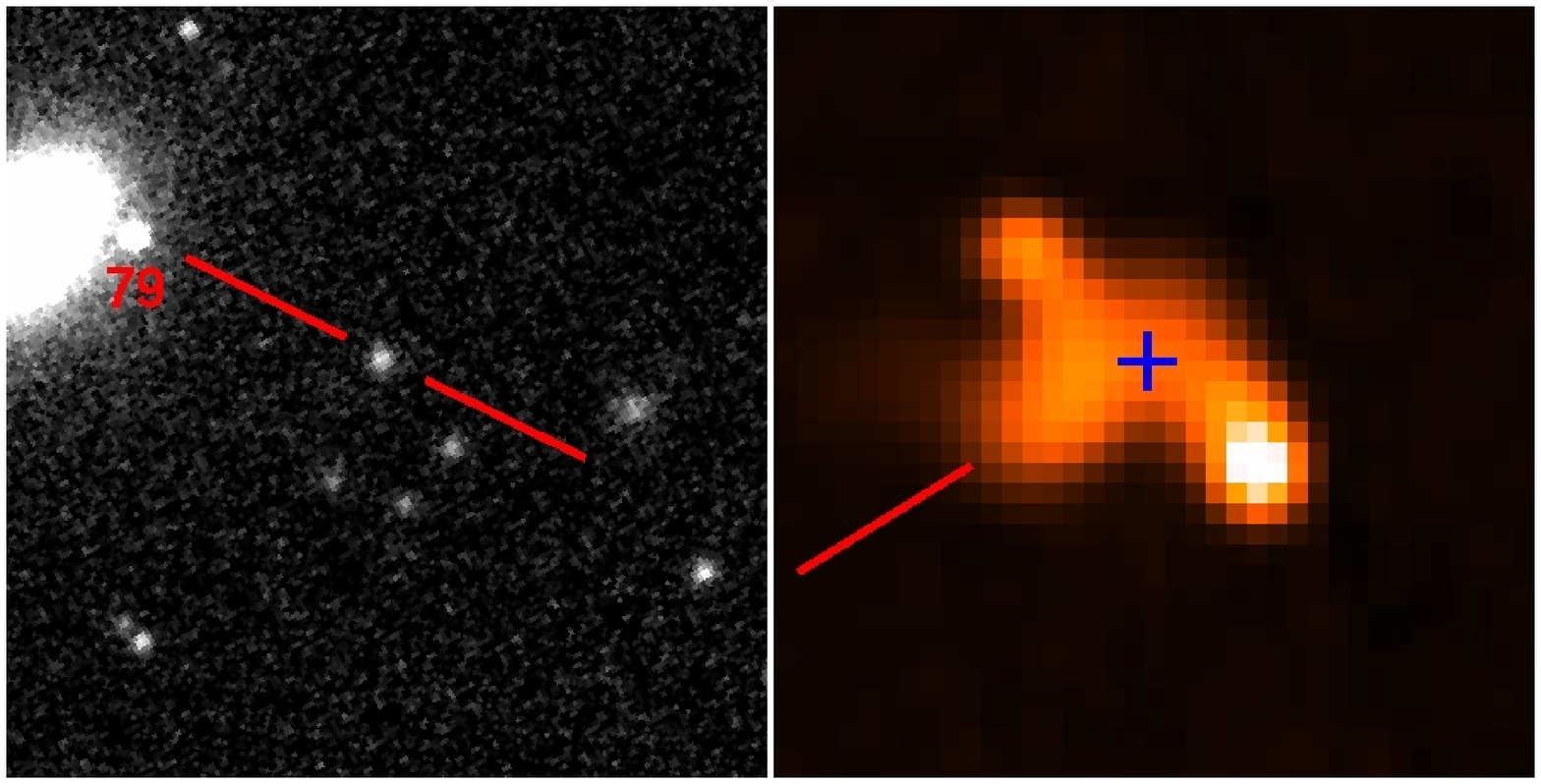}}
\centerline{
\includegraphics[width=78mm,angle=0]{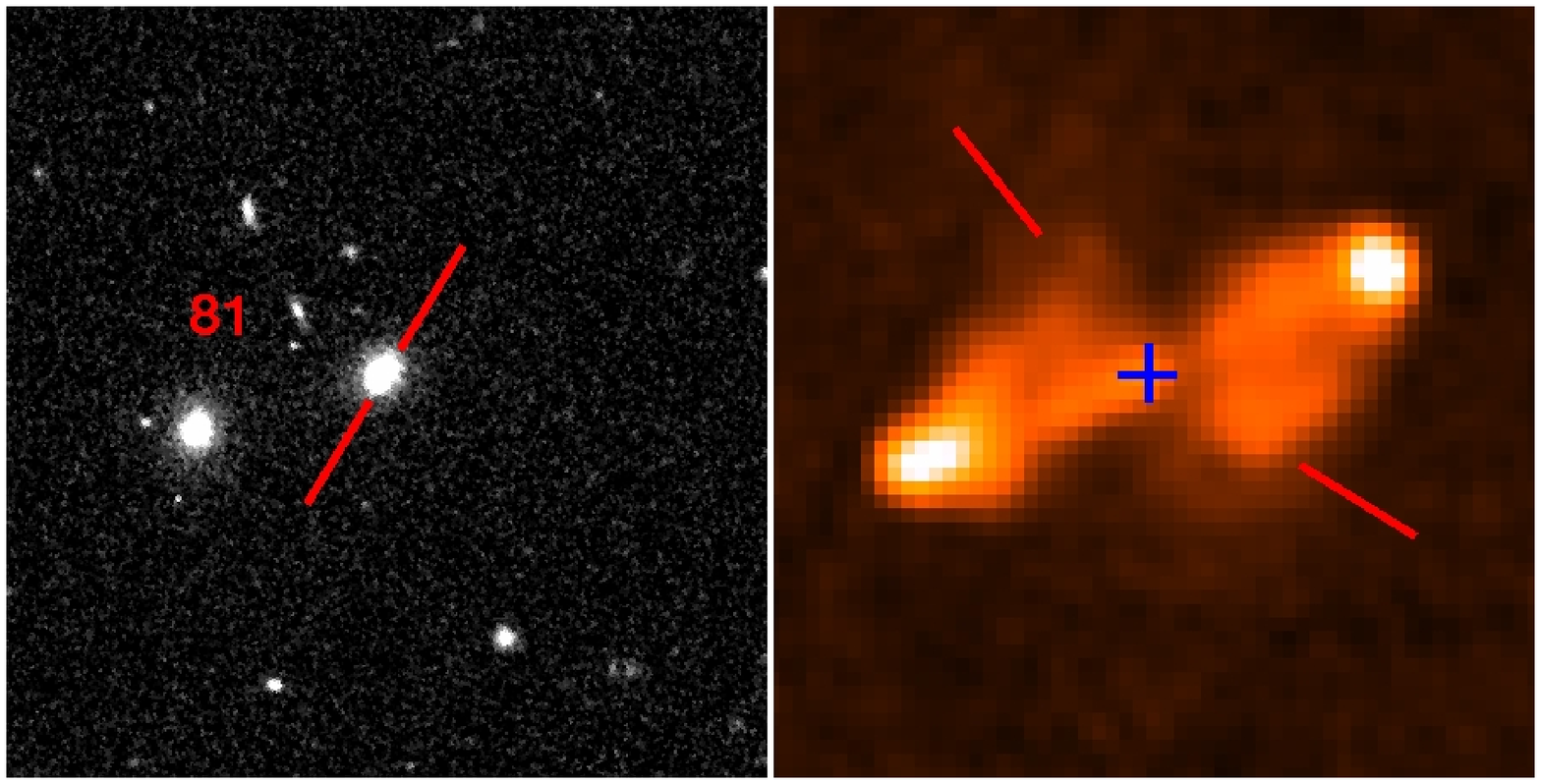}
\includegraphics[width=78mm,angle=0]{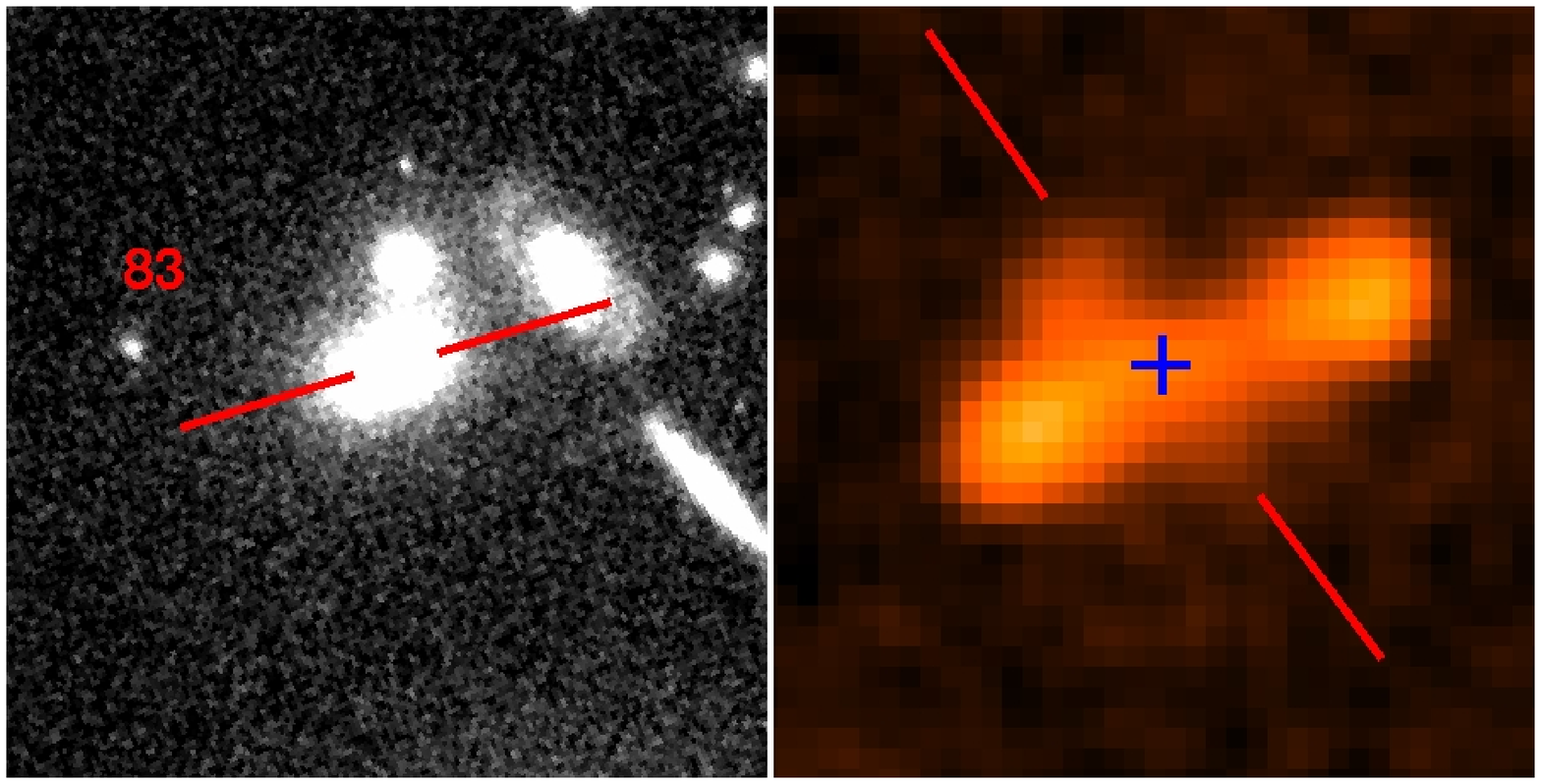}}
\centerline{
\includegraphics[width=78mm,angle=0]{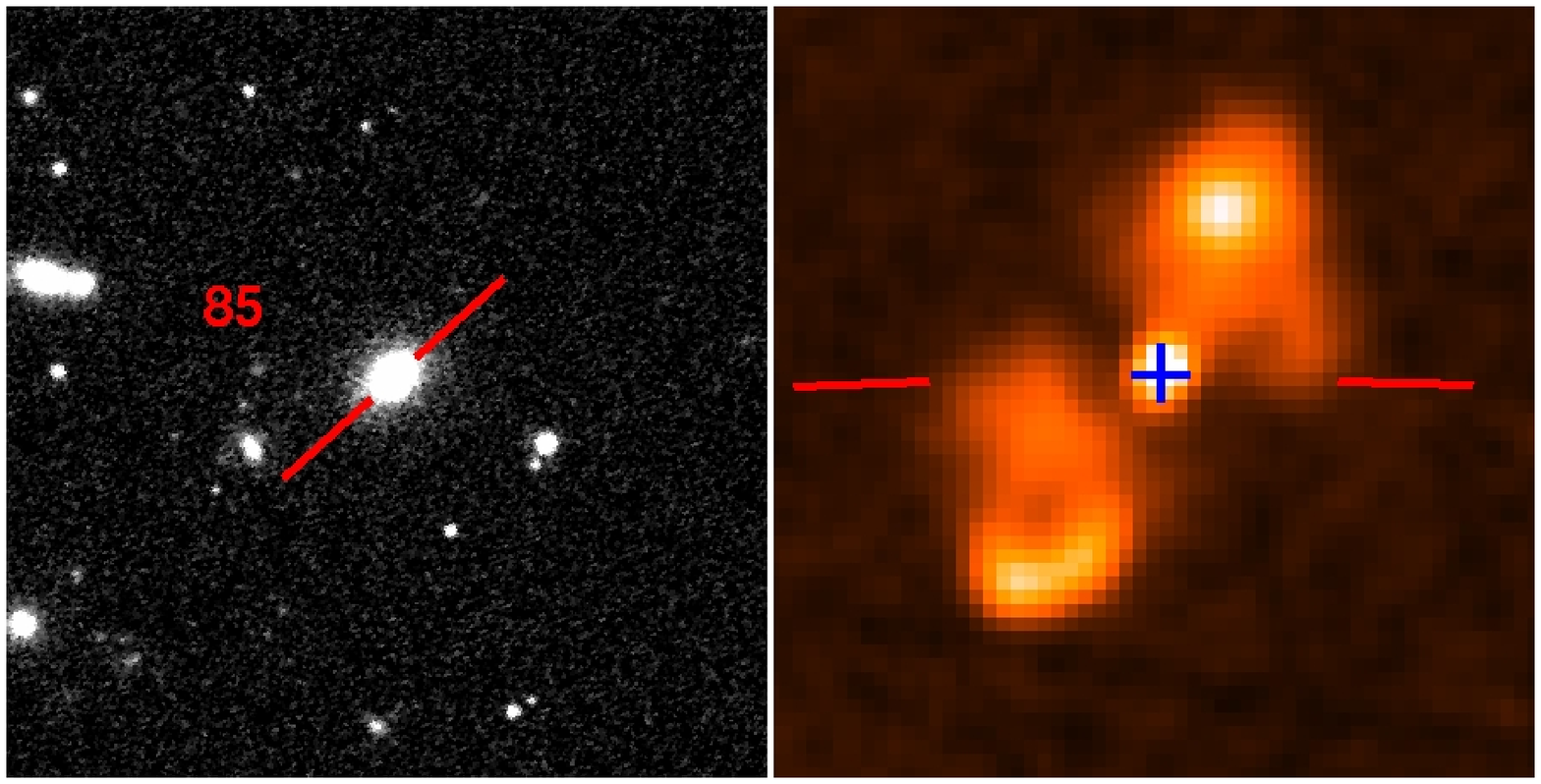}
\includegraphics[width=78mm,angle=0]{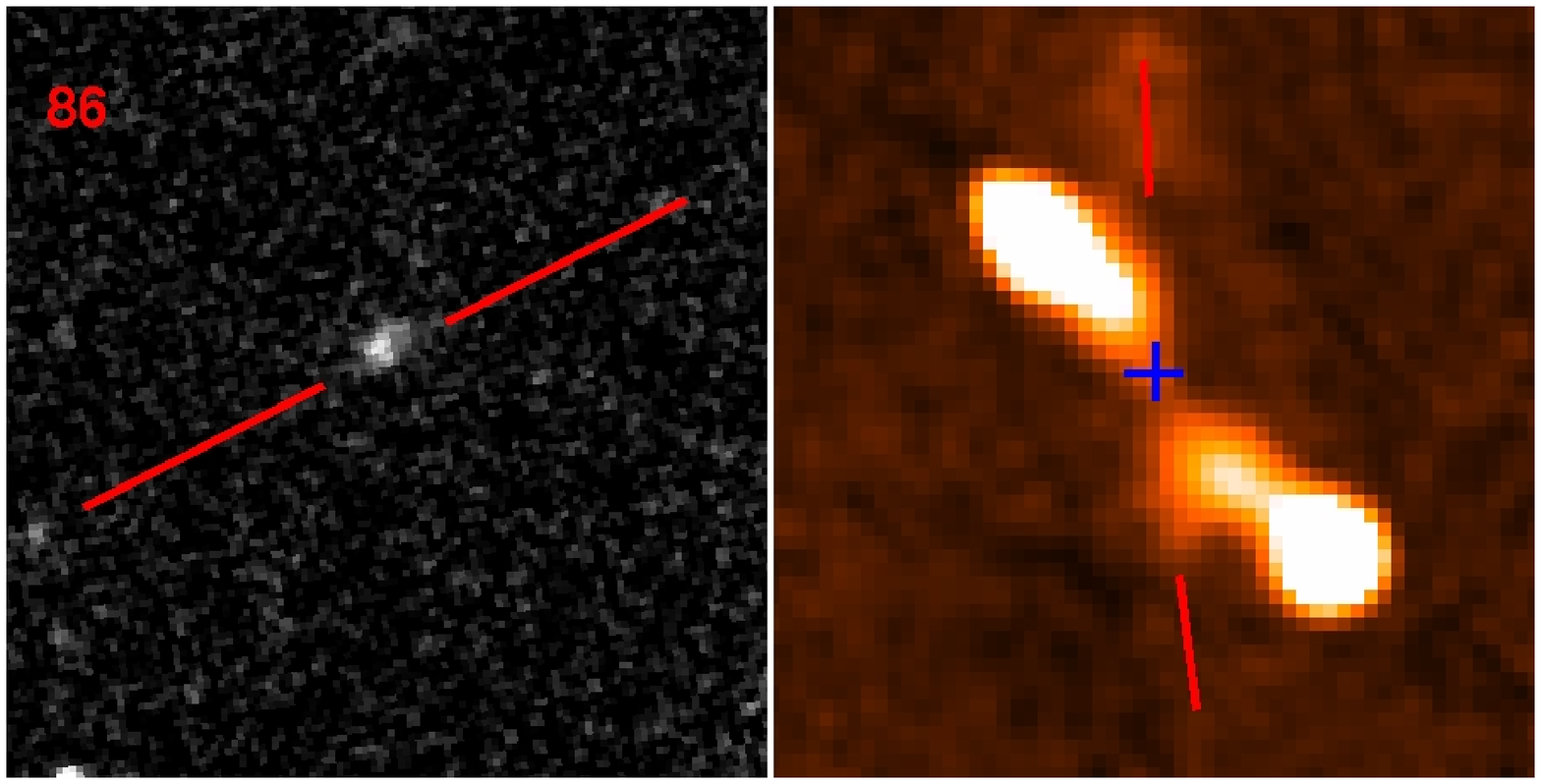}}
\centerline{
\includegraphics[width=78mm,angle=0]{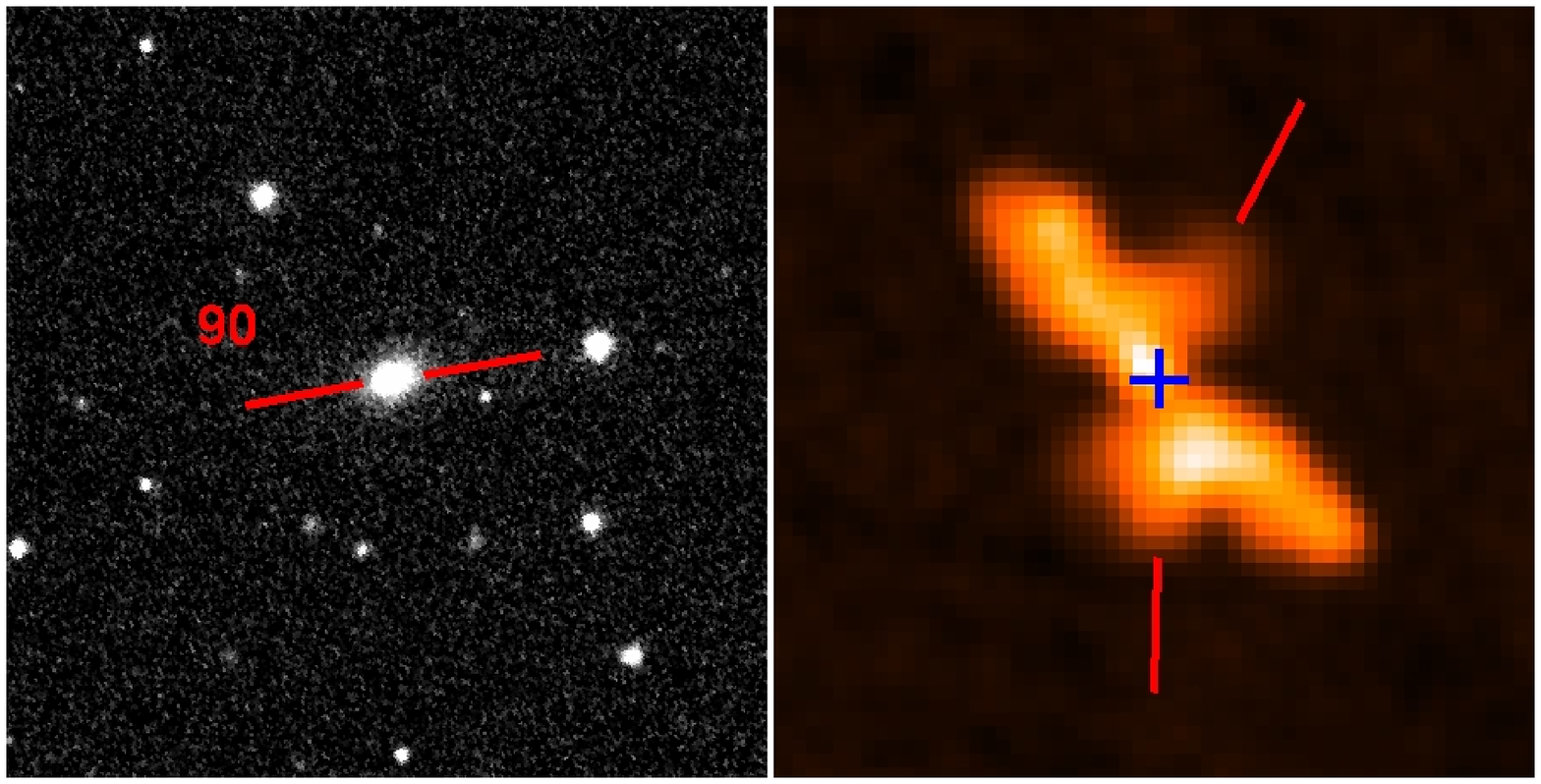}
\includegraphics[width=78mm,angle=0]{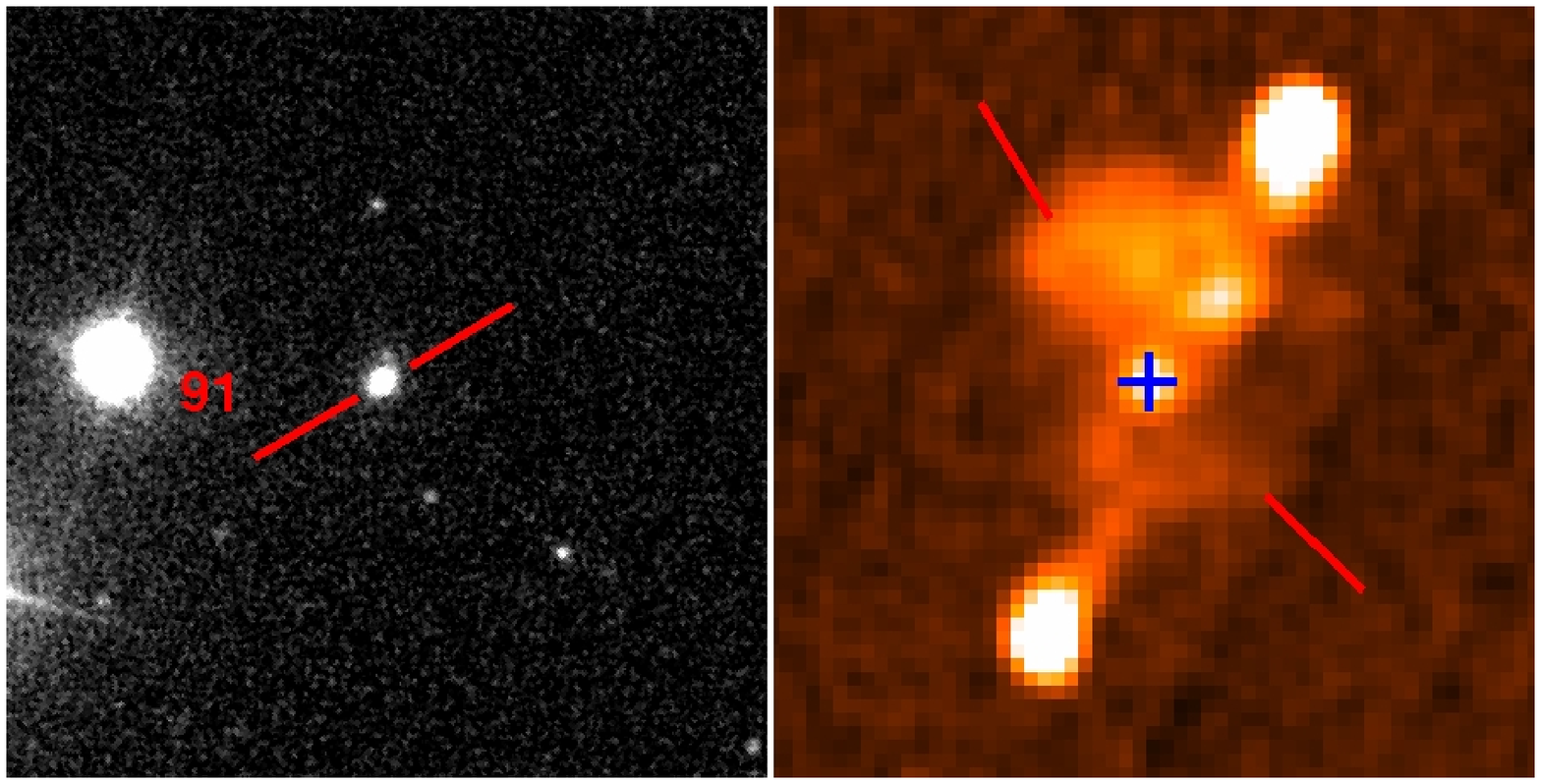}}
\centerline{
\includegraphics[width=78mm,angle=0]{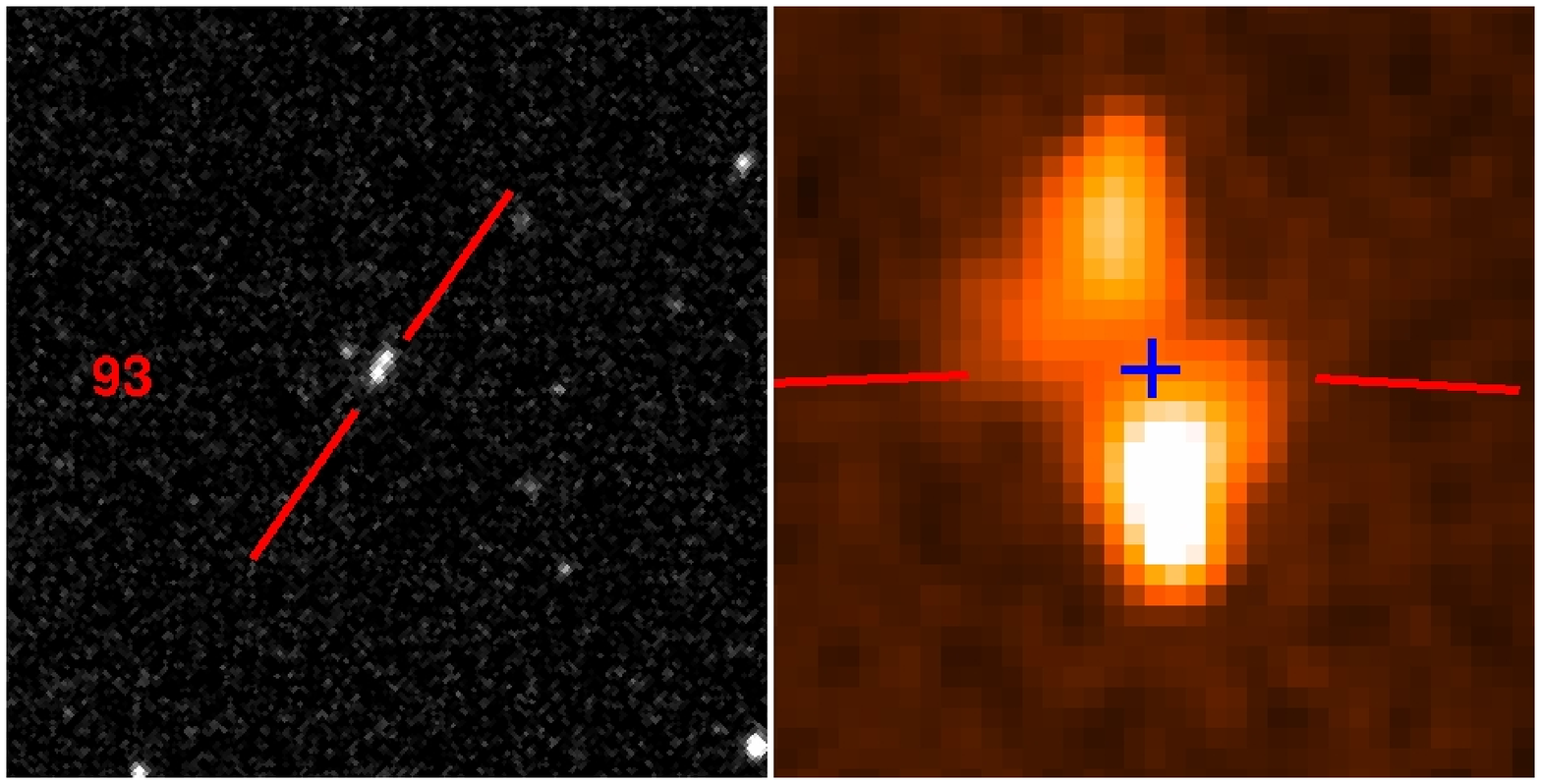}
\includegraphics[width=78mm,angle=0]{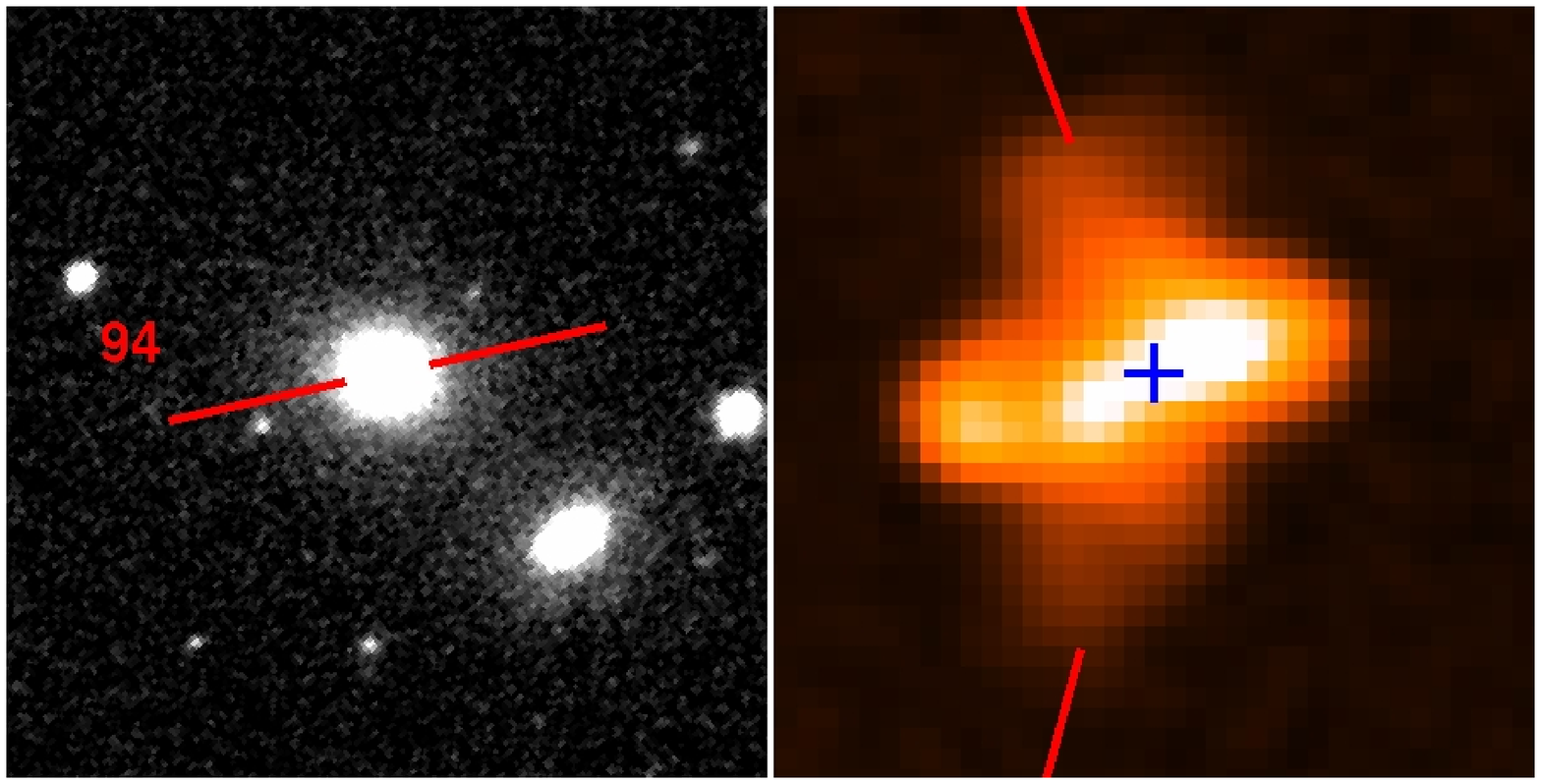}
}
\caption{(continued)}
\end{figure*}

\begin{figure*}
\centerline{
\includegraphics[width=65mm,angle=0]{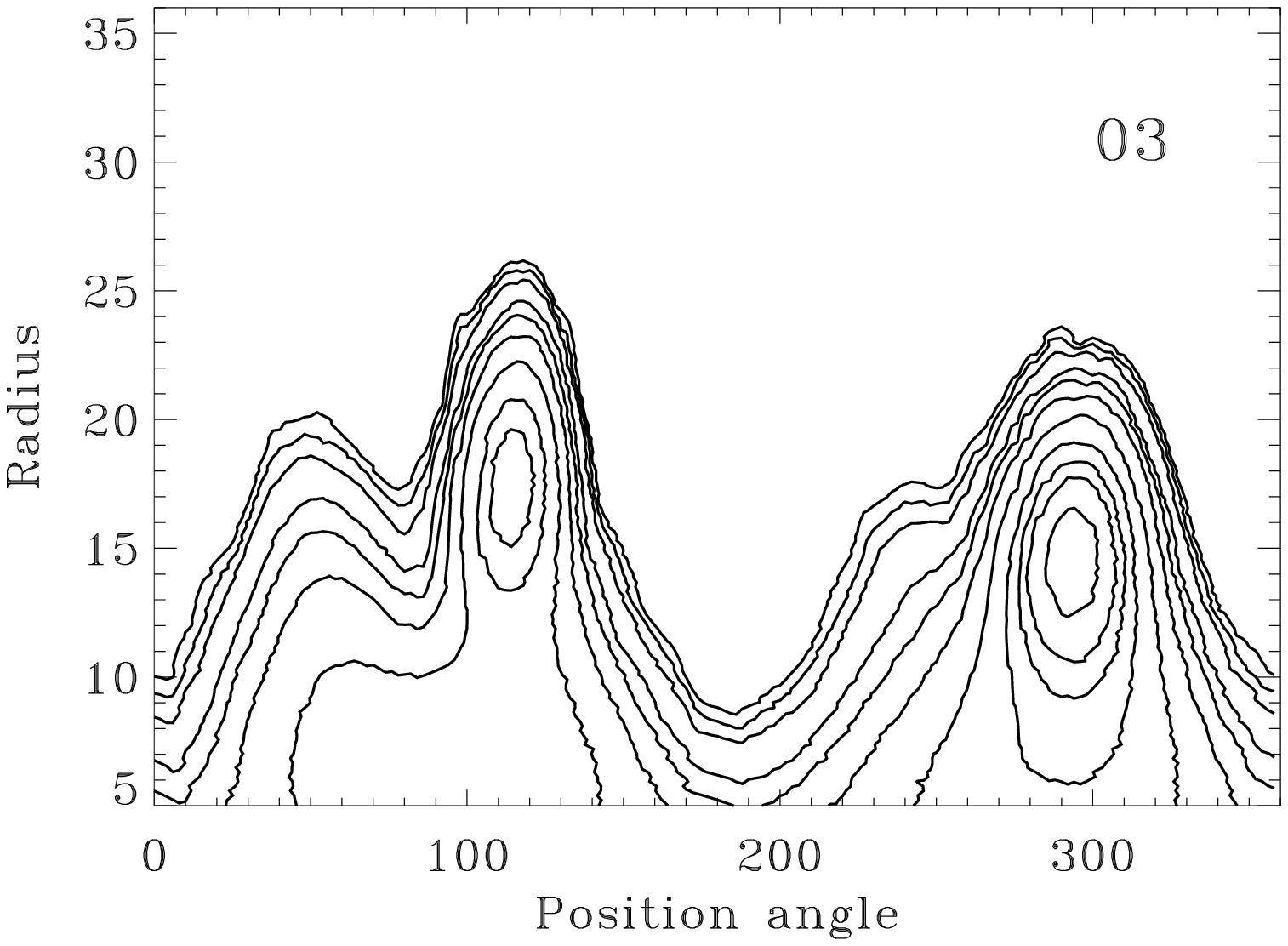}
\includegraphics[width=65mm,angle=0]{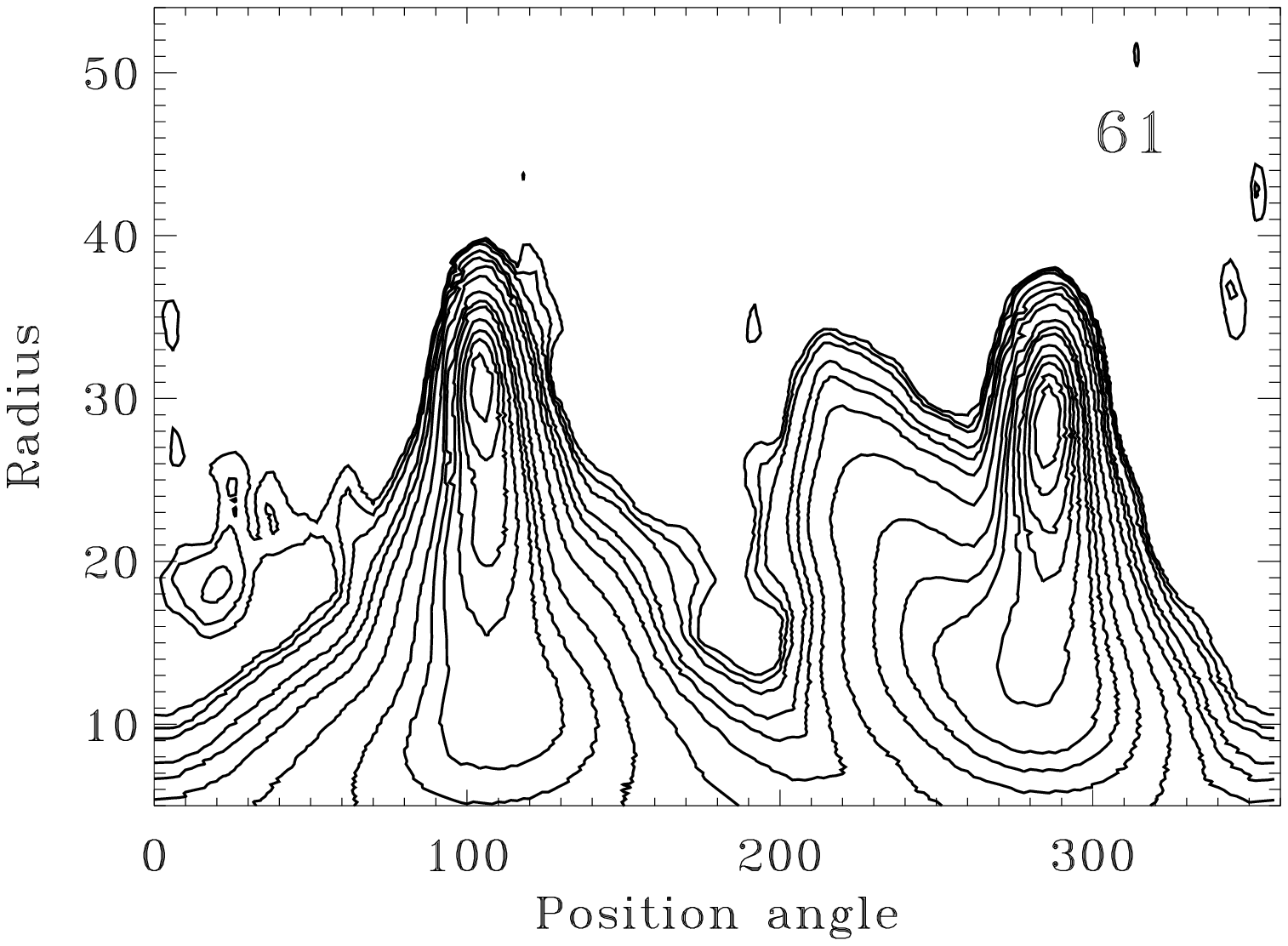}
\includegraphics[width=65mm,angle=0]{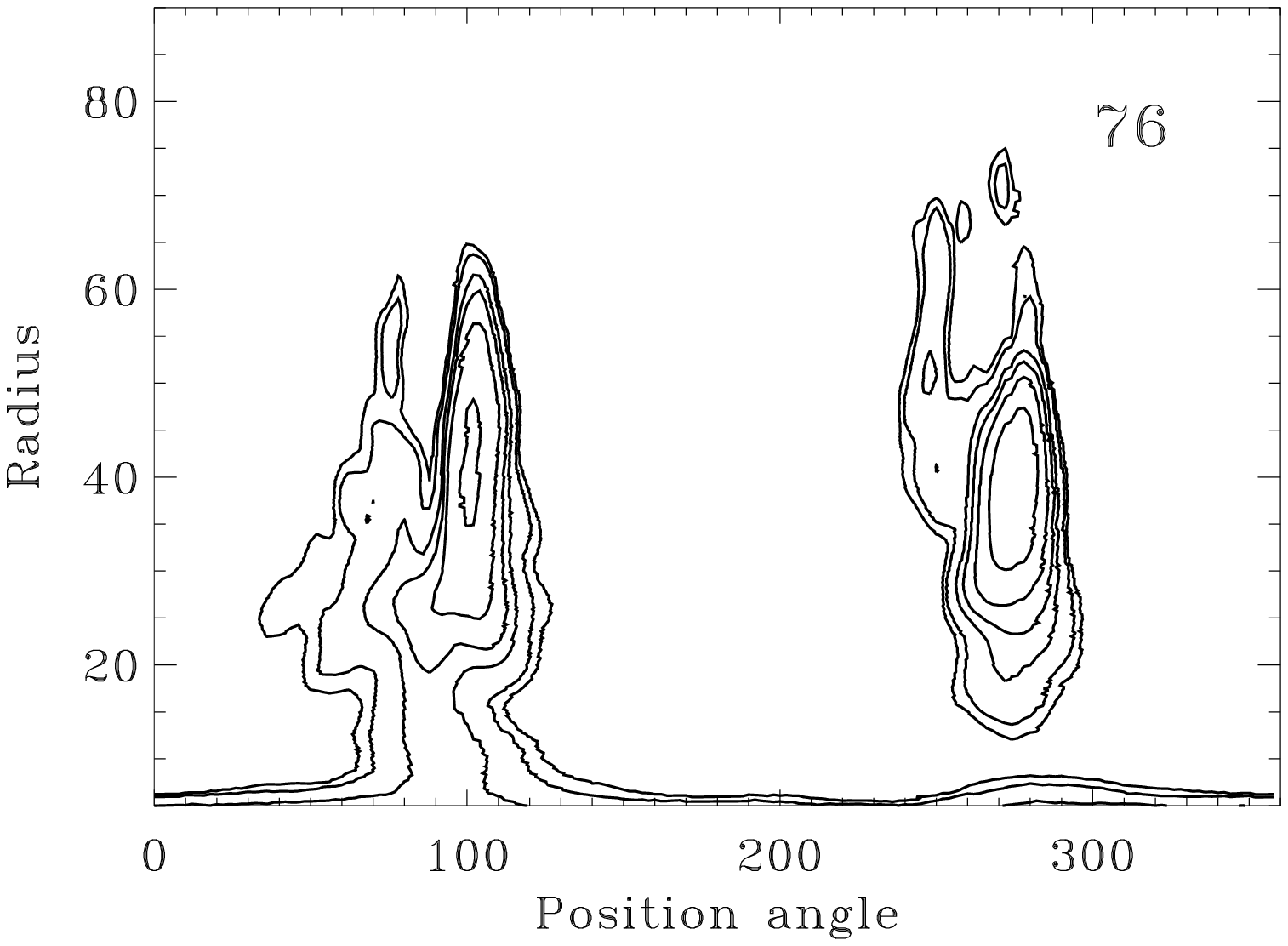}}
\caption{Radio contours of the FIRST images in polar coordinates for 3
  representative XRSs. The contours start at three times the local
    rms of the images, i.e., usually at $\sim$0.5 mJy/beam.}
\label{rteta}
\end{figure*}

\section{Optical versus radio-wings axis}
\label{axis}
Having measured the optical PA, we can proceed to  estimate the radio
PA of the wings for the 22 XRSs. For each one, we produced a polar
diagram of the radio emission by using the host galaxy as origin. Contours of
the FIRST images in polar coordinates for three representative XRSs are shown in
Fig. \ref{rteta}. While the main peaks in the polar diagrams are associated
with the primary lobes and hot spots, the secondary ones are located at the
PA of the radio wings. In particular we show XRS~03, where the faint SW wing
is well visible as a secondary peak in the polar diagram, the single winged
XRS~61, and the extended and diffuse wings of XRS~76, which produces well defined
linear features in this figure. The resulting values of wing PA, 
  defined as the orientation of the wings where they reach the local 3$\sigma$
  level, are tabulated in Table \ref{tab3}. We note that in several cases, only one
of the wings is visible.

\begin{figure*}
\centerline{
\includegraphics[width=78mm,angle=0]{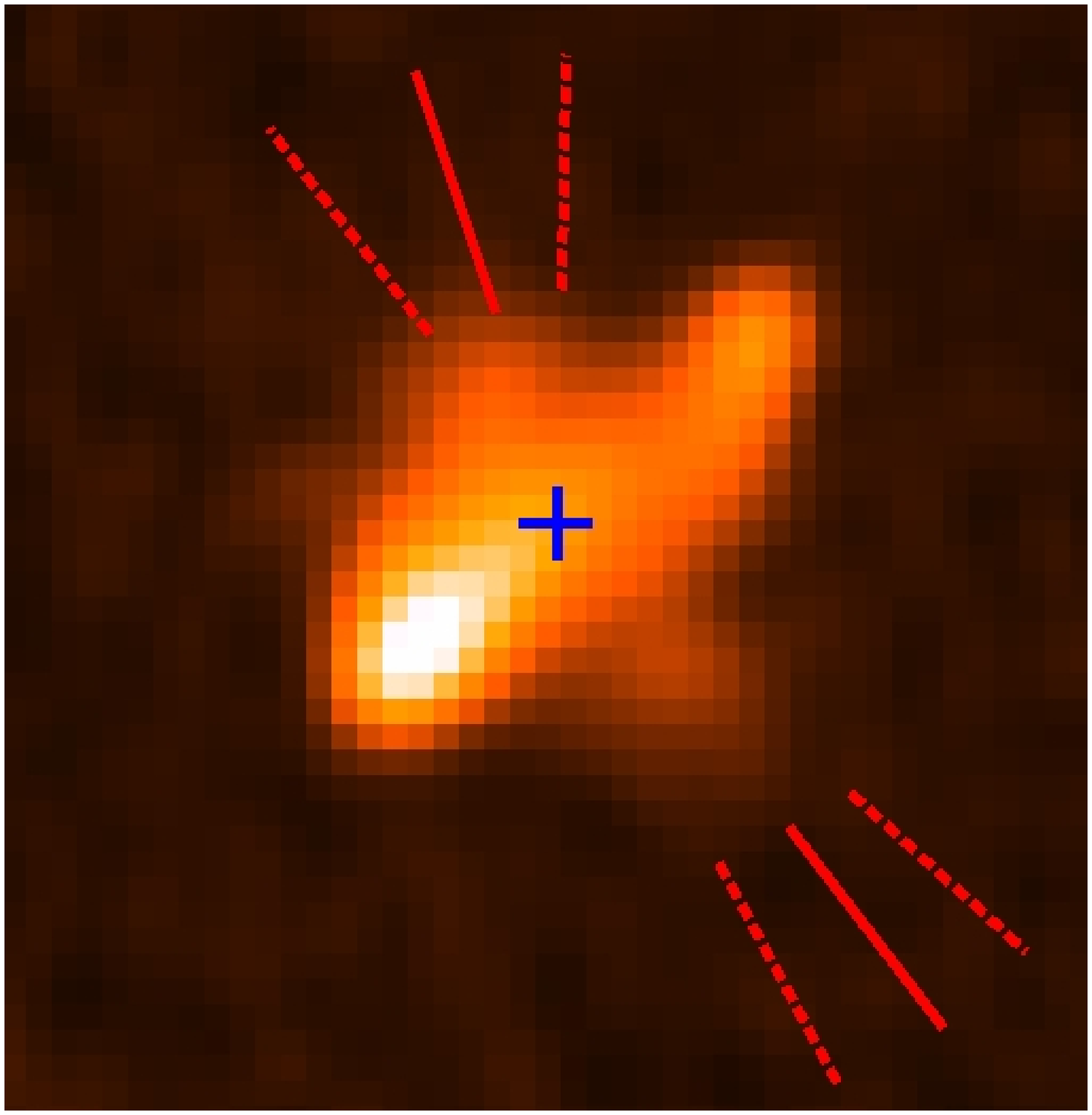}
\includegraphics[width=78mm,angle=0]{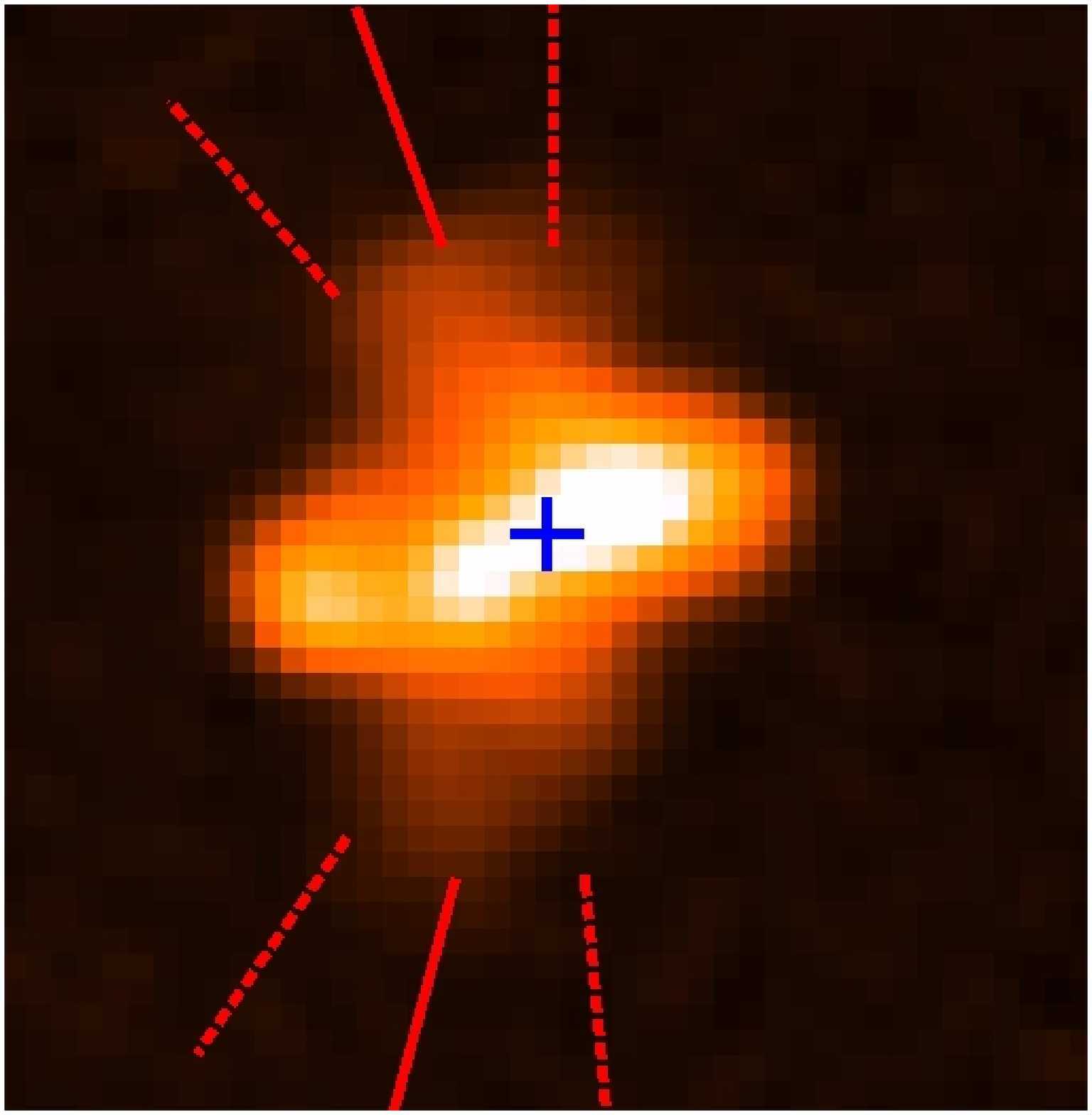}}
\includegraphics[width=85mm,angle=0]{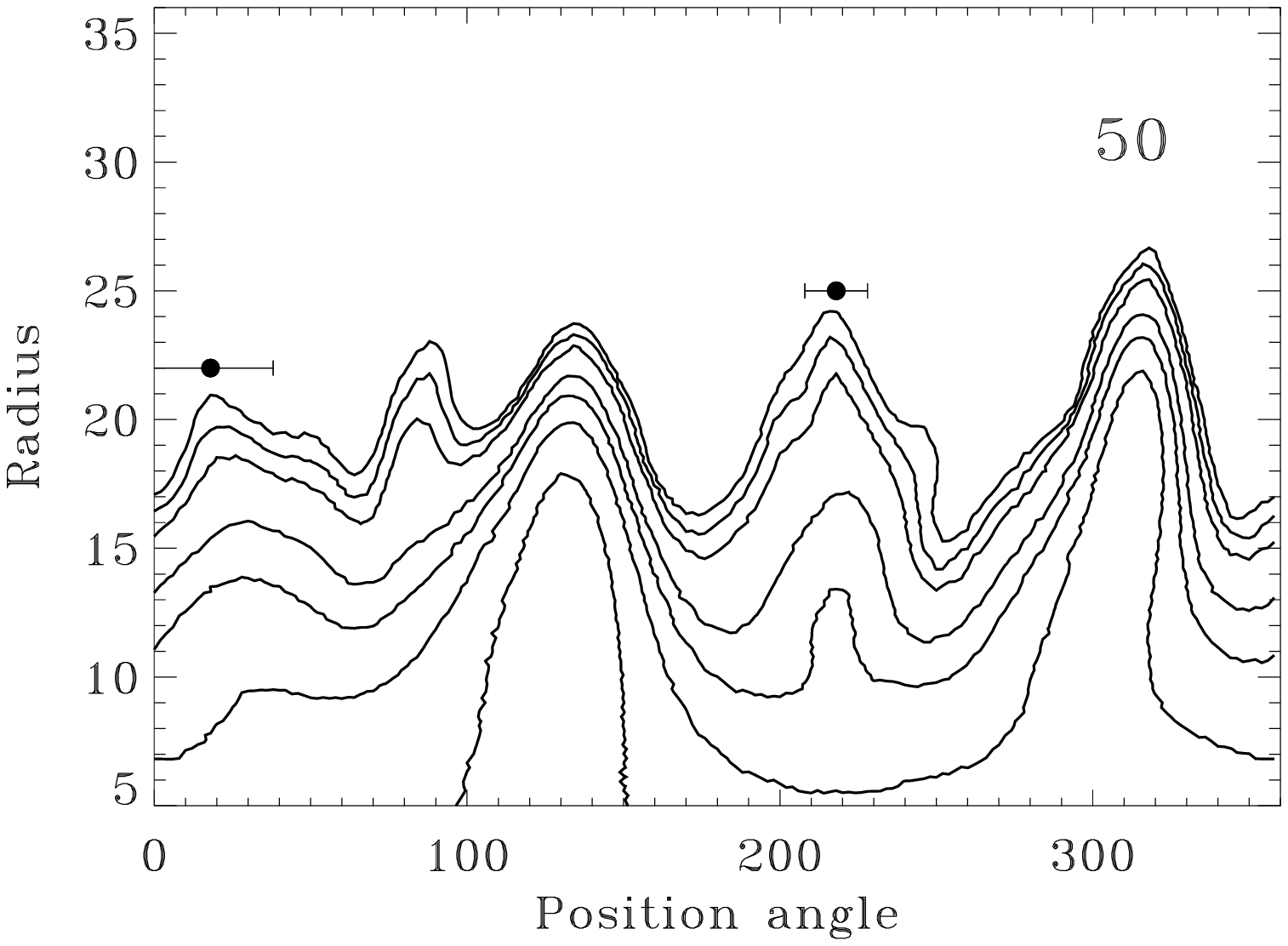}
\includegraphics[width=85mm,angle=0]{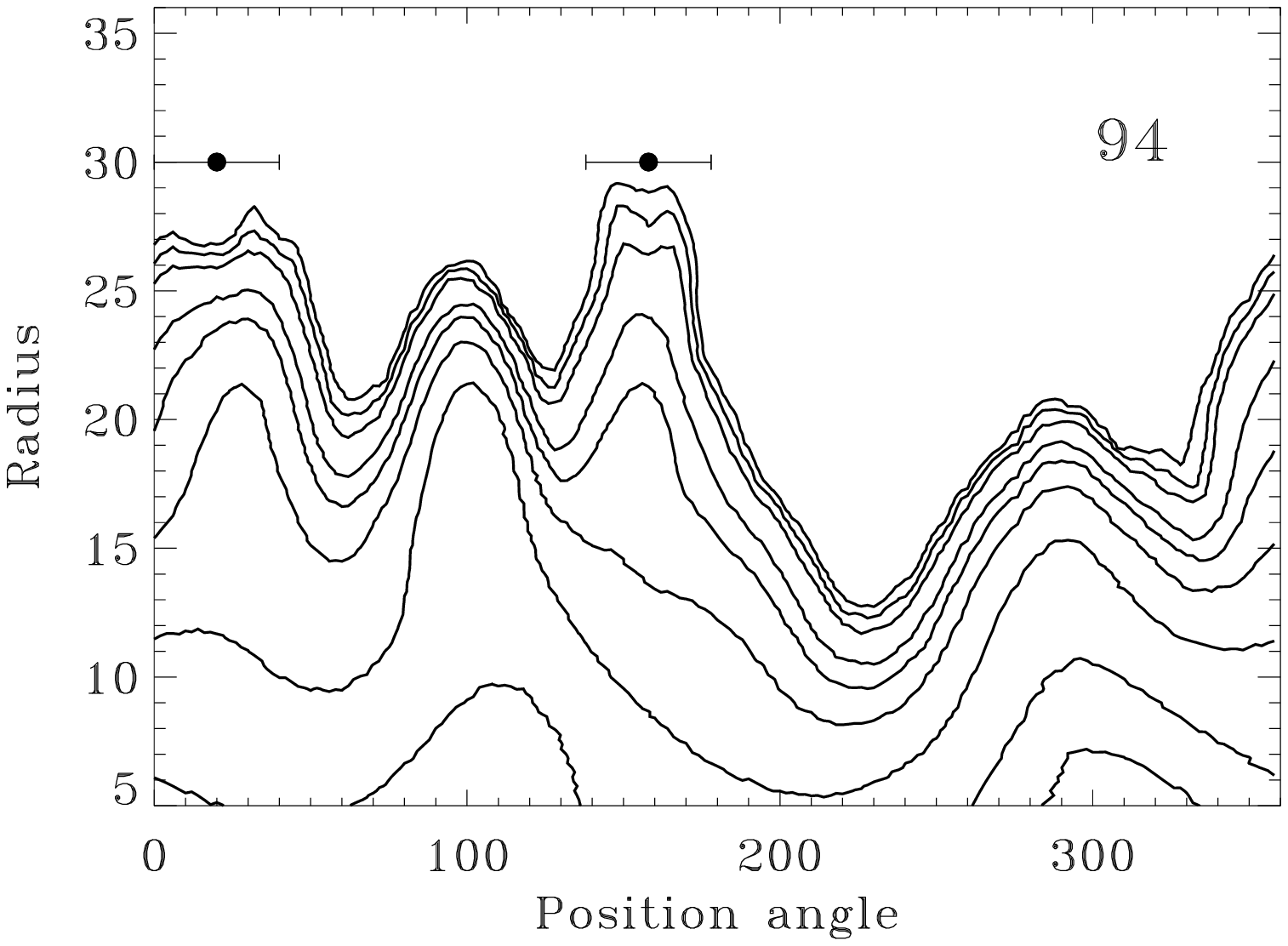}
\caption{Top panels: FIRST images of two XRSs (XRS~50 and XRS~94, on the left
  and right side, respectively) graphically showing the errors in the
  measurement of the wings P.A. The solid red line indicates the central value,
  the dashed ones differ by the estimated errors, i.e., $20^{\circ}$, except
  for the SW wing of XRS~50, where the error is $10^{\circ}$. Bottom panels:
  polar diagrams with  the wings PA values and errors indicated.}
\label{radioerror}
\end{figure*}

As a result of the complex structure of the radio wings, which often show
sub-structures and bendings, the wings' PA vary depending on the adopted
reference surface-brightness level. Thus an estimate of the error in their
axis can only obtained by visual inspection.  For each XRS we explored both
the polar diagram and the FIRST map to define the acceptable range of wing
orientations. Operatively, rather than defining a specific error to each
individual wing, we divided them into two groups to which we assigned PA
errors of 10$^\circ$ and 20$^\circ$, respectively. In Fig. \ref{radioerror} we
show graphically two examples of errors on the radio PA measurements. However,
as we show below, the precise value of the radio PA error does not have a
significant effect on our results concerning the relationship between optical
and radio structures.

\begin{table}
\begin{center}
 \caption{Optical and radio position (in degrees, from North to East) of
   X-shaped radio-galaxies with marks $\leq$ 2 and for which is is possible to
   measure the optical PA.}
\begin{tabular}{c|c|c|c}
\hline
\hline
Galaxy    & Opt. PA &  Radio PA & $\Delta$ PA \\
\hline
XRS~03    &  144 $\pm$   20  &    44$^a$ /233$^b$ & 80 / 89 \\ 
XRS~06    &  134 $\pm$    6  &   234$^b$ /---     & 80 / --- \\         
XRS~07    &  141 $\pm$   15  &    58$^b$ /257$^a$ & 83 / 64  \\         
XRS~11    &  134 $\pm$   20  &   119$^b$ /---     & 15 / ---  \\        
XRS~24    &   74 $\pm$   30  &     6$^b$ /195$^a$ & 68 / 59  \\         
XRS~44    &   12 $\pm$    2  &   108$^b$ /308$^a$ & 84 / 64  \\ 
XRS~49    &  152 $\pm$    4  &    59$^a$ /244$^b$ & 87 / 88 \\ 
XRS~50    &  120 $\pm$    8  &    18$^b$ /218$^a$ & 78 / 82  \\ 
XRS~61    &  100 $\pm$   37  &   213$^a$ /---     & 67 / --- \\ 
XRS~63    &   54 $\pm$   11  &   133$^b$ /309$^a$ & 79 / 75 \\ 
XRS~71    &   79 $\pm$    3  &   303$^a$ /---     & 44 / --- \\ 
XRS~76    &  127 $\pm$    5  &    76$^a$ /250$^a$ & 51 / 57  \\ 
XRS~77    &   96 $\pm$   19  &     2$^b$ /216$^a$ & 86 / 60  \\ 
XRS~79    &   64 $\pm$   19  &   121$^a$ /---     & 57 / --- \\ 
XRS~81    &  148 $\pm$   10  &    38$^a$ /239$^b$ & 70 / 89  \\ 
XRS~83    &  106 $\pm$    4  &    35$^a$ /217$^b$ & 71 / 69  \\ 
XRS~85    &  131 $\pm$    2  &    92$^b$ /268$^a$ & 39 / 43  \\ 
XRS~86    &  115 $\pm$   29  &     2$^a$ /187$^a$ & 67 / 72  \\ 
XRS~90    &  100 $\pm$    8  &   179$^a$ /333$^a$ & 79 / 53  \\ 
XRS~91    &  121 $\pm$    6  &    31$^a$ /226$^b$ & 90 / 75 \\ 
XRS~93    &  144 $\pm$    7  &    92$^a$ /267$^b$ & 52 / 57  \\ 
XRS~94    &  102 $\pm$    2  &    20$^b$ /158$^b$ & 82 / 56  \\ 
\hline
\end{tabular}
\label{tab3}

\end{center}
\medskip
\noindent
\small{Galaxy ID; optical position angle; radio position angle of the wing(s)
  with a code reporting the error on this measurement: 
  $^a=10^\circ$ and $^b=20^\circ$; relative offset between optical major axis and
  radio wings.}
\end{table}

In Fig. \ref{cfr} we show sideby side, the optical and radio image of the 22
XRSs for which the optical position angle can be measured.

\begin{figure}
\centerline{ \includegraphics[width=95mm,angle=0]{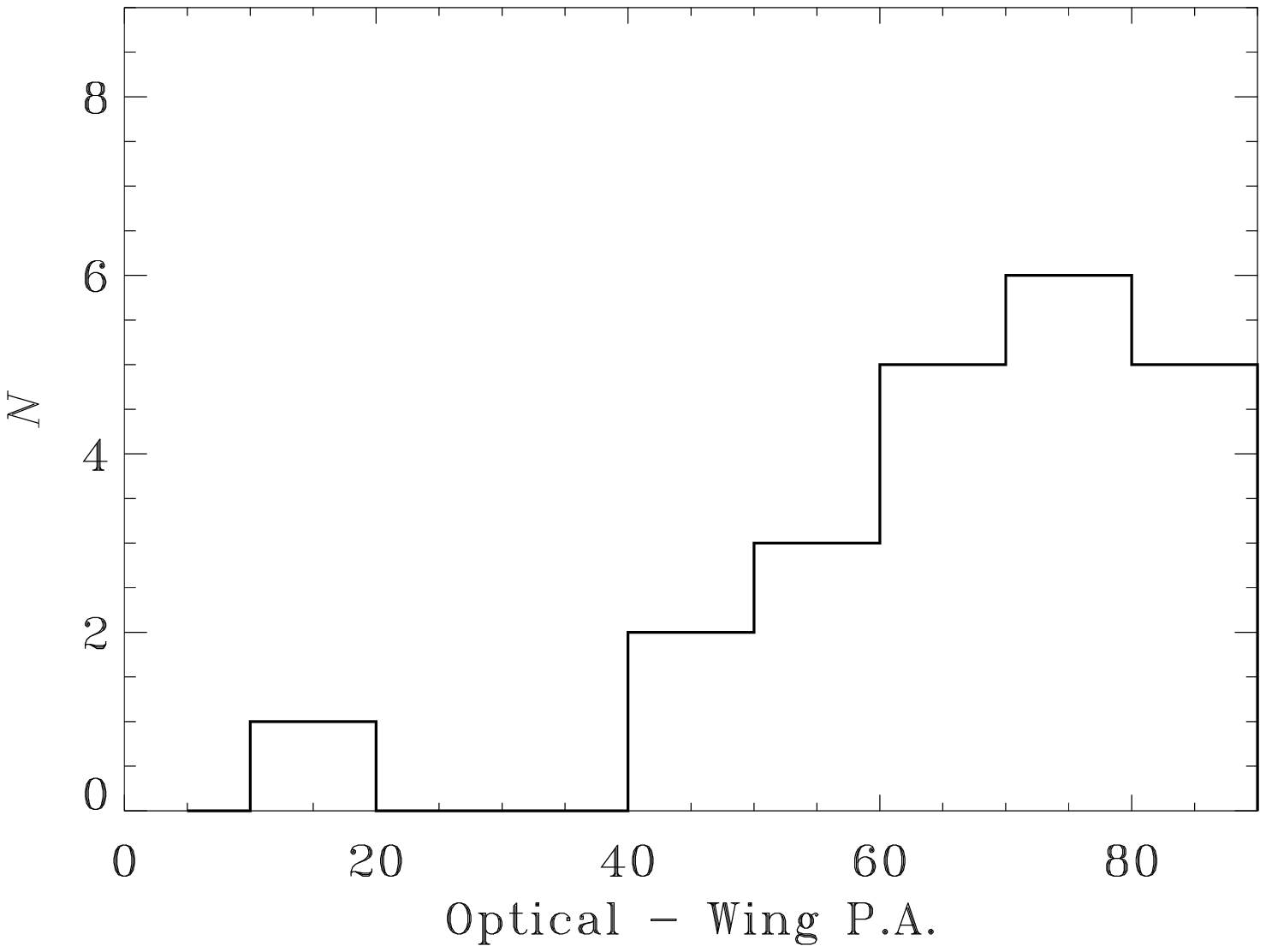}}
\caption{Difference between optical and wing PA (in degrees) for the 22
  XRSs for which such a comparison is possible. When two wings are present, we
  averaged the two values.}
\label{dpahist}
\end{figure}

The difference between the optical and radio axis are given in Table \ref{tab3}
and shown as a histogram in Fig. \ref{dpahist}. When both radio-wings are
detected we adopted the average offset. There is a clear preference in favor of
values higher than 40$^\circ$, with only one object (namely XRS~11) below this
value. The probability, obtained with the Kolmogoroff-Smirnov test, that this is
compatible with a uniform distribution is $P = 0.9 \times 10^{-4}$.

Interestingly, \citet{battye09} found that red elliptical galaxies of
relatively low radio-loudness show a highly significant alignment between the
radio major axis and the optical minor axis; however this effect is  far
less pronounced with respect to what we found for the radio wings, and they
interpreted this as due to a bias for the spin axis of the central engine being
aligned with the minor axis of the host.

At this stage we consider the effects of the inclusion/exclusion of the
sources suggested by the inspection of the higher resolution radio images from
\citet{roberts15}. The two sources for which an X-shape morphology is less
secure are XRS 05 and 12, but no measurement of the optical PA was possible in
these objects. The objects that might instead be included are XRS 27, 32, 62,
69, and 92). We perform the same analysis as for the other XRSs and in only 2
(XRS 27 and 92) we could measure the optical PA (154$^\circ$ and 135$^\circ$,
respectively). Since the radio-optical PA differences are 44$^\circ$ and
75$^\circ$,\footnote{For J1606+0000 we used the \citet{hodgeskluck10} radio
  images.} respectively, the inclusion of these two sources would strengthen
the statistical significance of our results.

\begin{figure}
\centerline{
\includegraphics[width=95mm,angle=0]{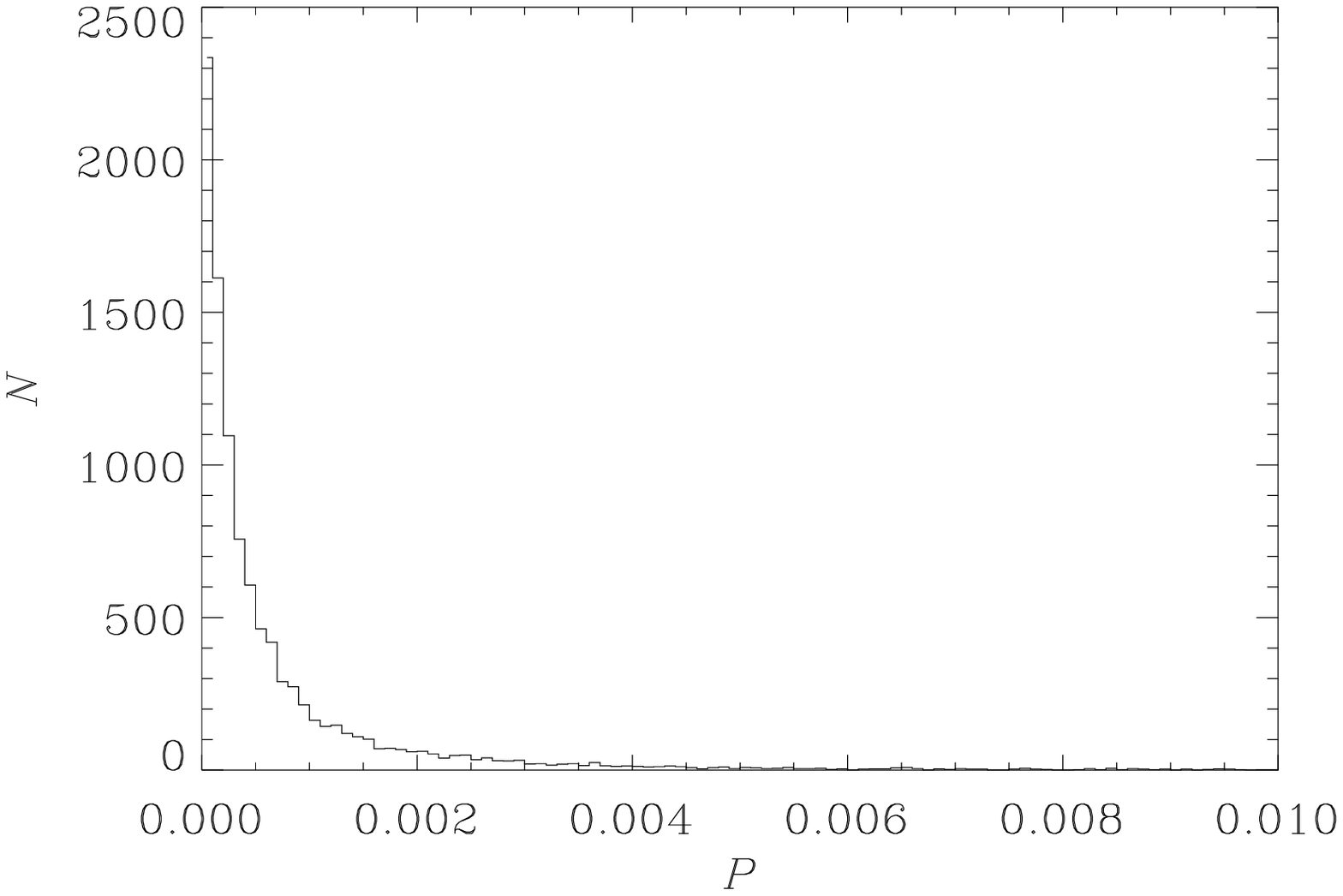}}
\caption{Distribution of probability resulting from a Kolmogoroff-Smirnov test
  that the distribution of offsets between optical and radio-wing axis is
  compatible with a uniform distribution. This is obtained with the bootstrap
  method from 10,000 realizations that result from randomly varying both axis
  around the central values with an amplitude given by the measurements
  errors.}
\label{probhist}
\end{figure}

To account for the measurement errors in both the optical and radio
PA, we again used  the bootstrap method. We extracted 10000 times 22 pairs of
PA values with a normal distribution centered on the best value and with a
width given by the PA error. We then estimated the radio-optical
offsets. The histogram of the resulting probability is shown in
Fig. \ref{probhist}. The average value is $P = 7 \times 10^{-4}$. To
account for the uncertainties in our estimates of the error in the wings PA,
we repeated the analysis doubling all values, finding that the average
probability increases only to $P = 5 \times 10^{-3}$. 

\medskip
We conclude that   a highly significant connection exists between the
optical and the radio-wings axis, with the wings preferentially aligned with
the host's minor axis. The same result has been obtained by
\citet{butterflies} from the study of a sample of nine XRSs selected from the
literature. Thanks to the larger size of the sample considered here, the
statistical significance is strongly improved. Since the XRSs samples in these
two studies do not have objects in common, the overall probability that XRS
wings are randomly oriented with respect to their hosts is even lower than the
$\sim10^{-3}$ probability quoted here. \citet{butterflies} also find that
wings form in galaxies with a larger ellipticity with respect to a reference
sample of non-winged FR~II radio-galaxies. We cannot confirm the overall
validity of this finding since the 22 XRSs for which we can measure the
optical axis are at a much larger distance than those studied by
\citeauthor{butterflies}, with a median redshift of the two samples being 0.31
and 0.085, respectively. Generally, the larger distance compromises  an
accurate measurement of the ellipticity of XRSs hosts. However, if we focus on
the four XRSs with $z<0.1$ (namely, XRS~44, XRS~49, XRS~76, and XRS~83), they
have an ellipticity in the range 0.21-0.36, which is similar to the XRSs considered by
\citeauthor{butterflies} and larger than the median ellipticity of FR~II
radio-galaxies ($\sim$0.13).

The relationship between the optical axis and wings orientation suggests the
existence of causal connection between the presence of radio wings (and thus
the origin of XRSs) and the properties of their host. Our results strengthen
the interpretation proposed by \citeauthor{butterflies} that XRSs naturally
form when a jet propagates in a non-spherical gas distribution. In this case,
the cocoon expansion along the direction of maximum pressure gradient (the
galaxy minor axis) occurs at a speed comparable to that of the advance of the
jet head along the radio source main axis, producing the X-shaped morphology.

\begin{figure}
\centerline{
\includegraphics[width=88mm,angle=0]{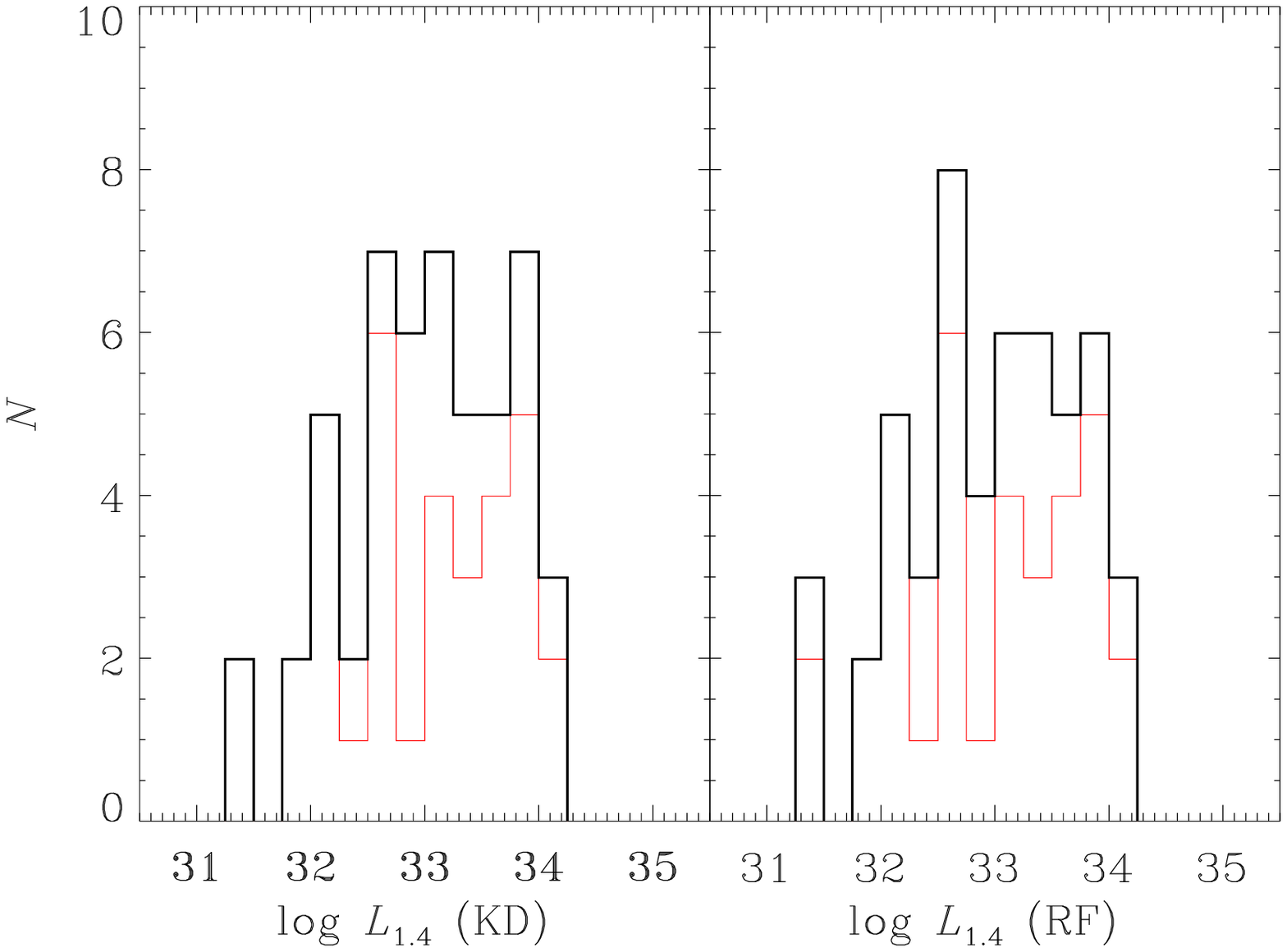}}
\label{zlum}
\caption{Radio luminosity distribution of XRSs at 1.4 GHz; the luminosities
  are estimated, when no spectroscopic redshift is available by adopting the
  KD and RF photometric redshifts, in the left and right panel,
  respectively. The red histograms represent the fraction of objects with
  spectroscopic redshifts.}
\end{figure}

\section{Redshift and radio power of XRSs}
\label{redshift}
We collected the spectroscopic redshifts for our sources from \citet{cheung09}
and \citet{landt10}, see Table \ref{tab2}. When no spectroscopic redshift is
available, we adopt the photometric redshifts provided by the SDSS database, which were estimated with the kd-tree nearest neighbor (KD, \citealt{csabai07}) or the
random forests (RF, \citealt{carliles10}) methods. Only in three cases was no
redshift estimate  available. 

The comparison between the spectroscopic and photometric redshifts for the 26
XRSs where these two quantities are available indicates that the average
relative differences are $|z_{\rm spec}-z_{\rm phot}|/z_{\rm spec}=0.13$ for
both the KD and RF methods. When only the two photo-z estimates are available
they are, with only two exceptions, consistent with each other within the
errors. This indicates that the redshift estimates (and the corresponding
radio luminosities) are generally robust. In Fig. \ref{zlum} we present the
radio luminosity histograms of XRSs, having taken the 1.4 GHz from the FIRST
catalog as published by \citet{cheung09}. In the two panels the luminosities
have been estimated, when no spectroscopic redshift is measured, by adopting
the KD and RF values, respectively. The red histograms represent the fraction
of objects with spectroscopic redshifts.  The radio luminosity distribution
spans the range $L \sim10^{32} - 10^{34}$ $\ergsHz$ at 1.4 GHz, regardless of
the photo-z adopted.

XRSs extend by a factor of 10 below the least luminous FR~II of the 3CR sample
\citep{buttiglione09}, but this effect has already been noted by
\citet{kozie11} from the analysis of FR~II that was extracted from the
SDSS. More interesting is the presence of a high power cut-off for XRSs. We
cannot firmly conclude that this is not related to some selection bias that is
due, for example, to the detectability of the wings and of the hosts in higher
redshift/power objects; indeed, the objects in which we could measure the
optical PA have a median redshift of 0.30, against 0.43 for the sources where
this estimate was unsuccessful; alternatively this result can be interpreted
in the framework of the lateral cocoon expansion model presented above. In
fact, high luminosity radio galaxies that are associated with jets of greater
power are less affected by the properties of the surrounding medium. They
might escape the region where the cocoon expansion in influenced by the
asymmetric host before the lateral wings can develop.

\section{Spectral analysis}
\label{spectral}
The aim of this section is to derive information on the spectroscopic properties
of XRSs from the point of view of their stellar population and emission lines.
This analysis can falsify the proposed model that poses that XRSs are
``normal'' FR~II whose radio morphology is only caused by the jet propagation
in an asymmetric ISM. In fact, no differences are expected between the XRS's
properties and those of non-winged FR~II. However, we note that the competing
model based on jet re-orientation (due, e.g., to a galaxies merger) does not
require an enhanced star-formation activity to be observed in XRSs, since this is
expected only when one of the coalescing galaxies is gas rich; no effect on
the stellar population is expected from a ``dry'' merger. Furthermore, there
might be a substantial temporal gap between the various phases of the merger
event, i.e., the black hole coalescence and reorientation, the phase of high
star formation, and the onset of the AGN activity (e.g., \citealt{blecha11}).

We collected the available 28 SDSS spectra for the selected sub-sample of 53
XRSs. For two XRSs (namely, XRS~48 and 73) we do not proceed with this
analysis since they are quasars with no visible starlight. The resulting list
of 26 objects is presented in Table \ref{spectra}. Each spectrum is modeled by
using the Gandalf software \citep{sarzi06} from which we obtain a
decomposition of the stellar light into 39 templates and the measurements of
their emission lines. The different emission lines were all fixed to the same
velocity by using, as an initial guess, the host redshift.

\subsection{Emission line properties}
\label{el}

\begin{figure*}
\centerline{
\includegraphics[width=170mm,angle=0]{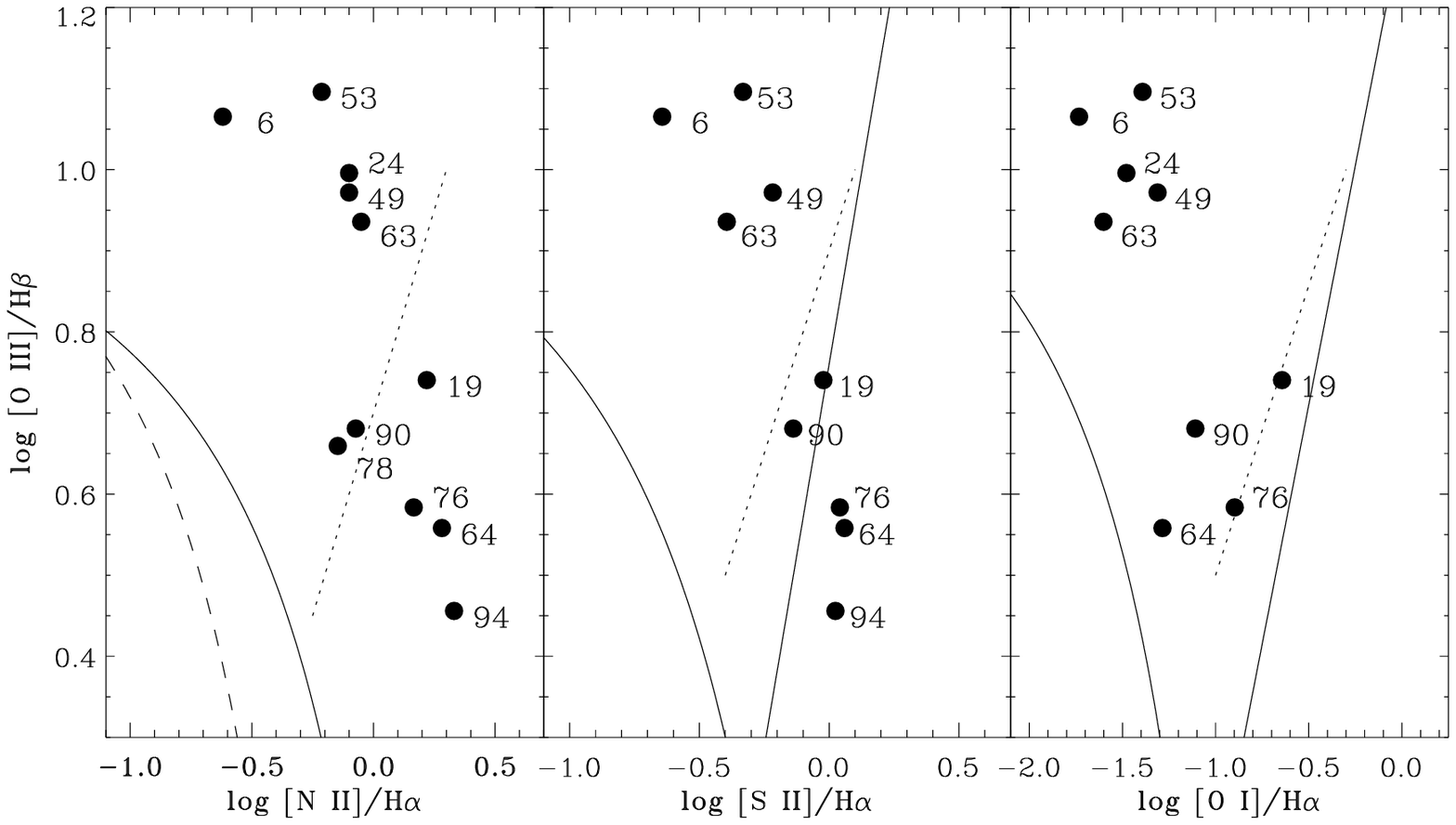}}
\caption{Spectroscopic diagnostic diagrams: (left) [O~III]/\Hb\ vs
  [N~II]/\Ha\ ratios, (center) [O~III]/\Hb\ vs [S~II]/\Ha\ ratio, (right)
  [O~III]/\Hb\ vs [O~I]/\Ha. The curves divide AGN (above the solid curved
  line) from star-forming galaxies. In the left panel, between the long-dashed
  and solid curve  are the composite galaxies \citep{kewley06}. In the
  middle and right panels, the straight solid line divides Seyferts (upper
  left region) from LINERs (right region). The dashed lines mark the
  approximate boundaries between HEGs and LEGs derived from the 3CR
  radio-galaxies in each diagram, from \citet{buttiglione10}, Fig. 7.}
\label{dd}
\end{figure*}

The emission line measurements of the 26s XRSs are used to build the diagnostic
diagram  to derive a spectroscopic classification. This is
obtained by comparing the ratio of the intensity between [O~III] and H$\beta$
with those of the [N~II], [S~II], and [O~I] line against H$\alpha$, following
the analysis presented by \citet{kewley06}. In eight cases the H$\beta$ line (and
often also the [O~III] line) is not detected; furthermore, the H$\alpha$ line
falls outside six SDSS spectra because of their high redshift. We are then left with
11 objects that can be located in the diagnostic diagrams (see
Fig. \ref{dd}). We classify a source as HEG when it falls above the dashed
lines that mark the boundaries between HEGs and LEGs, as derived from the 3CR
radio-galaxies \citep{buttiglione10}. Two objects (namely XRS~78 and 90) fall
into the HEGs region in the left and right panels, while, in the LEGs
region, they are only  in the middle one: we consider them as HEGs (a conclusion that
will be strengthened by their location in the diagram that compares line and
radio luminosity, see below).  We find seven HEGs and four LEGs.

Considering again the six high redshift sources with no information on the
H$\alpha$ line, in all cases the [O~III]/\Hb\ is larger than eight. Since no LEG
in the 3CR is found above this ratio, we consider them as HEGs.

\begin{table*}
\begin{center}
\caption{Spectral properties of the 26 X-shaped RG with SDSS/BOSS spectra.}
\begin{tabular}{c|c|c|c|r|c|l}
\hline
\hline
ID      & redshift & survey & Plate MJD Fiber &  $L_{\rm[O~III]}$   & Sp. type & Method     \\
\hline                      
XRS~03  &0.591  & BOSS &  4222  55444   554  &   42.61  &    HEG   & [O~III]/H$\beta$         \\    
XRS~05  &0.304  & SDSS &  0394  51913   364  &   40.48  &    LEG   & [O~III]/L$_{\rm r}$    \\       
XRS~06  &0.281  & SDSS &  1499  53001   522  &   42.81  &    HEG   & D.D.                      \\    
XRS~13  &0.659  & BOSS &  4392  55833   757  &   42.58  &    HEG   & [O~III]/H$\beta$     \\       
XRS~19  &0.376  & BOSS &  4445  55869   202  &   41.05  &    LEG   & D.D.                 \\       
XRS~20  &0.213  & SDSS &  1209  52674   339  &   39.98  &    LEG   & [O~III]/L$_{\rm r}$     \\       
XRS~24  &0.520  & BOSS &  5301  55987   885  &   42.40  &    HEG   & D.D.                      \\    
XRS~25  &0.591  & BOSS &  4795  55889   717  &   42.15  &    HEG   & [O~III]/H$\beta$               \\       
XRS~26  &0.227  & SDSS &  1201  52674   639  &$<$40.13  &    LEG   & [O~III]/L$_{\rm r}$      \\       
XRS~30  &0.166  & SDSS &  1744  53055   069  &$<$39.82  &    LEG   &  [O~III]/L$_{\rm r}$     \\       
XRS~44  &0.081  & SDSS &  1225  52760   366  &   39.74  &    LEG   &  [O~III]/L$_{\rm r}$          \\    
XRS~49  &0.079  & SDSS &  2089  53498   176  &   41.46  &    HEG   & D.D.                      \\    
XRS~53  &0.424  & BOSS &  5848  56029   983  &   42.45  &    HEG   & D.D.                \\       
XRS~63  &0.447  & BOSS &  5977  56098   073  &   42.56  &    HEG   & D.D.                      \\    
XRS~64  &0.183  & SDSS &  0910  52377   437  &   40.45  &    LEG   & D.D.                 \\       
XRS~72  &0.641  & BOSS &  4038  55363   846  &   41.79  &    ---   & ---                 \\       
XRS~76  &0.037  & SDSS &  2131  53819   533  &   40.03  &    LEG   & D.D.                      \\    
XRS~77  &0.367  & SDSS &  1326  52764   048  &   42.53  &    HEG   & [O~III]/H$\beta$          \\    
XRS~78  &0.504  & BOSS &  4024  55646   127  &   41.53  &    HEG   & D.D.                  \\       
XRS~81  &0.188  & SDSS &  1396  53112   073  &   40.01  &    LEG   & [O~III]/L$_{\rm r}$           \\    
XRS~83  &0.084  & SDSS &  1647  53531   096  &$<$39.36  &    LEG   & [O~III]/L$_{\rm r}$           \\    
XRS~84  &0.536  & BOSS &  6024  56088   833  &$<$40.48  &    LEG   & [O~III]/L$_{\rm r}$     \\       
XRS~86  &0.659  & BOSS &  5481  55983   295  &   42.32  &    HEG   & [O~III]/H$\beta$          \\    
XRS~90  &0.174  & SDSS &  2173  53874   431  &   41.39  &    HEG   & D.D.                      \\    
XRS~93  &0.556  & BOSS &  6034  56103   675  &   41.79  &    HEG   & [O~III]/H$\beta$          \\    
XRS~94  &0.107  & BOSS &  5008  55744   852  &   39.84  &    LEG   & D.D.                      \\    
\hline
\end{tabular}
\label{spectra}
\medskip

\end{center}

\noindent
\small{1) Galaxy ID, 2) redshift, 3) survey for which the data have been
  obtained, 4) Plate, Julian Day and Fiber identifying the spectrum, 
5) logarithm of the [O~III] luminosity in $\ergs$, 6 and 7) spectral type and method used to derive it. }
\end{table*}

\subsection{Radio and emission line properties}
\label{rel}

\begin{figure}
\centerline{
\includegraphics[width=90mm,angle=0]{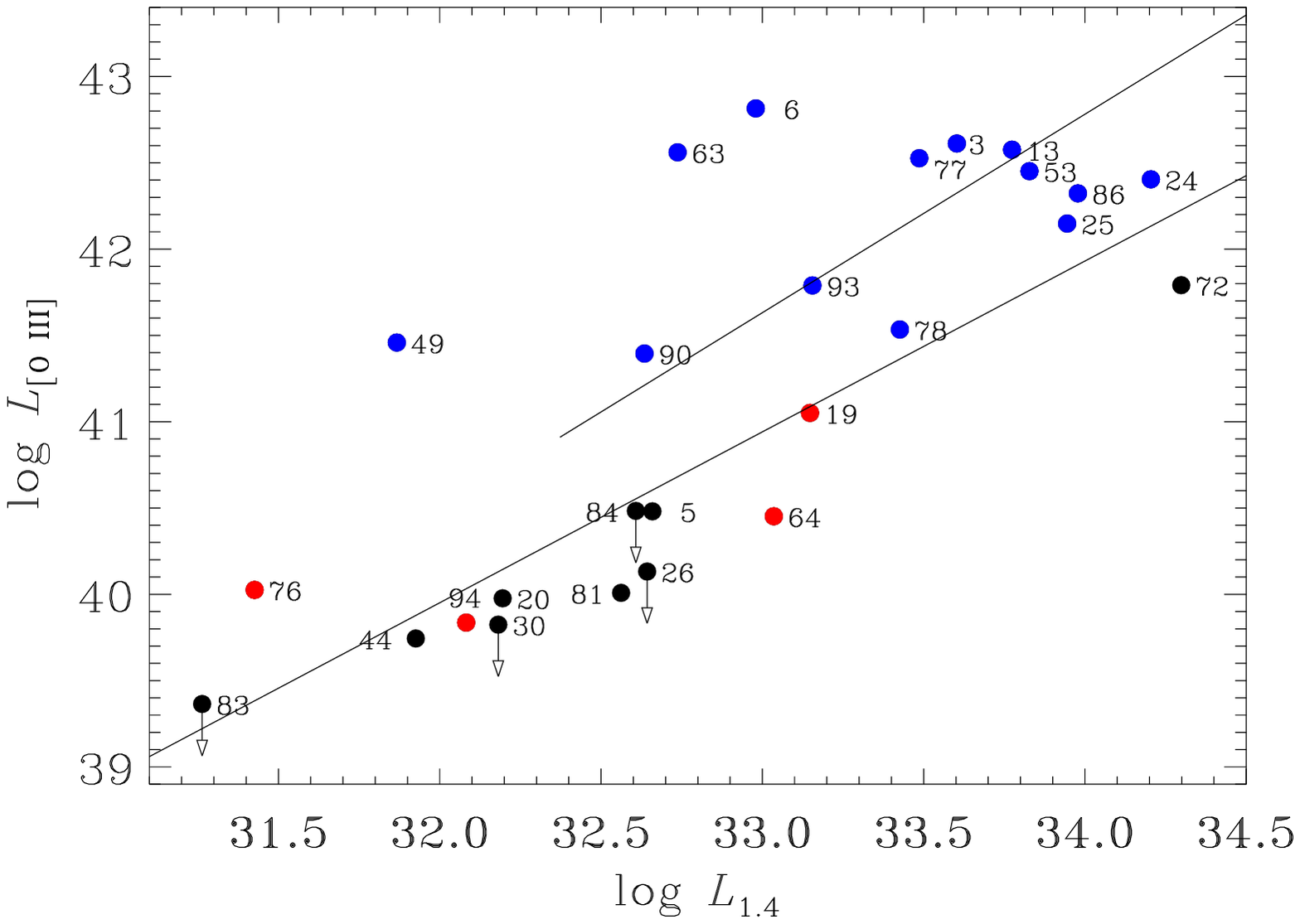}}
\caption{Comparison of the luminosity of the [O~III] line and the radio power
  at 1.4 GHz (in $\ergs$ and $\ergsHz$ units, respectively). The XRSs are
  color-coded based on their classification: blue = HEG, red = LEG, black =
  unclassified.}
\label{lo3l178}
\end{figure}

We now consider the relation between emission lines and radio luminosity. In
3C radio-galaxies the two spectral types follow different correlations between
$L_{\rm[O~III]}$ and $L_r$, with HEGs having
larger line luminosity at given radio power.  In Fig. \ref{lo3l178} we compare
emission lines and radio luminosity, color-coding the different source
types. Both HEGs and LEGs among the XRSs are located close to the respective
correlations defined by the 3CR RGs. 

As already noticed in the 3C sample, several galaxies of uncertain spectral
classification can be  classified robustly based on their location in this
diagram. In fact, with the sole exception of XRS~72, all unclassified sources
fall very closely along the $L_{\rm[O~III]}$ - $L_r$ relation defined by 3CR
  LEGs. With this strategy we are able to define a spectral type for all but
  one of the XRSs.

\subsection{Stellar populations}
\label{population}

\begin{figure*}
\centerline{
\includegraphics[width=180mm,angle=0]{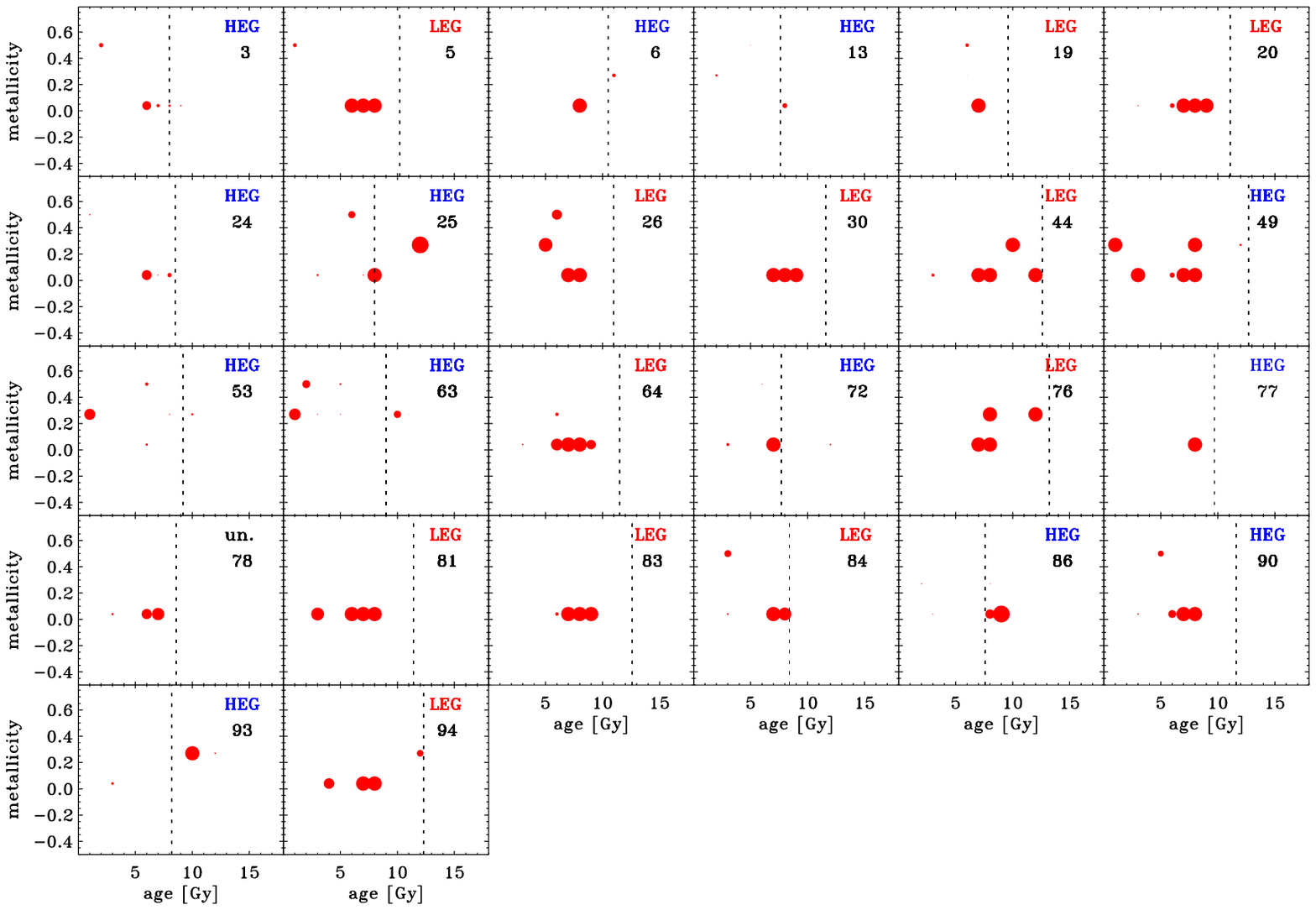}}
\caption{Decomposition of the starlight emission in various templates of
  stellar populations obtained with Gandalf. For each source we show the
  contribution of the 39 models used, formed by combining set with three
  different metalicity (Y axis) and 13 different ages (X axis). The area of
  the symbols show the relative contribution of each template. The dashed
  vertical line marks the age of the Universe at the redshift of each source.}
\label{age}
\end{figure*}

The Gandalf software provides us also with a decomposition of the starlight
among the 39 templates of varying metallicity (from solar to thrice solar)
  and age (up to 12 Gyr) from \citet{bruzual03}. In Fig. \ref{age} we
show the result of this analysis.  For each source we show the contribution
of the 39 models used, formed by combining three different metallicities (Y
axis) and 13 different ages (X axis). The area of the symbols show the
relative contribution of each template to the XRS's starlight.

HEGs are known to show almost invariably blue colors, an indication of recent
star formation, while in LEGs, the fraction of actively star-forming objects is
not enhanced with respect to quiescent galaxies \citep{baldi08,smolcic09}. A
quantitative comparison between these results and those presented here cannot
be performed since the indicators of young stars are rather different. 
  Furthermore, the precise decomposition in the various stellar populations is
  rather uncertain, being affected by, e.g., the age-metallicity
  degeneracy. Nonetheless, a prominent young stellar population (with an age
$\le 3$ Gyr and a fractional contribution of at least 10\%) is found in eight
(out of 13) HEGs and in only one (out of 12) LEGs among the XRSs.

\medskip
Apparently, both the emission lines and the stellar population properties of
XRSs do not differ from those of 3C FR~II. This is what is expected if the
appearance of the radio wings is due to the hosts asymmetries.

\section{Summary and conclusions}
\label{summary}

The aim of this paper is to explore the properties of XRSs, from
the point of view of the geometrical relationship between radio and optical
structures, of their emission lines, and of their stellar population. Starting
with a sample of XRSs identified in the FIRST survey and covered by the SDSS,
we restrict our analysis to the 53 objects with the clearest X-shaped
morphology. The identification of the host of the XRSs in the SDSS images is,
in most cases, straightforward since we found a single optical source located
close to the midpoint of the line that joins the peaks of the two main radio
lobes. Only in five sources is the identification  not secure, because we
either have more than one plausible association or no optical source is found
at the center of the radio source.

Spectroscopic redshifts are available for 35 sources and for all others (except
three) there are photometric redshift measurements provided by the SDSS
database. The median redshift is $z=0.31$. At this distance, the typical core
size of RG's hosts covers only 1$\arcsec$. Indeed, most of the XRS's hosts are
rather compact sources in the SDSS images: the classical approach for estimating
the position angle of the hosts major axis, based on fitting elliptical
isophotes to the optical images, cannot generally be used.

Consequently, we adopt a different strategy: for each galaxy, we projected its
optical image onto an axis. The orientation of the galaxy's major axis is
defined as the angle at which the width of the projected profile reaches its
maximum. With this method, we also derive the host's size and axial ratio. The
bootstrap technique was used to estimate the errors of these quantities. While
some galaxies are too faint, too compact, or not sufficiently elongated, we
obtain a robust measurement of the PA of the major axis for 22 galaxies.

We find that the radio-wing axis of all (but one) of the XRSs form an angle larger
than 40$^\circ$ with the optical major axis, with a probability that this is
compatible with a uniform distribution of only $P = 0.9 \times 10^{-4}$.  A
highly significant connection between the optical and the radio-wing axis
emerges, with the wings preferentially aligned with the host's minor axis,
This confirms the previous findings but, because of the larger size of the sample
studied, with a strongly improved statistical significance.  We also confirm
that wings develop in highly elliptical galaxies, although this conclusion can
only be obtained  from the study of the four nearest XRSs. 

The relationship between the optical axis and wing orientation indicate that
the formation of the XRSs is intimately related to the host's geometry. These
results strengthen the interpretation that the X-shaped morphology in
radio-sources has an hydro-dynamical origin, rather than being the result of a
jet-reorientation following a black hole coalescence. In particular XRSs
naturally form when a jet propagates in a non-spherical gas distribution: the
rapid cocoon expansion along the direction of the host's minor axis produces
the X-shaped morphology.

In this framework, it is possible to further progress in our understanding of
the origin of XRSs along two lines. Firstly, we used the optical imaging data
obtained from the SDSS. This survey provides us with a homogeneous dataset,
but of limited depth; a targeted optical survey can lead to a higher number
of radio galaxies for which it is possible to measure the geometrical
parameters. Finally, the planned JVLA surveys will produce deeper and higher
resolution images, enabling an even larger sample of bona-fide XRSs to be built.

We have also performed a study of the properties of the emission lines (and
their connection with the spectroscopic types and radio luminosity) as well as
of their stellar population. This analysis can falsify the proposed model that
poses that XRSs are ``normal'' FR~II, whose radio morphology is only caused by
an asymmetric ISM. In fact, in this case, no differences are expected between
the XRS's properties and those of non-winged FR~II. Conversely, any anomaly of
XRSs with respect to non-winged FR~II may indicate that they are the result of
a specific evolutionary path, such as, one that involves the occurrence of a
merger. This is one of the requirements of the jet re-orientation scenario, an
alternative to the hydro-dynamical model we propose. A proper classification
into the various spectroscopic classes is essential for a proper comparison
with non-winged radio galaxies. We do not find any difference between the
properties of XRSs and the general FR~II population.

\bibliographystyle{aa}

\end{document}